\documentclass[fleqn,usenatbib]{mnras}
\usepackage{amsmath}    
\usepackage{newtxtext,newtxmath}
\usepackage[T1]{fontenc}
\usepackage{ae,aecompl}
\usepackage{graphicx}	
\usepackage{amssymb}
\usepackage{geometry}
\usepackage{float}
\usepackage{caption}
\usepackage{subcaption}
\usepackage{dblfloatfix}
\usepackage{url}

\newcommand{\hinvMsun}{h^{-1}{\rm M}_\odot}
\newcommand{\hinvMpc}{h^{-1}{\rm Mpc}}
\newcommand{\Vmax}{$V_{\rm max}$}
\newcommand{\Vpeak}{$V_{\rm peak}$}
\newcommand{\ahalf}{$a_{\rm M/2}$}
\newcommand{\Rvir}{$R_{\rm vir}$}
\newcommand{\Mh}{M_{\rm h}}
\newcommand{\Mnl}{M_{\rm nl}}
\newcommand{\Meff}{M_{\rm eff}}

\title[Dependence of Halo Bias and Kinematics on Assembly Variables]{Dependence of Halo Bias and Kinematics on Assembly Variables}

\author[Xu \& Zheng]{
Xiaoju Xu\thanks{E-mail: xiaoju.xu@utah.edu}
and Zheng Zheng\thanks{E-mail: zhengzheng@astro.utah.edu}\\
\\
Department of Physics and Astronomy, University of Utah, 115 South
1400 East,
Salt Lake City, UT 84112, USA
}

\date{Accepted XXX. Received YYY; in original form ZZZ}

\pubyear{2017}

\begin{document}
\label{firstpage}
\pagerange{\pageref{firstpage}--\pageref{lastpage}}
\maketitle

\begin{abstract}

Using dark matter haloes identified in a large $N$-body simulation, we 
study halo assembly bias, with halo formation time, peak maximum circular 
velocity, concentration, and spin as the assembly variables. Instead of 
grouping haloes at fixed mass into different percentiles of each assembly 
variable, we present the joint dependence of halo bias on the {\it values} 
of halo mass and each assembly variable. In the plane of halo mass and one 
assembly variable, the joint dependence can be largely described as halo 
bias increasing outward from a global minimum. We find it unlikely to have 
a combination of halo variables to absorb all assembly bias effects. We 
then present the joint dependence of halo bias on two assembly variables 
at fixed halo mass. The gradient of halo bias does not necessarily follow 
the correlation direction of the two assembly variables and it varies with 
halo mass. Therefore in general for two correlated assembly variables one 
cannot be used as a proxy for the other in predicting halo assembly 
bias trend. Finally, halo assembly is found to affect the kinematics of 
haloes. Low-mass haloes formed earlier can have much higher pairwise velocity 
dispersion than those of massive haloes. In general, halo assembly leads to 
a correlation between halo bias and halo pairwise velocity distribution, 
with more strongly clustered haloes having higher pairwise velocity and 
velocity dispersion. However, the correlation is not tight, and the kinematics 
of haloes at fixed halo bias still depends on halo mass and assembly variables.

\end{abstract}


%%%%%%%%%%%%%%%%% BODY OF PAPER %%%%%%%%%%%%%%%%%%

\section{Introduction}
\label{sec:intro}

Large-volume galaxy redshift surveys over a range of redshifts,
including the Sloan Digital Sky Survey (SDSS; \citealt{York00}),
the Two-degree Field Galaxy Redshift Survey (2dFGRS; \citealt{Colless99}),
the SDSS-III \citep{Eisenstein11} and SDSS-IV \citep{Dawson16}, and
the WiggleZ Dark Energy Survey \citep{Blake11}, have transformed
the study of large-scale structure, producing detailed distribution
of galaxies in the universe as a function of their properties and resulting
in galaxy clustering measurements with high precision.
Galaxy clustering data from such surveys play an important role in
understanding galaxy formation and evolution and in learning about
cosmology, in particular in constraining dark energy and testing 
gravitational theories. 

The formation of galaxies involves poorly understood baryonic processes,
causing a hurdle to interpret galaxy clustering data. In contrast, the
formation and evolution of dark matter haloes are dominated by gravitational
interactions and their properties are well understood with analytic models
and $N$-body simulations %\citep[e.g.][]{Press74,Bond91,Mo96,Jing98a,Sheth99,Sheth01a,Jenkins01,Tinker08}. 
\citep[e.g.][]{Press74,Mo96,Tinker08}. 
Over the past two decades, 
an informative way to interpret galaxy clustering has been developed and 
made wide applications, which is to link galaxies to the underlying dark 
matter halo population. The two commonly adopted descriptions of the 
galaxy-halo connection are the halo occupation distribution (HOD) and 
conditional luminosity function (CLF) frameworks \citep[e.g.][]{Berlind02,Yang03,Zheng05},
which have been successfully applied 
to galaxy clustering data to infer the relation between galaxy properties and halo mass 
%\citep[e.g.][]{Bullock02,Bosch03,Zheng04,Yang05,Zehavi05,Cooray06,Hamana06,Lee06,Phleps06,White07,White11,Zheng07,Blake08,Brown08,Zheng09,Zehavi11,Coupon12,Guo14,Skibba15,Guo15a,Guo15c,Guo16a,Xu16,Xu18}.
\citep[e.g.][]{Zehavi05,Zehavi11}.

The HOD/CLF framework chooses the halo mass as the halo variable, and the
implicit assumption is that the statistical properties of galaxies inside
haloes only depend on halo mass, not on halo environment or growth history.
This assumption of an environment-independent HOD/CLF is partly motivated
by the excursion set theory \citep{Bond91}.
However, $N$-body simulations show that the
clustering of haloes depends on not only halo mass, but also 
halo assembly history \citep[e.g.][]{Gao05,Zhu06,Jing07}, 
halo structure \citep[e.g.][]{Wechsler06,Gao07,Faltenbacher10,Paranjape17a}, and halo environment \citep[e.g.][]{Harker06,Salcedo18},
%\citep[e.g.][]{Sheth04,Gao05,Wechsler06,Zhu06,Jing07}, environment %\citep[e.g.][]{,Harker06,Salcedo18}, and halo structure %\citep[e.g.][]{Gao07,Faltenbacher10,Paranjape17a}
%Bett07,Li08,Lazeyras17,Paranjape18,Villarreal17,Han18,Musso18,
which is termed as {\it halo assembly bias}.
Halo properties that correlate with halo environment or assembly history
are broadly referred to here as halo assembly variables, such as halo formation
time, halo concentration, maximum circular velocity of halo, and halo spin. 
Studying the origin of halo assembly bias remains active ongoing effort 
\citep[e.g.][]{Borzyszkowski17}. %{Sandvik07,Dalal08,Hahn07,Ludlow09,Wang09,Borzyszkowski17}. 
%,Diemand07,Wang11,Shi15,Hahn09,
For massive haloes (with mass higher than the nonlinear mass $\Mnl$ for 
collapse), the assembly bias appears to be a generic feature in the Extended
Press-Schechter theory, related to the difference in the curvature of 
peaks of the same height in a Gaussian density field \citep[e.g.][]{Dalal08}.
For low mass haloes, the assembly bias is proposed to be originated from the
strong influence of the environment, especially the tidal field, on the 
evolution of haloes \citep[e.g.][]{Hahn07,Hahn09,Wang11,Shi15}. 

Whether halo assembly bias is inherited by galaxies is still under investigation %in modelling \citep[e.g.][]{Zentner16,Zehavi18,Zu18} , 
in both simulations \citep[e.g.][]{Chaves16,Busch17,Garaldi18} and observations
\citep[e.g.][]{Lin16,Miyatake16,Guo17,Zu17a}. 
%,Vakili16,
If galaxy properties are closely tied to halo growth history, the inherited
assembly bias from the host haloes would require a modification of the 
current halo model of galaxy clustering to include such an effect. 
Otherwise, we would infer incorrect galaxy-halo 
connections and introduce possible systematics in cosmological constraints
(e.g. \citealt{Zentner14,Hearin15,Zentner16}; but see \citealt{McEwen16}). 
Along the path, a better characterisation of halo assembly bias is necessary, 
which motivates the work in this 
paper. In most previous studies, halo
assembly bias is inferred by grouping each halo assembly variable at fixed halo mass 
into certain percentiles. While this can reveal the trend of assembly bias, 
it does not provide a clear view on the multivariate dependence of the halo bias, 
which needs the measurement of halo bias at certain values, not certain percentiles, 
of a given assembly variable. If halo assembly effect is to be incorporated into the 
halo modelling of
galaxy clustering, a natural route is to describe halo bias in terms of
multiple halo properties.
We will present the multivariate dependence of halo bias 
and study whether there is a combination of halo variables to minimise assembly
bias. In addition, we also investigate the scale dependence of halo assembly
bias and the assembly effect on halo kinematics.

The structure of the paper is as follows. In Section 2, we describe the 
simulation data used in this study and the measurement of halo bias.
In Section 3, we present the joint dependence of halo bias on halo mass and 
each halo assembly variable, study an effective halo variable to minimise halo
assembly bias, show the scale dependence of halo assembly bias, and describe 
the dependence of halo bias on two halo assembly variables.
In Section 4, we analyse the assembly effect on halo kinematics.
Finally, in Section 5, we summarise our results and discuss their implications.

%2
\section{Simulation and Halo Bias Calculation}
\label{sec:sim}

For the work presented in this paper, we make use of the MDR1 MultiDark 
simulation \citep[][]{Prada12,Riebe13}\footnote{\url{https://www.cosmosim.org/cms/simulations/mdr1/}}.
The MDR1 $N$-body simulation adopts a spatially flat $\Lambda$CDM cosmology, 
with $\Omega_{\rm m}=0.27$, $\Omega_{\rm b}=0.0469$, $h=0.70$, 
$n_{\rm s}=0.95$, and 
$\sigma_8=0.82$. The simulation has $2048^3$ particles in a box of 1$h^{-1}$Gpc on one side, with particle mass $8.721\times 10^9\hinvMsun$. Dark matter haloes
are identified with the Rockstar algorithm \citep[][]{Behroozi13}, which finds
haloes through adaptive hierarchical refinement of friends-of-friends groups 
using the six-dimensional phase-space coordinates and the temporal 
information. 

While the MDR1 simulation is adopted for presenting most of the results, we also make use of the Bolshoi simulation for some investigations that need high completeness in certain regions of the parameter space. The  Bolshoi simulation \citep{Klypin2011}\footnote{\url{https://www.cosmosim.org/cms/simulations/bolshoi/}} with a $250 h^{-1}{\rm Mpc}$ box adopts the same cosmology as MDR1, but with 64 times higher mass resolution and 7 times higher force resolution.

We use the $z=0$ Rockstar halo catalogue. In addition to halo mass $\Mh$, 
 
for which we adopt $M_{\rm 200b}$ (i.e. the mean density of a halo being 200 times the background density of the universe), we 
consider four other halo properties: \Vpeak, peak maximum circular
velocity of the halo over its accretion history; $\lambda$, halo spin 
parameter that characterises its angular momentum; \ahalf, cosmic scale 
factor when the halo obtains half of its 
current ($z=0$) total mass, which characterises halo formation time; 
$c$, halo concentration parameter, which is the
ratio of the halo virial radius \Rvir\ to the scale radius $r_{\rm s}$. All the 
four quantities are related to the halo assembly history and therefore are dubbed as halo assembly variables.
 
We measure halo bias for haloes above $\sim 4\times 10^{11}\hinvMsun$ (about 50 particles).
The large simulation box also serves well our purpose of studying the joint 
dependence of halo bias on multiple halo properties, as it enables halo bias
in fine bins of halo properties to be measured. To further reduce the 
uncertainty in the halo bias measurements, we derive halo bias from the 
two-point halo-mass cross-correlation function $\xi_{\rm hm}$ and the
two-point mass auto-correlation function $\xi_{\rm mm}$ 
\citep[e.g.][]{Jing99,Jing07}, i.e.
\begin{equation}
b(r)=\frac{\xi_{\rm hm}(r)}{\xi_{\rm mm}(r)}.
\end{equation}
In practice, for a given sample of haloes and a pair separation bin 
$r\pm dr/2$, we count the number of halo-mass particle pairs, 
$N_{\rm hm}(r\pm dr/2)$, 
and that of mass particle-particle pairs, $N_{\rm mm}(r\pm dr/2)$. With the
periodic boundary condition, the corresponding pair counts from 
randomly distributed haloes and mass particles can be simply calculated as 
$N_{\rm hm, ran}(r\pm dr/2) = n_{\rm h} n_{\rm m} V d^3r$ and 
$N_{\rm mm, ran}(r\pm dr/2) = n_{\rm m}^2 V d^3r/2$, where $n_{\rm h}$ and 
$n_{\rm m}$ are the
halo and mass particle number densities in the simulation box of volume $V$. 
The two-point correlation functions are then computed as
\begin{equation}
\xi_{\rm hm}(r) = \frac{N_{\rm hm}(r\pm dr/2)}{N_{\rm hm, ran}(r\pm dr/2)} -1
\end{equation}
and
\begin{equation}
\xi_{\rm mm}(r) = \frac{N_{\rm mm}(r\pm dr/2)}{N_{\rm mm, ran}(r\pm dr/2)} -1.
\end{equation}

%3
\section{Joint Dependence of Halo Bias on Mass and Assembly Variables}
\label{sec:joint_dep}

In this section, 
we first present the results of the joint dependence of halo bias on both 
mass and one of the assembly variables. Then we investigate whether an 
effective halo mass can be constructed to absorb the assembly effect in halo 
clustering. We also discuss how the halo assembly affects the scale dependence 
of halo bias. Finally, we present the dependence of halo bias on two halo assembly
variables at fixed halo mass.

%3.1
\subsection{Dependence of halo bias on mass and one assembly variable}
\label{sec:bias_M_A}

\begin{figure*}
	\centering
	\begin{subfigure}[h]{0.48\textwidth}
		\includegraphics[width=\textwidth]{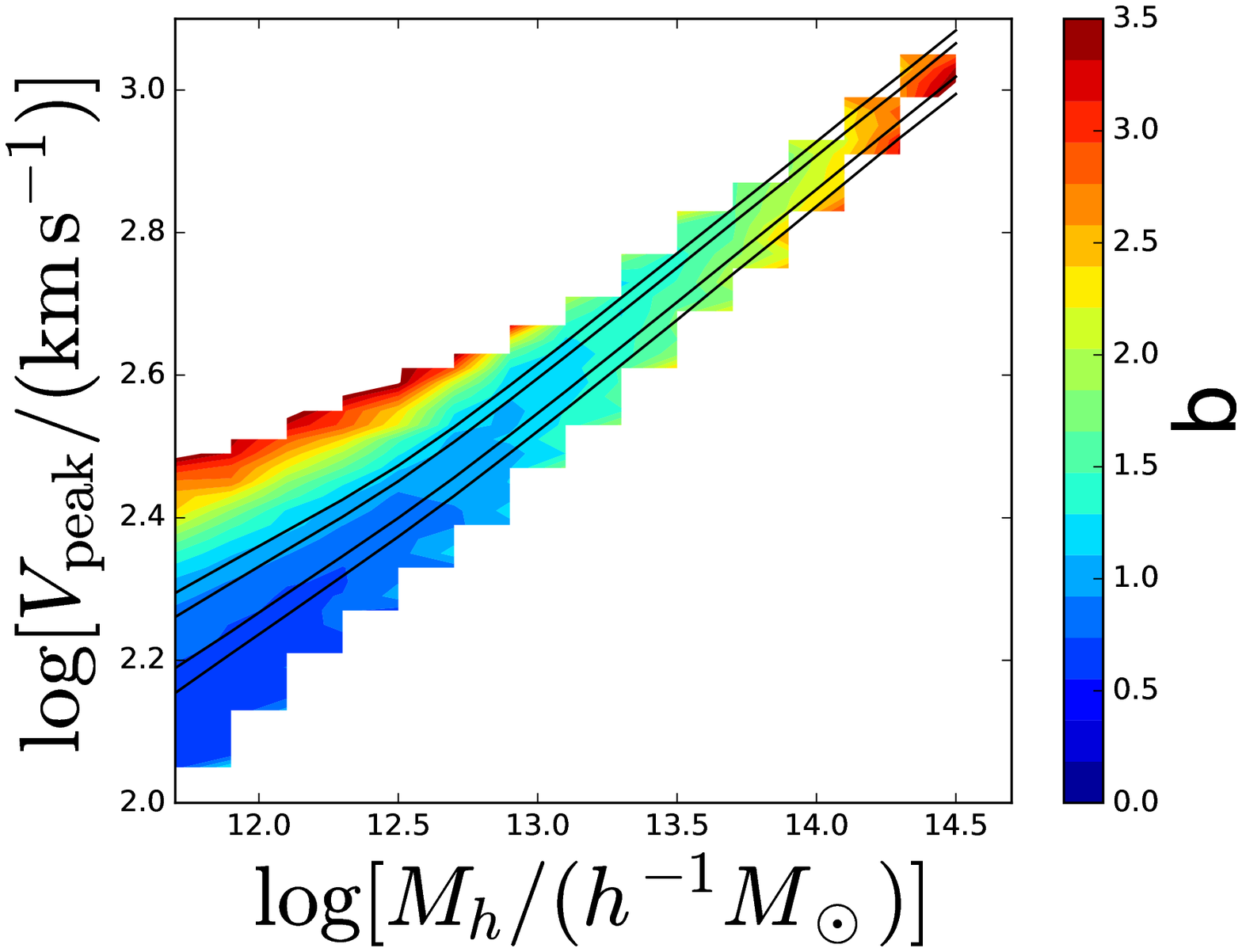}
	\end{subfigure}
	\hfill
	\begin{subfigure}[h]{0.48\textwidth}
                \includegraphics[width=\textwidth]{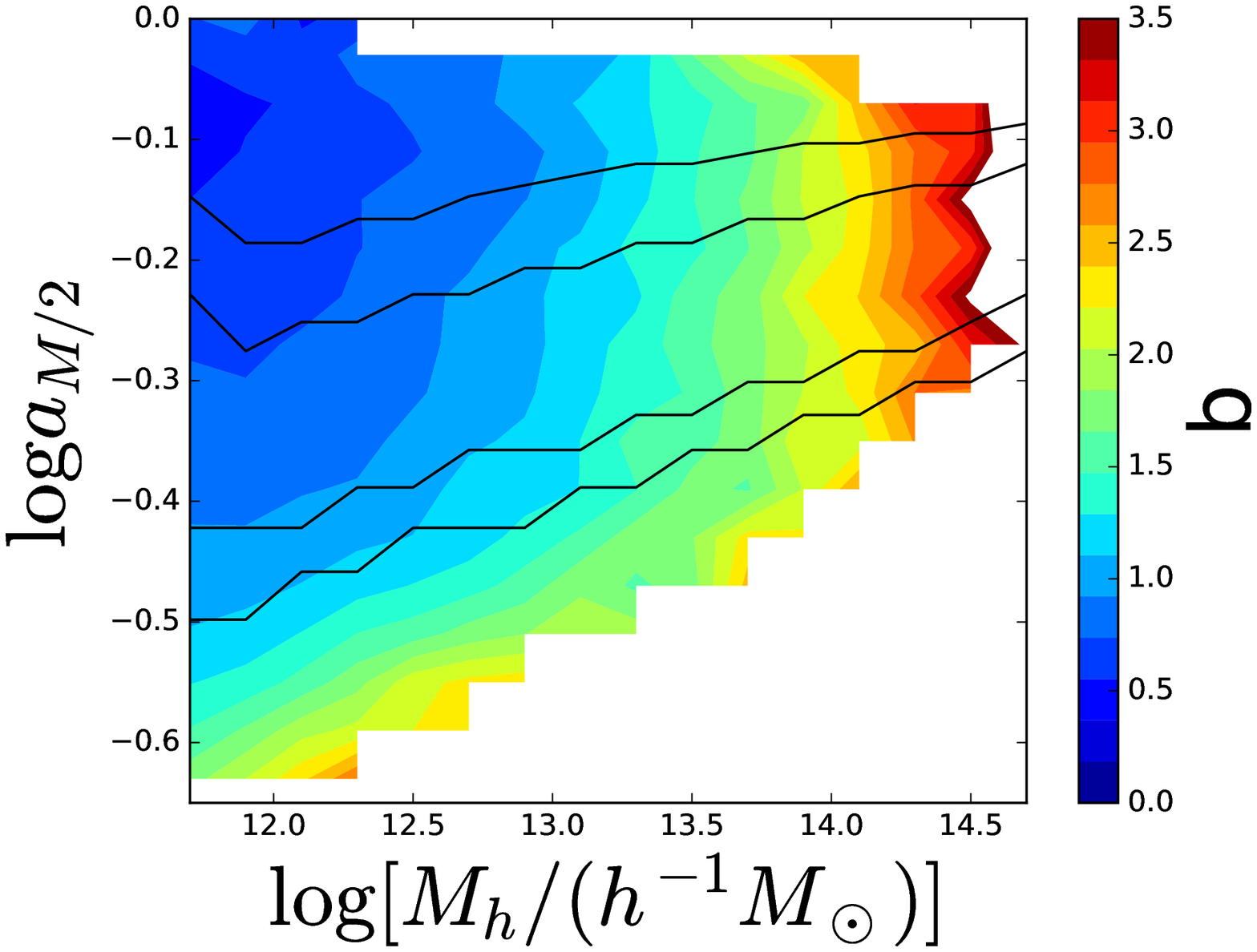}
	\end{subfigure}
	\hfill
	\begin{subfigure}[h]{0.48\textwidth}
                \includegraphics[width=\textwidth]{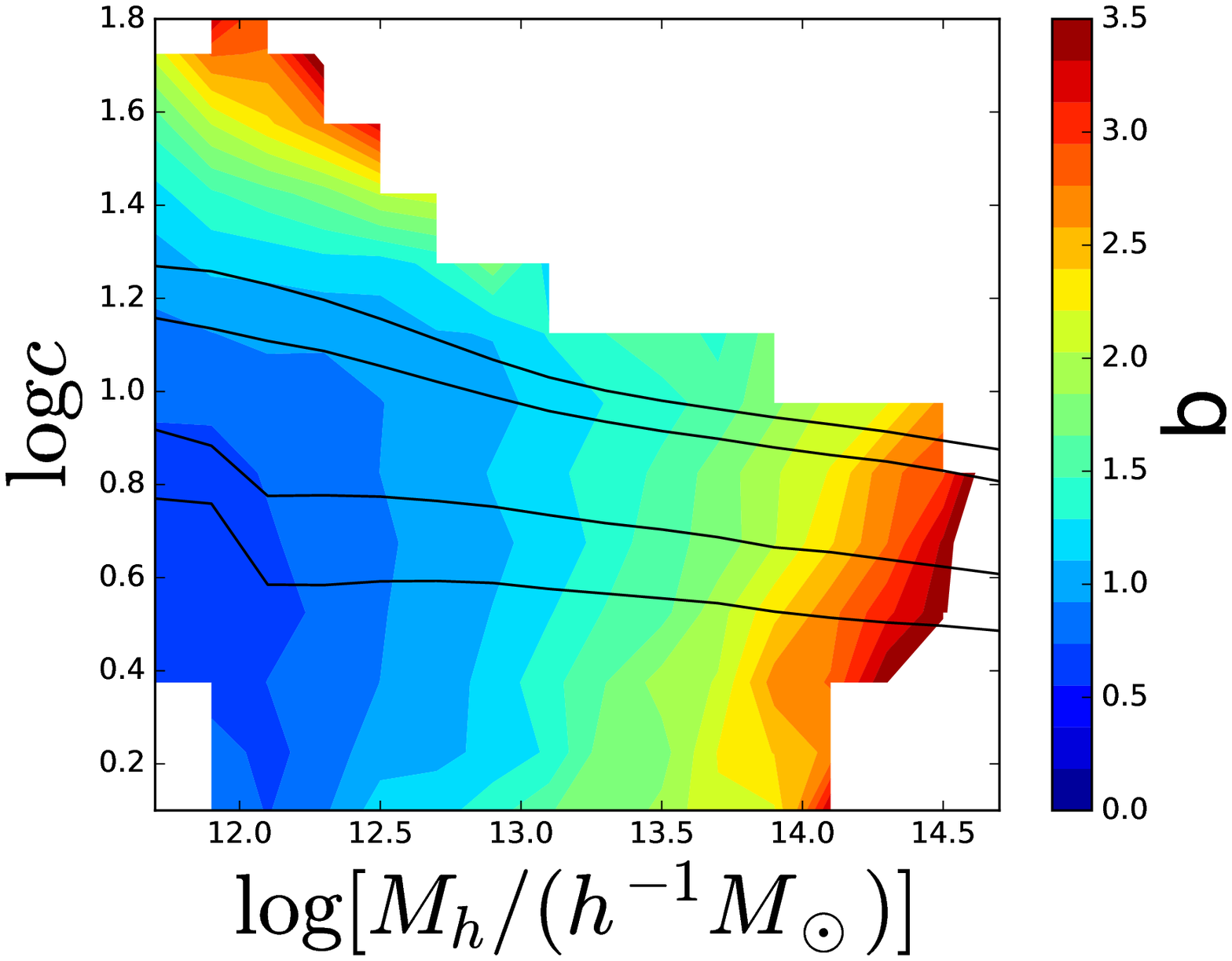}
	\end{subfigure}
	\hfill
	\begin{subfigure}[h]{0.48\textwidth}
                \includegraphics[width=\textwidth]{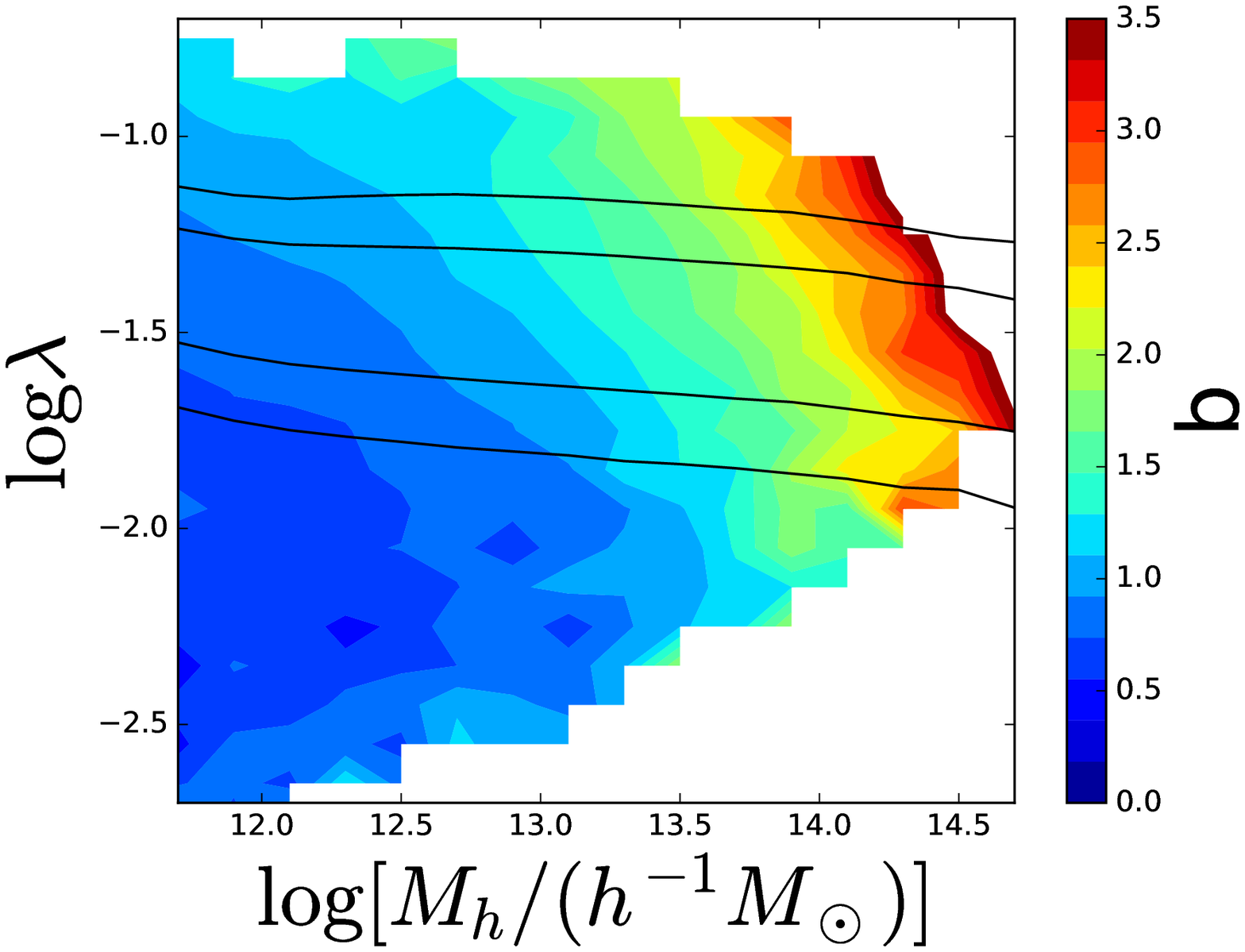}
	\end{subfigure}
	\hfill
\caption{Joint dependence of halo bias on halo mass and each assembly variable. The assembly variables in the four panels are peak maximum circular velocity
\Vpeak, halo formation scale factor \ahalf, halo concentration $c$, and 
halo spin $\lambda$, respectively. The four black curves in each panel 
mark the central 50 and 80 per cent of the distribution of the 
corresponding assembly variable as a function of halo mass.
}
\label{fig:joint_dep}
\end{figure*}

To separate the effects of mass dependence and assembly dependence, we first 
choose the halo mass bin to have a narrow width, $\Delta\log\Mh=0.2$. For each
halo mass bin, we further divide haloes into bins of each assembly variable, 
and typically there are about 20 assembly variable bins.
We measure halo bias for haloes in each bin of mass and assembly variable.

We take the average above $10\hinvMpc$ as the large-scale halo bias factor, where the ratio between $\xi_{\rm hm}$ 
and $\xi_{\rm mm}$ has becomes flat.

To
maintain reasonable signal-to-noise ratios, we only consider bins with more 
than 100 haloes.  

The joint dependences of halo bias on mass and each assembly variable are
shown in Fig.~\ref{fig:joint_dep}.\footnote{

While the plot does not show the level of uncertainties in the contours, it is clear that the main trend is not affected by noise. The contribution of noise distorts the otherwise smooth contours, showing up as jags in the contour curves or alterations of contour levels (such as the small fluctuation seen in the lower part of the bottom-right panel of Fig.~\ref{fig:joint_dep}). Rather than providing a corresponding plot of the noise levels, we suggest to use the fluctuations in the map as an estimate of the magnitude of noise. This rule of thumb also applies to curve plots, and in general the deviation from a smooth curve gives us the level of noise. However, we do add error bars for a few curves in the curve plots of this paper, and in most cases the noise level is low and the trend is not affected by the noise.
}
The black curves in each panel delineate
the central 50 and 80 per cent of the distribution of the corresponding 
assembly variable as a function of halo mass.

The lowest mass bin in the investigation is $\sim 4\times 10^{11}\hinvMsun$, corresponding to about 50 particles. \citet{Paranjape17a} show that for haloes with less than 400 particles, the distribution of halo concentration is not converged (also see \citealt{Trenti10}). The spin distribution is also affected for haloes with low number of particles \citep[e.g.][]{Trenti10,Benson17}.
To see how the resolution effect changes the results, we perform a test (see Appendix~\ref{sec:appendix}) of replacing haloes of $\log[\Mh/(\hinvMsun)]<12.8$ with those in the Bolshioi simulation, which ensures that each of the lowest mass haloes in our analysis contains more than 3000 particles. We find that the overall trend shown in Fig.~\ref{fig:joint_dep} still holds. Given the higher signal-to-noise ratio and larger range in assembly variables with the MDR1 simulation, we present the halo bias results using the MDR1 simulation except for the analysis in Section~\ref{sec:Meff}, where extending to much lower halo mass is needed. For halo pairwise velocity statistics, a caveat is made in Section~\ref{sec:pairwise_vel}.

The top-left panel shows the joint dependence of halo bias on halo mass
$\Mh$ and peak maximum circular velocity \Vpeak. At low halo masses halo
bias increases as \Vpeak increases, while at high halo masses the trend 
is reversed. The transition mass is around $\log[\Mh/(\hinvMsun)]=13.0$,
which is about three times the nonlinear mass for collapse 
($\log[\Mnl/(\hinvMsun)]=12.5$ for the adopted cosmology). Haloes with low 
mass and high \Vpeak have a bias factor comparable to that of high mass 
haloes, indicating that they could originate from highly stripped high mass 
haloes \citep[e.g.][]{Wang11}.

In the top-right panel, it can be seen that at fixed mass haloes formed 
earlier (i.e. with lower \ahalf) are more strongly clustered. The trend becomes
weak for haloes above $\sim 10^{14}\hinvMsun$, manifested by the almost 
vertical contours. Extremely old haloes (the bottom boundary in the panel) 
at all masses are all as strongly clustered as massive haloes.

At fixed mass, old haloes tend to be more concentrated, i.e. with higher 
concentration parameters $c$. The dependence of halo bias on 
concentration (bottom-left panel) follows a similar trend as with formation 
time (see \S~\ref{sec:2var} for the condition for this to be valid). 
The dependence becomes weak around $3\Mnl$, which is consistent with
previous work 

of direct measurements by binning halo concentration \citep[e.g.][]{Wechsler06,Jing07}
and high precision inference with a lognormal model and Separate Universe technique \citep{Paranjape17a}. We note that the latter shows concentration-dependent assembly bias as a function of the actual value of concentration, instead of percentiles of concentration. 

However, at higher halo 
masses, halo bias decreases with increasing halo concentration, clearly 
different from that in the \ahalf\ case. This difference implies that
halo concentration $c$ can depend on quantities other than formation time
(\ahalf), like the mass accretion rate \citep[e.g.][]{Wechsler02}. 
%\textbf{\citet{Paranjape17a} presented more detailed analysis on concentration, they constructed an analytical model for linear and quadratic Eulerian bias, with the assumption that concentration is a lognormal distributed parameter. Together with the Separate Universe technique, their result largely consistent with the ones traditional method.}

In the bottom-right panel, the joint dependence of halo bias on halo mass
and spin is shown. The distribution of halo spin (indicated by the black
curves) is only weakly dependent on halo mass. Overall the contours show an
ordered pattern (except for the extremely low spin tail
of low mass haloes), with haloes of higher spin more strongly clustered. 
Unlike the cases in the other three panels, the trend persists over all halo
masses, which is consistent with the result using percentiles of spin 
distribution \citep[][]{Gao07}.

Compared to studies of halo bias for haloes selected from percentiles of 
assembly variables, our results show clearly the joint dependence of halo
bias as a function of halo mass and the value of each assembly variable.
In the $\Mh$--$A$ ($A$ being one of the assembly variable) plane, the common
description of our results can be that halo bias increases outward from a
point of global minimum. Such a description is not obvious if the investigation
is limited to the main distribution range of the assembly variable. The 
previous results can all be understood by considering slices in the 
$\Mh$--$A$ plane and by noticing the shape and orientation of the contours.

With the above description motivated by our calculation, it would be 
interesting to ask whether we can define a new halo variable to account for 
both the halo mass and assembly history dependences of halo bias. We do such
an exercise below.

%3.2
\subsection{Effective mass to absorb the assembly bias effect?}
\label{sec:Meff}

\begin{figure*}
\centering
	\begin{subfigure}[h]{0.48\textwidth}
	\centering
	\includegraphics[width=\textwidth]{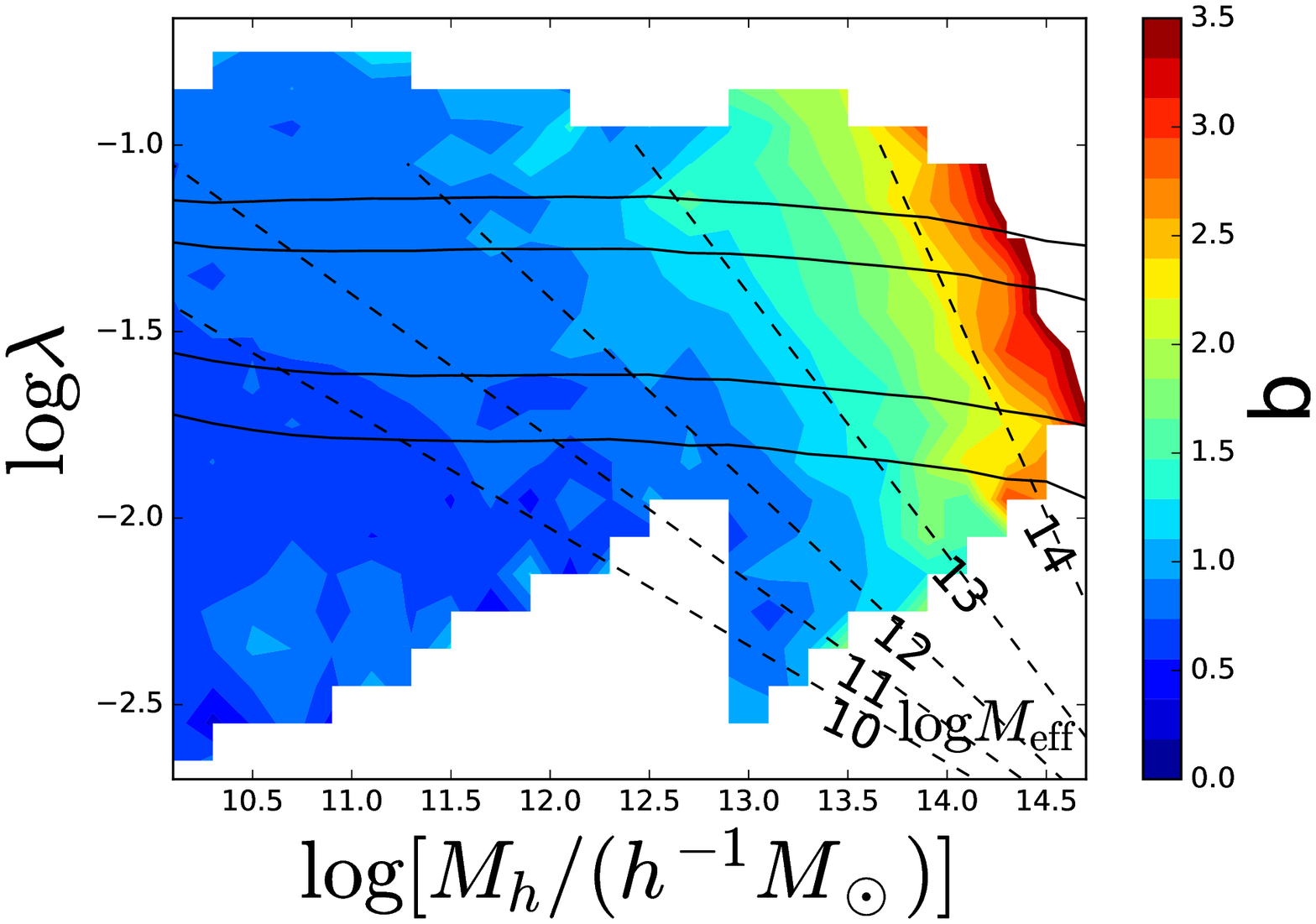}\hfill
	\end{subfigure}
	\begin{subfigure}[h]{0.48\textwidth}
	\centering
	\includegraphics[width=\textwidth]{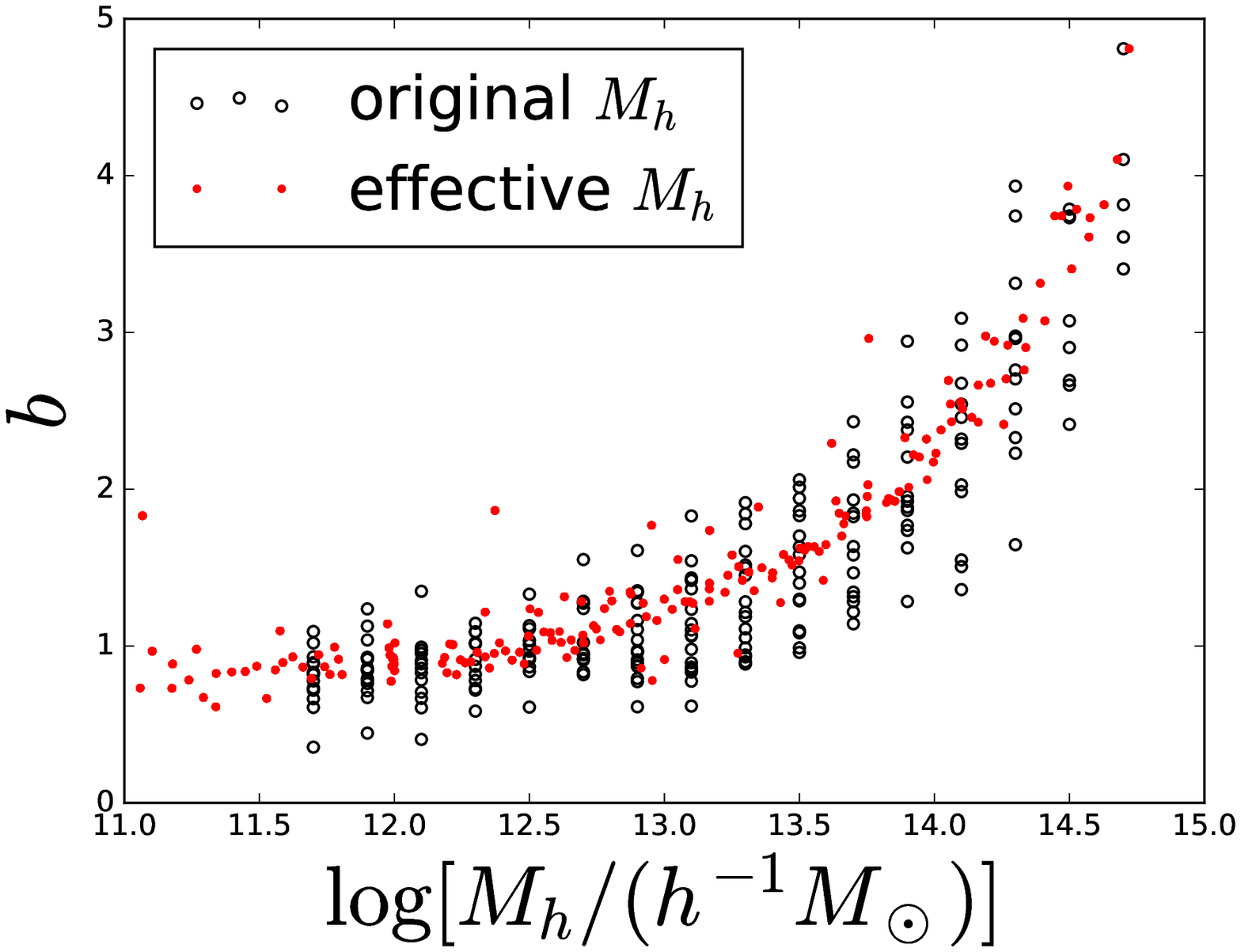}\hfill
	\end{subfigure}
\caption{Effective halo mass to absorb assembly bias effect with halo spin. 
The left panel is the same as the bottom-right panel of 
Fig.~\ref{fig:joint_dep}, and dashed lines are added to illustrate how the
effective mass is defined (see text). In the right panel, the black points
show the halo bias as a function of halo mass, with the scatter at fixed halo
mass coming from haloes with different spin parameters (assembly bias effect
with spin). The red points show the dependence of halo bias on effective
halo mass, and the scatter caused by the assembly effect with spin 
is substantially reduced.
}
\label{fig:Meff}
\end{figure*}

\begin{figure*}
\centering
	\begin{subfigure}[h]{0.48\textwidth}
	\centering
	\includegraphics[width=\textwidth]{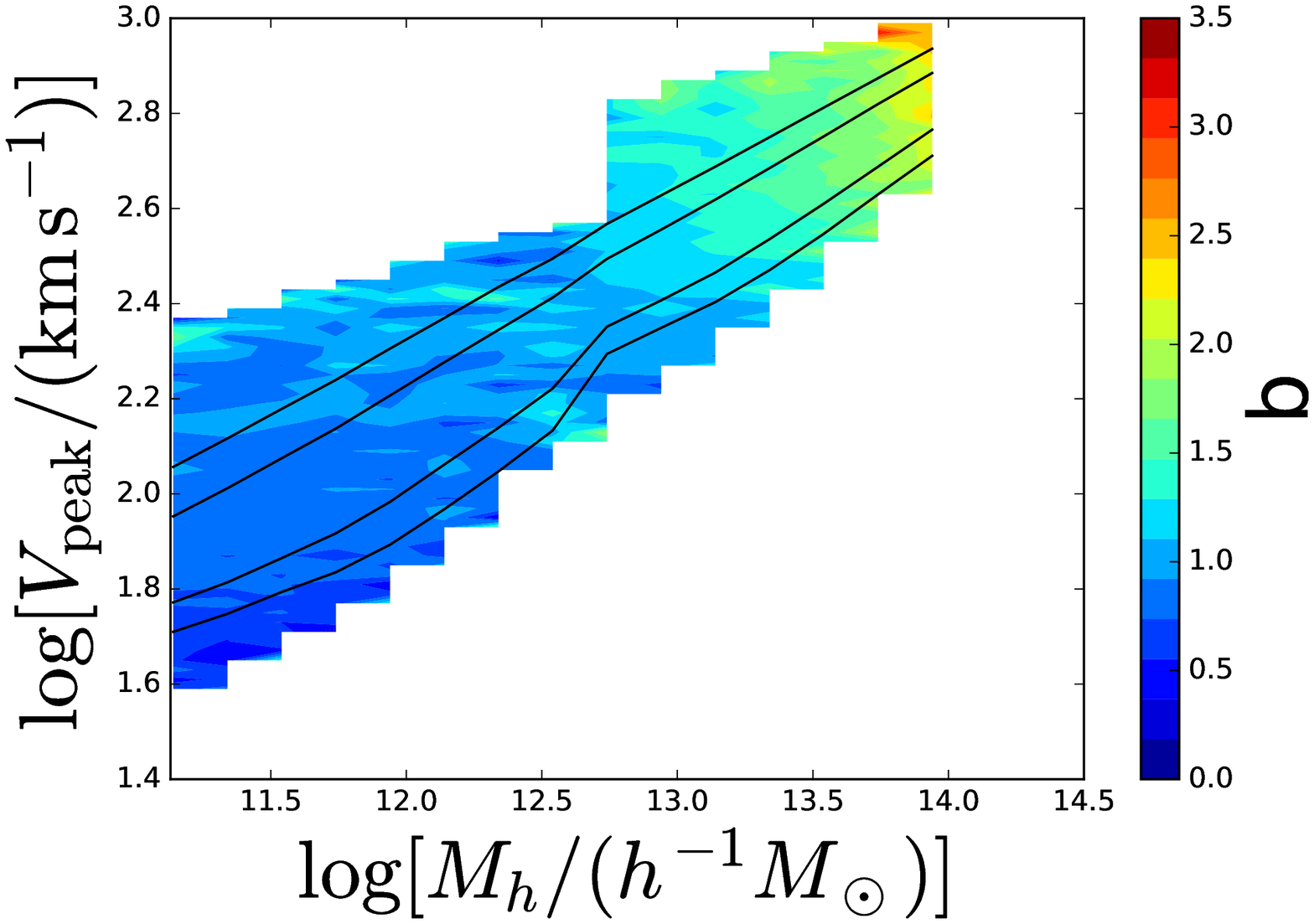}\hfill
	\end{subfigure}
	\begin{subfigure}[h]{0.48\textwidth}
	\centering
        \includegraphics[width=\textwidth]{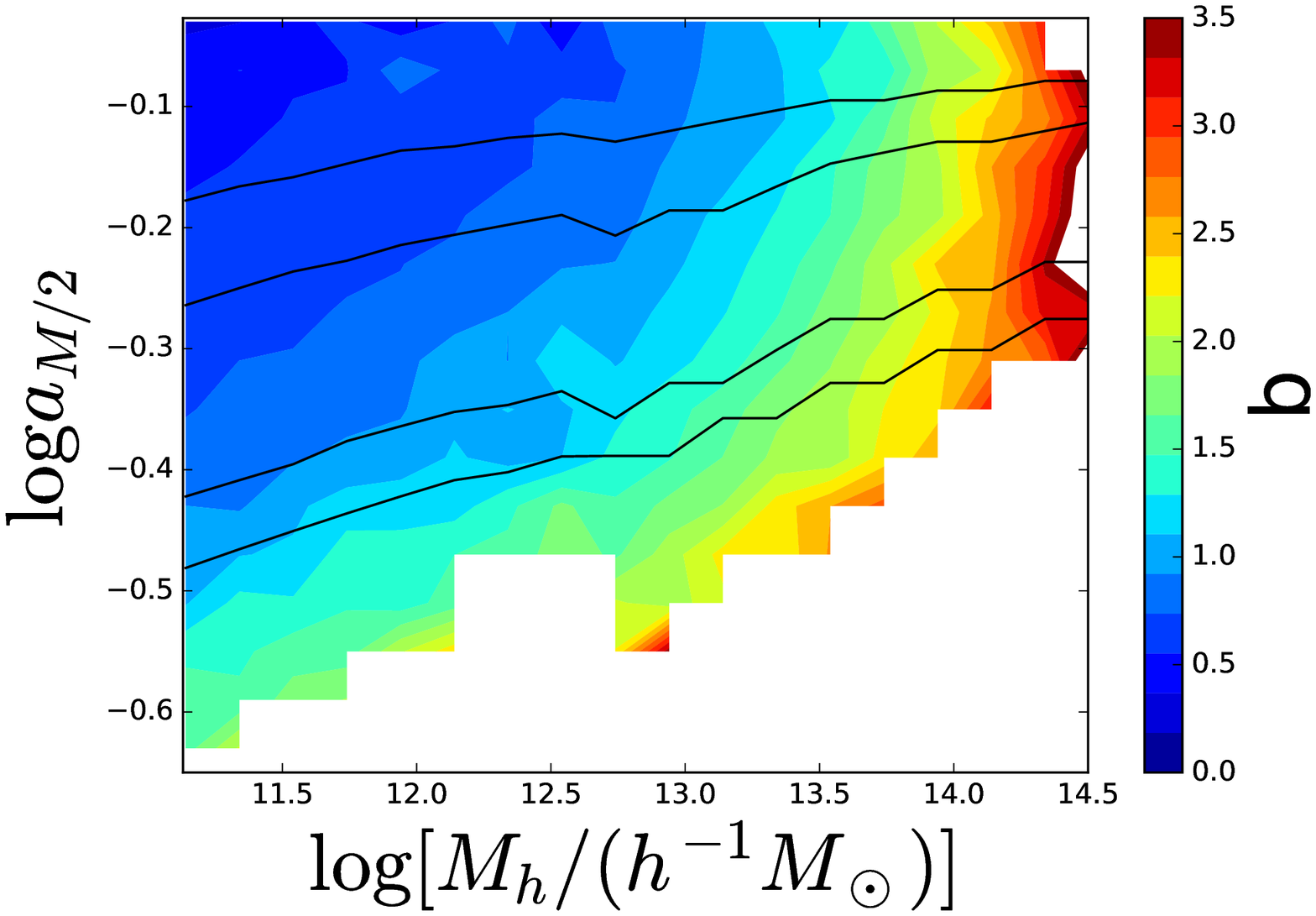}\hfill
	\end{subfigure}
	\begin{subfigure}[h]{0.48\textwidth}
	\centering
        \includegraphics[width=\textwidth]{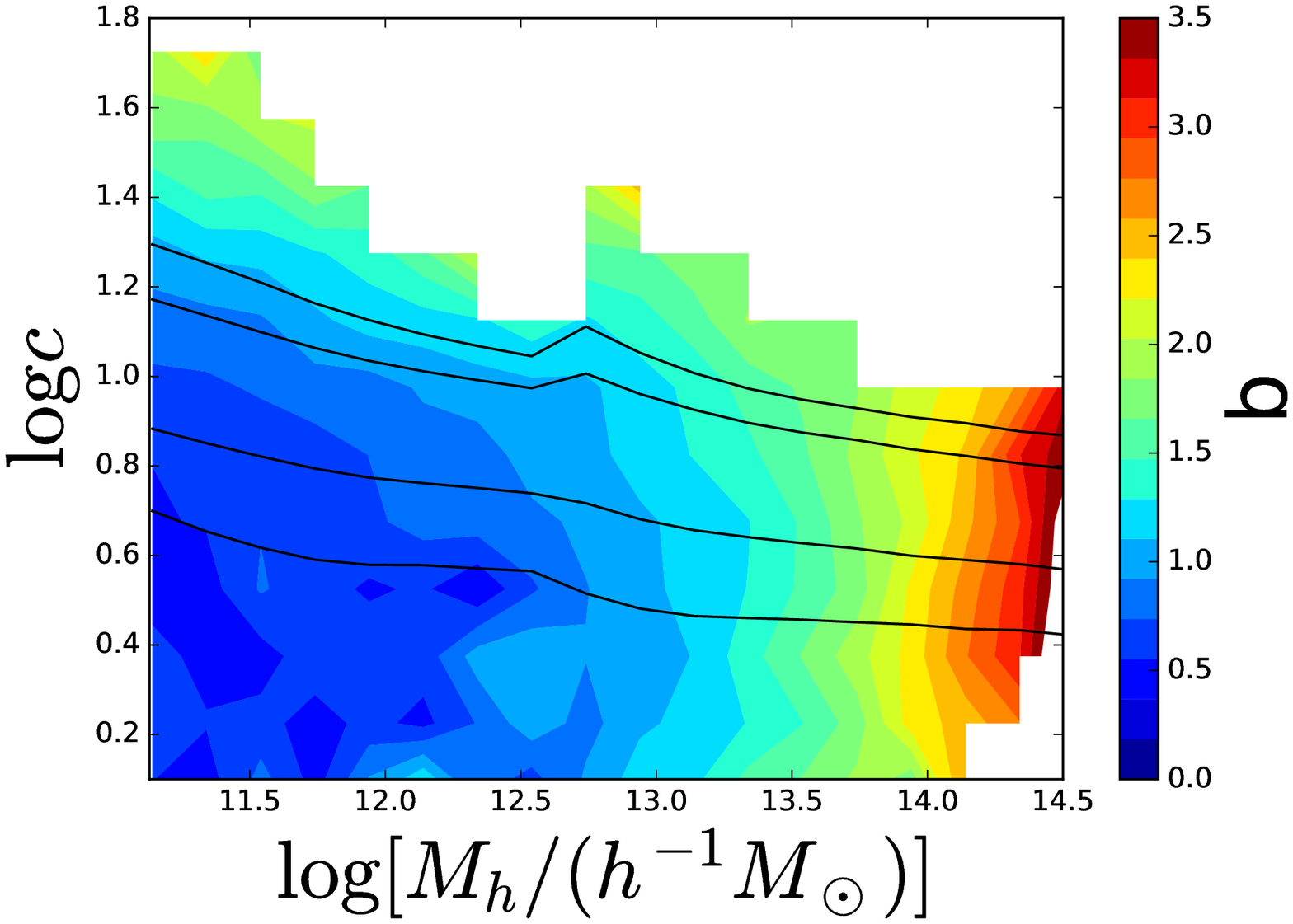}\hfill
	\end{subfigure}
	\begin{subfigure}[h]{0.48\textwidth}
	\centering
        \includegraphics[width=\textwidth]{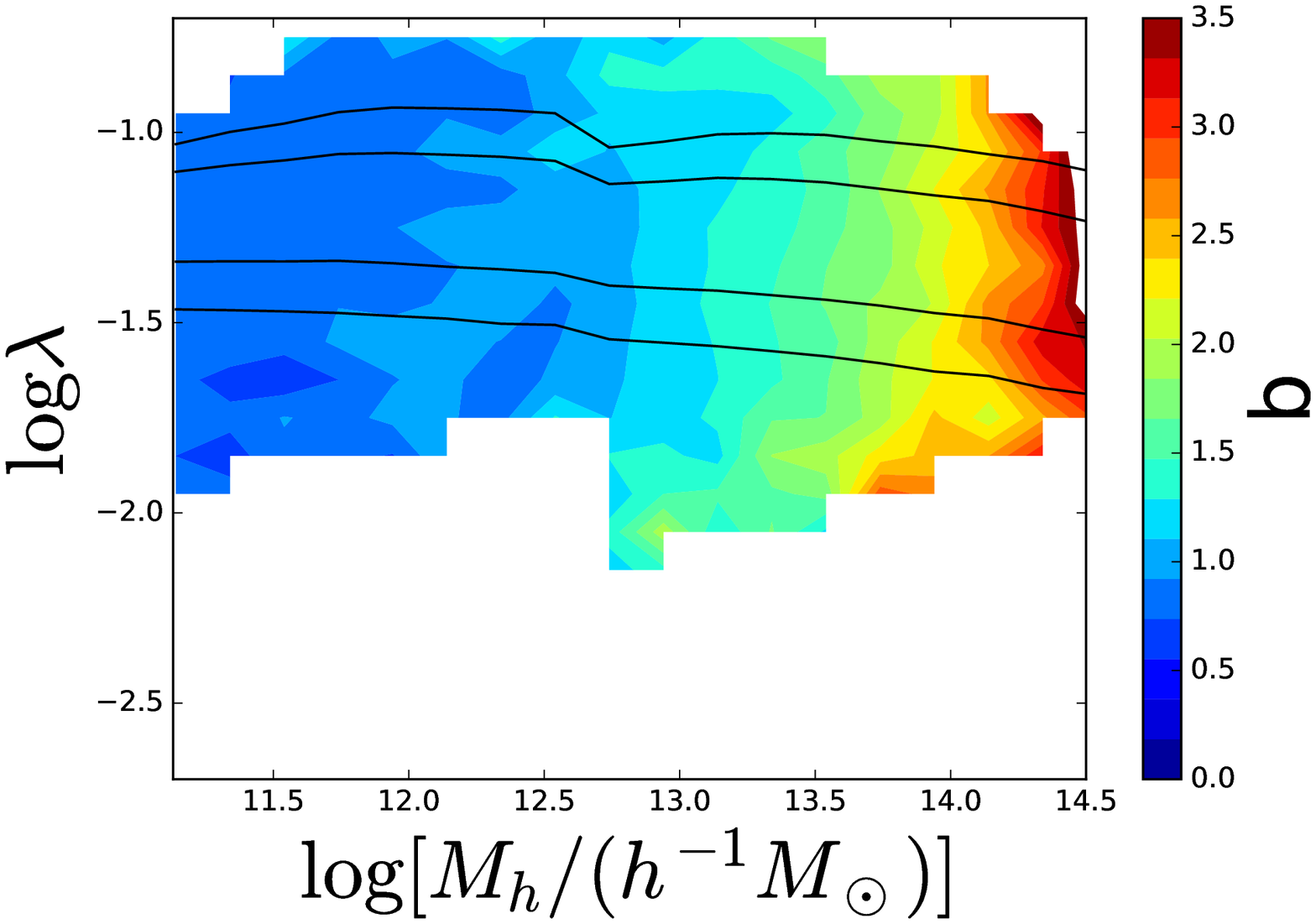}\hfill
	\end{subfigure}
\caption{Same as Fig.~\ref{fig:joint_dep}, but with original halo mass replaced by 
effective halo mass. The effective halo mass absorbs the halo assembly bias
with halo spin, manifested by the vertical contours in the bottom-right
panel. However, assembly bias effect still exists with other assembly 
variables, as seen in the other three panels.
}
\label{fig:bMeff}
\end{figure*}

\begin{figure*}
\centering
	\begin{subfigure}[h]{0.48\textwidth}
	\centering
	\includegraphics[width=\textwidth]{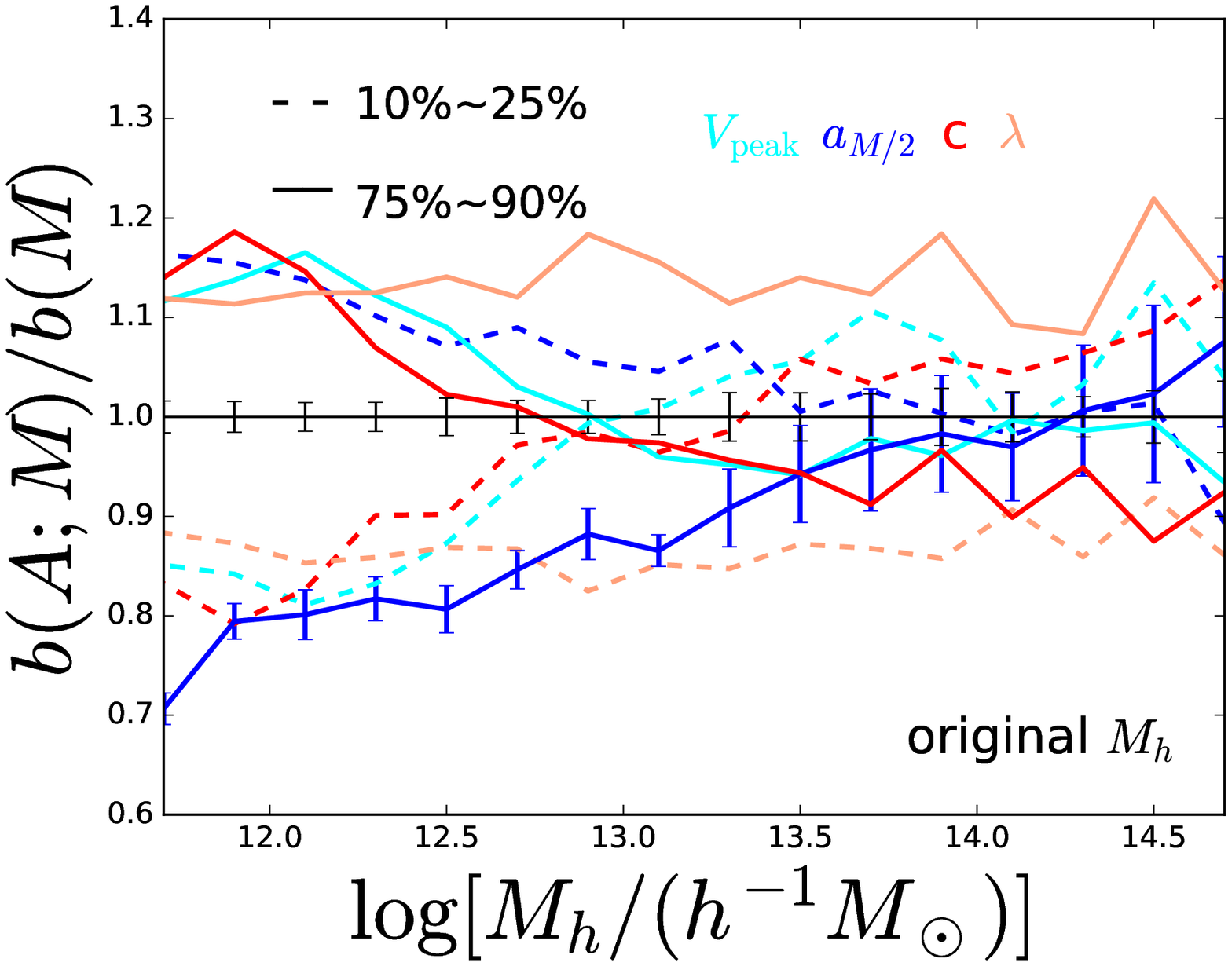}\hfill
	\end{subfigure}
	\begin{subfigure}[h]{0.48\textwidth}
	\centering
	\includegraphics[width=\textwidth]{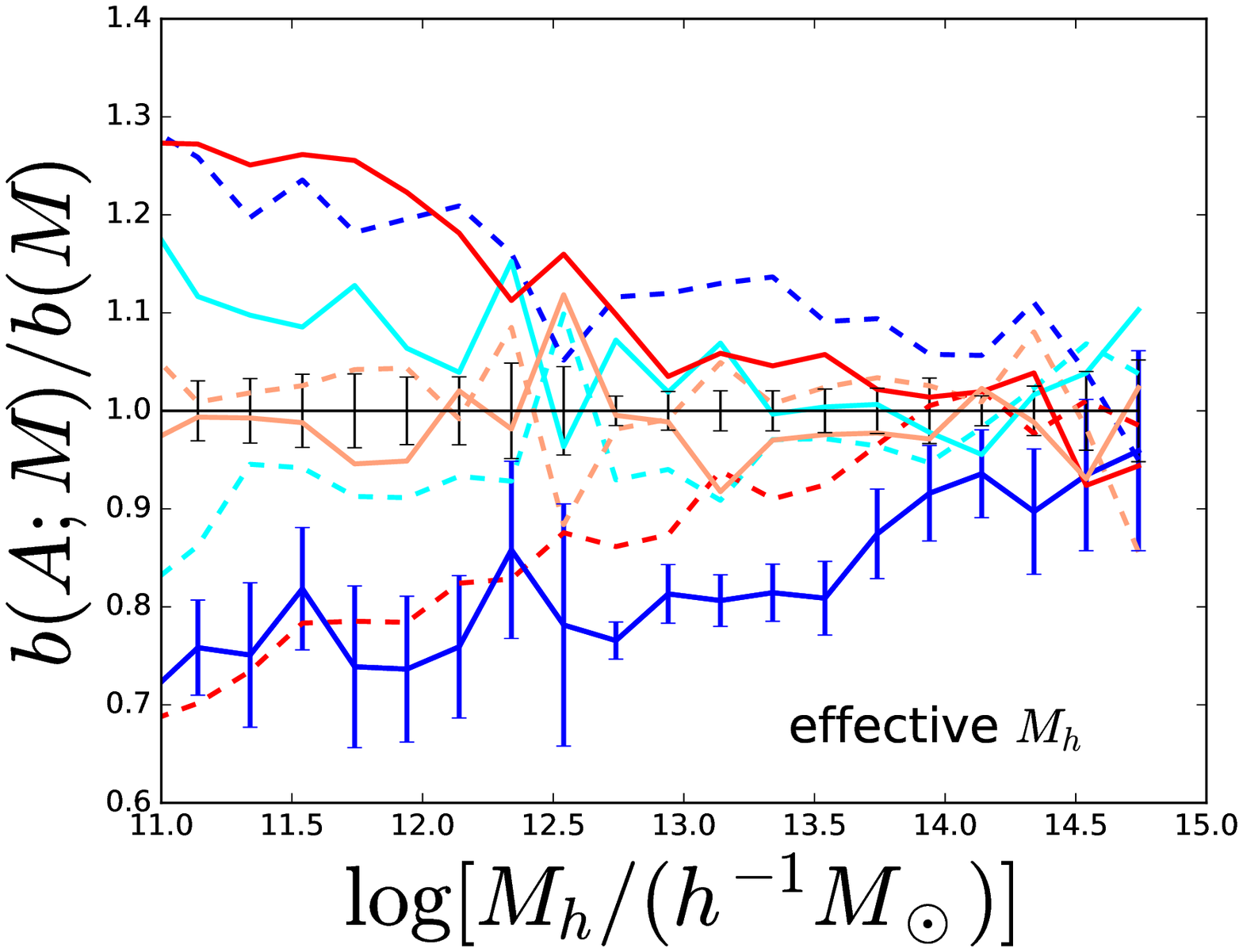}\hfill
	\end{subfigure}
\caption{
Comparison between assembly bias effects under original halo mass (left) 
and effective halo mass (right). At fixed original or effective halo mass, 
haloes are grouped according to the percentiles of each assembly variable, and 
their bias values normalised by the mass-only dependent bias are shown as curves 
in each panel (dashed for the 10--25 per cent percentile and solid for the 75--90 
per cent percentile).
}
\label{fig:bpercentile}
\end{figure*}

To see whether we can construct a halo variable from combinations of halo mass and
assembly variables to minimise the assembly history effect on halo bias, we 
investigate an example with halo mass $\Mh$ and spin $\lambda$. As shown in 
Fig.~\ref{fig:joint_dep}, the distribution of halo spin is almost independent 
of halo mass and the joint dependence of halo bias on $\Mh$ and $\lambda$ has
a pattern easy to describe.

Within the most part of the parameter space extended by 
$\log\Mh$ and $\log\lambda$, a given value of halo bias factor approximately 
corresponds to a straight line (see the dashed lines in Fig.~\ref{fig:Meff} 
as examples). We find that the set of
straight lines can be well described as converged to a single point
($\log M_{\rm h,0}$, $\log\lambda_0$), located towards the bottom-right corner 
of the left panel of Fig.~\ref{fig:Meff}. The slope 
$(\log \Mh - \log M_{\rm h,0}) / (\log\lambda - \log\lambda_0)$ of each line
can be defined as a new halo variable, which is a combination of 
halo mass $\Mh$ and spin $\lambda$. Halo bias is monotonically dependent on
this variable. Equivalently, for each line, we can evaluate the halo mass
at a fixed spin value and use this halo mass as the new variable. We choose
the fixed spin value to be $\log\lambda_{\rm eff}=-1.4$, roughly the value 
averaged over all haloes, and define the corresponding halo mass as the
effective halo mass $\Meff$ through
\begin{equation}
\log\Meff = \log M_{\rm h, 0} + \frac{\log\lambda_{\rm eff} - \log\lambda_0}{\log\lambda - \log\lambda_0}(\log \Mh -\log M_{\rm h,0}).
\end{equation}

To cover a large range in effective halo mass, we perform the analysis using a combination of haloes in both the MDR1 and Bolshoi simulations. As seen from Fig.~\ref{fig:Meff}, for an effective mass of $\log[\Meff/(\hinvMsun)]\sim$ 11--12.5, we have contributions from haloes of mass below $\log[\Mh/(\hinvMsun)]\sim 11.6$ with high spin. Furthermore, there is numerical effect for haloes with less than a few hundred particles \citep[e.g.][]{Paranjape17a,Trenti10,Benson17} in the MDR1 simulation. To include low mass haloes and minimise the potential numerical effect, we calculate effective halo masses for both MDR1 and Bolshoi haloes, then present the results using MDR1 haloes for $\log[\Meff/(\hinvMsun)]\geq 12.7$ and Bolshoi haloes for $\log[\Meff/(\hinvMsun)]< 12.7$.

%\textbf{ In this section we combine MDR1 and Bolshoi simulation together, the reason is that effective mass catalog calculated only using MDR1 haloes are incomplete. For example, haloes with mass lower than $4\times 10^{11}\hinvMsun$ and high spin could also have a high effective mass, but these low mass haloes would not be included in MDR1. To include these haloes, we calculate effective mass for both MDR1 and Bolshoi haloes, then use MDR1 for $\log\Meff>12.7$ since it can cover the whole range of spin parameter in MDR1, and use Bolshoi for $\log\Meff<12.7$.}

To verify that the effect of assembly bias in spin is absorbed into the 
effective halo mass, we measure halo bias in fine bins of $\log\Meff$
and $\log\lambda$. In the right panel of Fig.~\ref{fig:Meff}, the black 
circles show the dependence of halo bias on the original halo mass $\Mh$, 
and the
spread at fixed mass reflects the assembly bias effect in halo spin. The
red dots represent the dependence of halo bias on the effective halo mass
$\Meff$. At fixed $\Meff$ the scatter is much smaller than that seen in the
$b$--$\Mh$ dependence, demonstrating that the assembly bias effect in spin 
has been well absorbed into the effective mass.

Fig.~\ref{fig:bMeff} is similar to Fig.~\ref{fig:joint_dep}, but for the 
joint dependence of halo bias as a function of effective mass $\Meff$ and
each of the assembly variable. The fact that the assembly bias effect in
spin has been absorbed into $\Meff$ shows up as largely vertical contours
in the $\log\Meff$--$\log\lambda$ plane (bottom-right panel).

However, in the other three panels of Fig.~\ref{fig:bMeff}, where the joint 
dependence of halo bias on $\Meff$ and the other assembly variables are shown,
assembly bias still exists at fixed $\Meff$. The assembly bias effect
becomes weaker at the high $\Meff$ end. This is not surprising, given that
the effect is weak at high $\Mh$ in Fig.~\ref{fig:joint_dep} and the $\lambda$
dependence appears to be the strongest one. 

In Fig.~\ref{fig:bpercentile}, we show the assembly bias effect as a function 
of the original halo mass (left panel) and the effective halo mass (right panel), 
by grouping haloes into 10--25 per cent and 75--90 per cent percentiles of 
different assembly variables. Error bars are calculated with the jackknife method. For clarity, we only plot error bars for $b(M)$ (black solid) and \ahalf\ (dark blue). Under the effective halo mass, the assembly bias 
in halo spin almost disappears, and the curves of bias from the two percentile 
subsamples fall on top of each other (on the line of unity, corresponding to 
the average bias as a function of effective mass), while under the original halo 
mass the assembly effect in halo spin is at the level of $\sim$15 per cent.
Above $\log[\Meff/(\hinvMsun)]\sim 13.5$, the assembly effects in \Vpeak
and $c$ are also reduced, but not in \ahalf. However, towards lower $\Meff$, 
the assembly bias effect in other assembly variables becomes enhanced, compared
to the case with the original halo mass. Therefore, the effective mass does not 
work all the way to reduce assembly bias effect in other variables and over the 
full mass range.

With the joint dependence of halo bias on $\Meff$ and assembly variables, in 
principle we can perform a similar variable transformation to largely reduce 
the assembly effect in another chosen variable, and the analysis can be 
repeated until the assembly effect in all the assembly variables of interest 
is recursively absorbed. It may be possible to find a principal direction or locus 
in the multidimensional space spanned by halo mass and assembly variables such 
that halo bias has the strongest dependence on the corresponding variable 
combination. However, there is a limit on how far we would like to go following
such a path. If the ultimate goal is to replace halo mass with the putative 
variable combination in halo modelling of galaxy clustering, it is desirable 
that the variable combination has a tight correlation with galaxies properties. 
It has been established that halo mass plays the dominant role in shaping galaxy 
formation and evolution and in determining the main properties of galaxies 
\citep[e.g.][]{White78,Birnboim03,Keres05}. Although the combination variable 
$\Meff$ absorbs the assembly bias effect in halo spin, a fixed $\Meff$ spans 
a large range in halo mass. For example, haloes of effective mass 
$\Meff=10^{12}\hinvMsun$, which corresponds to the middle dashed line in the left 
panel of Fig.~\ref{fig:Meff}, can come from haloes of original mass 
$\Mh\sim 10^{11}\hinvMsun$ with high spin or those of $\Mh\sim 10^{13}\hinvMsun$ 
with low spin. The large range in original halo mass suggests
that $\Meff$ would not be a good variable to choose for a tight correlation with 
galaxy properties. Together with the results in Fig.~\ref{fig:bMeff} and
Fig.~\ref{fig:bpercentile}, it implies that it is unlikely to find the right 
combination of halo variables to completely absorb the assembly bias effect 
in every assembly variable and at the same time to keep a close connection to 
galaxy properties.

%3.3
\subsection{Scale dependence of halo bias with assembly variable}
\label{sec:scale_dep}

\begin{figure*}
    \centering
    \begin{subfigure}[h]{0.99\textwidth}
        \centering
        \includegraphics[width=\textwidth]{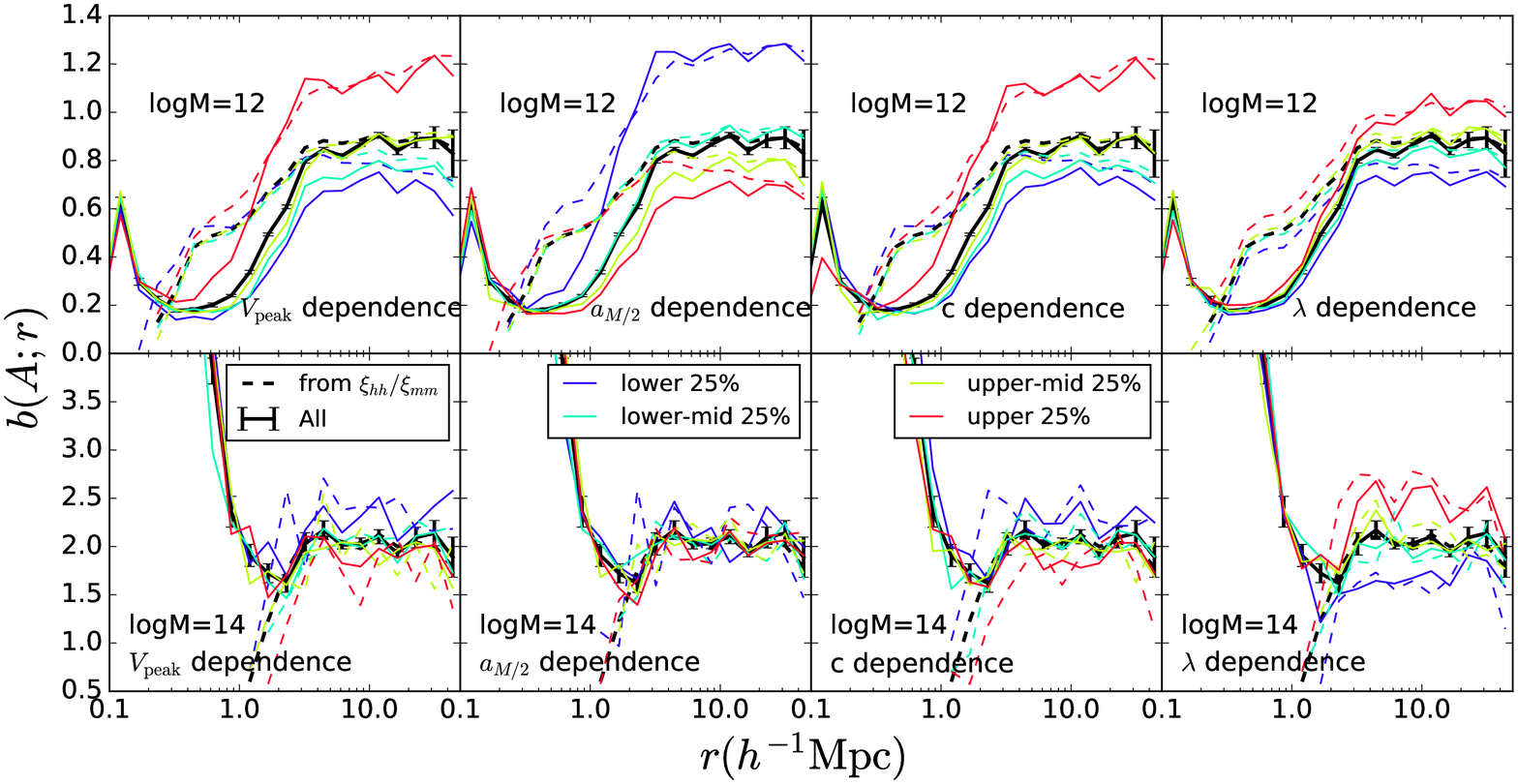}
    \end{subfigure}
    \hfill
    \begin{subfigure}[h]{0.99\textwidth}
        \centering
        \includegraphics[width=\textwidth]{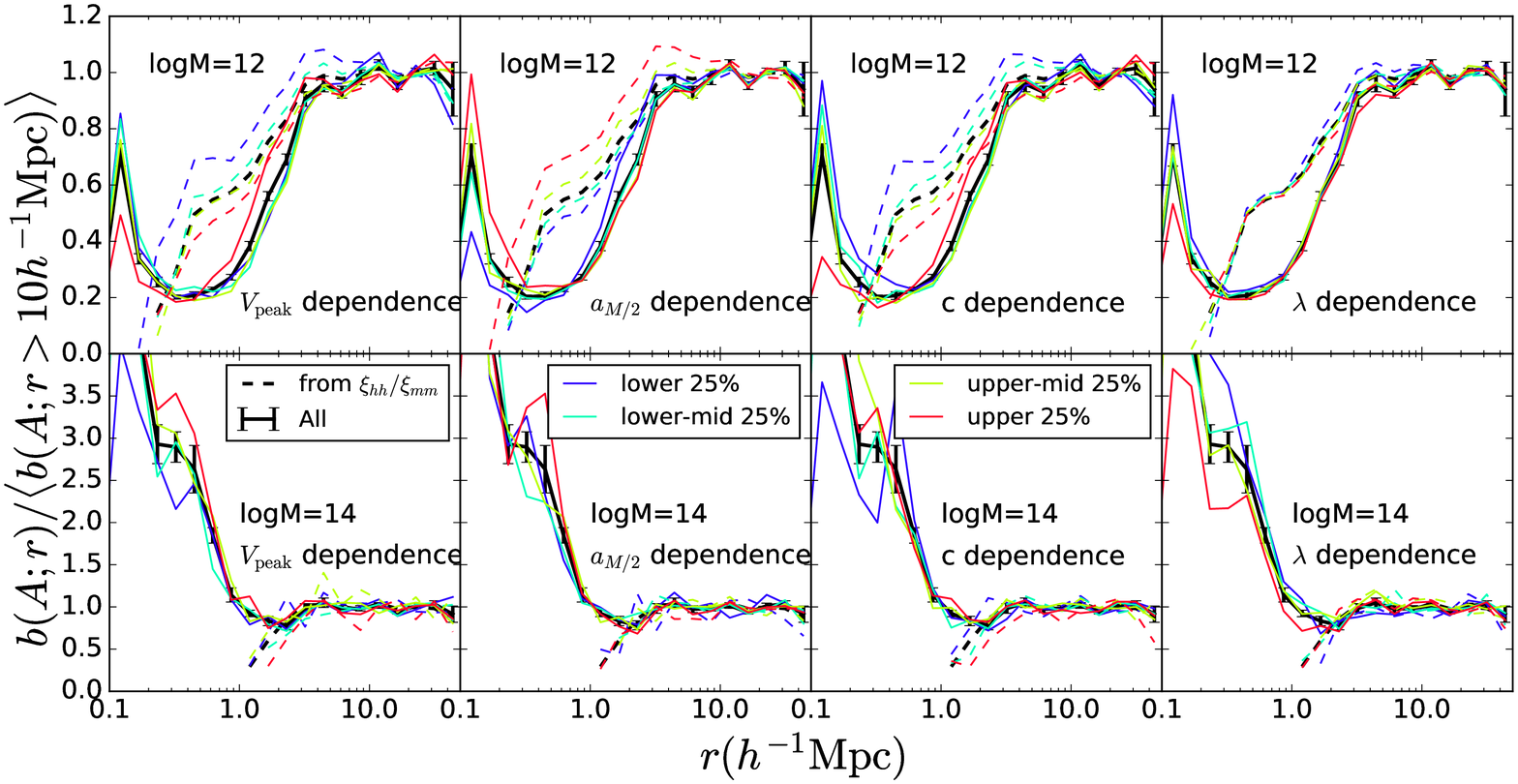}
    \end{subfigure}
\caption{
Scale dependence of assembly bias. The upper subplot shows the scale-dependent 
halo bias factor for different assembly variables. In each panel, 
solid curves are calculated through halo-mass cross-correlation functions 
and dotted curves through halo-halo auto-correlation functions. Each set
of five curves are for haloes in different percentiles of each assembly
variable: lower 25 per cent (blue), lower-middle 25 per cent (cyan), 
upper-middle 25 per cent (green), upper 25 per cent (red), and all haloes 
in the mass bin (black). Top and bottom panels are for haloes of 
$\log[\Mh/(\hinvMsun)]$=12 and 14, respectively. Bottom subplot is similar, 
but each halo bias curve is normalised by the corresponding large-scale value
(average on scales larger than $10\hinvMpc$). 
}
\label{fig:scale_dep}
\end{figure*}

While the assembly history of haloes affects large-scale clustering, it can 
also have an effect on small-scale clustering. In Fig.~\ref{fig:scale_dep},
the scale dependences of halo bias on assembly variables are shown at two 
halo masses, $\log[\Mh/(\hinvMsun)]$=12 and 14, respectively. The top eight 
panels show the original bias factors $b(A;r)$ as a function of $r$, calculated 
by grouping haloes into various percentiles of each assembly variable $A$.
To emphasise the shape difference on small scales, in the bottom eight panels 
we also show halo bias factors normalised by the corresponding large-scale 
value (average values above $10\hinvMpc$), i.e.
$b(A;r)/\langle b(A;r>10\hinvMpc)\rangle$.  
For clarity, in each panel, error bars are only plotted for the curve corresponding to all haloes to indicate the uncertainties.
%The differences between the curves of different percentiles are not caused by noise.}
%a result of is n for the jackknife method are added to all the black lines one each panel to give a sense of uncertainty, this can show that the discrepancy between different percentiles can not be explained by uncertainty.}

In the non-normalised case, for $\Mh=10^{12}\hinvMsun$ haloes, all halo bias 
factors (solid curves in the top panels) start to drop at $\sim 3\hinvMpc$ 
towards small scales, which means that the two-point halo-mass cross-correlation
function is shallower than the mass auto-correlation function on such scales. 
Note that the drop occurs on scales larger than the halo virial radius 
(\Rvir$\sim 0.21\hinvMpc$). The panels show that halo assembly history 
influences how steep the bias factor drops towards small scale. Except for 
the halo spin case, all other cases have a similar trend that can be related
through the correlation between halo assembly variables. Haloes that form 
earlier (smaller \ahalf), thus on average with higher halo concentration 
$c$ and higher \Vpeak, have a steeper drop in halo bias towards smaller 
scales. The dotted curves in each panel are the measurements of halo bias using 
halo-halo auto-correlation functions, and the trend is similar to  that derived 
from halo-mass cross-correlation functions.

For the case with halo spin, the overall shape difference is smaller than seen
in other three cases. As halo spin shows an anti-correlation with halo 
concentration (e.g. \citealt{Maccio07}; also Fig.~\ref{fig:bon2_mb1} and
Fig.~\ref{fig:bon2_mb2}), one would expect the trend to be that 
the scale-dependent bias for haloes with lower spin drops more steeply
towards smaller scales. However, the result is opposite (see \S~\ref{sec:2var} 
for an interpretation).

The bottom panels of the non-normalised case in Fig.~\ref{fig:scale_dep} 
show the results with $\Mh=10^{14}\hinvMsun$ haloes. As the virial radius 
\Rvir\ is $\sim 0.98 \hinvMpc$, the rise on small scales reflects the 
mass distribution inside haloes. For all cases, the dotted curves for halo 
biases from halo-halo auto-correlation functions have a cutoff around 
$r\sim 2$\Rvir, a manifestation of the halo exclusion effect. Above 
$\sim 2$\Rvir, the shape of scale-dependent halo bias factors do not 
show a strong dependence on assembly variables (shown more clearly in
the corresponding panels for the normalised case).

Overall the shape of the scale dependence of halo bias at small scales depend on the value of the assembly variable in consideration, as shown in the normalised case. The dependence is not strong, but it is discernible for low mass haloes. For high mass haloes, the dependence becomes weak (except for the spin case). In such a situation, the scale-dependent assembly bias at a given halo mass $\Mh$ can be well approximated as $b(\Mh,A;r)=\langle b(\Mh,A)\rangle f(\Mh;r)$, where $\langle b(\Mh,A)\rangle$ is the large-scale bias with the value of assembly variable being $A$ and $f(\Mh;r)$ characterise the shape (being unity on large scales by construction).

\citet{Sunayama16} study the scale dependence of halo assembly bias for the 
case of \Vmax, the maximum circular velocity at $z=0$. They find that for
low-mass haloes ($\Mh\la \Mnl$) the ratio of biases between high \Vmax\ and low 
\Vmax\ haloes exhibits a pronounced scale dependence at 0.5--5$\hinvMpc$, 
and the scale dependence becomes weak towards higher halo mass. Our results,
if put in a similar format, are in broad agreements with their findings. They 
also show that the scale dependence can be partially attributed to haloes
previously residing in higher mass haloes but ejected to become host haloes
at the epoch of interest, which is consistent with the proposed origin of 
assembly bias for low mass haloes \citep[e.g.][]{Wang09,Wetzel14}

If galaxy properties track the halo assembly history, the scale-dependent bias
from the halo assembly effect could be a way to reveal the assembly bias in 
galaxy clustering. However, the results in Fig.~\ref{fig:scale_dep} show that 
the scale dependence from assembly effect is not strong, and the overall shape 
does not vary significantly with the magnitude of any assembly variables, at 
a level of $\sim$10 per cent on scales around $3\hinvMpc$ for low mass haloes
(see the normalised curves in Fig.~\ref{fig:scale_dep}). Also on scales where 
the scale dependence has a relatively large amplitude, the one-halo contribution
to the galaxy correlation function would tend to mask it. Therefore, in practice,
it is probably difficult to discern the assembly effect from the scale dependence
in galaxy clustering.

%3.4
\subsection{Dependence of halo bias on two assembly variables}
\label{sec:2var}

\begin{figure*}
    \centering
    \begin{subfigure}[h]{0.32\textwidth}
        \centering
        \includegraphics[width=\textwidth]{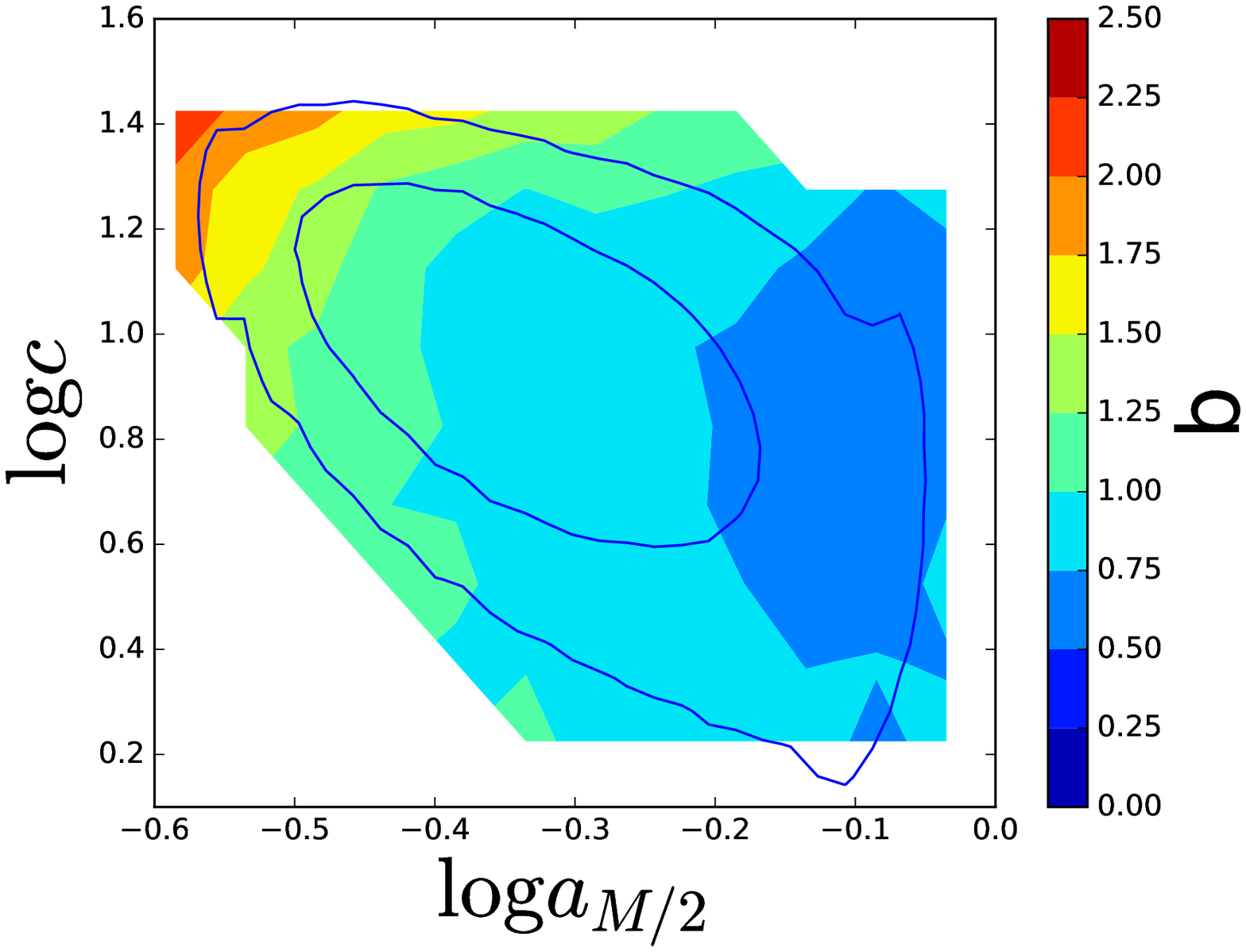}
    \end{subfigure}
    \hfill
    \begin{subfigure}[h]{0.32\textwidth}
        \centering
        \includegraphics[width=\textwidth]{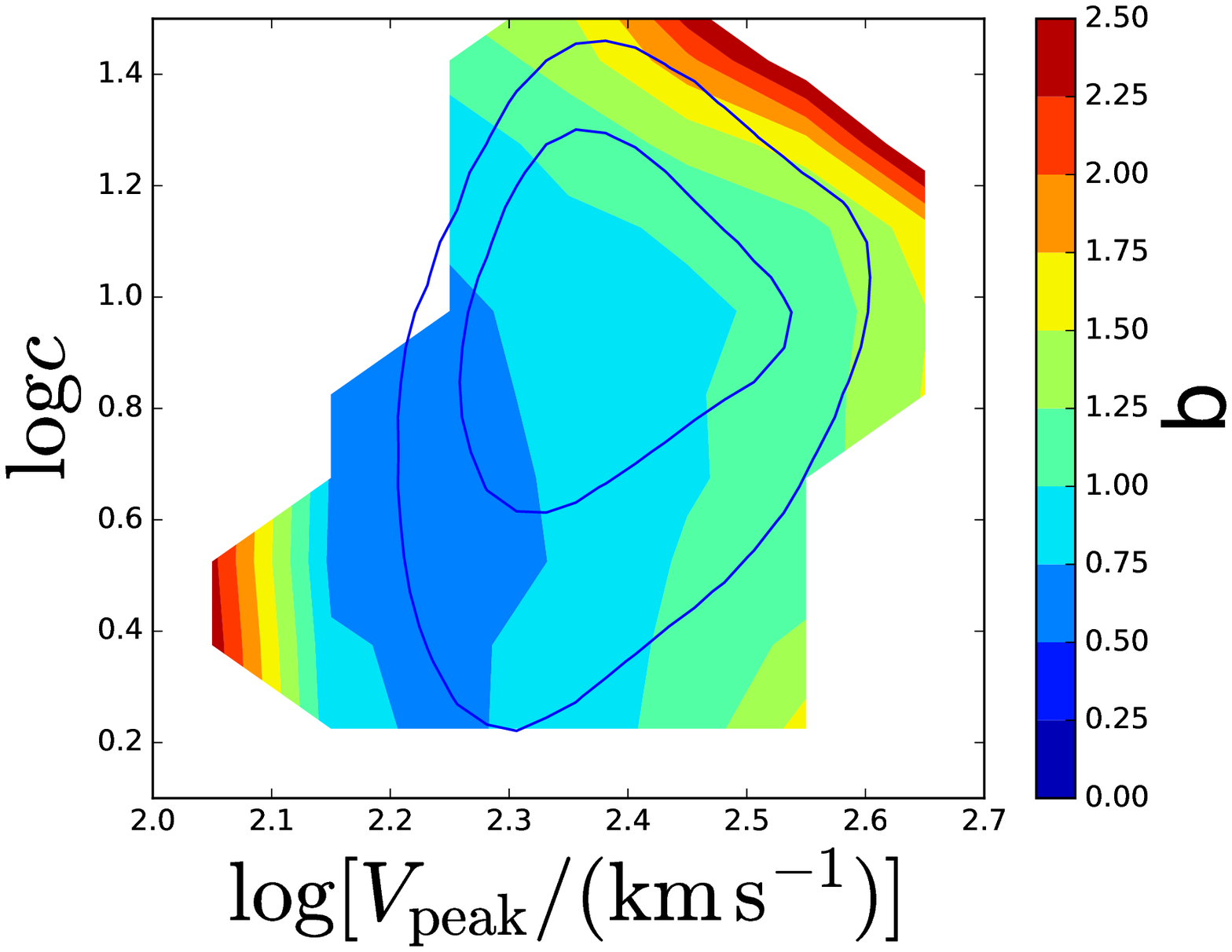}
    \end{subfigure}
	\hfill
    \begin{subfigure}[h]{0.32\textwidth}
        \centering
        \includegraphics[width=\textwidth]{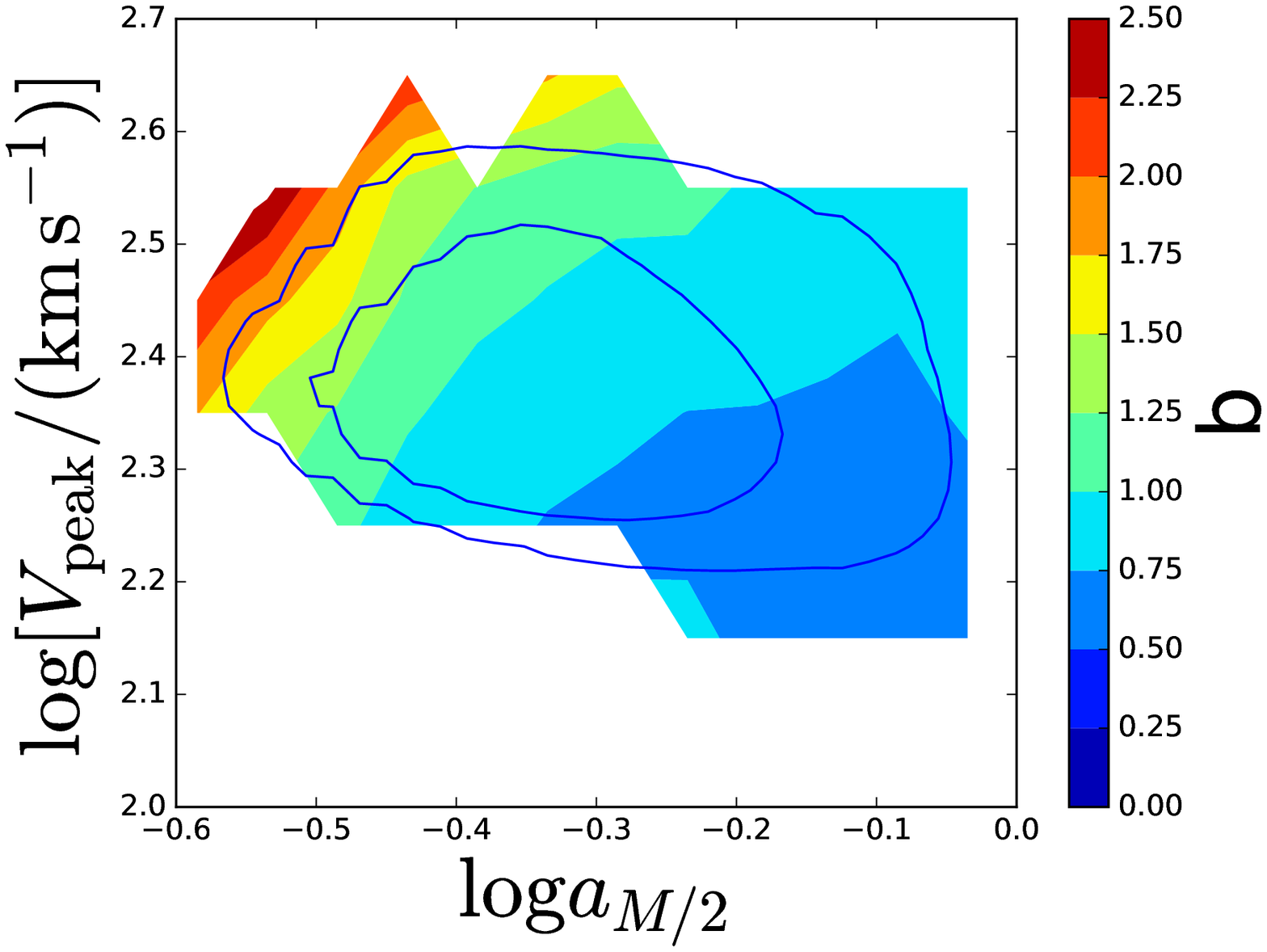}
    \end{subfigure}
	\hfill
    \begin{subfigure}[h]{0.32\textwidth}
        \centering
        \includegraphics[width=\textwidth]{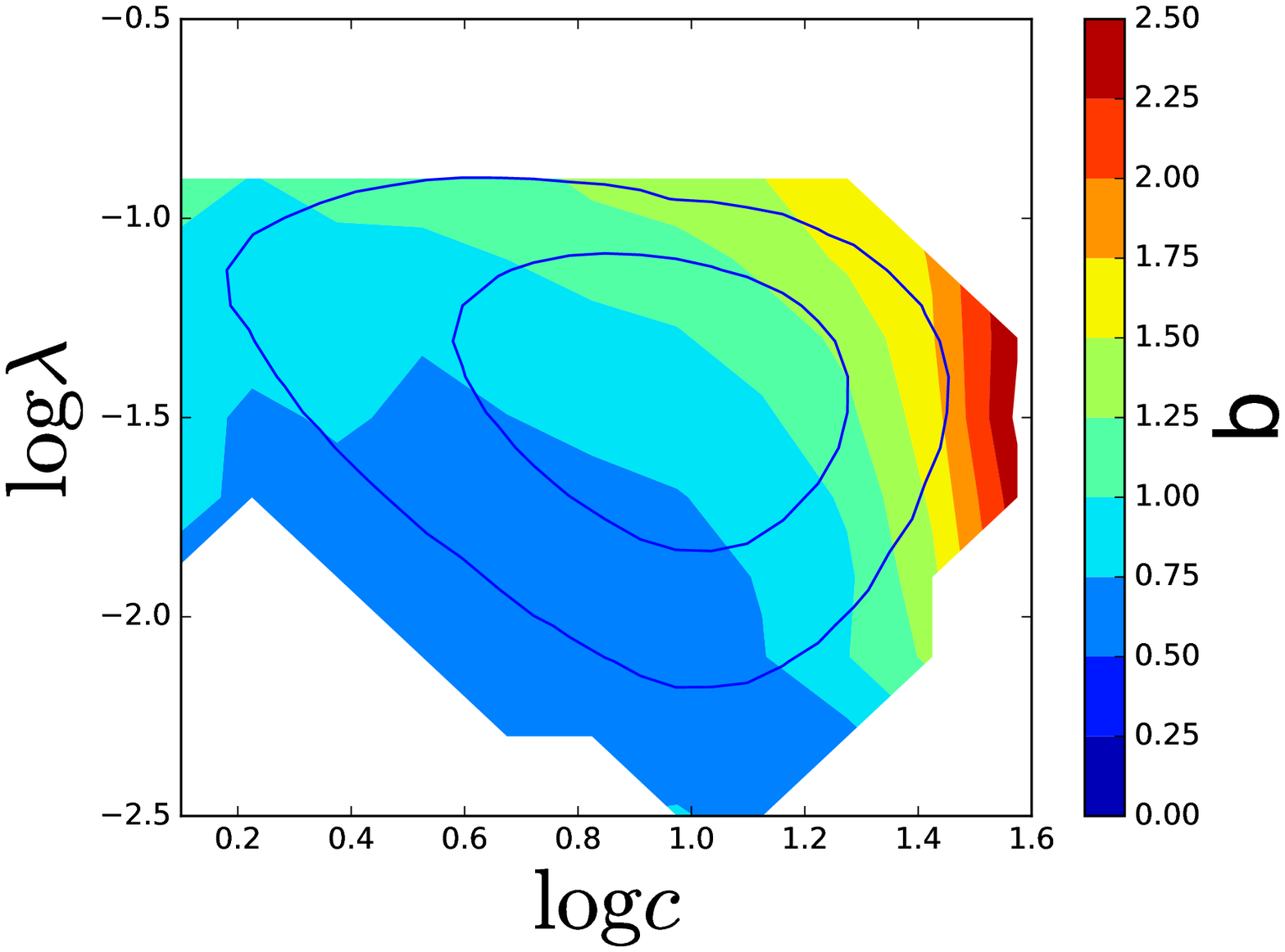}
    \end{subfigure}
	\hfill
    \begin{subfigure}[h]{0.32\textwidth}
        \centering
        \includegraphics[width=\textwidth]{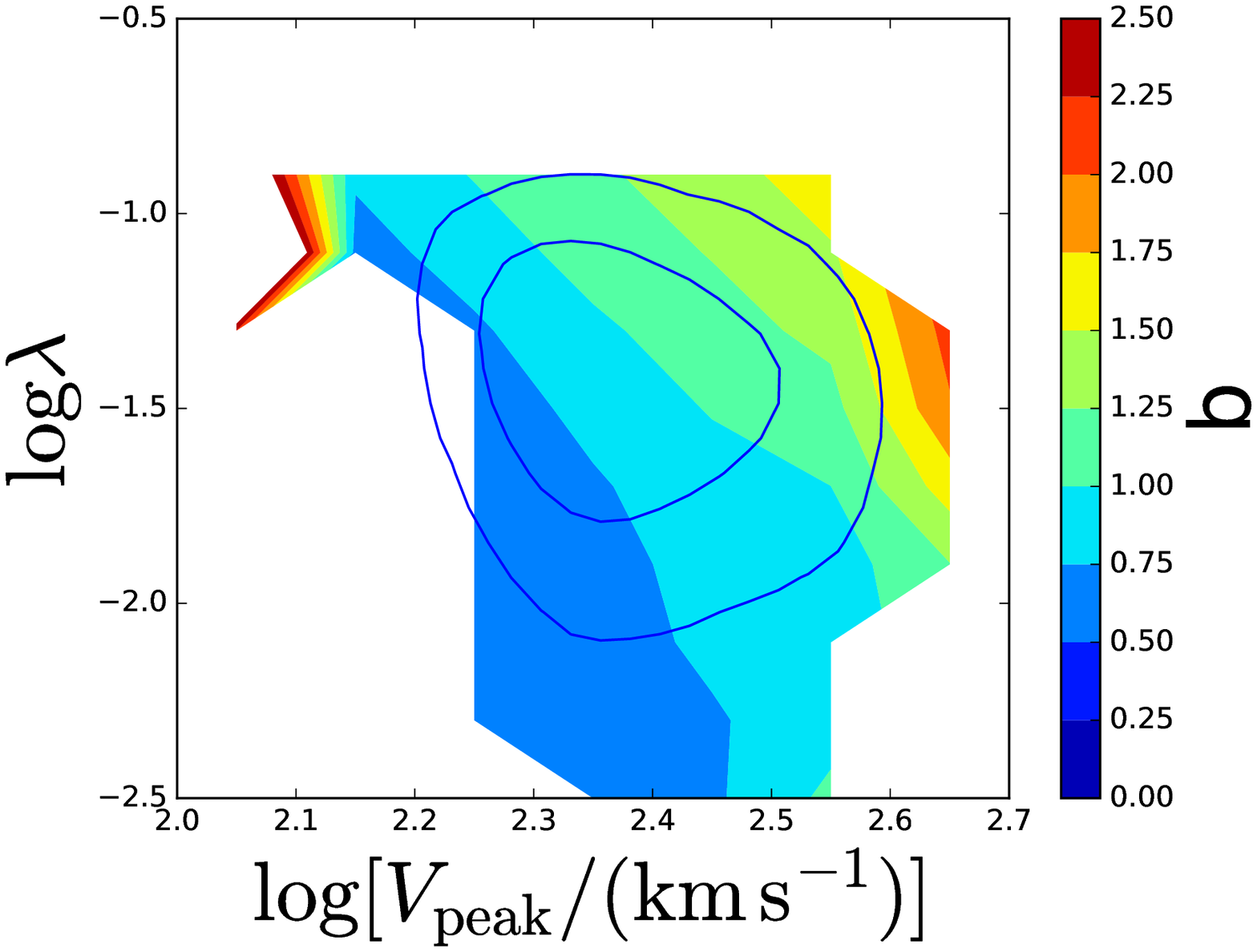}
    \end{subfigure}
	\hfill
    \begin{subfigure}[h]{0.32\textwidth}
        \centering
        \includegraphics[width=\textwidth]{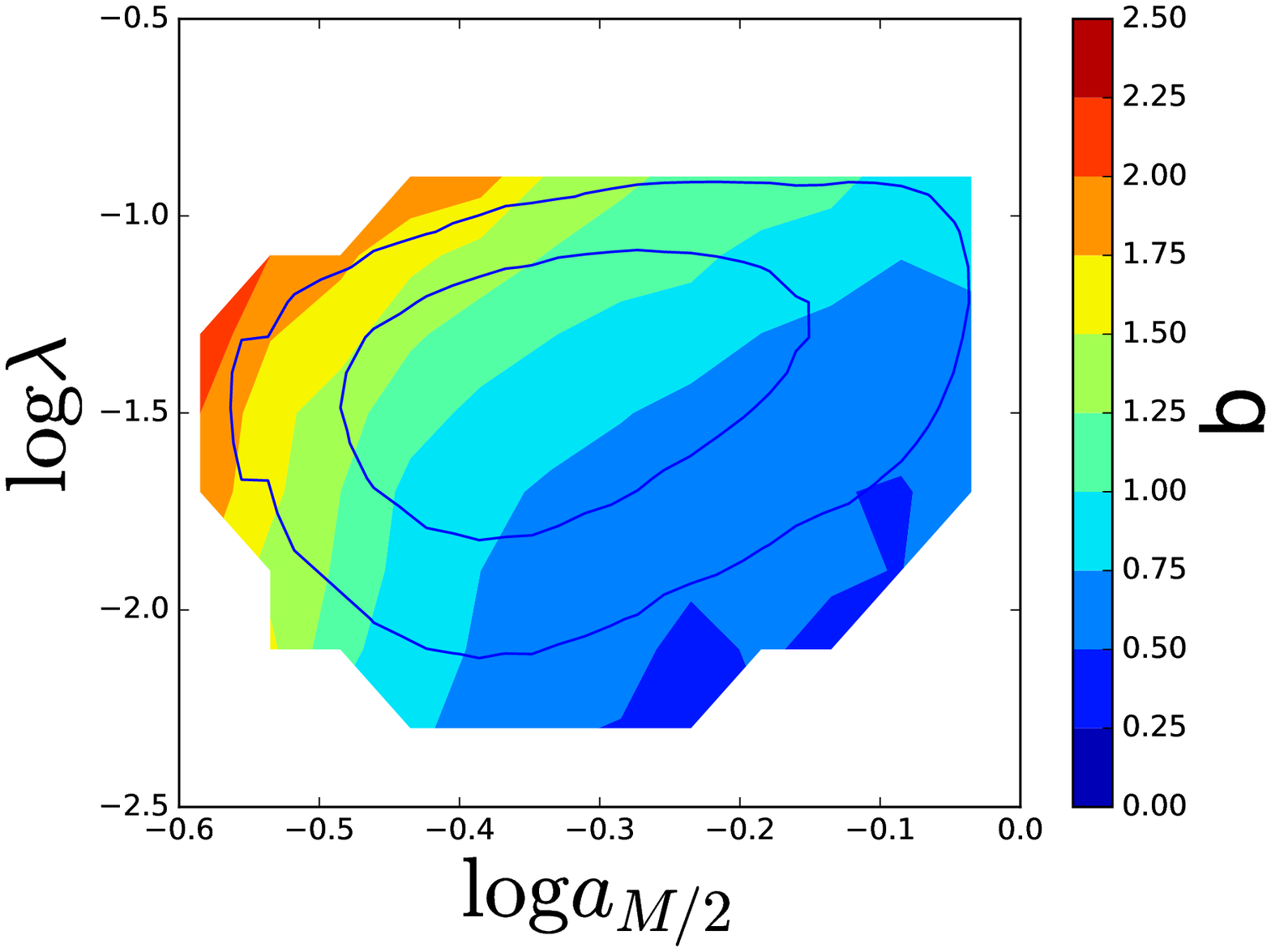}
    \end{subfigure}
\caption{Halo bias as a function of two assembly variables for haloes 
in the mass range of $\log[\Mh/(\hinvMsun)]=12.5\pm 0.5$. In each panel,
the contours mark the 68 per cent and 95 per cent distribution of the
two assembly variables and the colour scale shows the value of halo bias.}
\label{fig:bon2_mb1}
\end{figure*}

\label{sec:2var_dep} 
\begin{figure*}
    \centering
    \begin{subfigure}[h]{0.32\textwidth}
        \centering
        \includegraphics[width=\textwidth]{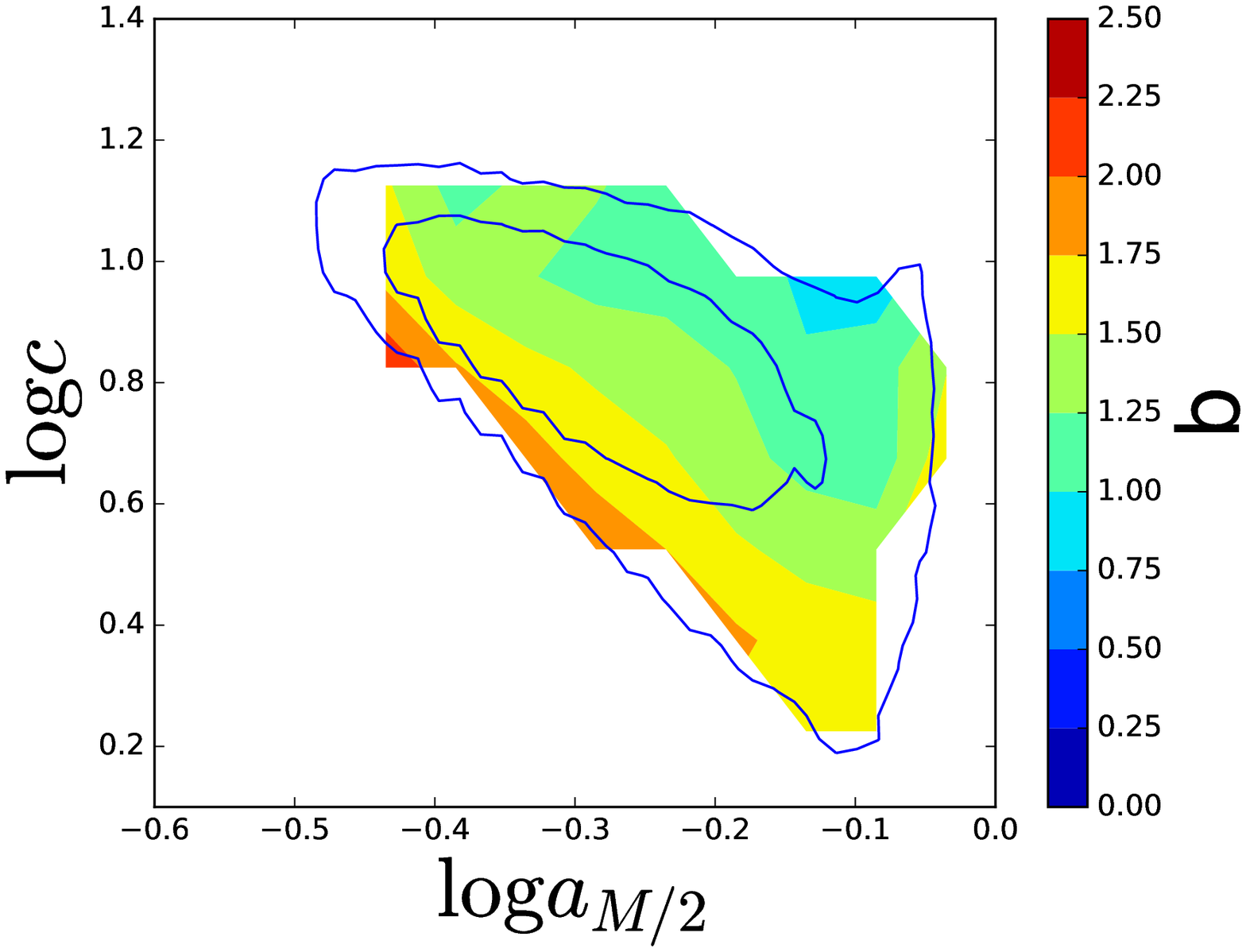}
    \end{subfigure}
    \hfill
    \begin{subfigure}[h]{0.32\textwidth}
        \centering
        \includegraphics[width=\textwidth]{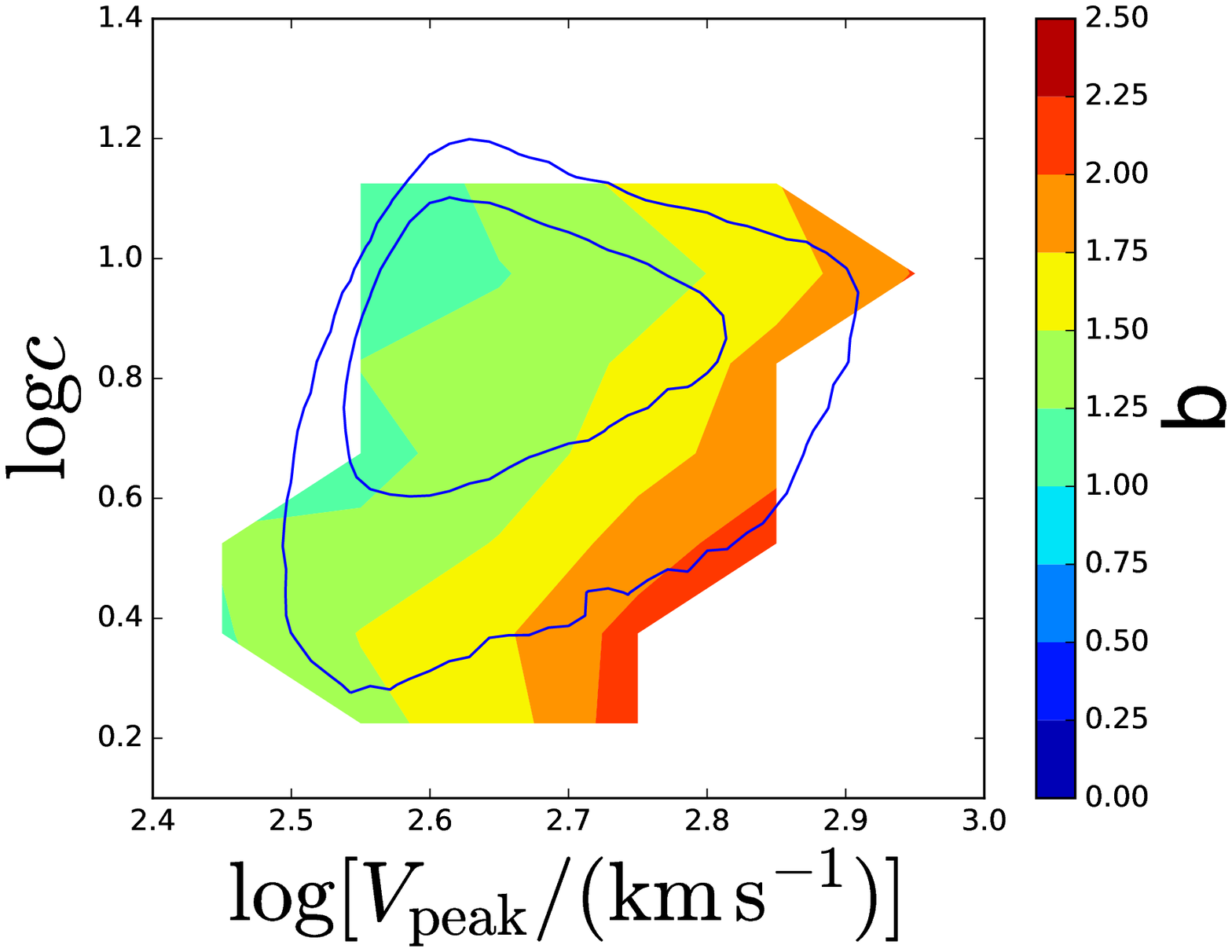}
    \end{subfigure}
	\hfill
    \begin{subfigure}[h]{0.32\textwidth}
        \centering
        \includegraphics[width=\textwidth]{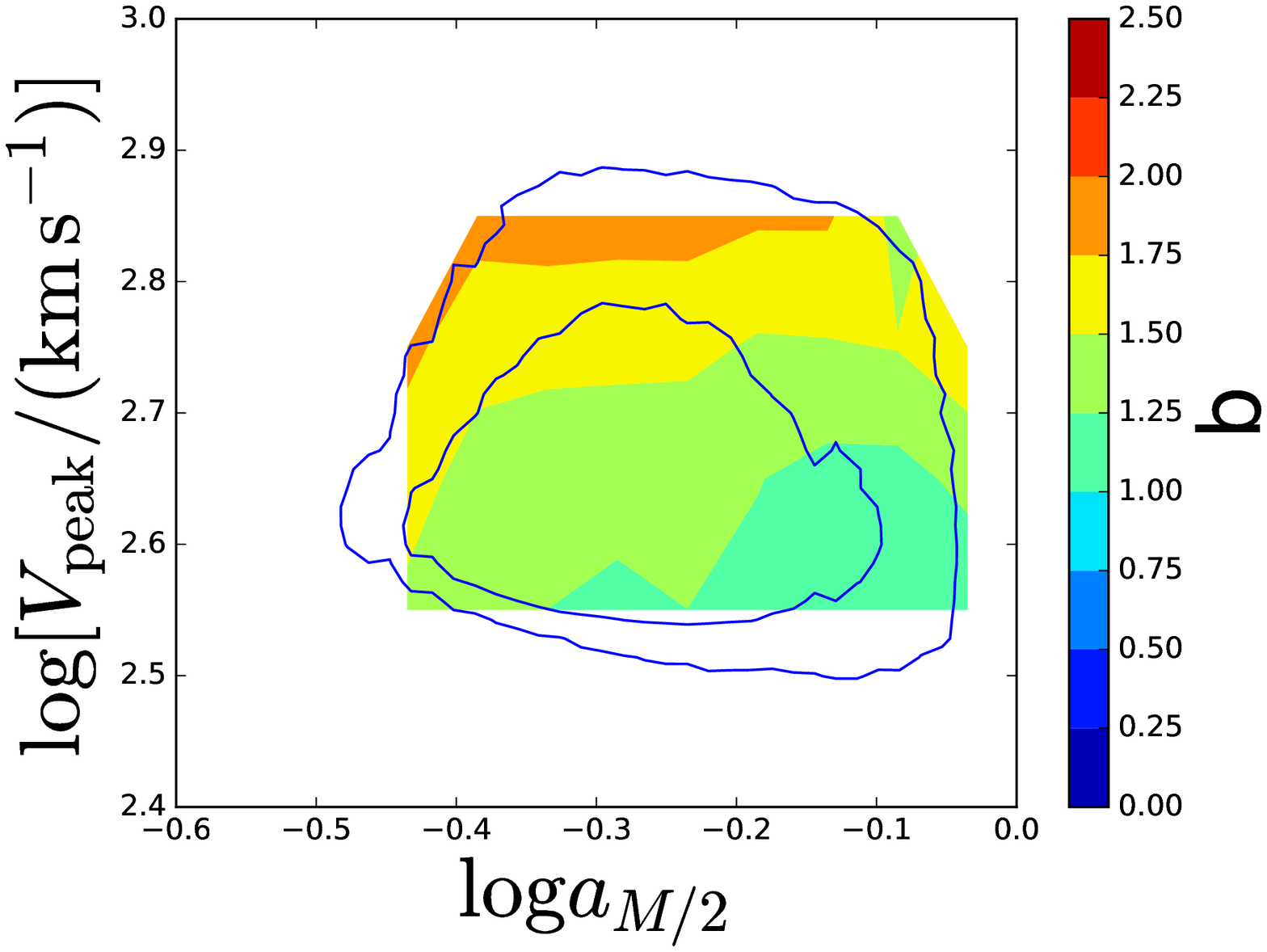}
    \end{subfigure}
	\hfill
    \begin{subfigure}[h]{0.32\textwidth}
        \centering
        \includegraphics[width=\textwidth]{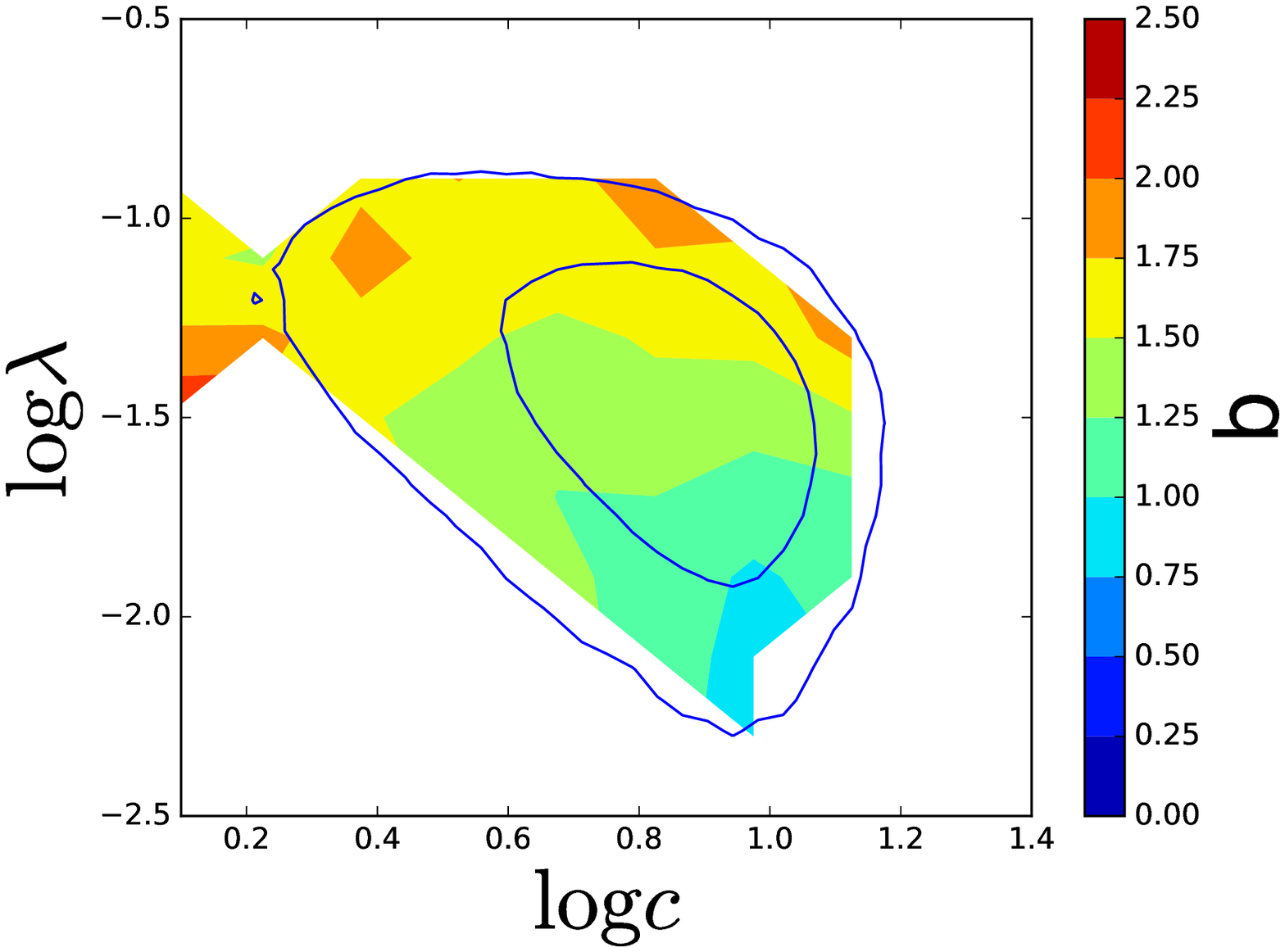}
    \end{subfigure}
	\hfill
    \begin{subfigure}[h]{0.32\textwidth}
        \centering
        \includegraphics[width=\textwidth]{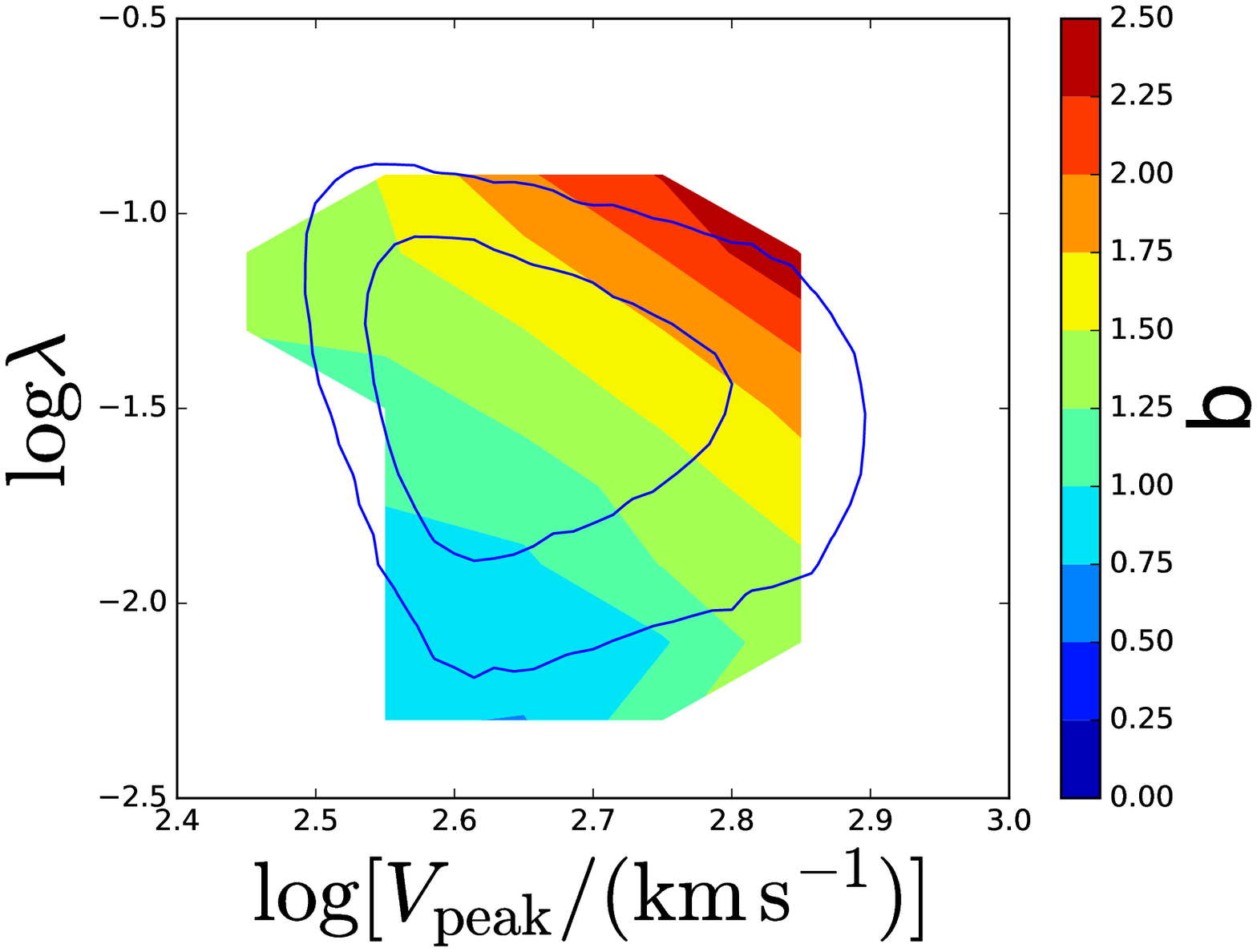}
    \end{subfigure}
	\hfill
    \begin{subfigure}[h]{0.32\textwidth}
        \centering
        \includegraphics[width=\textwidth]{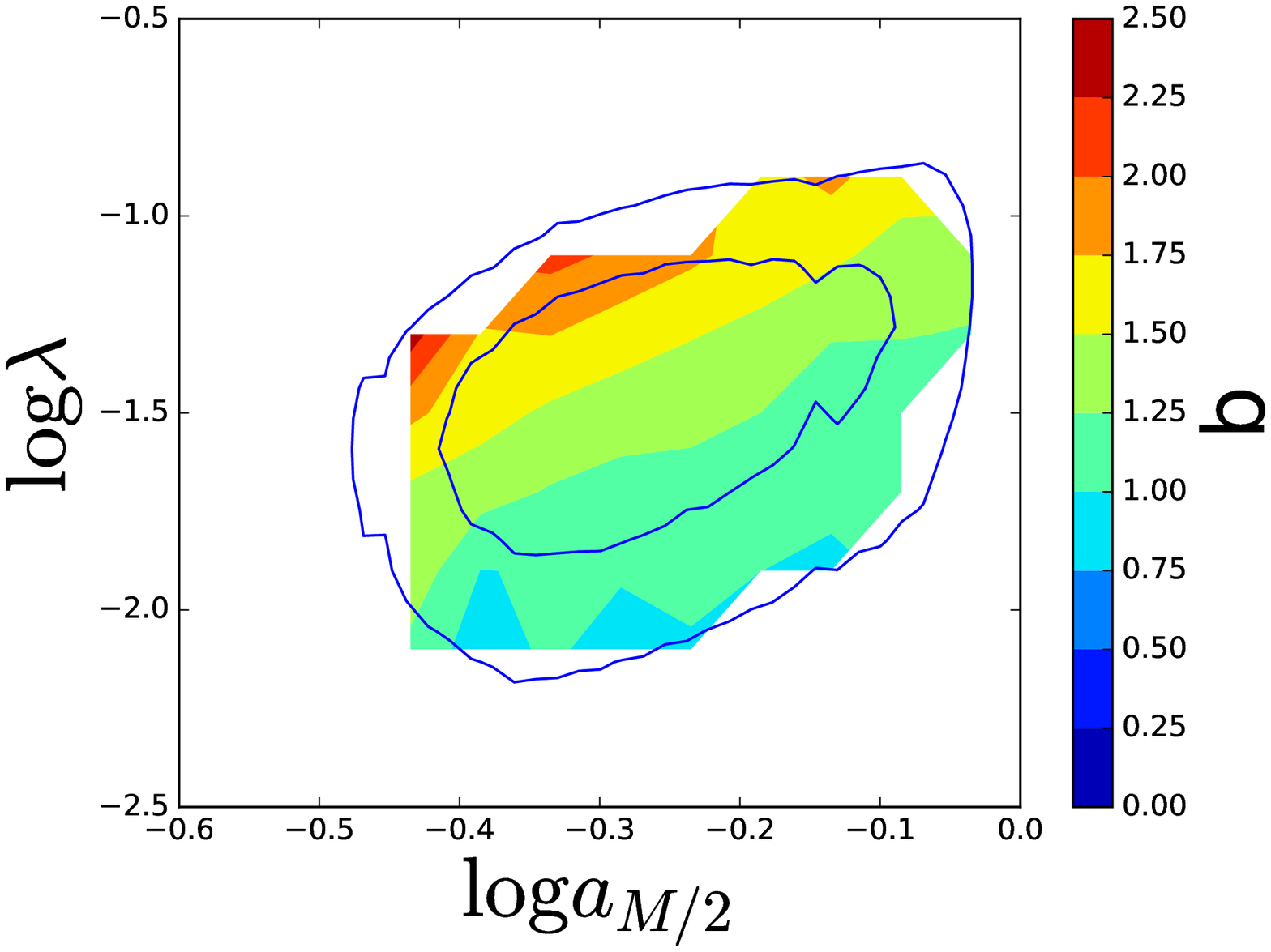}
    \end{subfigure}
\caption{Same as Fig.~\ref{fig:bon2_mb1}, but for haloes in the mass range of $\log[\Mh/(\hinvMsun)]=13.5\pm 0.5$.}
\label{fig:bon2_mb2}
\end{figure*}

At fixed halo mass, Fig.~\ref{fig:joint_dep} shows the dependence of halo bias
on each assembly variable. While in general assembly variables are correlated with 
each other, there are differences in the assembly bias trends with different assembly 
variables. To better understand the trends, we present the correlation between
any two assembly variables and how halo bias depends on them at fixed halo mass.

Each panel of Fig.~\ref{fig:bon2_mb1} shows halo bias as a function 
of two assembly variables, measured with haloes of mass in the range 
$\log[\Mh/(\hinvMsun)]=12.5\pm 0.5$ (i.e. about $\Mnl$). The contours enclose the
68 and 95 per cent distribution of the two assembly variables, respectively, which
also reveal the correlation between them. As can be seen in the top-left panel, halo
concentration $c$ and formation scale factor \ahalf\ are anti-correlated, i.e. older
haloes are more concentrated. The gradient of the halo bias follows the correlation 
direction. In such a case, the trend of dependence of halo bias on one variable can 
be used to predict that on the other variable. For example, at the above mass, given 
that more concentrated haloes are more strongly clustered and generally older, we 
can infer that older haloes are more strongly clustered than younger haloes. This is
indeed what is seen in Fig.~\ref{fig:joint_dep}. Besides $c$--\ahalf, the bias 
dependences on $c$--\Vpeak\ (top-middle panel) and \Vpeak--\ahalf\ (top-right panel)
also follow a similar behaviour. 

However, as halo spin $\lambda$ becomes involved, the above picture changes. For 
example, in the bottom-right panel, halo spin $\lambda$ is positively correlated with 
\ahalf, meaning that haloes of lower spin are generally older. Unlike the previous cases,
now the gradient of the halo bias is not along the correlation direction but about 
perpendicular to it, and the correlation can no longer be 
used to predict the bias dependence trend. That is, the fact that older haloes are 
more strongly clustered and have a lower spin does not imply that haloes of lower 
spin are more strongly clustered. In this example, the contrary is true (haloes of 
lower spin are less clustered as seen in the panel and in Fig.\ref{fig:joint_dep}).

The difference between the halo bias gradient and the correlation direction with two 
assembly variables can help resolve apparent puzzles in assembly bias trend with other 
variables, e.g. the dependence on halo age and the number of subhaloes (above some mass 
threshold). The occupation number of subhaloes is found to correlate with halo age
in the sense that older haloes have fewer subhaloes \citep[e.g.][]{Gao04}, resulting from
dynamical evolution and destruction of subhaloes. As in the mass range considered here
older haloes are more strongly clustered, the correlation would imply haloes of fewer
subhaloes are more strongly clustered. However, the dependence on subhalo occupation
number is found to be opposite to such a naive expectation \citep[e.g.][]{Croft12}.

Fig.~\ref{fig:bon2_mb2} shows halo bias as a function of two assembly variables for 
massive haloes ($\log[\Mh/(\hinvMsun)]=13.5\pm 0.5$, above $\Mnl$). Compared to the
case with low mass haloes (Fig.~\ref{fig:bon2_mb1}), while the correlation between
each two assembly variables does not change much, the direction of the gradient of 
halo bias can be substantially different. For example, the gradient direction in the
$c$--\ahalf\ panel now becomes perpendicular to the correlation direction. As a 
consequence, the trend in assembly bias effect and its relation to the correlation 
also changes. For cases involving halo spin, the gradient direction has only a mild 
change, indicating that the origin of spin bias is different from others 
\citep[e.g.][]{Salcedo18}.

With the Separate Universe technique, \citet{Lazeyras17} present the dependence of halo bias on two assembly variables in two halo mass bins, although the dependence is not compared with the correlation of each two variables. They show that the dependence changes with halo mass and the trend is weak if halo shape is used as one variable. \citet{Han18} present analysis of the multidimensional dependence of bias on halo properties. Their results, if projected onto the space of two assembly variables at fixed halo mass, can be compared to ours.

Overall we see that the joint dependence of halo bias on two assembly variables does not necessarily follow the correlation between the variables. One should be cautious in inferring assembly bias trend in one variable based on its correlation to the other variable. A similar conclusion is reached by \citet{Mao18} for the so-called ``secondary bias'' with cluster-size haloes. The pattern of the joint dependence varies with halo mass, which can be characterised by a rotation (e.g. in terms of the halo bias gradient) as halo mass increases.

%\textbf{Overall we see that the joint dependence of halo bias on two assembly variables does not necessarily follow the correlation between the variables. The pattern of the joint dependence varies with halo mass, which can be characterised by a rotation (e.g. in terms of the halo bias gradient) as halo mass increases. One should be cautious in inferring assembly bias trend in one variable based on its correlation to the other variable. A similar conclusion is reached by \citet{Mao17} for the so-called ``secondary bias'' with cluster-size haloes.\citet{Lazeyras17} also performed same type of reasoning with other halo properties. On the other hand, similar conclusion is reached by \citet{Mao17} for the so-called ``secondary bias'' with cluster-size haloes. \citet{Han18} also present analysis of joint dependence of bias on two or more halo properties. They define a microscopic halo bias for each halo, with machine learning method, they predict bias estimator as function of one halo property with estimator constructed including other properties, non-redundancy are found for some properties.}  

%4
\section{Dependence of Pairwise Velocity and Velocity Dispersions on Halo Assembly }
\label{sec:pairwise_vel}

\begin{figure*}
    \centering
    \begin{subfigure}[h]{0.33\textwidth}
        \centering
        \includegraphics[width=\textwidth]{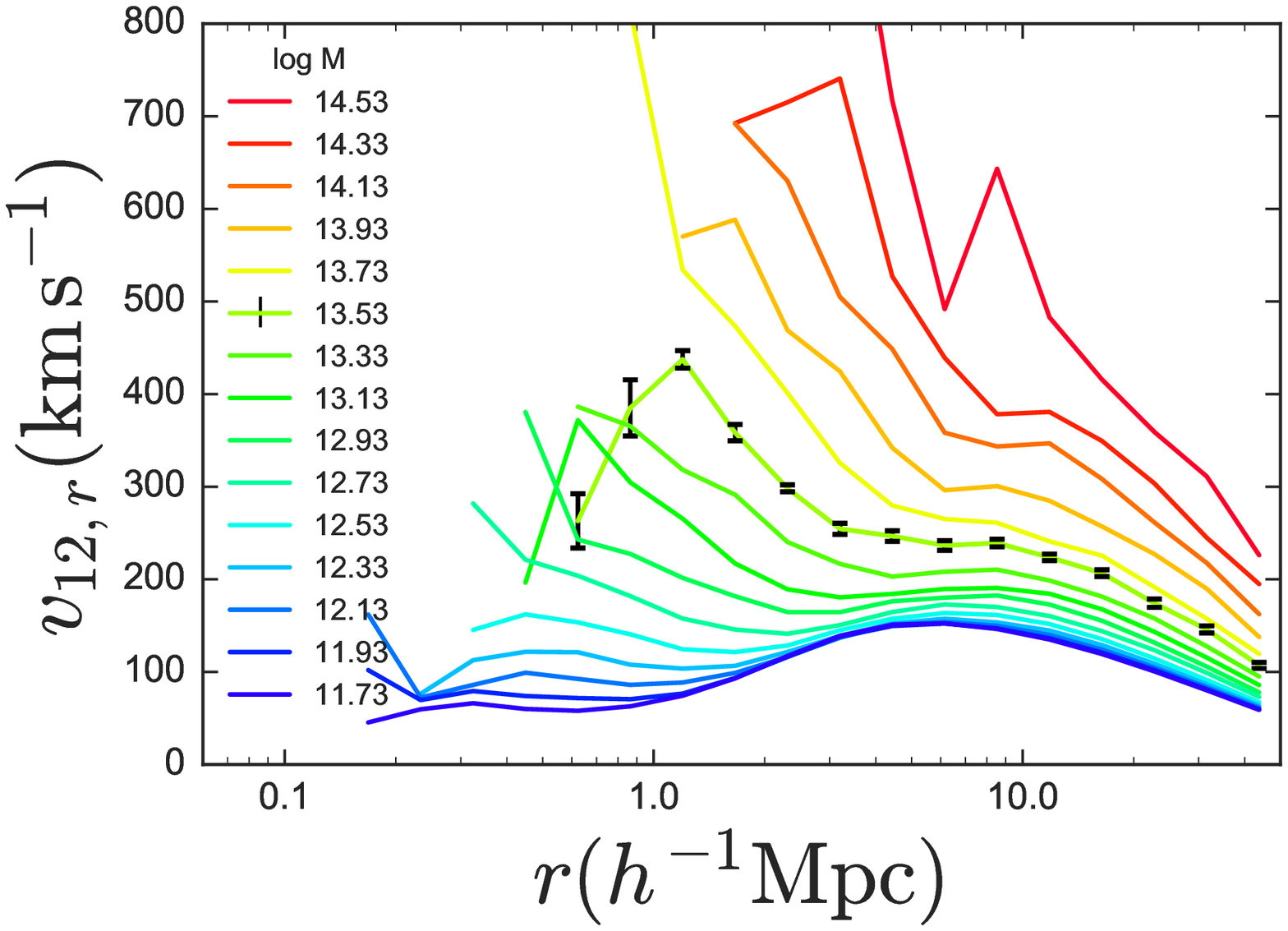}
    \end{subfigure}
    \hfill
    \begin{subfigure}[h]{0.33\textwidth}
        \centering
        \includegraphics[width=\textwidth]{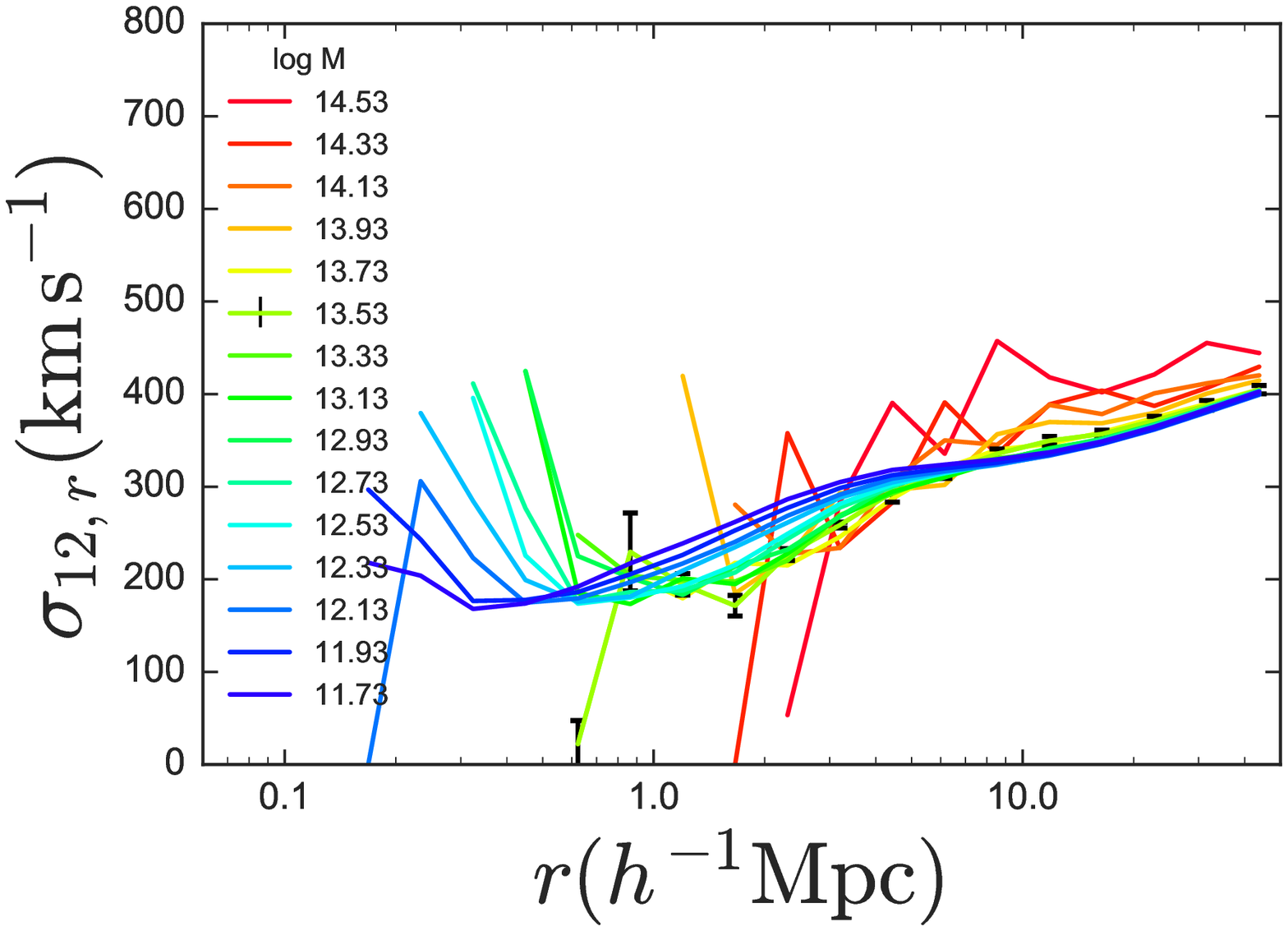}
    \end{subfigure}
    \hfill
    \begin{subfigure}[h]{0.33\textwidth}
        \centering
        \includegraphics[width=\textwidth]{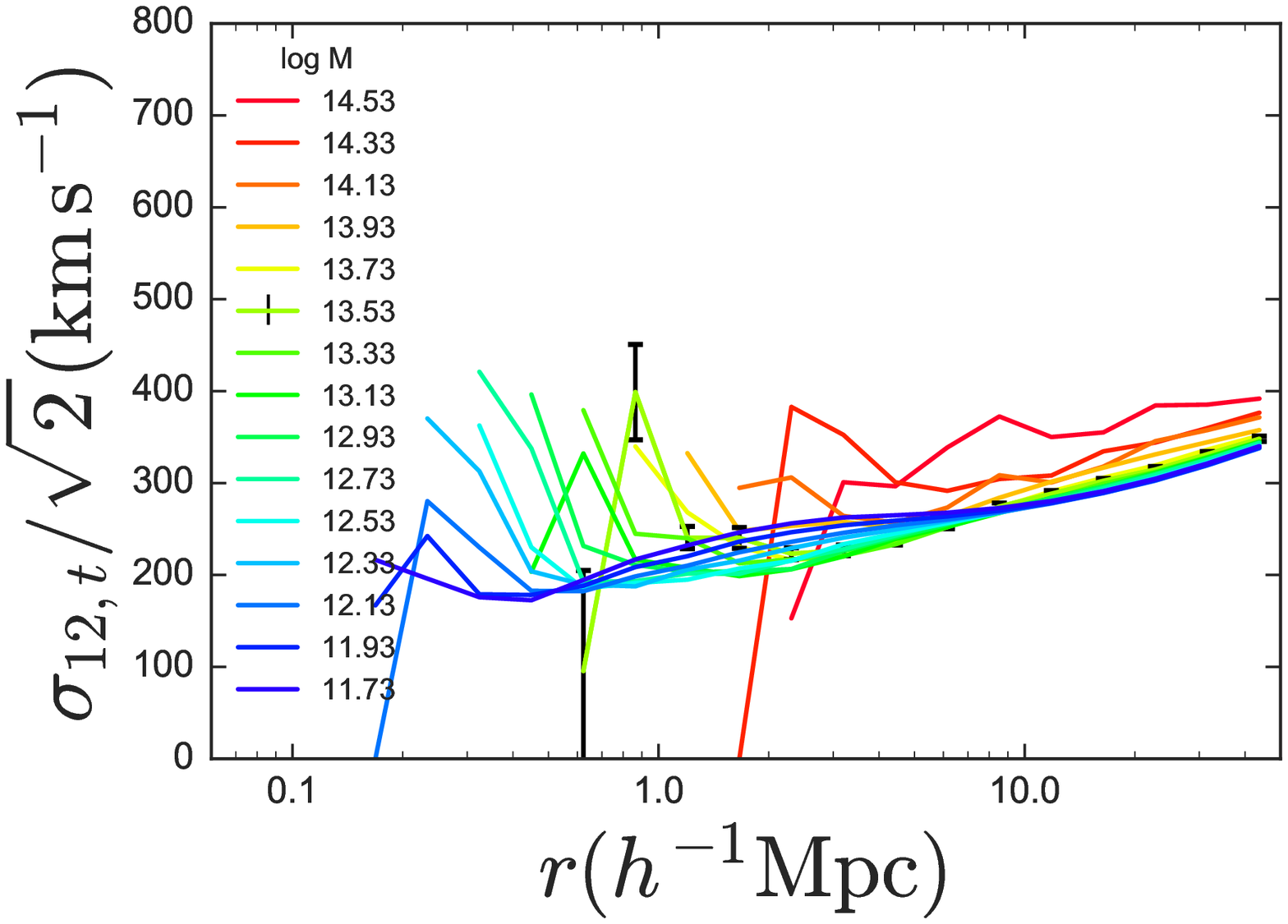}
    \end{subfigure}
\caption{
Pairwise velocity and velocity dispersions of haloes. The left panel shows
the scale-dependent pairwise radial velocity as a function of halo mass. The
middle and right panels are similar, but for pairwise radial and transverse
velocity dispersions, respectively. 
}
\label{fig:vel_mass}
\end{figure*}

\begin{figure*}
    \centering
    \begin{subfigure}[h]{0.24\textwidth}
        \centering
        \includegraphics[width=\textwidth]{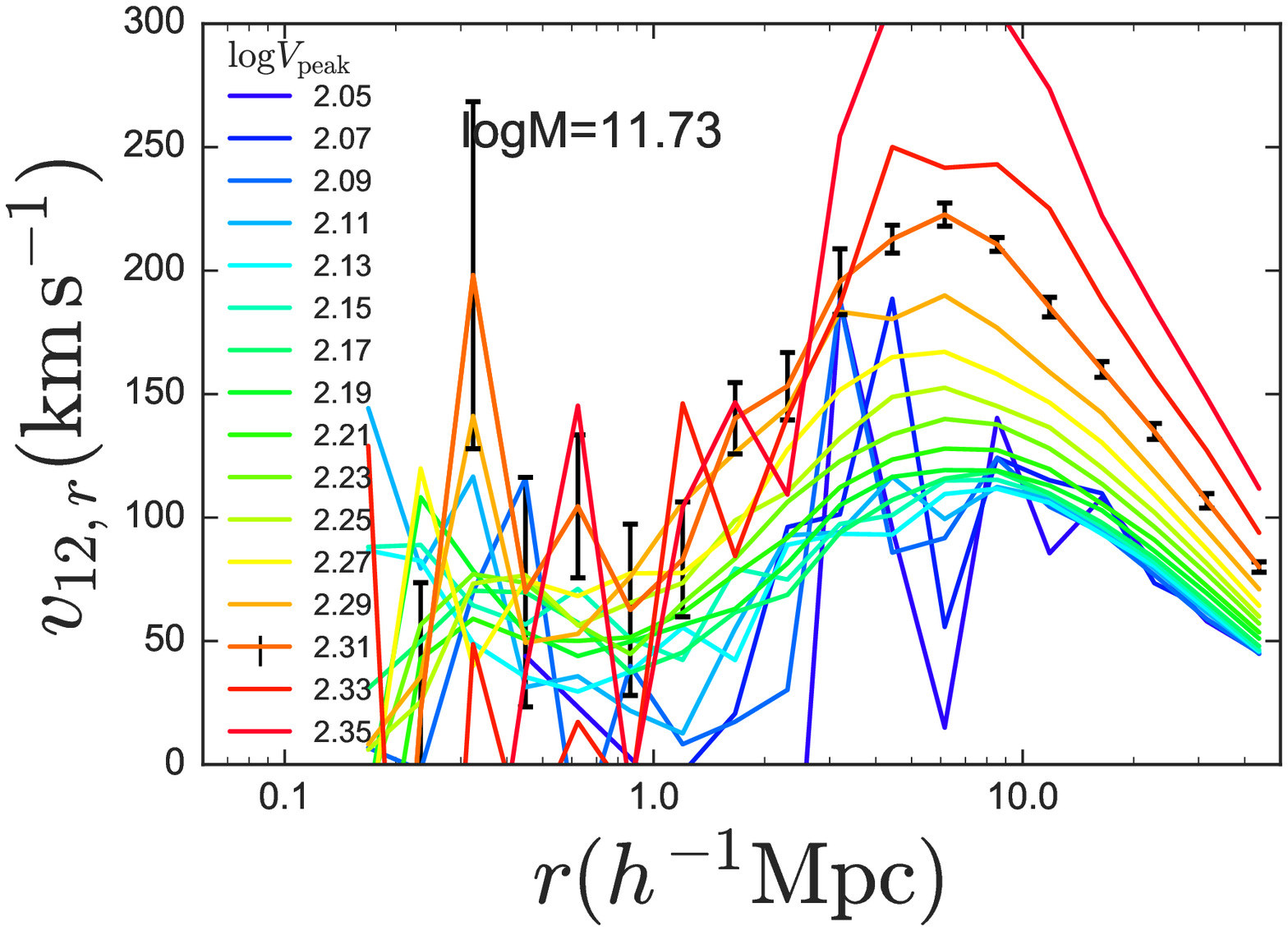}
    \end{subfigure}
    \hfill
    \begin{subfigure}[h]{0.24\textwidth}
        \centering
        \includegraphics[width=\textwidth]{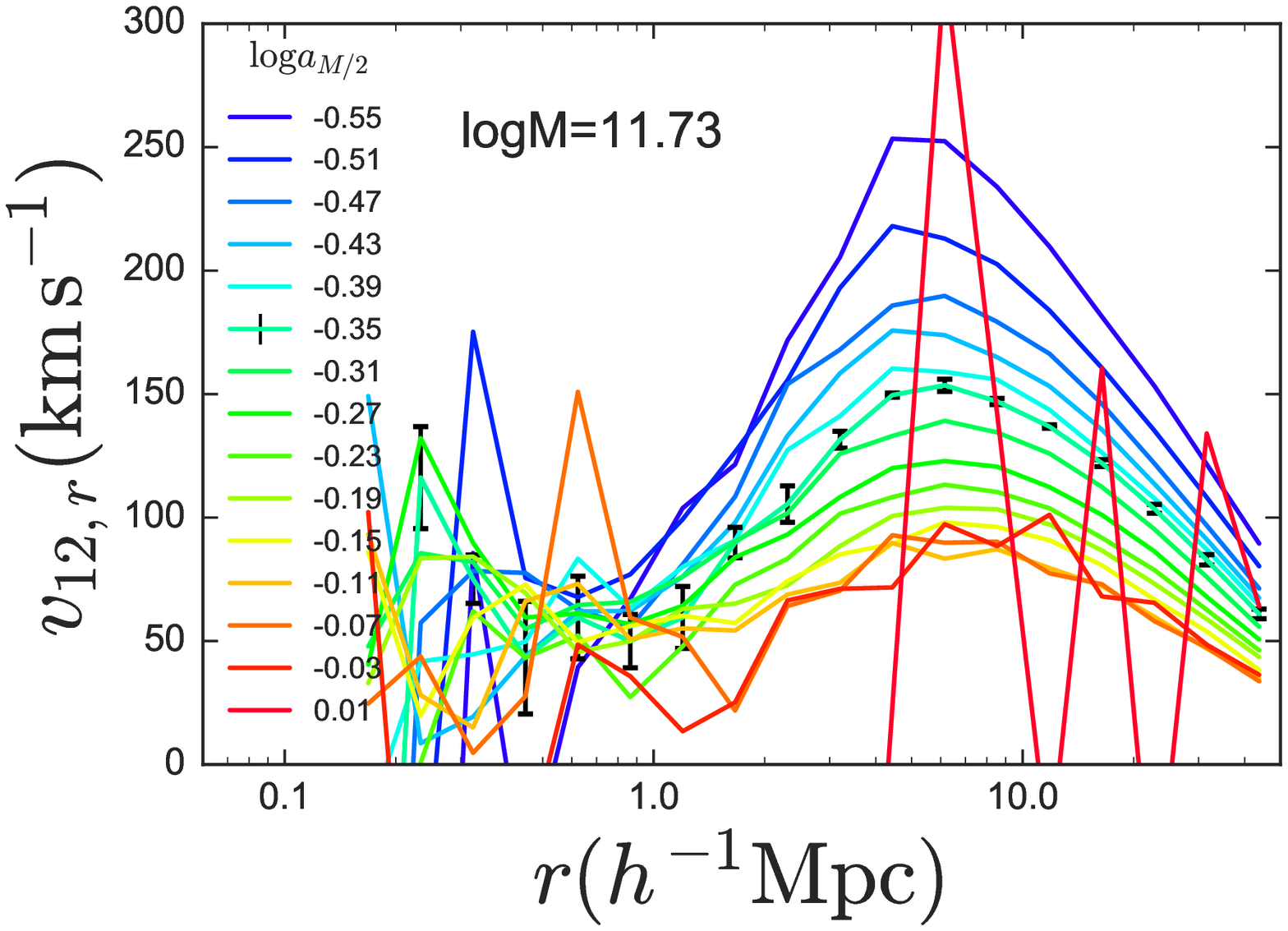}
    \end{subfigure}
	\hfill
    \begin{subfigure}[h]{0.24\textwidth}
        \centering
        \includegraphics[width=\textwidth]{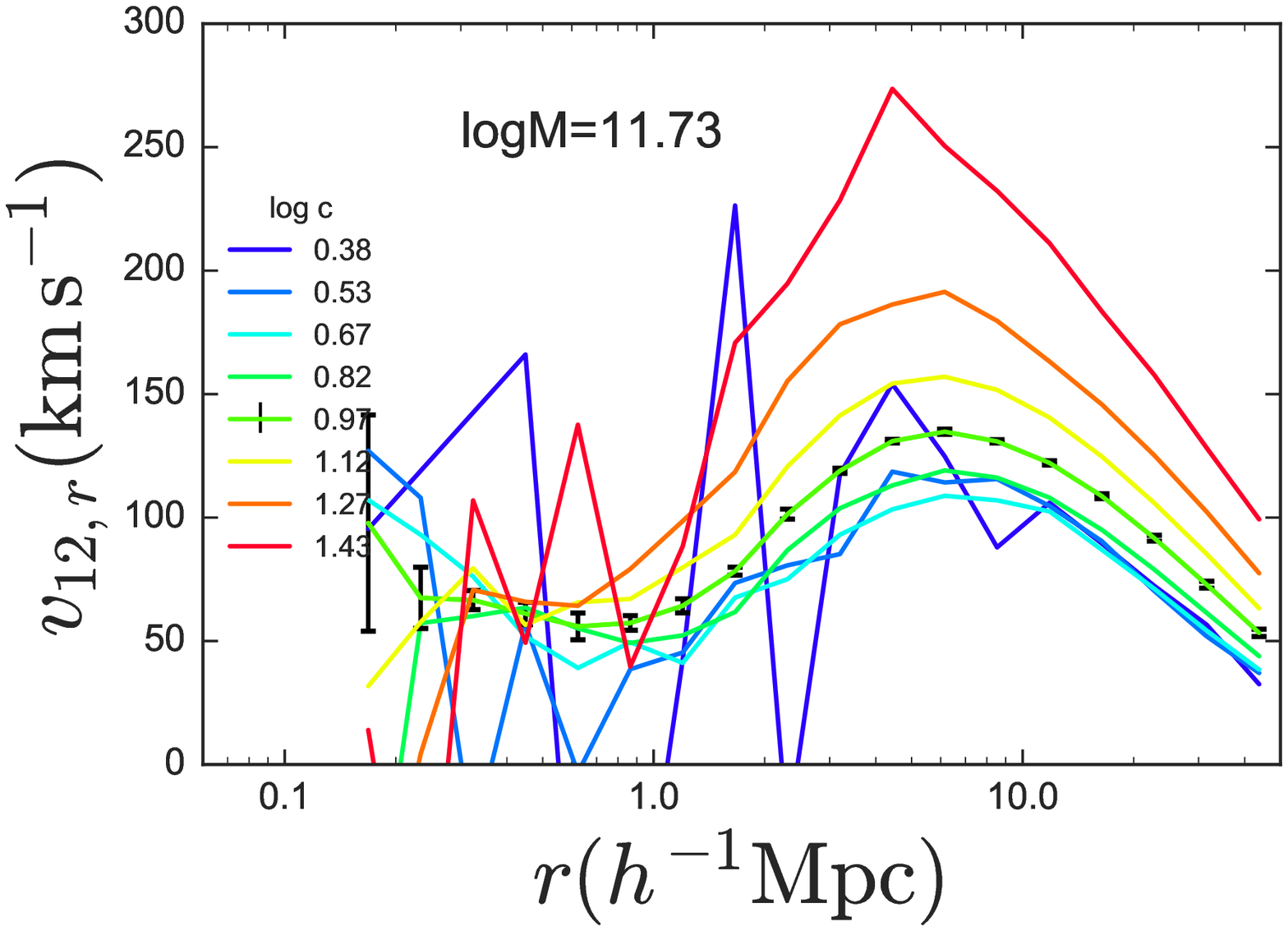}
    \end{subfigure}
	\hfill
    \begin{subfigure}[h]{0.24\textwidth}
        \centering
        \includegraphics[width=\textwidth]{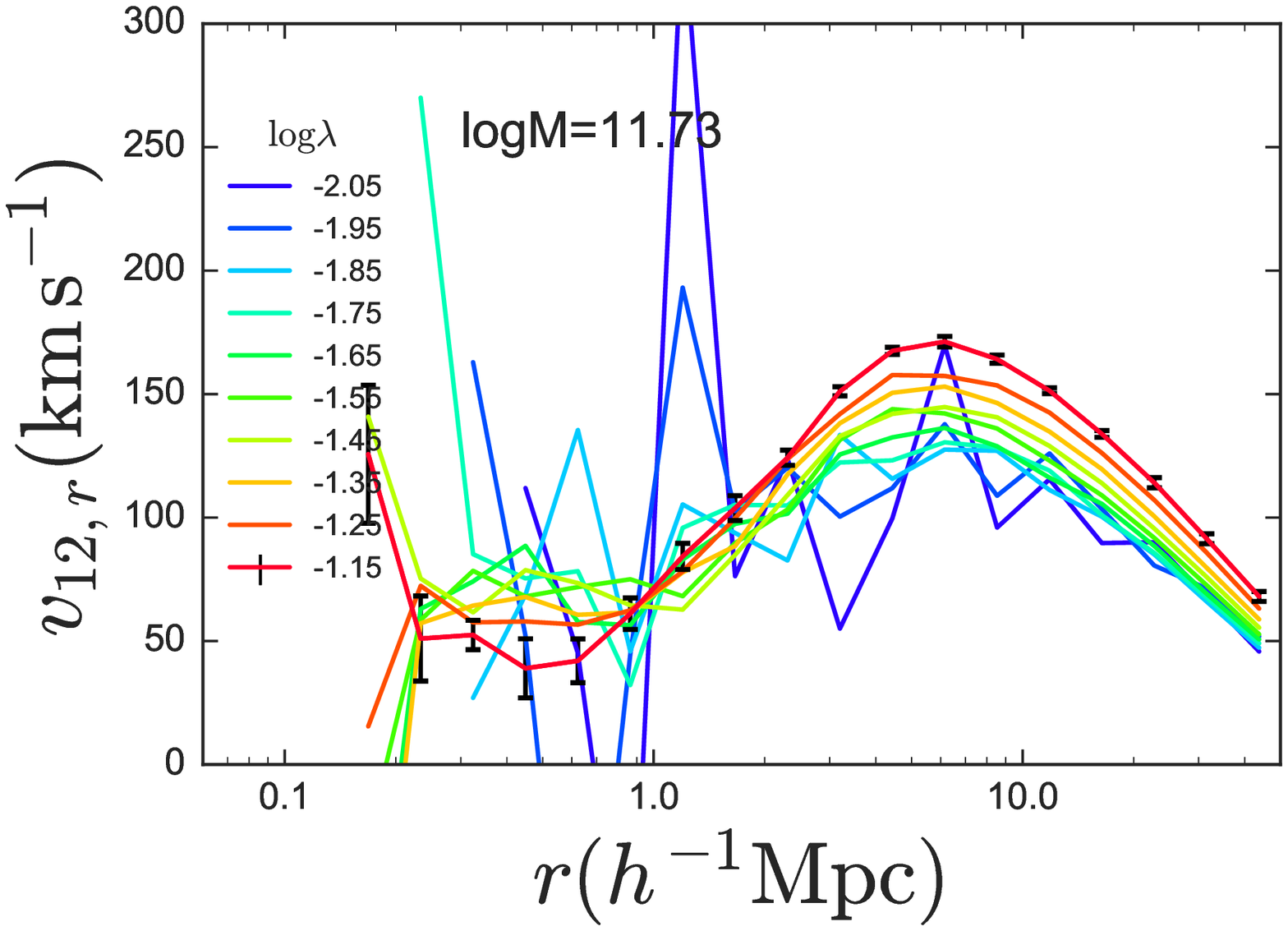}
    \end{subfigure}
	\hfill
    \begin{subfigure}[h]{0.24\textwidth}
        \centering
        \includegraphics[width=\textwidth]{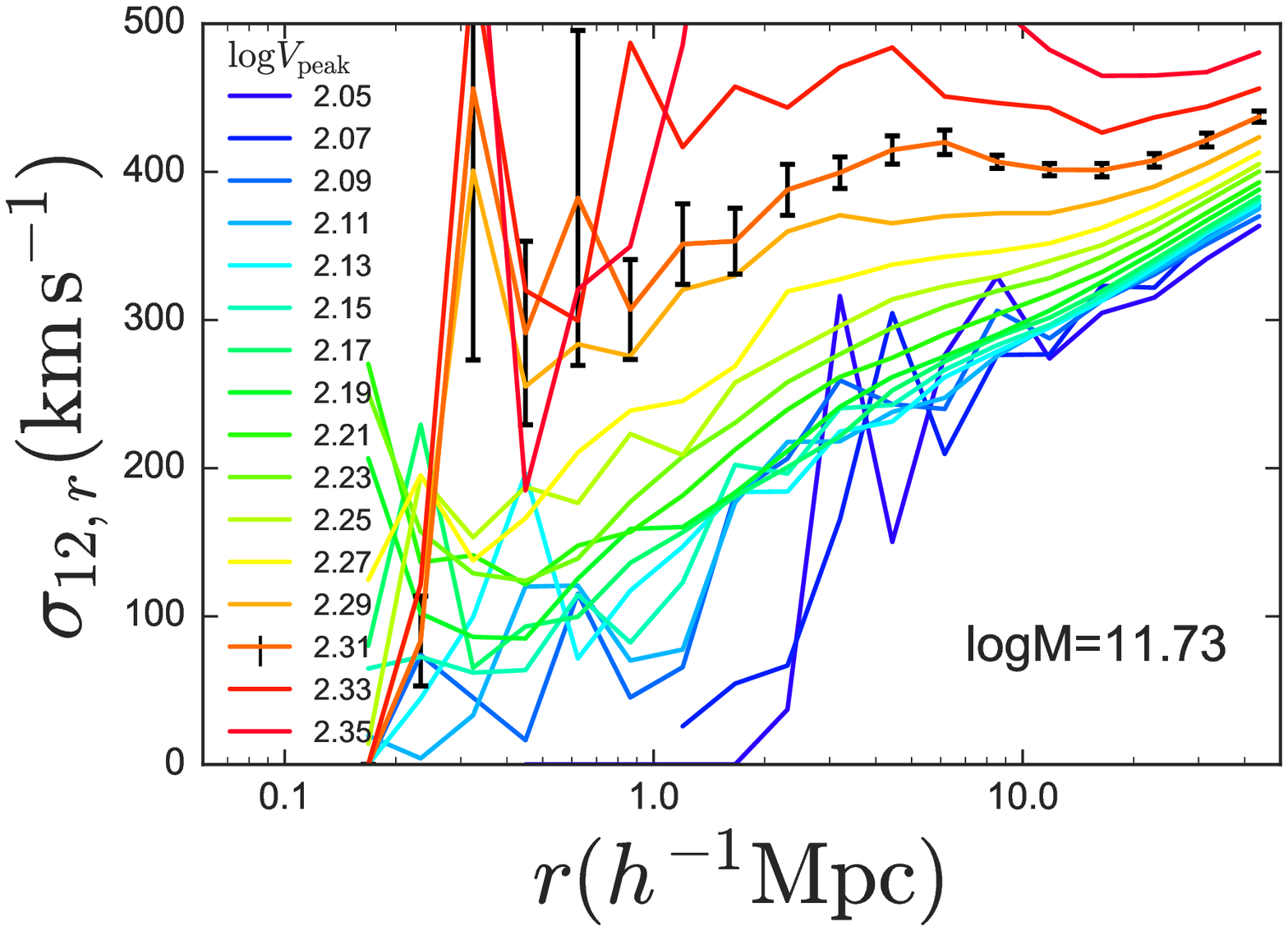}
    \end{subfigure}
	\hfill
    \begin{subfigure}[h]{0.24\textwidth}
        \centering
        \includegraphics[width=\textwidth]{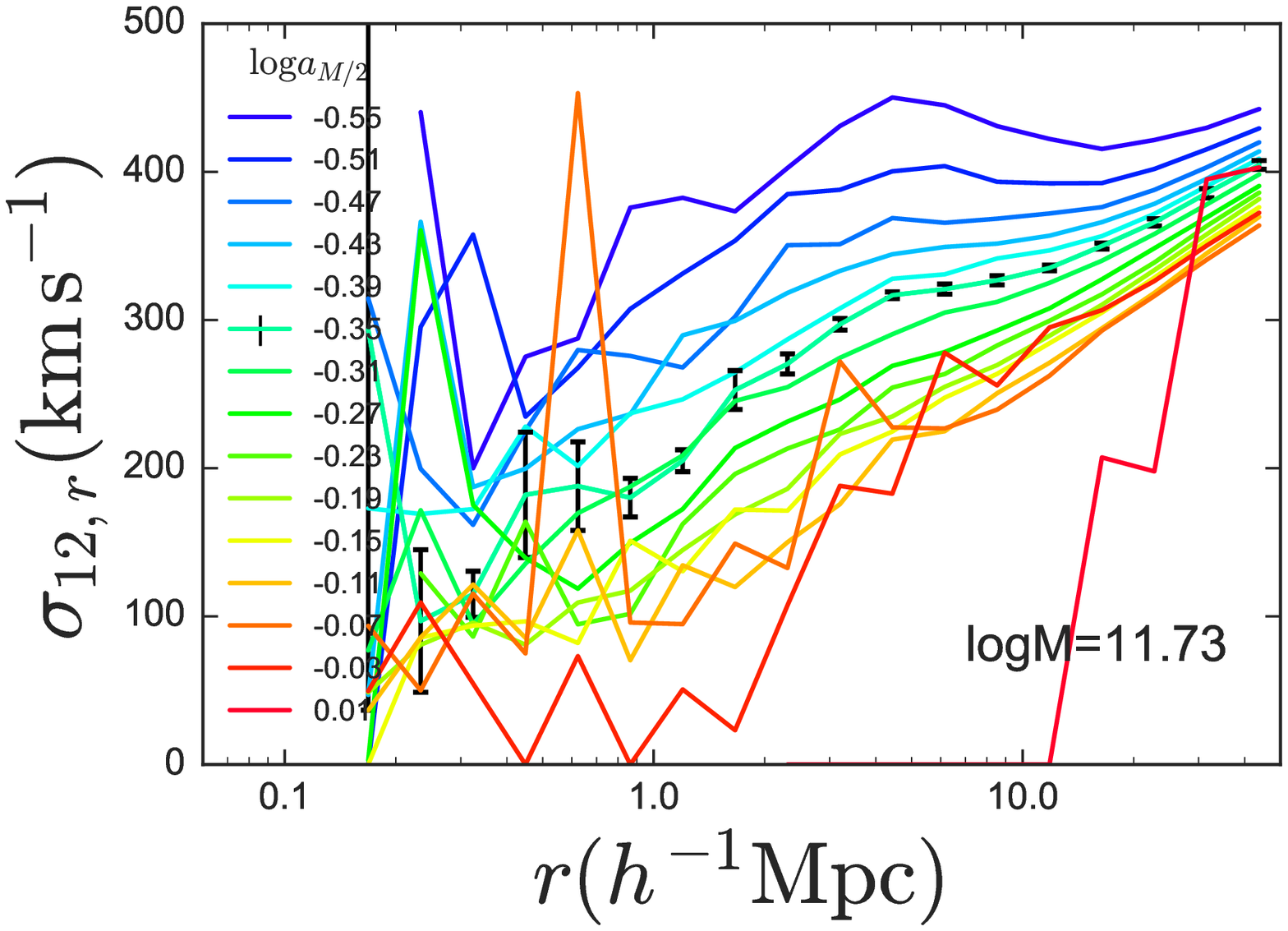}
    \end{subfigure}
	\hfill
    \begin{subfigure}[h]{0.24\textwidth}
        \centering
        \includegraphics[width=\textwidth]{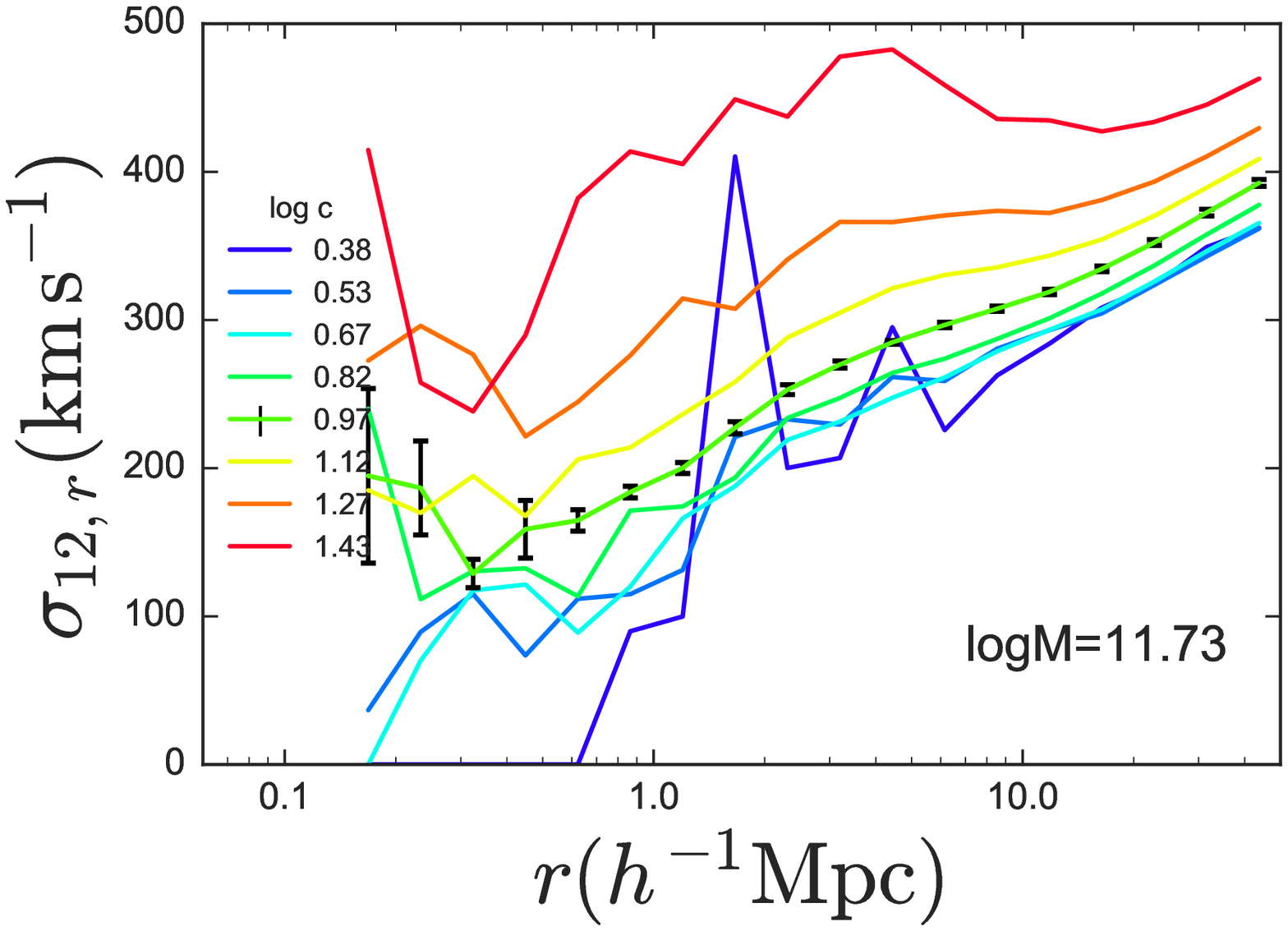}
    \end{subfigure}
	\hfill
    \begin{subfigure}[h]{0.24\textwidth}
        \centering
        \includegraphics[width=\textwidth]{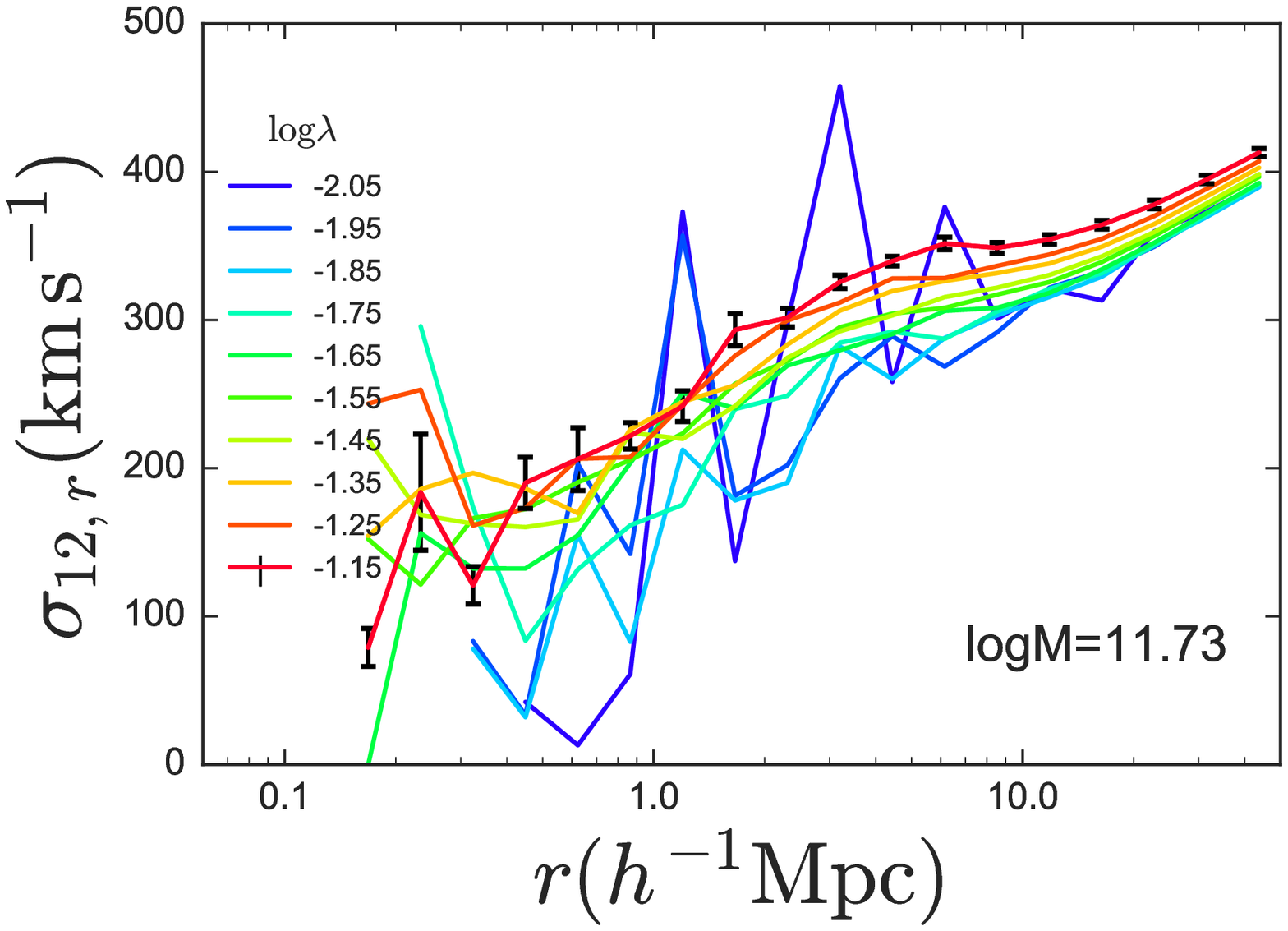}
    \end{subfigure}
    \begin{subfigure}[h]{0.24\textwidth}
        \centering
        \includegraphics[width=\textwidth]{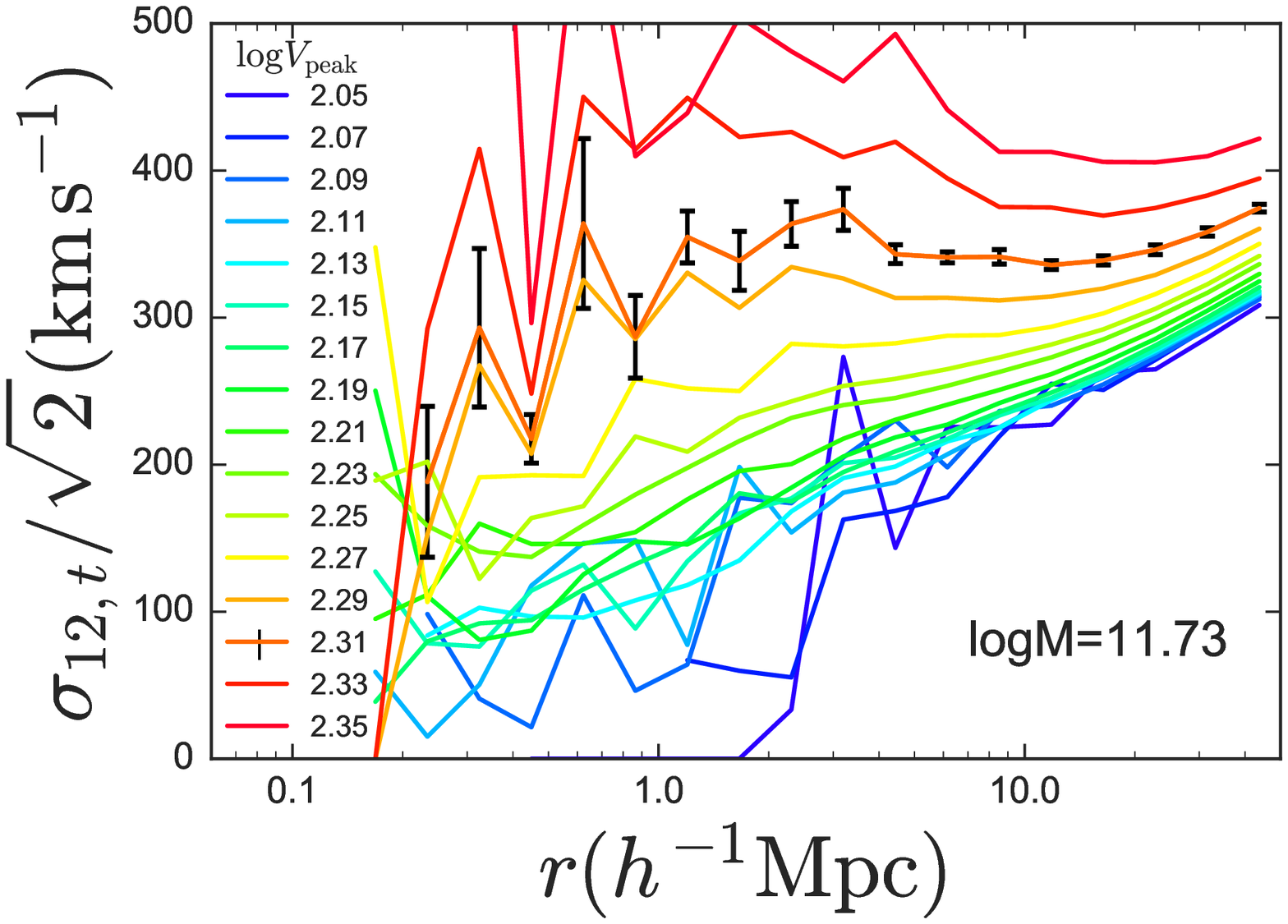}
    \end{subfigure}
    \hfill
    \begin{subfigure}[h]{0.24\textwidth}
        \centering
        \includegraphics[width=\textwidth]{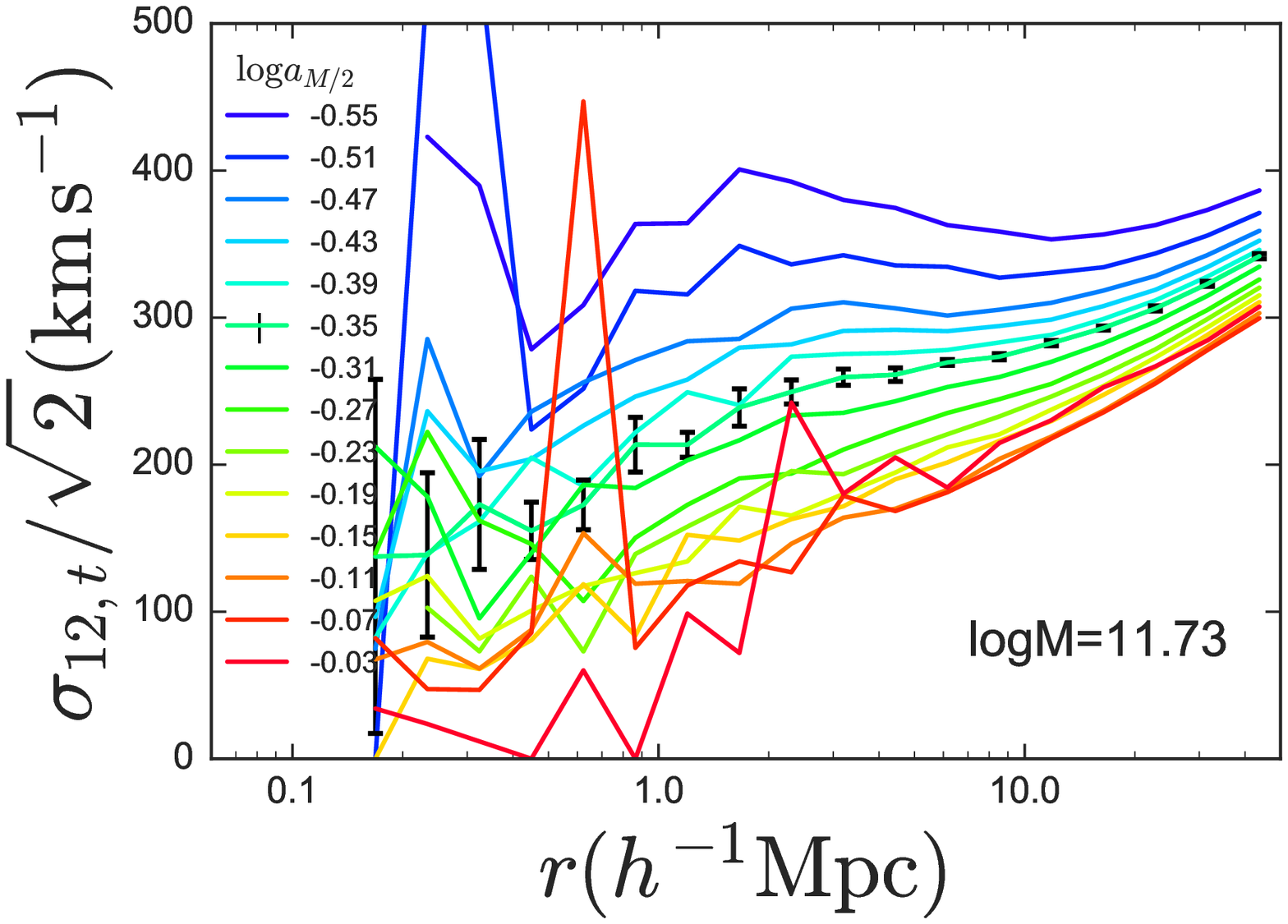}
    \end{subfigure}
	\hfill
    \begin{subfigure}[h]{0.24\textwidth}
        \centering
        \includegraphics[width=\textwidth]{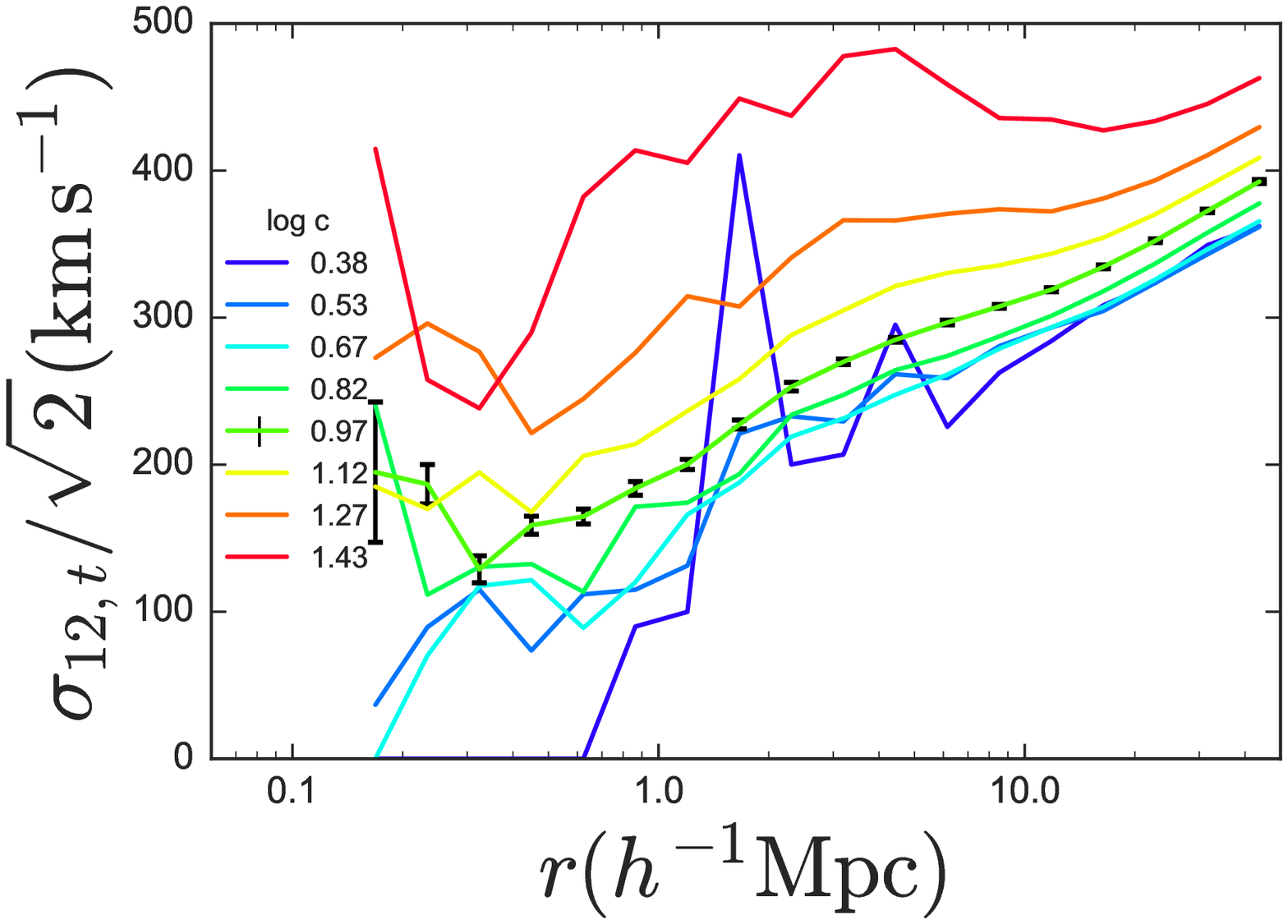}
    \end{subfigure}
	\hfill
    \begin{subfigure}[h]{0.24\textwidth}
        \centering
        \includegraphics[width=\textwidth]{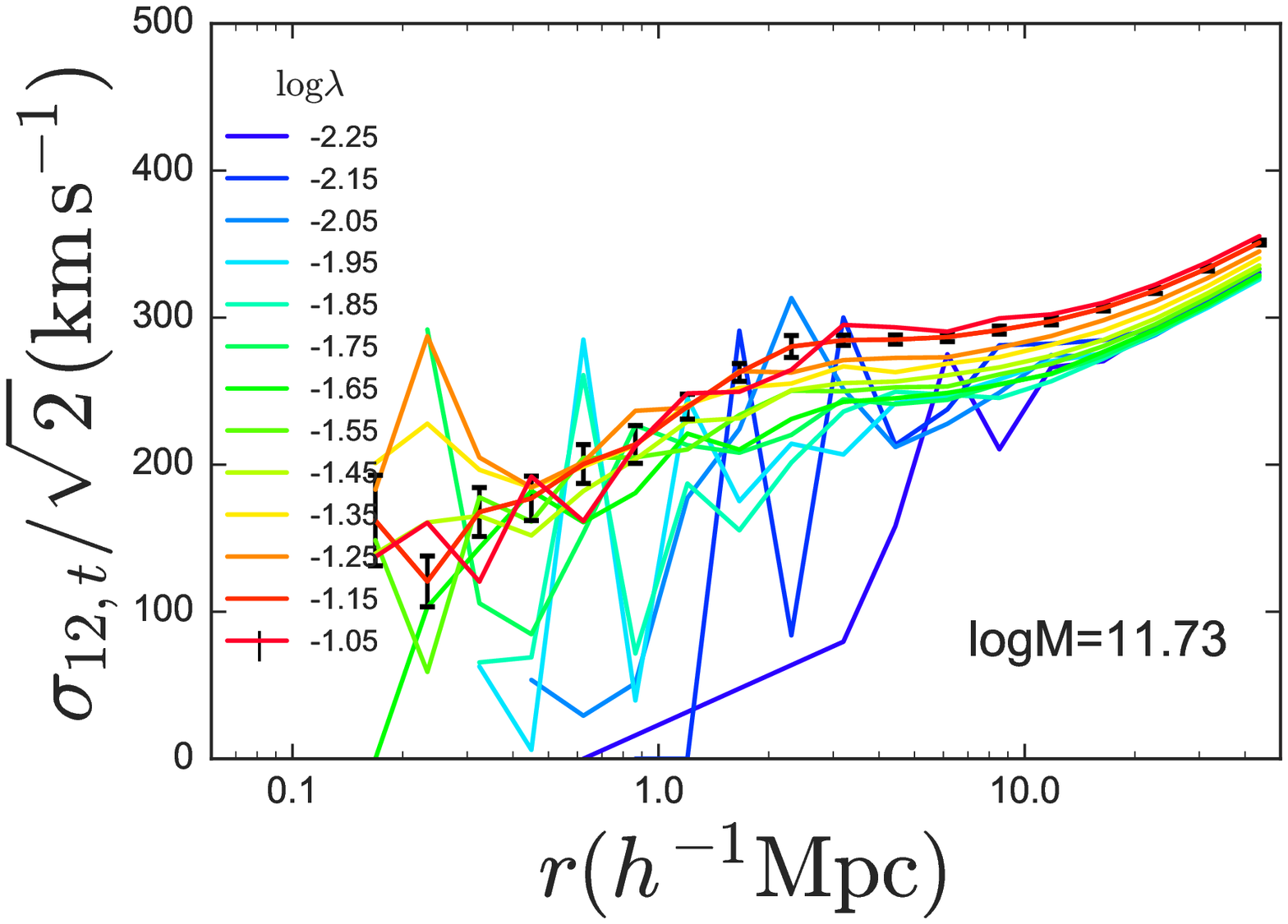}
    \end{subfigure}
\caption{
Dependence of halo pairwise velocity and velocity dispersions on 
assembly variables for low-mass haloes ($\log[\Mh/(\hinvMsun)]=11.73$). 
Top, middle, and bottom panels show pairwise radial 
velocity, pairwise radial velocity dispersion, and pairwise transverse 
velocity dispersion, respectively. In each panel, the curves are colour 
coded according to the value of the corresponding assembly variable.
}
\label{fig:vel_assembly}
\end{figure*}

\begin{figure*}
    \centering
    \begin{subfigure}[h]{0.24\textwidth}
        \centering
        \includegraphics[width=\textwidth]{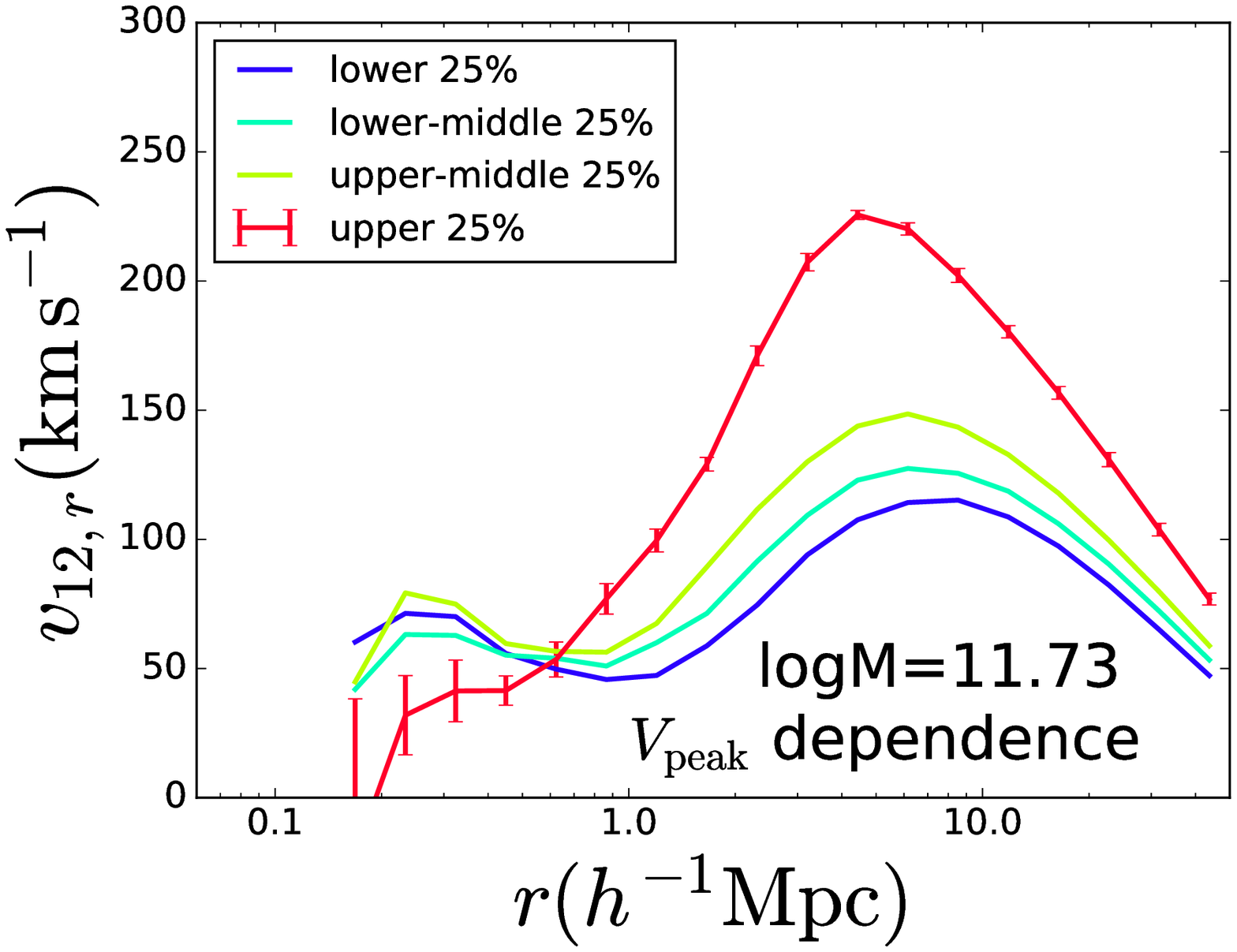}
    \end{subfigure}
    \hfill
    \begin{subfigure}[h]{0.24\textwidth}
        \centering
        \includegraphics[width=\textwidth]{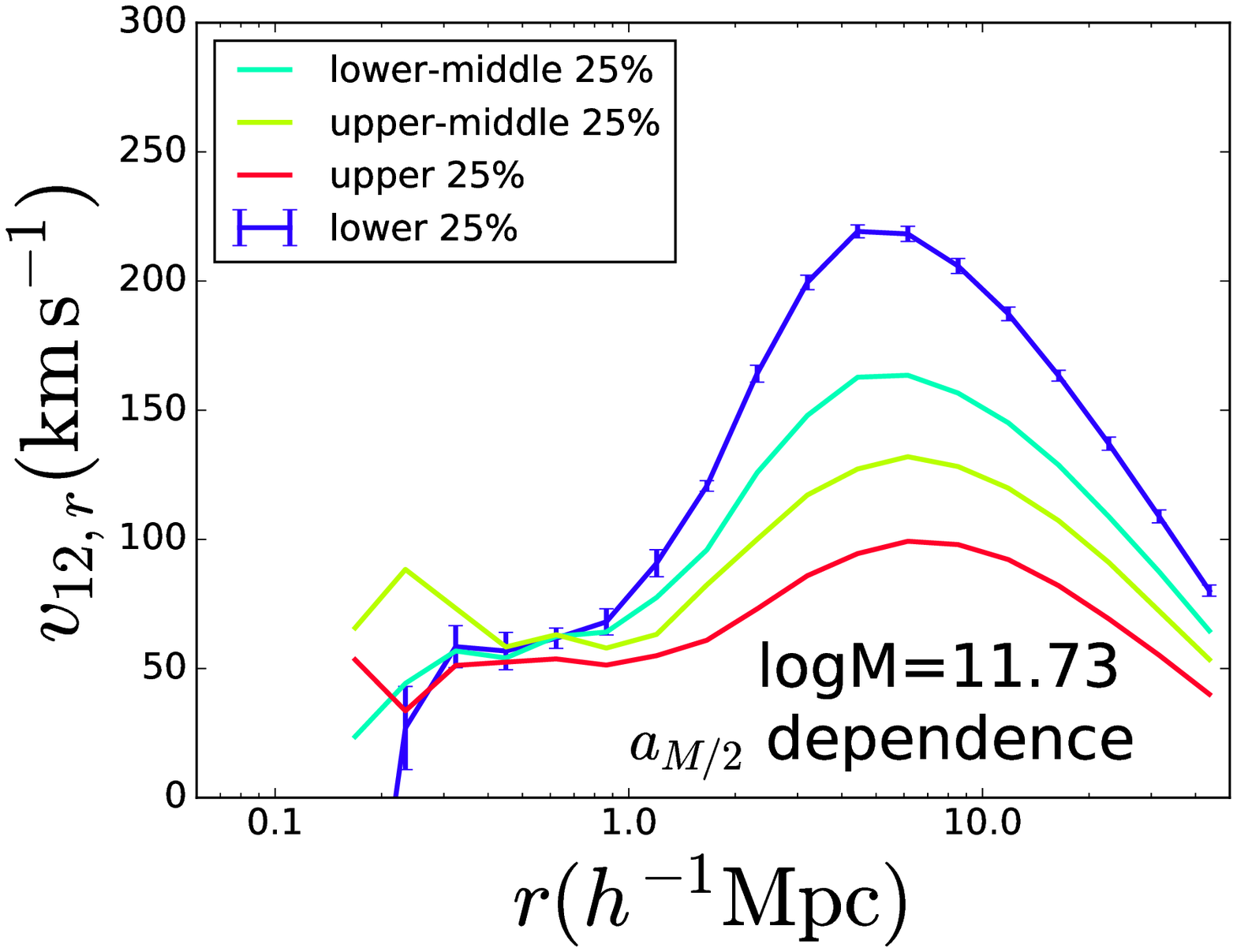}
    \end{subfigure}
	\hfill
    \begin{subfigure}[h]{0.24\textwidth}
        \centering
        \includegraphics[width=\textwidth]{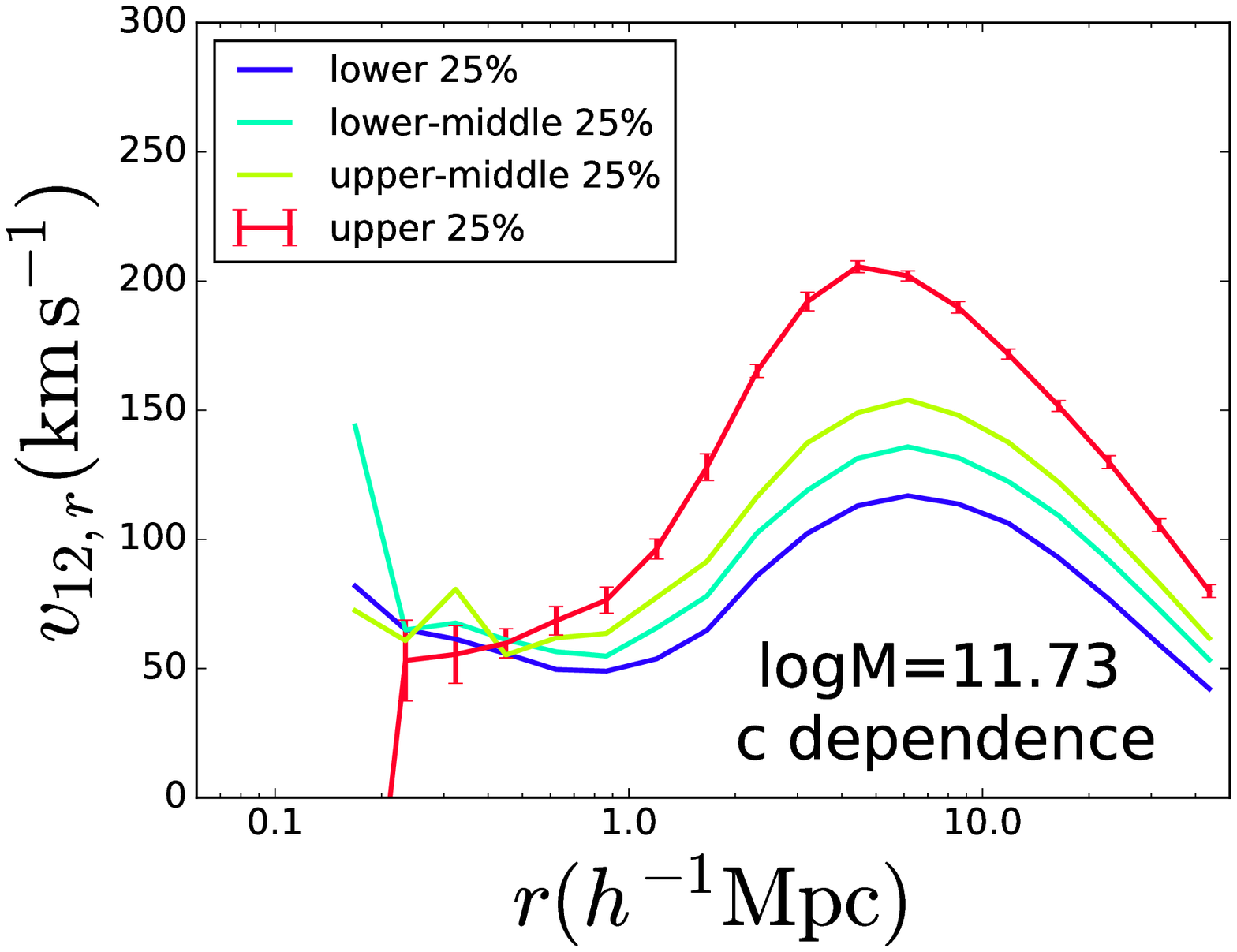}
    \end{subfigure}
	\hfill
    \begin{subfigure}[h]{0.24\textwidth}
        \centering
        \includegraphics[width=\textwidth]{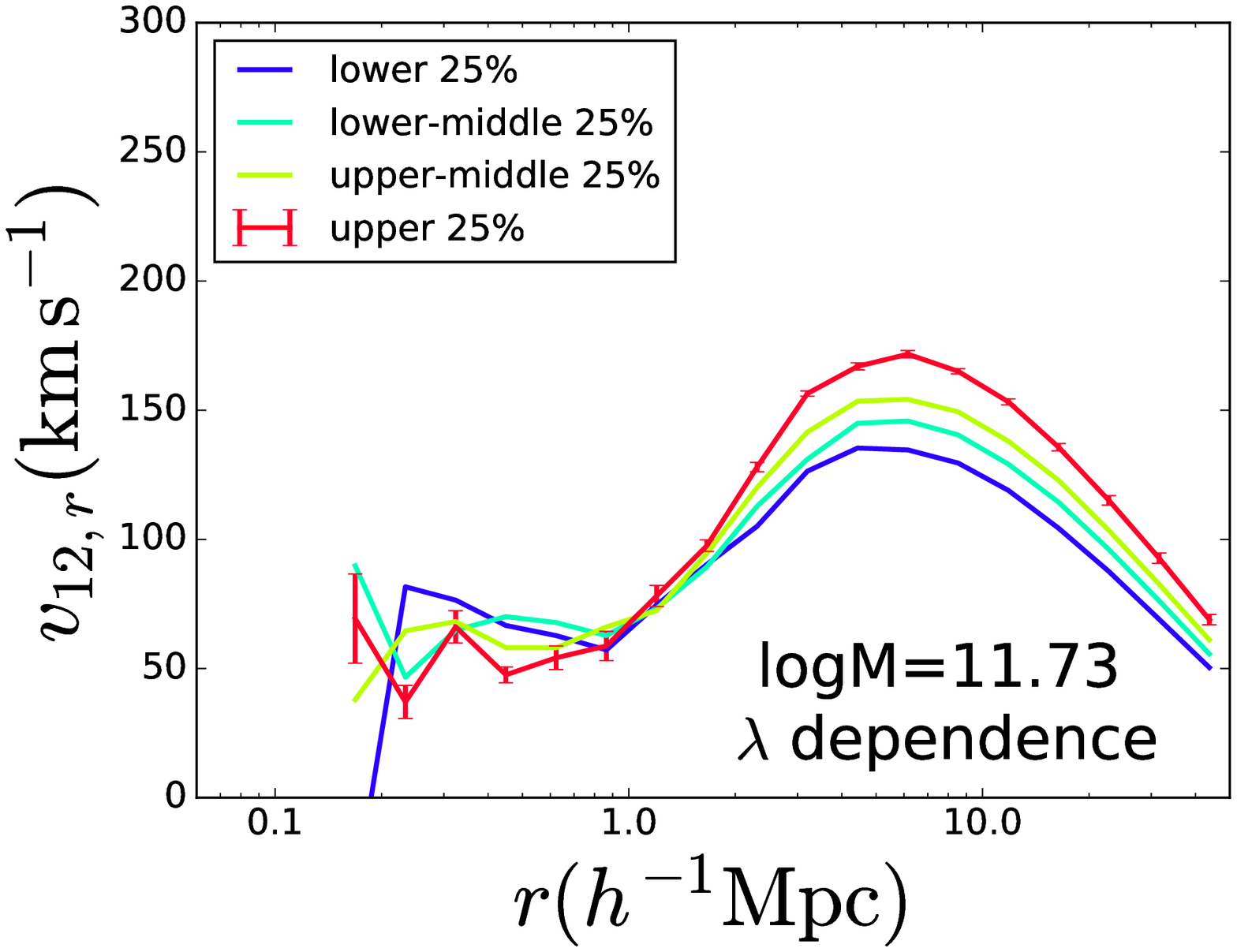}
    \end{subfigure}
	\hfill
    \begin{subfigure}[h]{0.24\textwidth}
        \centering
        \includegraphics[width=\textwidth]{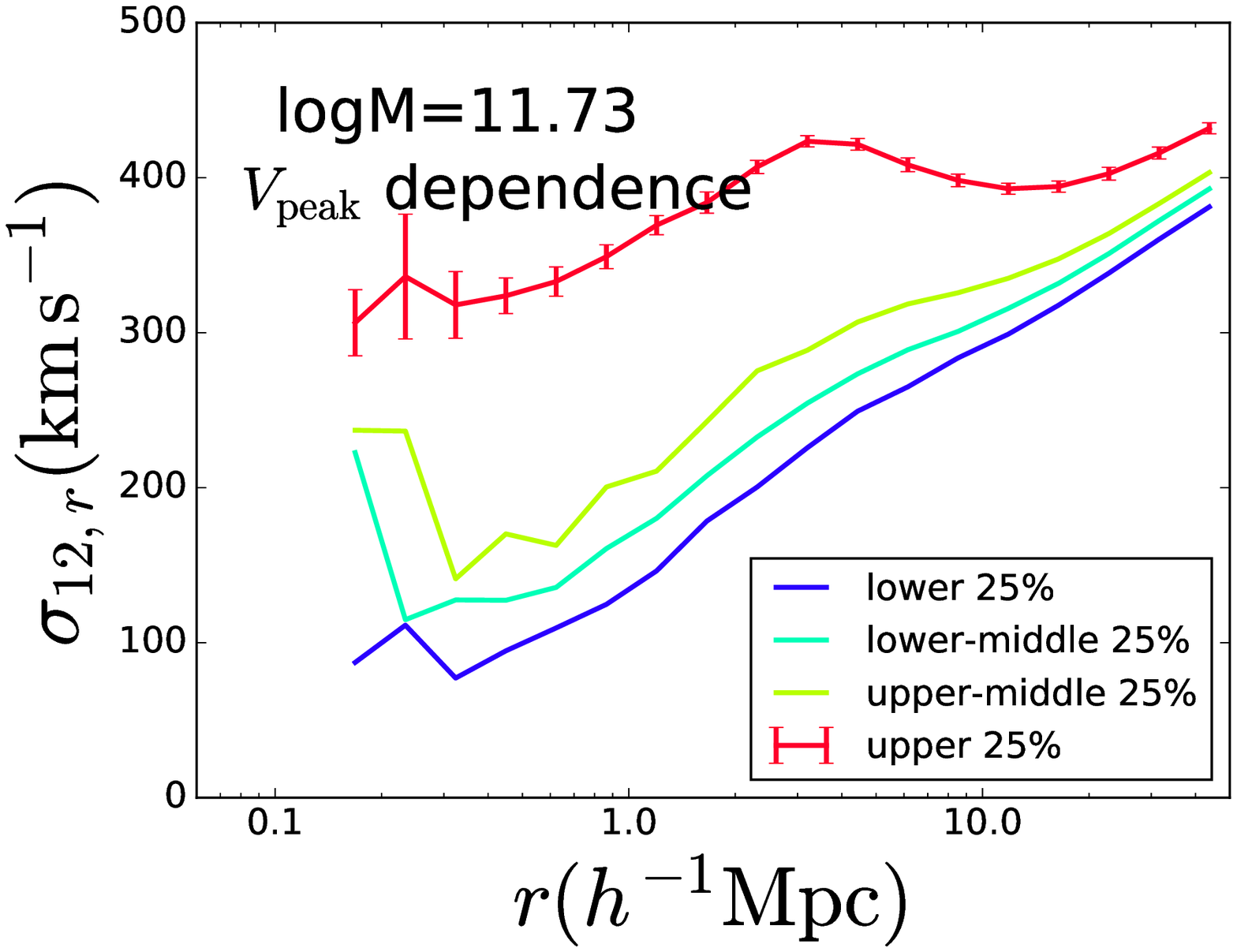}
    \end{subfigure}
	\hfill
    \begin{subfigure}[h]{0.24\textwidth}
        \centering
        \includegraphics[width=\textwidth]{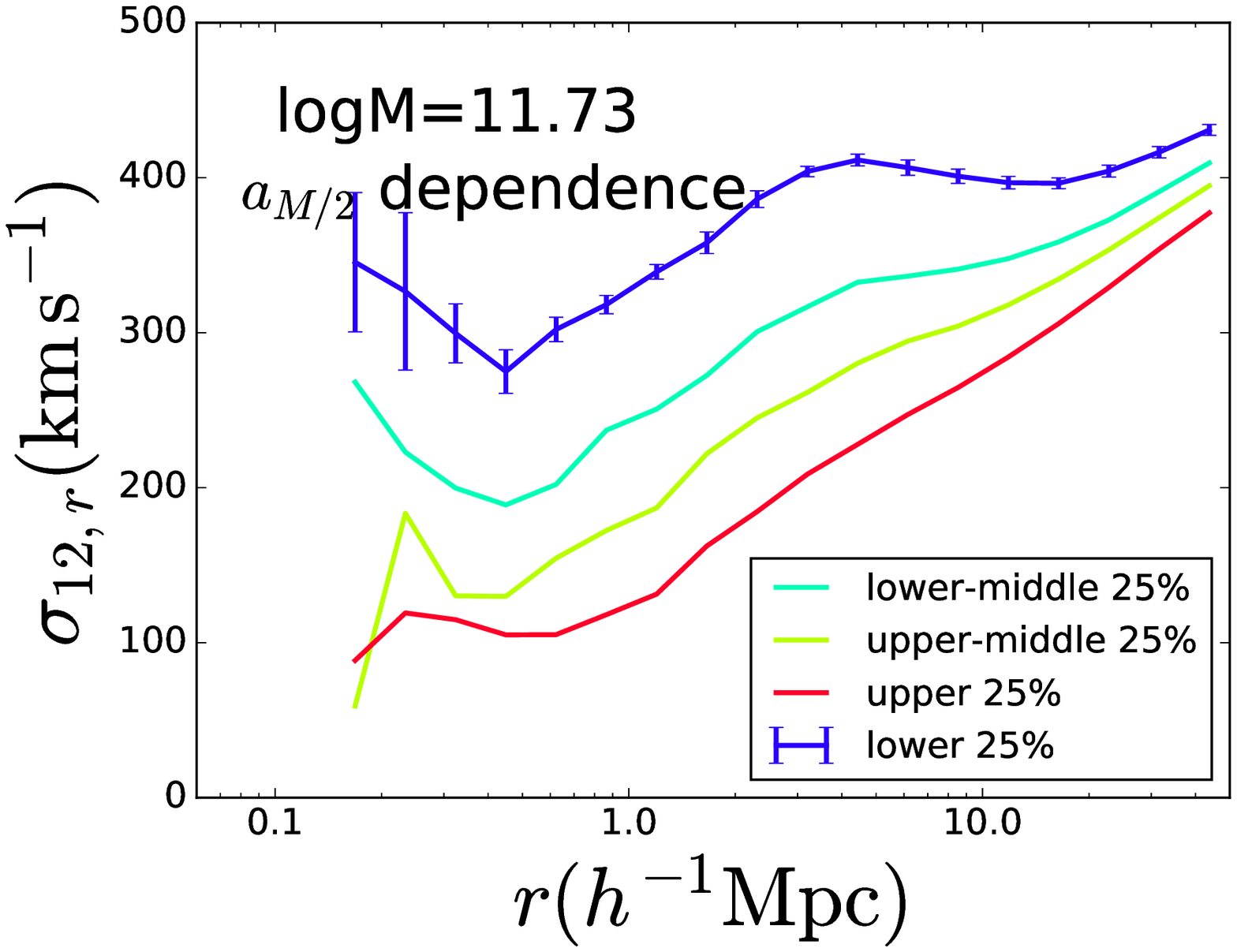}
    \end{subfigure}
	\hfill
    \begin{subfigure}[h]{0.24\textwidth}
        \centering
        \includegraphics[width=\textwidth]{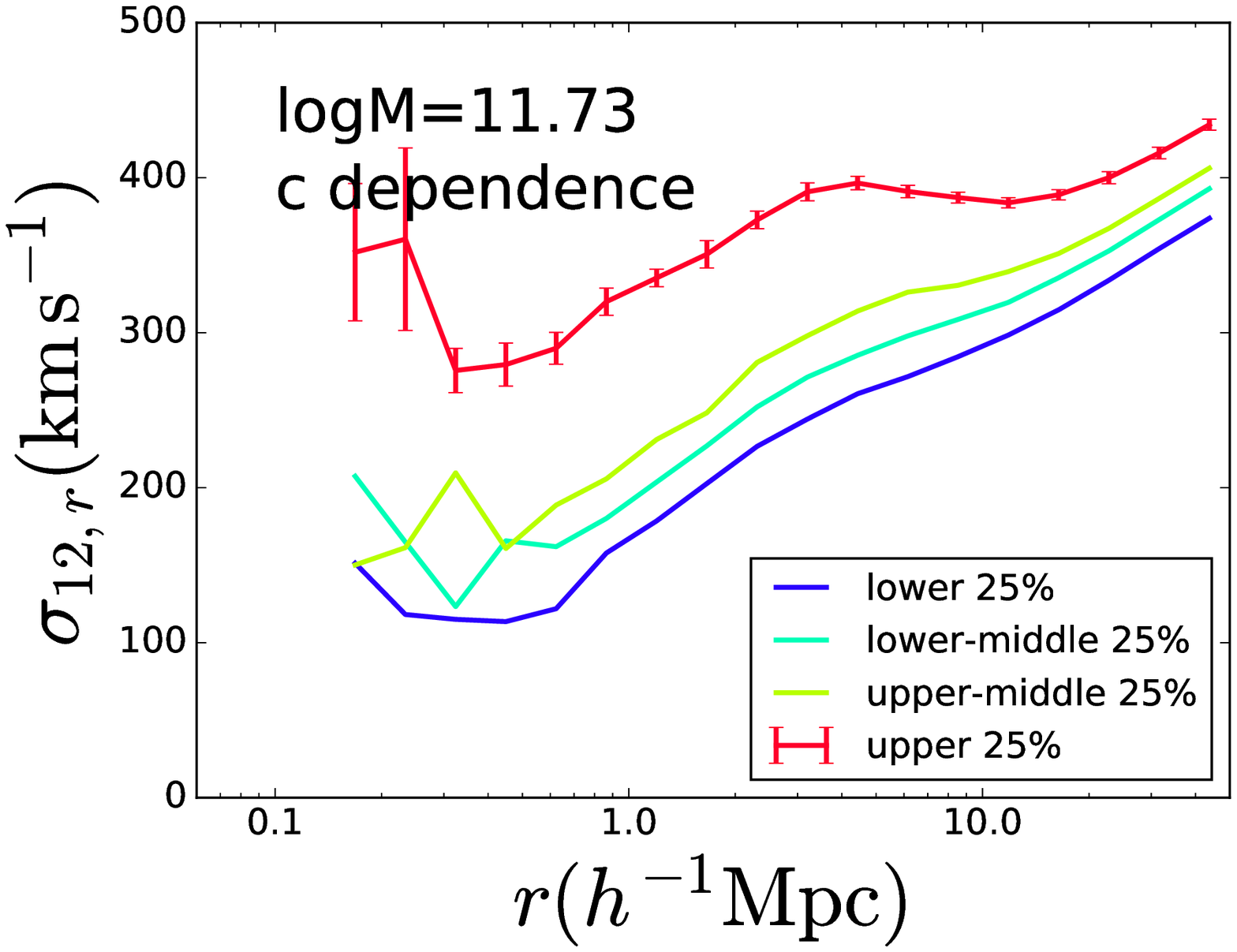}
    \end{subfigure}
	\hfill
    \begin{subfigure}[h]{0.24\textwidth}
        \centering
        \includegraphics[width=\textwidth]{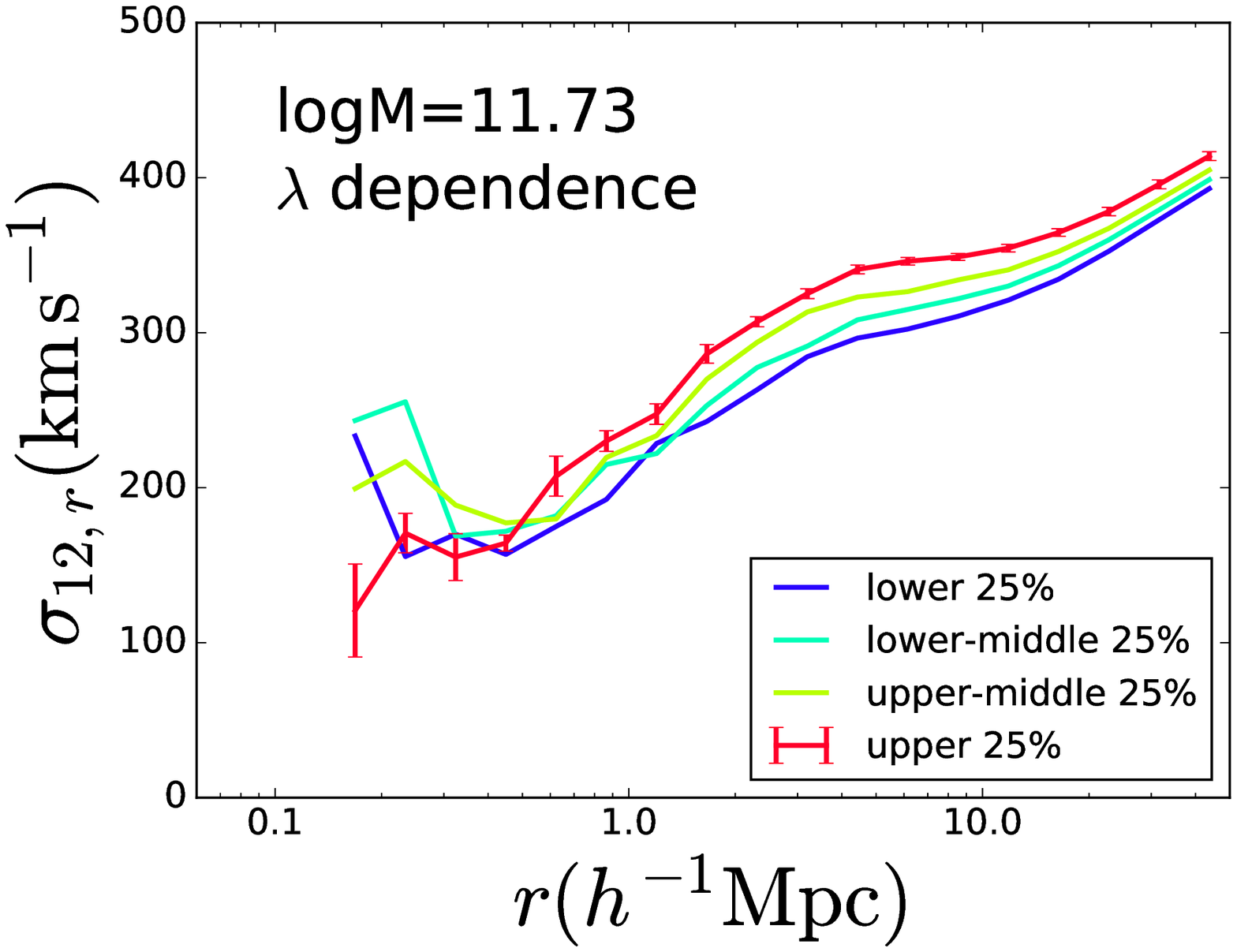}
    \end{subfigure}
    \begin{subfigure}[h]{0.24\textwidth}
        \centering
        \includegraphics[width=\textwidth]{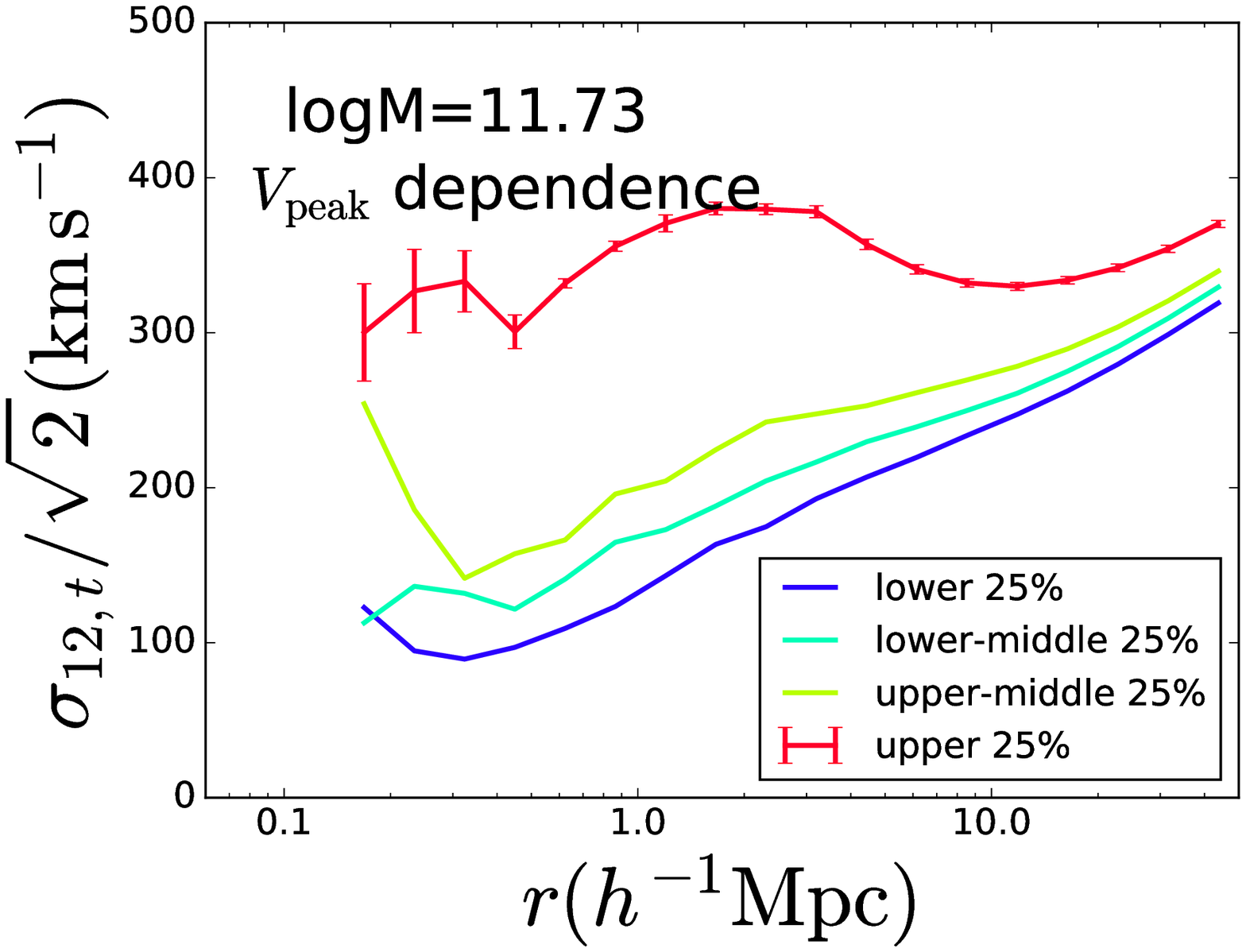}
    \end{subfigure}
    \hfill
    \begin{subfigure}[h]{0.24\textwidth}
        \centering
        \includegraphics[width=\textwidth]{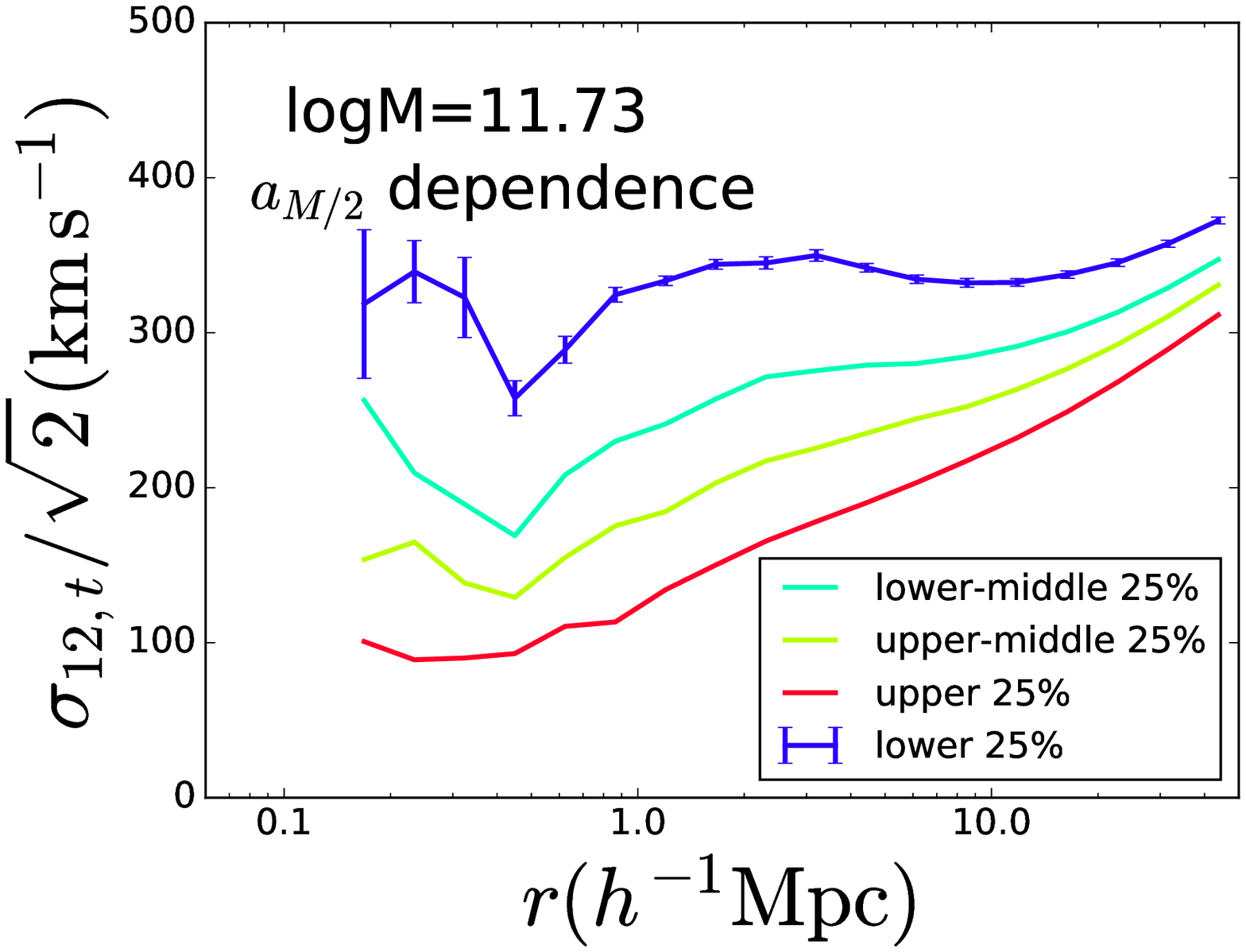}
    \end{subfigure}
	\hfill
    \begin{subfigure}[h]{0.24\textwidth}
        \centering
        \includegraphics[width=\textwidth]{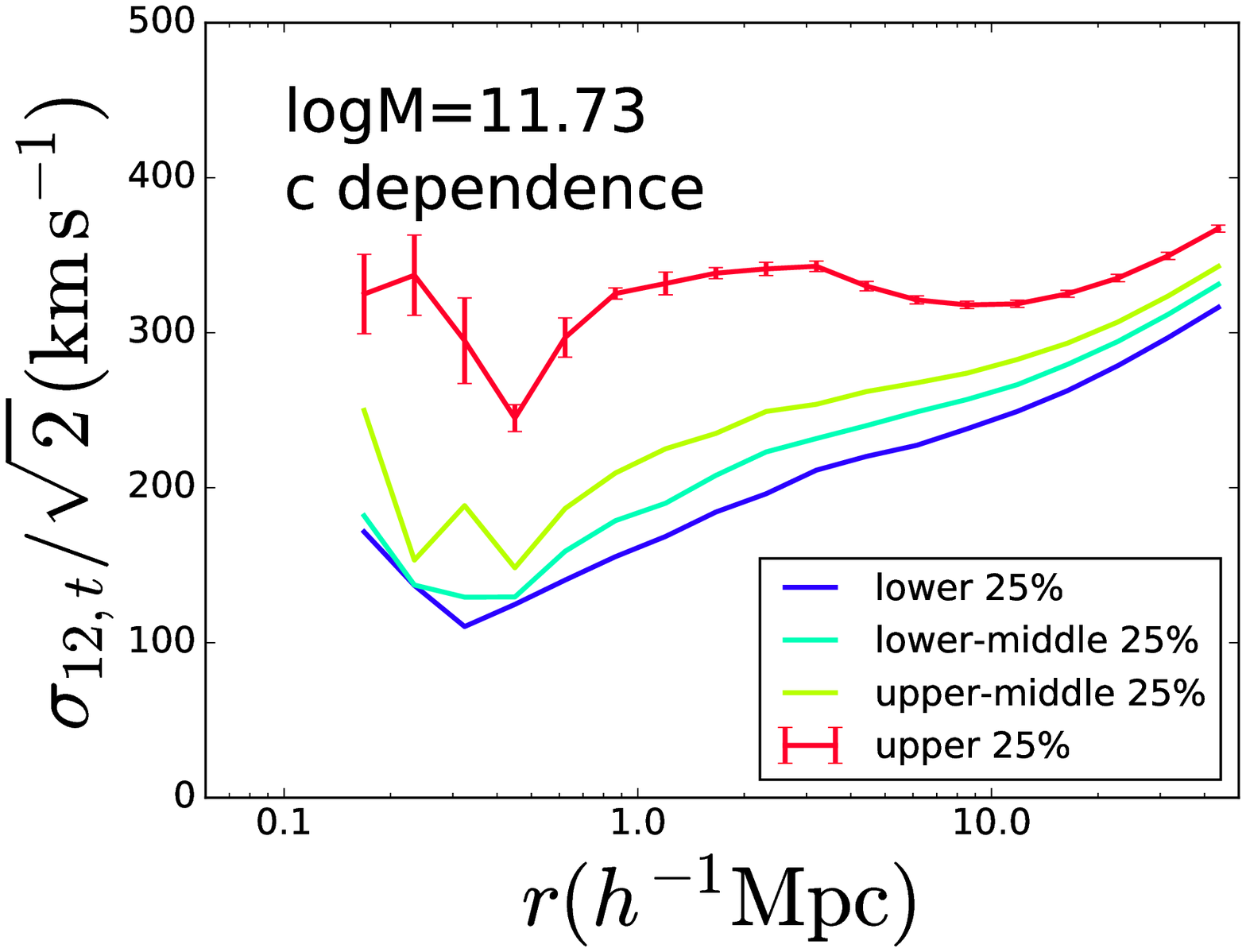}
    \end{subfigure}
	\hfill
    \begin{subfigure}[h]{0.24\textwidth}
        \centering
        \includegraphics[width=\textwidth]{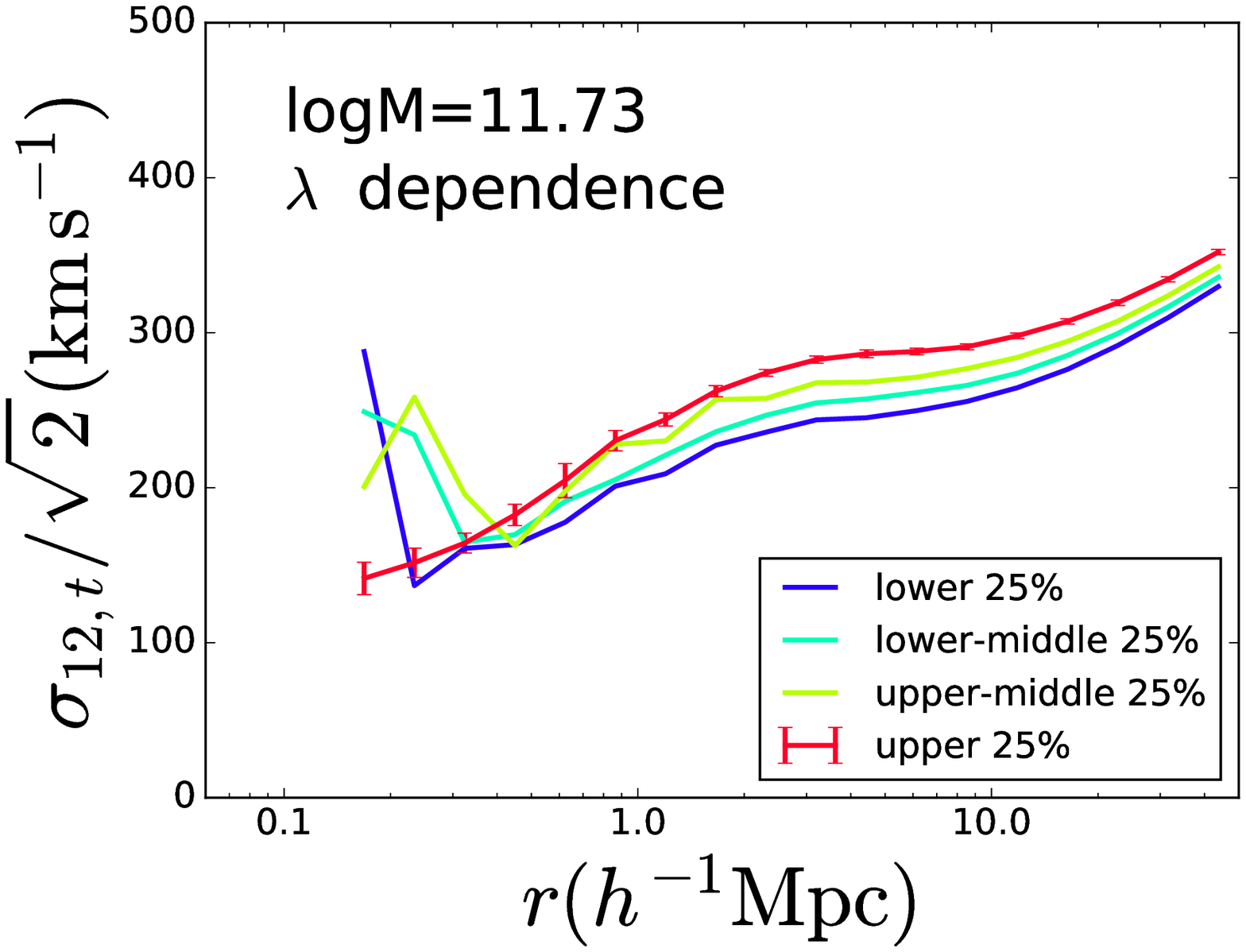}
    \end{subfigure}
\caption{
Same as Fig.~\ref{fig:vel_assembly}, but in each panel the curves are colour 
coded according to the percentile of the corresponding assembly variable.}
\label{fig:vel_assembly_percentile}
\end{figure*}

The assembly bias from the spatial distribution or clustering of haloes is 
related to halo environment, which should also affect the velocity distribution
of haloes. The velocity distribution of haloes is a major ingredient in 
modelling redshift-space clustering of galaxies. To yield insights on how halo assembly
may affect redshift-space clustering, we investigate the halo assembly effect on 
halo velocity distribution. 

We present the results
in terms of halo pairwise velocity,
$\boldsymbol{v}_{\rm 12}=\boldsymbol{v}_2 - \boldsymbol{v}_1$, 
as a function of halo pair separation, 
$r=|\boldsymbol{r}_{\rm 12}|=|\boldsymbol{r}_2 - \boldsymbol{r}_1|$, where 
$\boldsymbol{v}_{\rm i}$ and $\boldsymbol{r}_{\rm i}$ (${\rm i}$=1, 2) 
are the velocities
and positions of a pair of haloes. The pairwise radial and transverse velocities
are calculated as $\boldsymbol{v}_{\rm 12,r} = (\boldsymbol{v}_{\rm 12} \boldsymbol{\cdot} 
\boldsymbol{n}_{\rm 12}) \boldsymbol{n}_{\rm 12}$, and 
$\boldsymbol{v}_{\rm 12,t} = \boldsymbol{v}_{\rm 12} - \boldsymbol{v}_{\rm 12,r}$, where
$\boldsymbol{n}_{\rm 12}=\boldsymbol{r}_{\rm 12}/|\boldsymbol{r}_{\rm 12}|$ is the 
direction from one halo to the other of the pair. The corresponding velocity
dispersions are 
$\sigma_{\rm 12,r} = \langle v_{\rm 12,r}^2\rangle - \langle v_{\rm 12,r}\rangle^2$ 
and 
$\sigma_{\rm 12,t} = \langle v_{\rm 12,t}^2\rangle - \langle v_{\rm 12,t}\rangle^2$, 
with the average over all halo pairs at a given separation $r$.

Before moving on to present the results on pairwise velocity statistics, we point out the comparison between those from the Bolshoi and MDR1 simulations (see Appendix~\ref{sec:appendix}). In general, the MDR1 results in similar dependence patterns of the pairwise velocity statistics on assembly variables, except for halo spin.
The weak trend with spin seen in the MDR1 simulation does not show up in the Bolshoi simulation, which may be caused by noise in the Bolshoi simulation or unknown systematic effect in determining halo spin in low-resolution simulations. For consistency, we choose to present the MDR1 results with the above caveat for the spin results (more details in Appendix~\ref{sec:appendix}).

%4.1
\subsection{Assembly effect on pairwise velocity of haloes}
\label{sec:vel_A}
 
In Fig.~\ref{fig:vel_mass}, we first show the halo mass dependent pairwise 
velocity and velocity dispersions, as a function of halo pair separation.
Consistent with previous study \citep[e.g.][]{Zheng02}, the pairwise infall 
velocity $v_{\rm 12,r}$ (left panel) 
increases as halo mass increases, and the increase
is faster at smaller pair separation, reflecting the stronger gravitational 
influence from higher mass haloes. For high mass haloes (with mass above a 
few times $\Mnl$), the pairwise radial velocity continuously decreases toward
large separation. For low mass haloes, a bump in the pairwise radial velocity
shows up. The bump is around $r\sim 5\hinvMpc$ for haloes of the lowest mass
in the study and shifts to larger scales for haloes of higher mass. 
At large separation, the pairwise 
infall velocity tends to converge to an amplitude independent of halo mass,
which reflects the fact that all haloes feel the same large-scale gravitational 
potential field sourced by linear fluctuations. 

The pairwise radial velocity dispersion $\sigma_{\rm 12, r}$ 
(middle panel) decreases toward small halo pair separation. The dependence
on halo mass is weak -- in the range of $r\sim$ 1--5$\hinvMpc$, lower mass 
haloes have slightly higher $\sigma_{\rm 12, r}$. The dispersion 
$\sigma_{\rm 12, t}$
of the pairwise transverse velocity (right panel) has a similar trend. 
On large scales, the one-dimensional (1D) pairwise transverse velocity 
dispersion $\sigma_{\rm 12,t}/\sqrt{2}$ (right panel) is about 10 per cent
lower that the radial one (middle panel).

In Fig.~\ref{fig:vel_assembly}, the dependence of pairwise radial velocity 
(top panel) and the radial and transverse velocity dispersions (middle and 
bottom panels) on assembly variables are shown for haloes of 
$\log[\Mh/(\hinvMsun)]=11.73$. Below $\sim 1\hinvMpc$, the pairwise 
radial velocity does not show a strong dependence on any of the assembly 
variable. Above $\sim 1\hinvMpc$, the assembly effect becomes clear. 
Both radial and transverse pairwise velocity dispersions show a substantial 
dependence on assembly variables at all scales. To see the trend more clearly, Fig.~\ref{fig:vel_assembly_percentile} shows the dependences by grouping each 
assembly variable into four percentiles. For the radial 
pairwise velocity, the bump around $r\sim 5\hinvMpc$ seen in Fig.~\ref{fig:vel_mass}
shows up for each group, and the scatter caused by assembly effect also 
reaches maximum at this scale. The assembly effect in pairwise velocity with
halo spin is not as strong as with other assembly variables, opposite to
the trend seen in halo bias (e.g. left panel of Fig.~\ref{fig:bpercentile}).

For low-mass haloes presented here, similar to the spatial clustering, the 
trend of the dependence of pairwise velocity on assembly variables follows
the correlations among assembly variables, except for the 
case with halo spin. At fixed mass, haloes that on average form earlier 
(with lower \ahalf, higher $c$, or higher \Vpeak) have higher pairwise 
velocity and velocity dispersions. This is in line with the evolution of those 
low-mass haloes being influenced by the surrounding environment, especially 
the tidal field \citep[e.g.][]{Hahn07,Hahn09,Wang11,Shi15} and with some of them 
being ejected haloes around massive haloes \citep[e.g.][]{Wang09,Wetzel14}.
The deviation between the pairwise velocity trend with halo spin and the expectation 
from correlations with other assembly variables arises from the misalignment of
the gradient of pairwise velocity and the direction of the correlation in the
plane of spin and one other assembly variable, similar to what we discuss 
in \S~\ref{sec:2var} for halo bias.

\begin{figure*}
    \centering
    \begin{subfigure}[h]{0.24\textwidth}
        \centering
        \includegraphics[width=\textwidth]{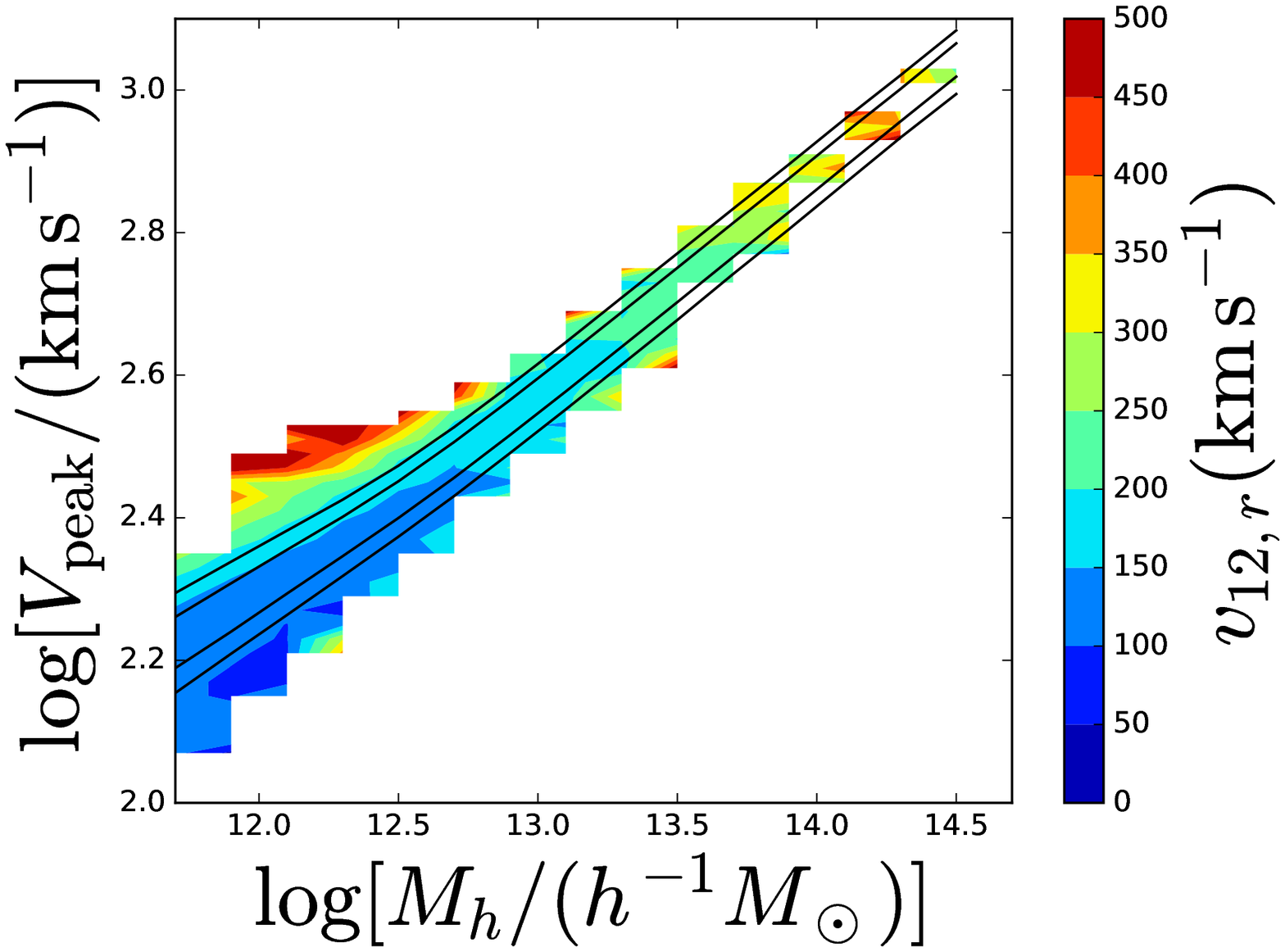}
    \end{subfigure}
    \hfill
    \begin{subfigure}[h]{0.24\textwidth}
        \centering
        \includegraphics[width=\textwidth]{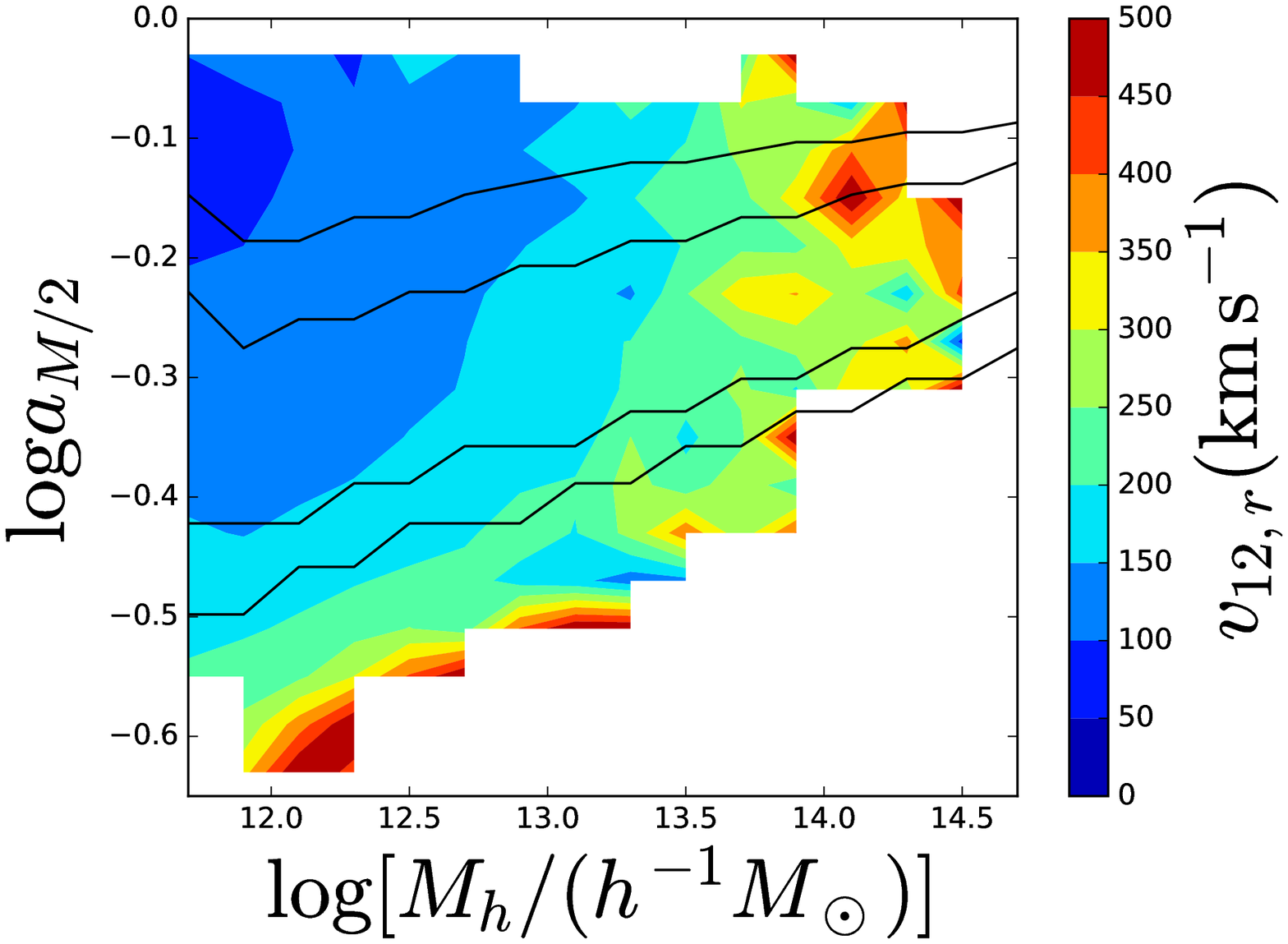}
    \end{subfigure}
    \hfill
    \begin{subfigure}[h]{0.24\textwidth}
        \centering
        \includegraphics[width=\textwidth]{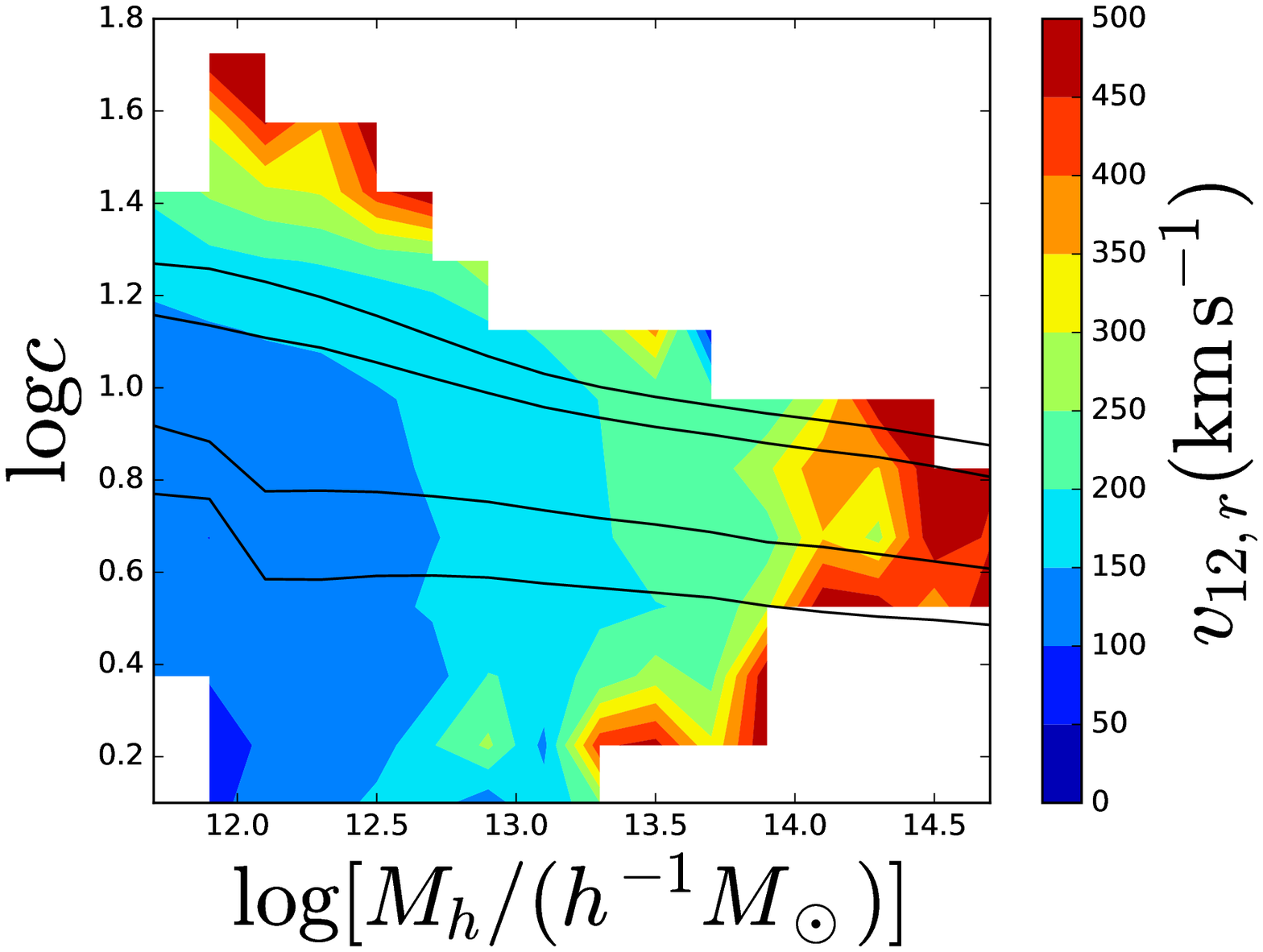}
    \end{subfigure}
    \hfill
    \begin{subfigure}[h]{0.24\textwidth}
        \centering
        \includegraphics[width=\textwidth]{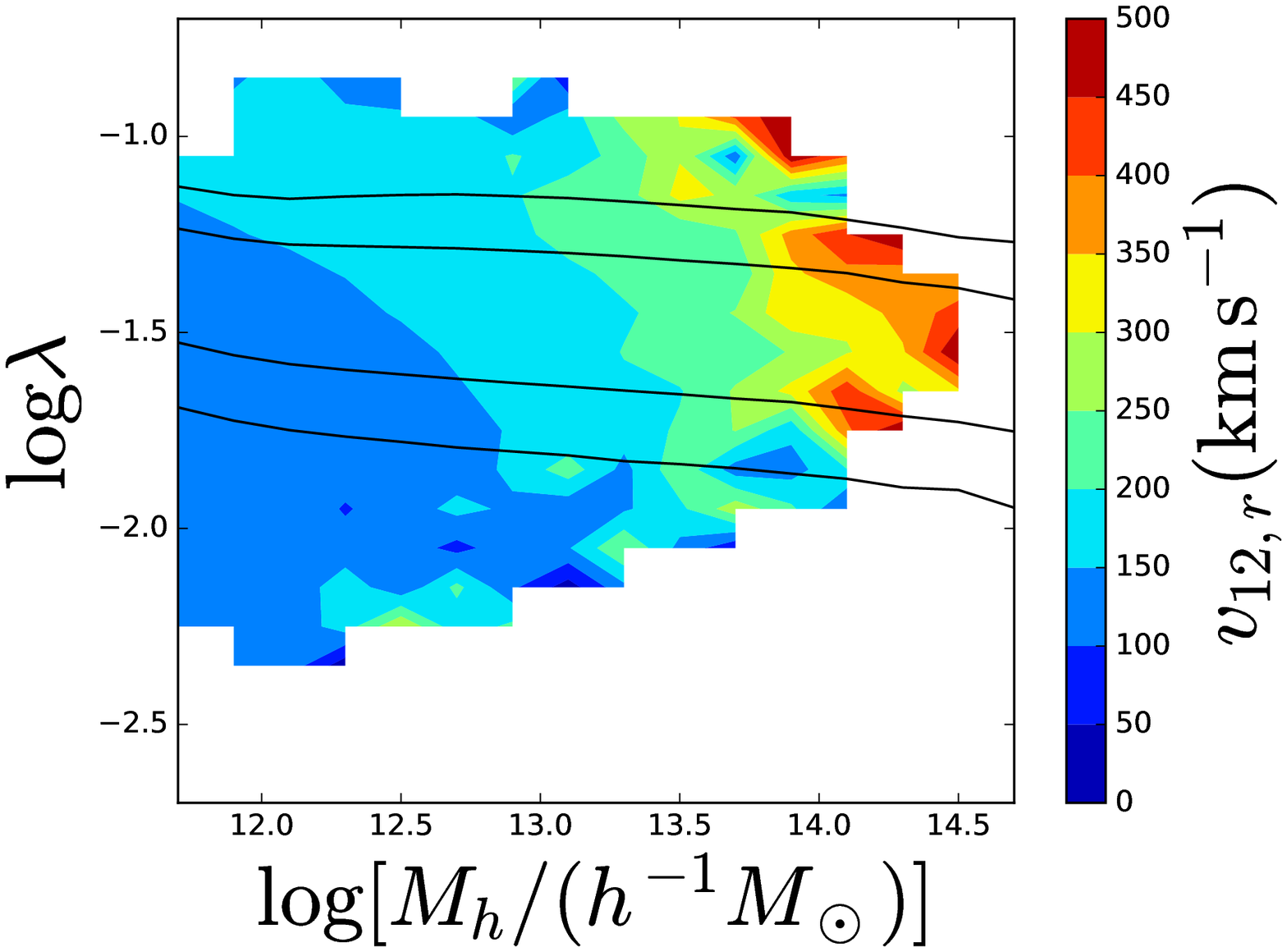}
    \end{subfigure}
    \hfill
    \begin{subfigure}[h]{0.24\textwidth}
        \centering
        \includegraphics[width=\textwidth]{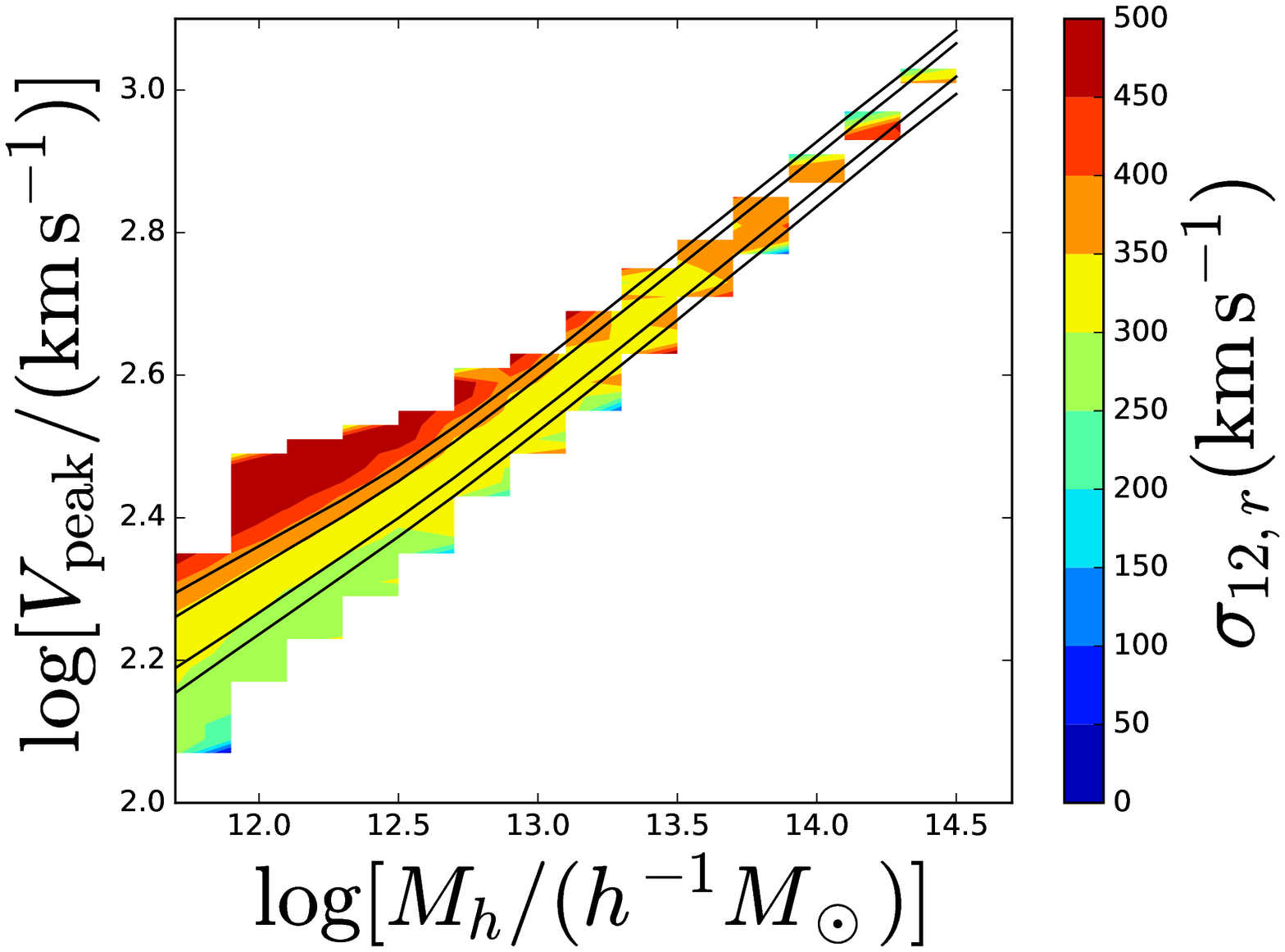}
    \end{subfigure}
    \hfill
    \begin{subfigure}[h]{0.24\textwidth}
        \centering
        \includegraphics[width=\textwidth]{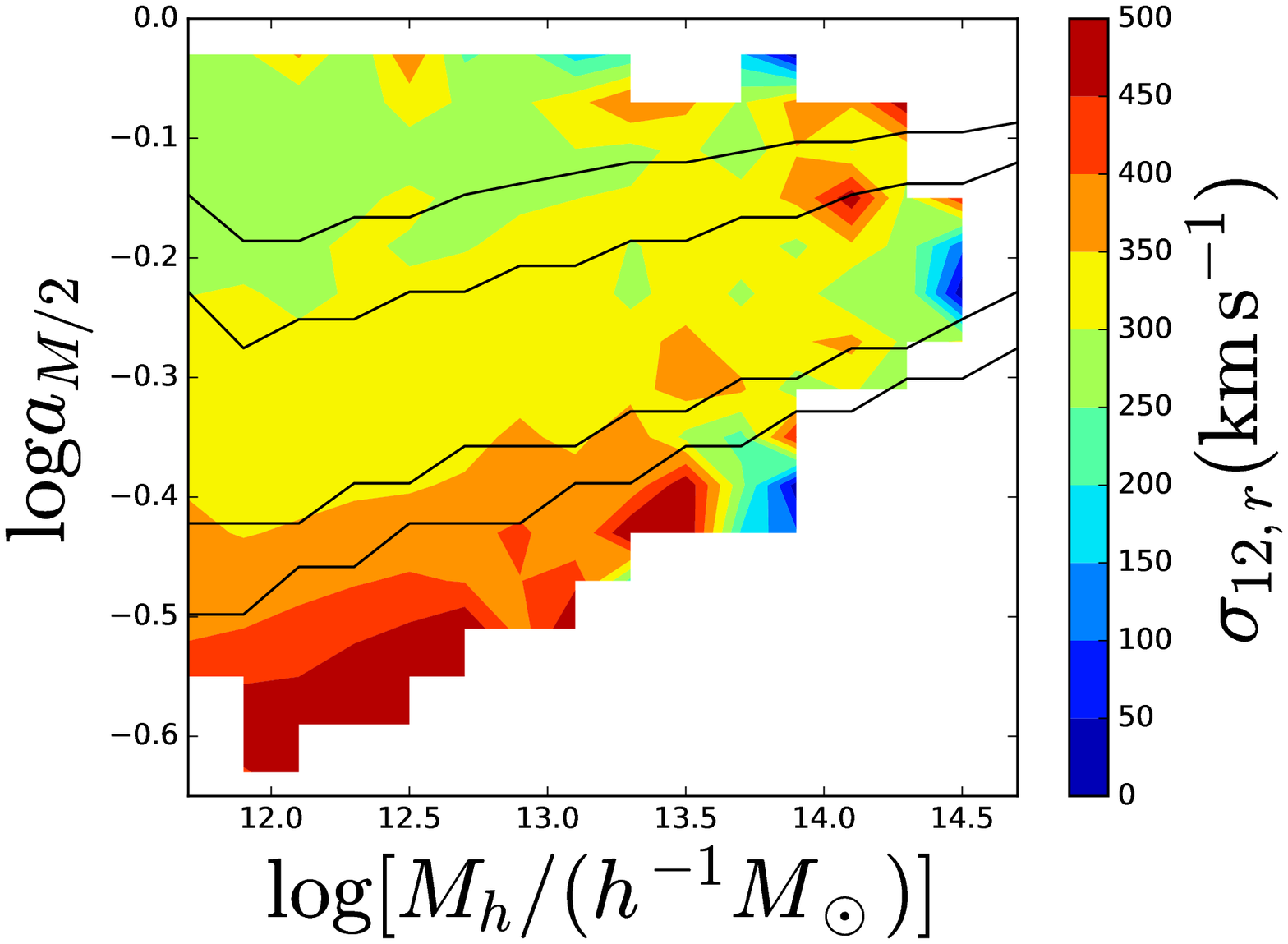}
    \end{subfigure}
    \hfill
    \begin{subfigure}[h]{0.24\textwidth}
        \centering
        \includegraphics[width=\textwidth]{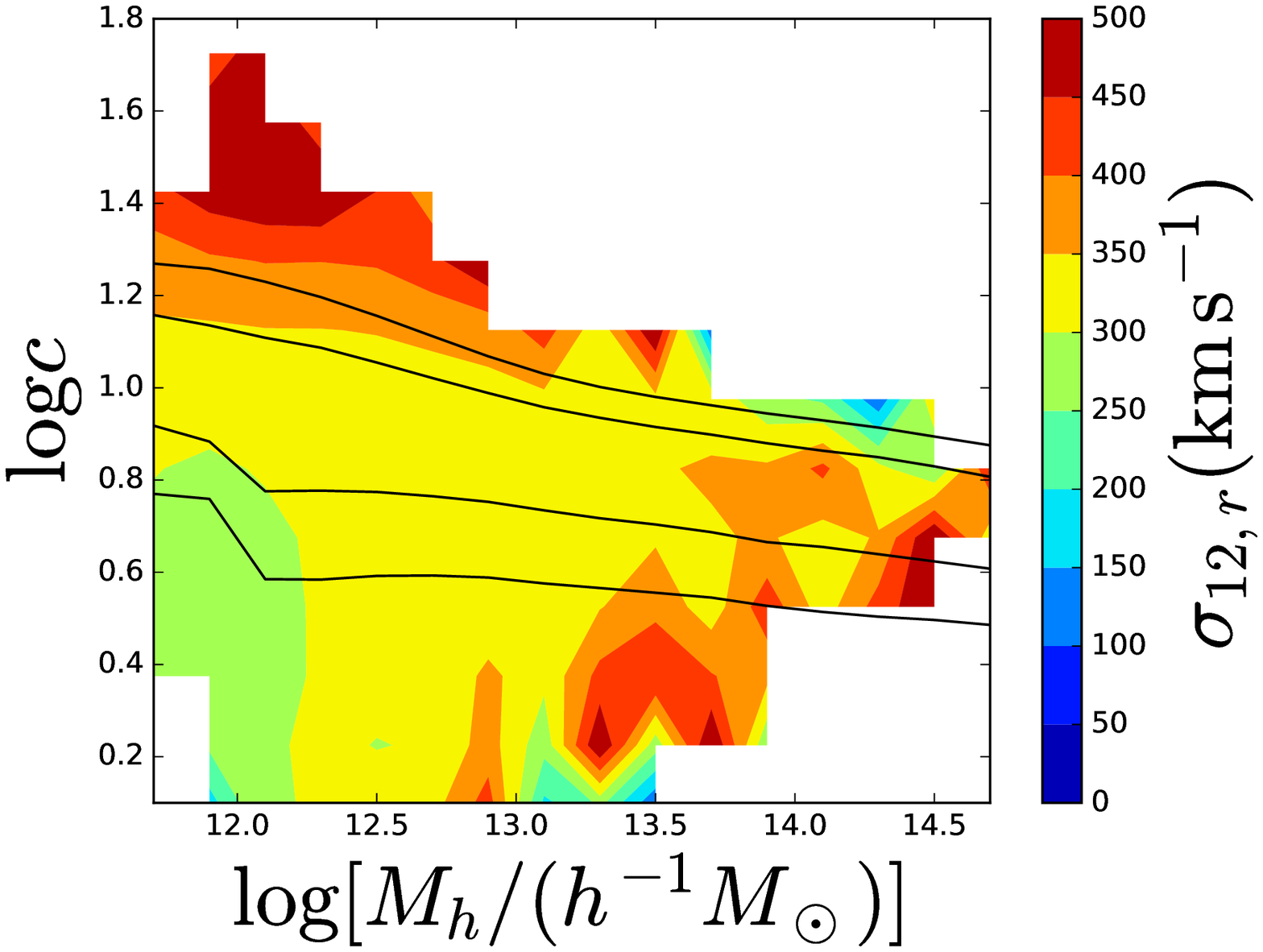}
    \end{subfigure}
    \hfill
    \begin{subfigure}[h]{0.24\textwidth}
        \centering
        \includegraphics[width=\textwidth]{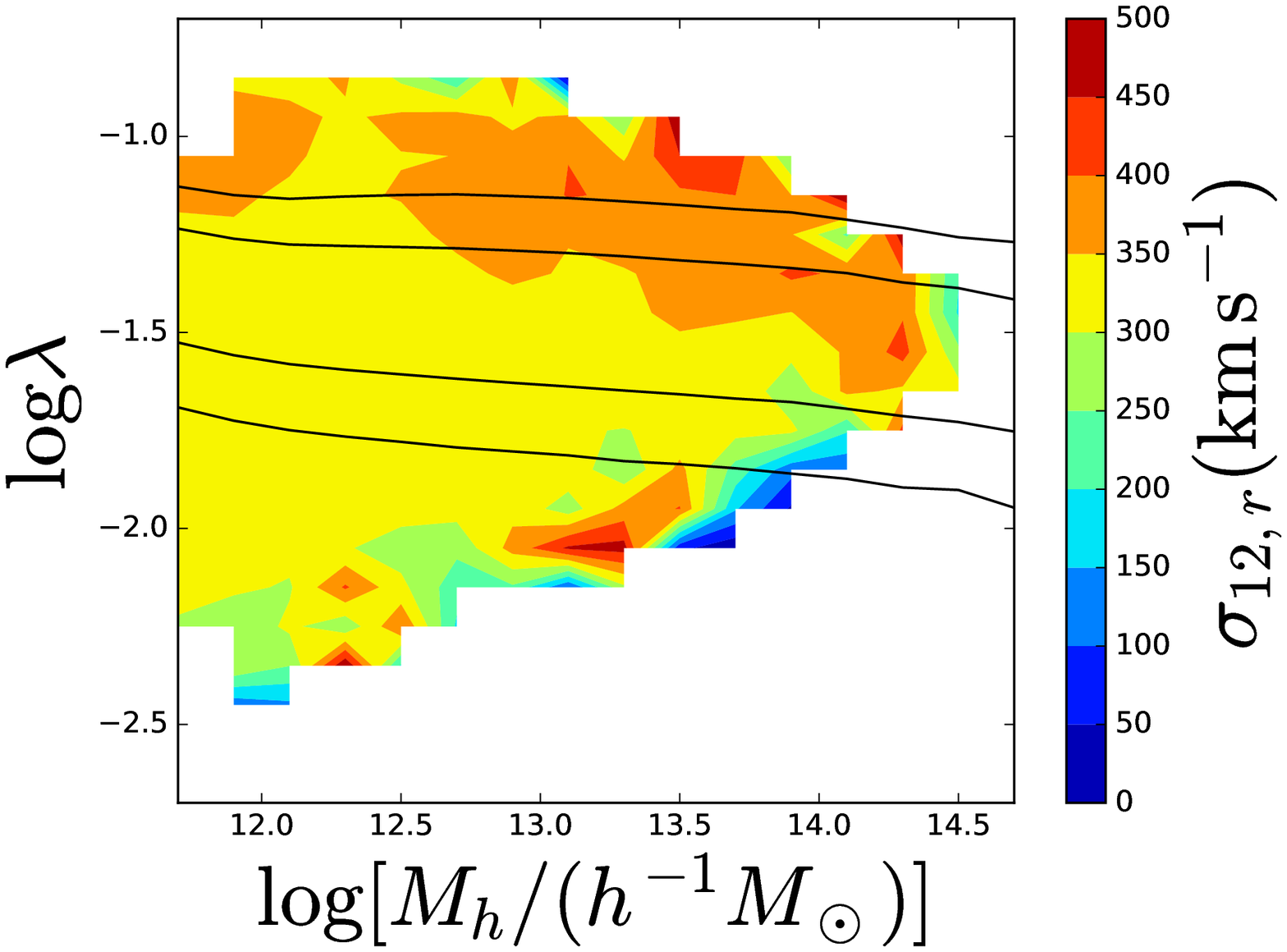}
    \end{subfigure}

    \hfill
    \begin{subfigure}[h]{0.24\textwidth}
        \centering
        \includegraphics[width=\textwidth]{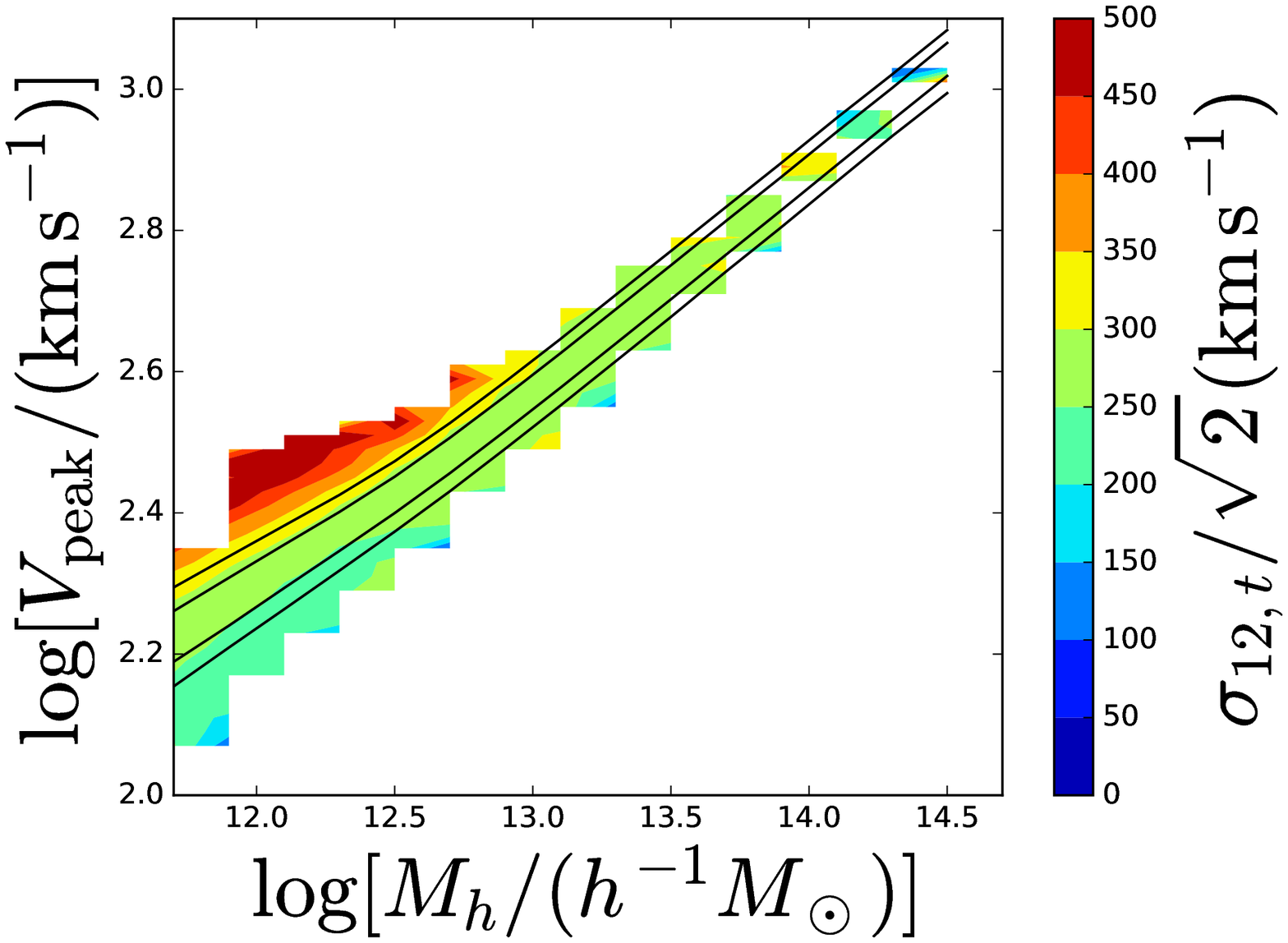}
    \end{subfigure}
    \hfill
    \begin{subfigure}[h]{0.24\textwidth}
        \centering
        \includegraphics[width=\textwidth]{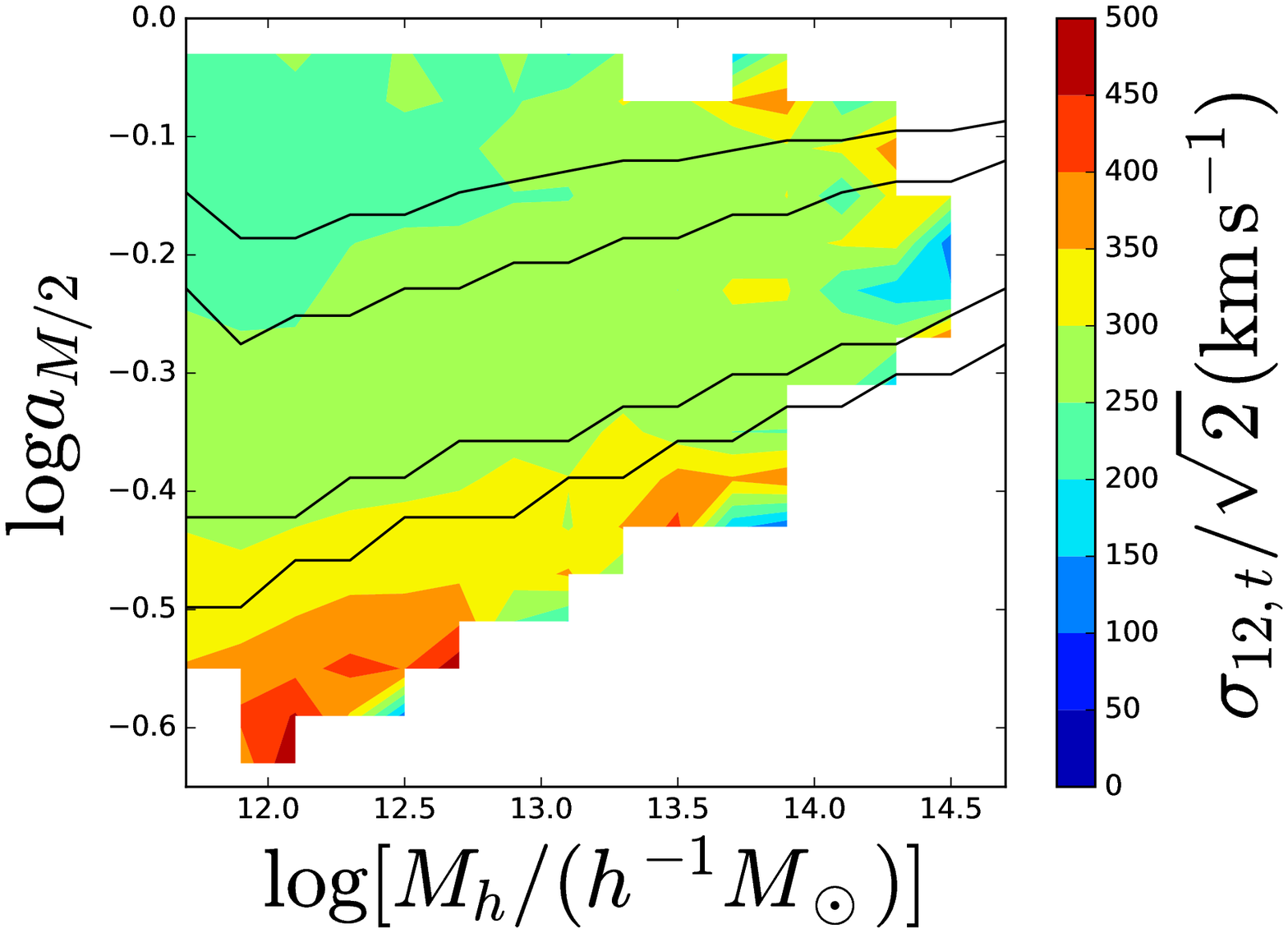}
    \end{subfigure}
    \hfill
    \begin{subfigure}[h]{0.24\textwidth}
        \centering
        \includegraphics[width=\textwidth]{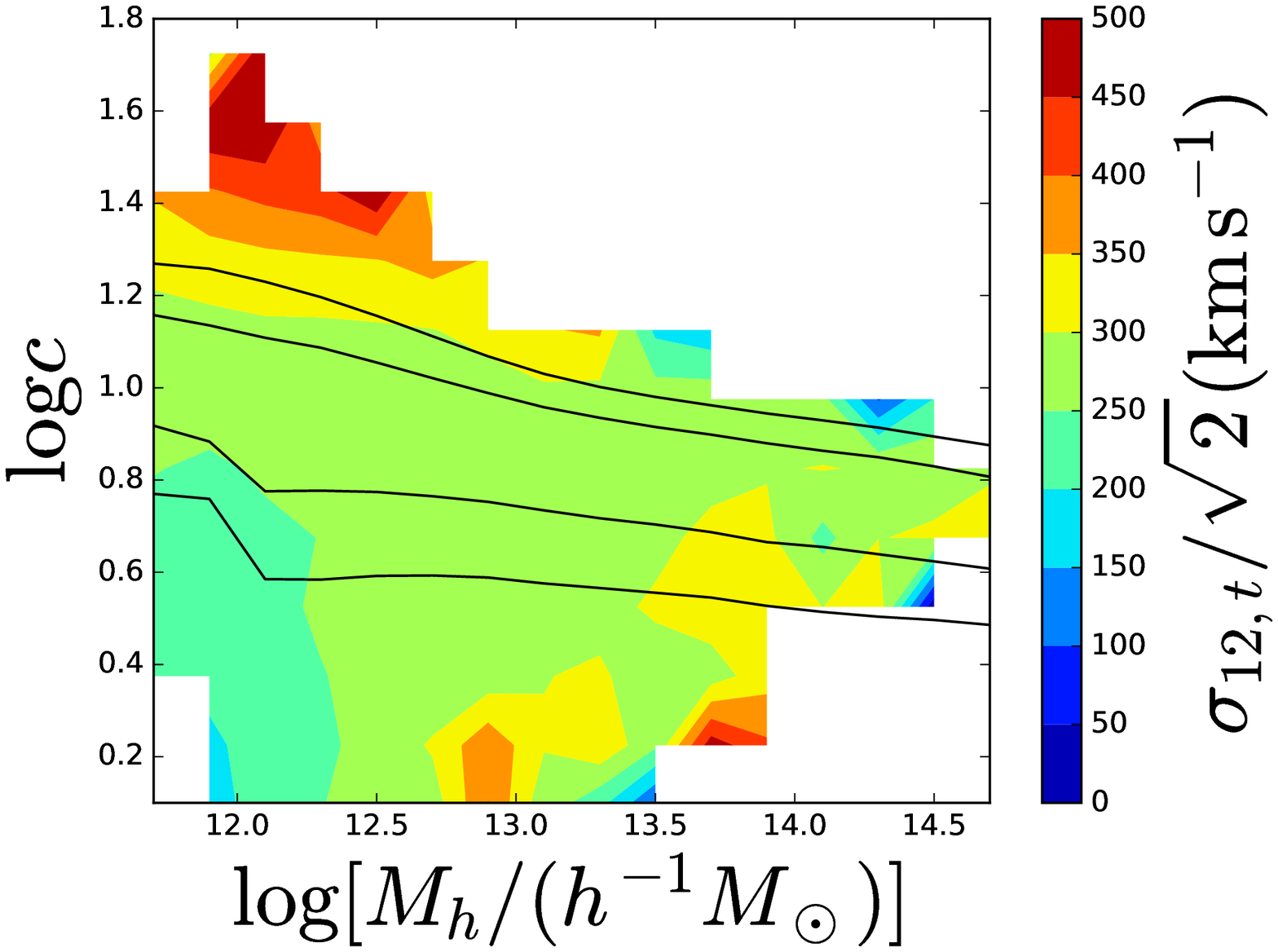}
    \end{subfigure}
    \hfill
    \begin{subfigure}[h]{0.24\textwidth}
        \centering
        \includegraphics[width=\textwidth]{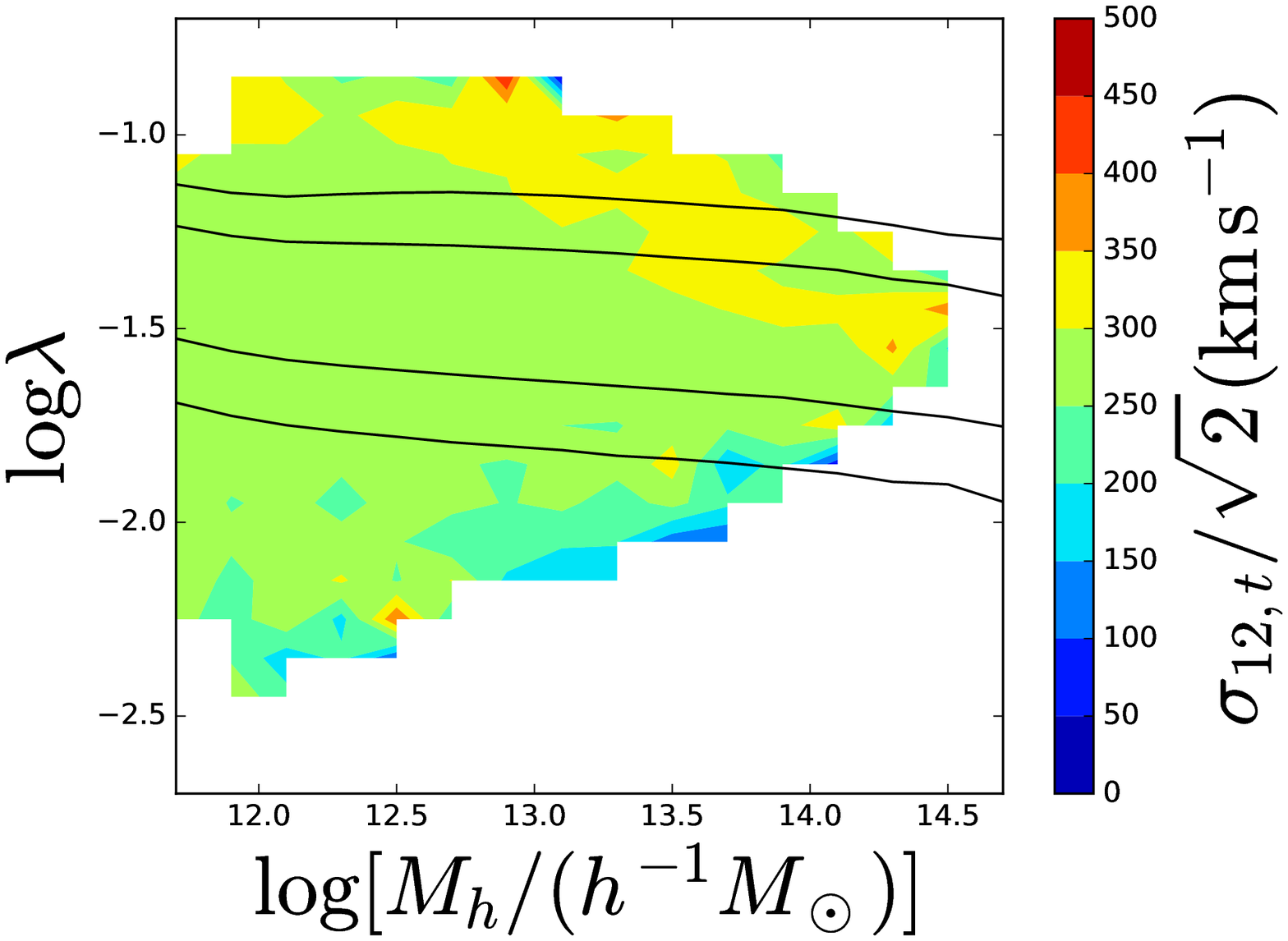}
    \end{subfigure}
\caption{
Dependence of halo pairwise velocity and velocity dispersions on 
halo mass and each of the assembly variables.
Top, middle, and bottom panels show pairwise radial 
velocity, pairwise radial velocity dispersion, and pairwise transverse 
velocity dispersion, respectively. The pairwise velocity and velocity 
dispersions are evaluated at $r\sim 10\hinvMpc$. The four black curves 
in each panel mark the central 50 and 80 per cent of the distribution 
of the corresponding assembly variable as a function of halo mass.
}
\label{fig:vel_m_A}
\end{figure*}

\begin{figure*}
    \centering
    \begin{subfigure}[h]{0.24\textwidth}
        \centering
        \includegraphics[width=\textwidth]{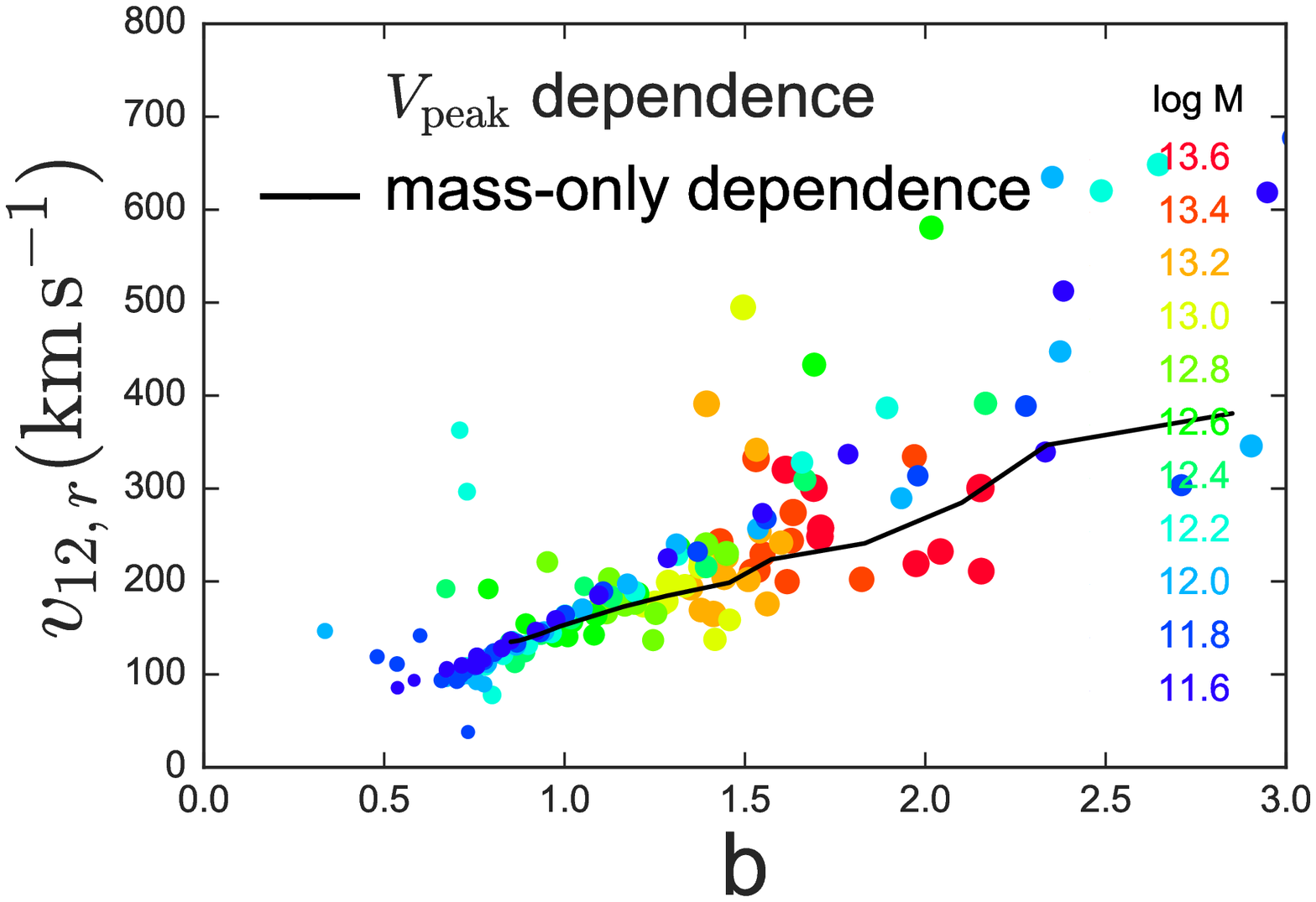}
    \end{subfigure}
    \hfill
    \begin{subfigure}[h]{0.24\textwidth}
        \centering
        \includegraphics[width=\textwidth]{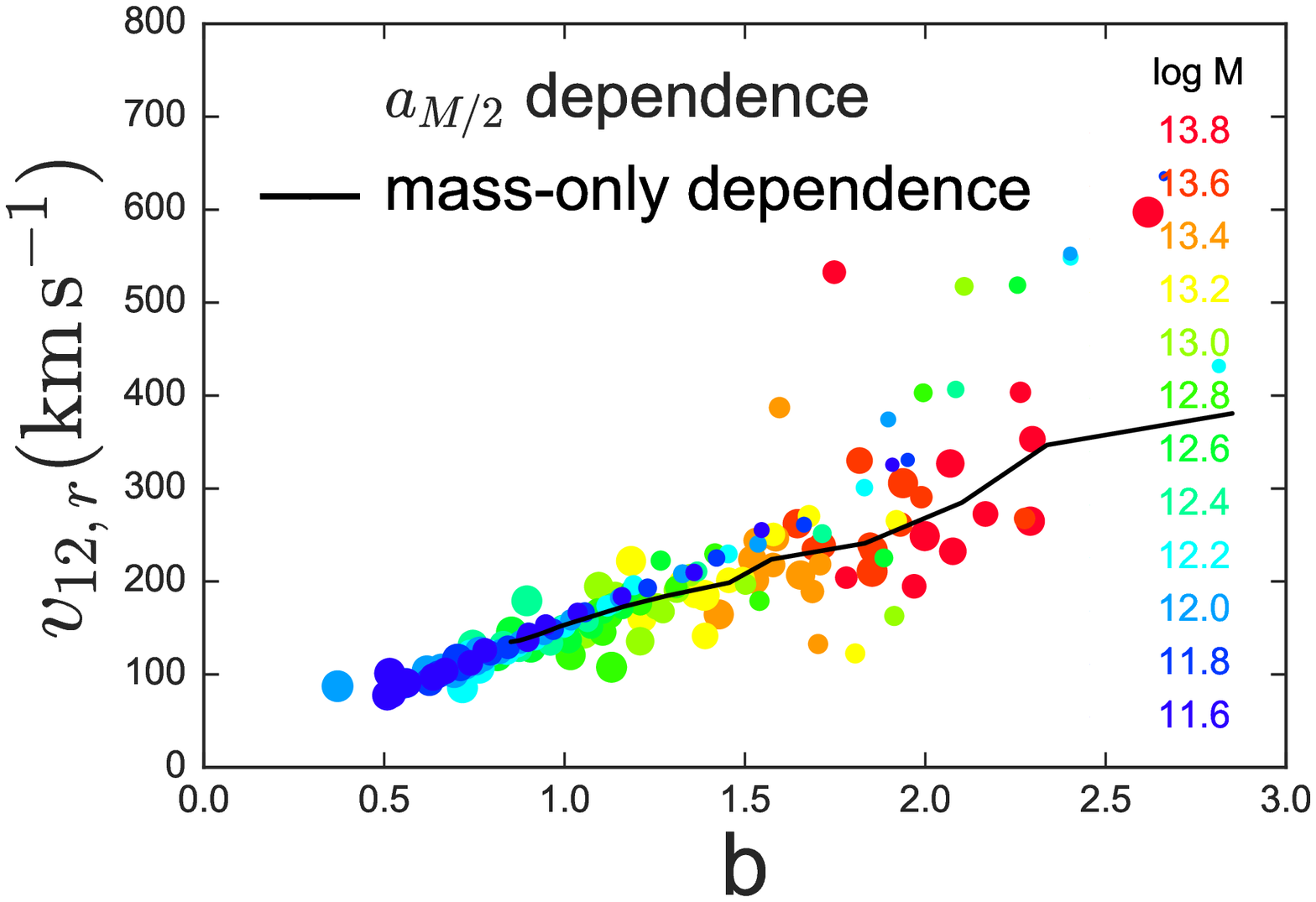}
    \end{subfigure}
    \hfill
    \begin{subfigure}[h]{0.24\textwidth}
        \centering
        \includegraphics[width=\textwidth]{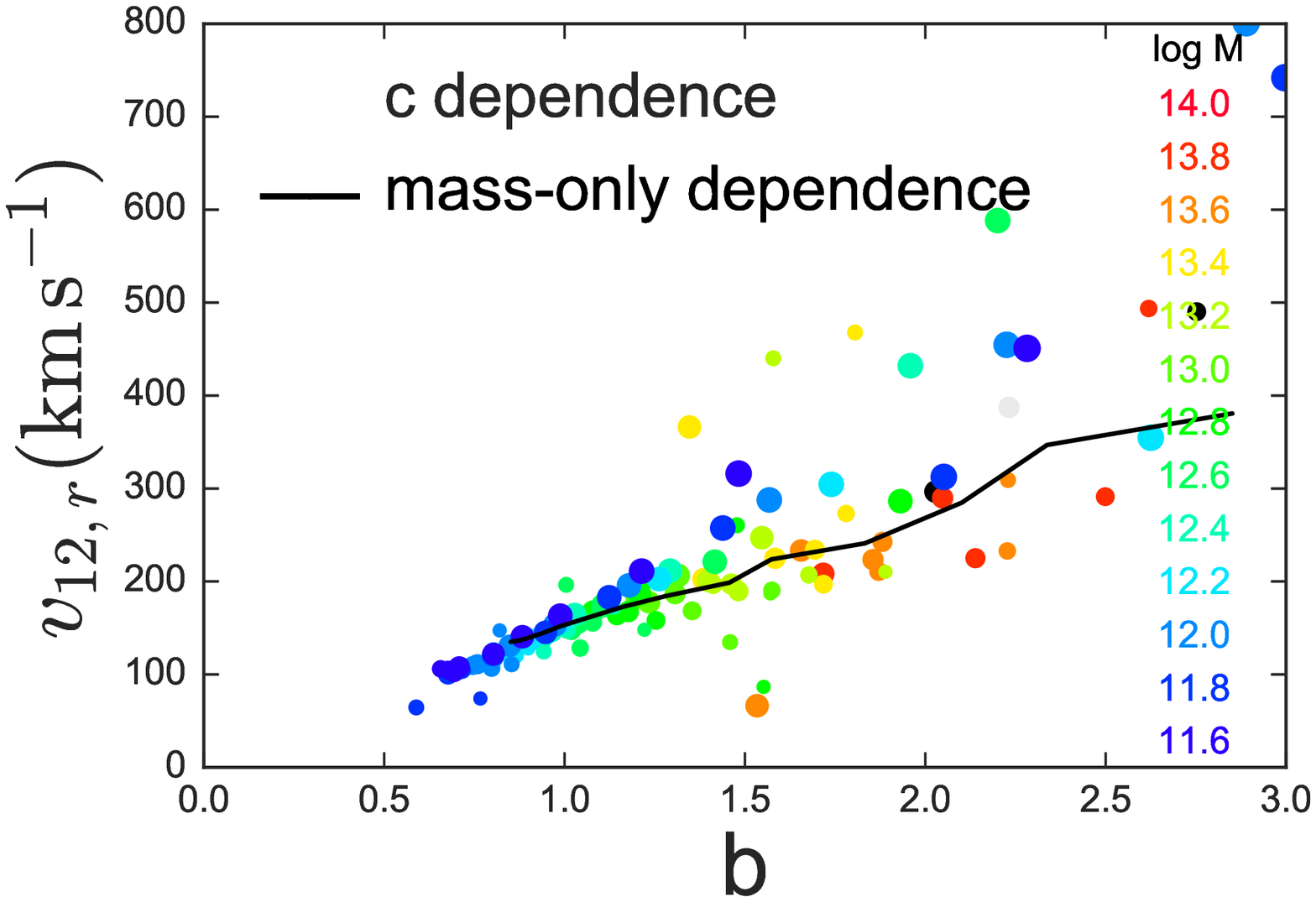}
    \end{subfigure}
    \hfill
    \begin{subfigure}[h]{0.24\textwidth}
        \centering
        \includegraphics[width=\textwidth]{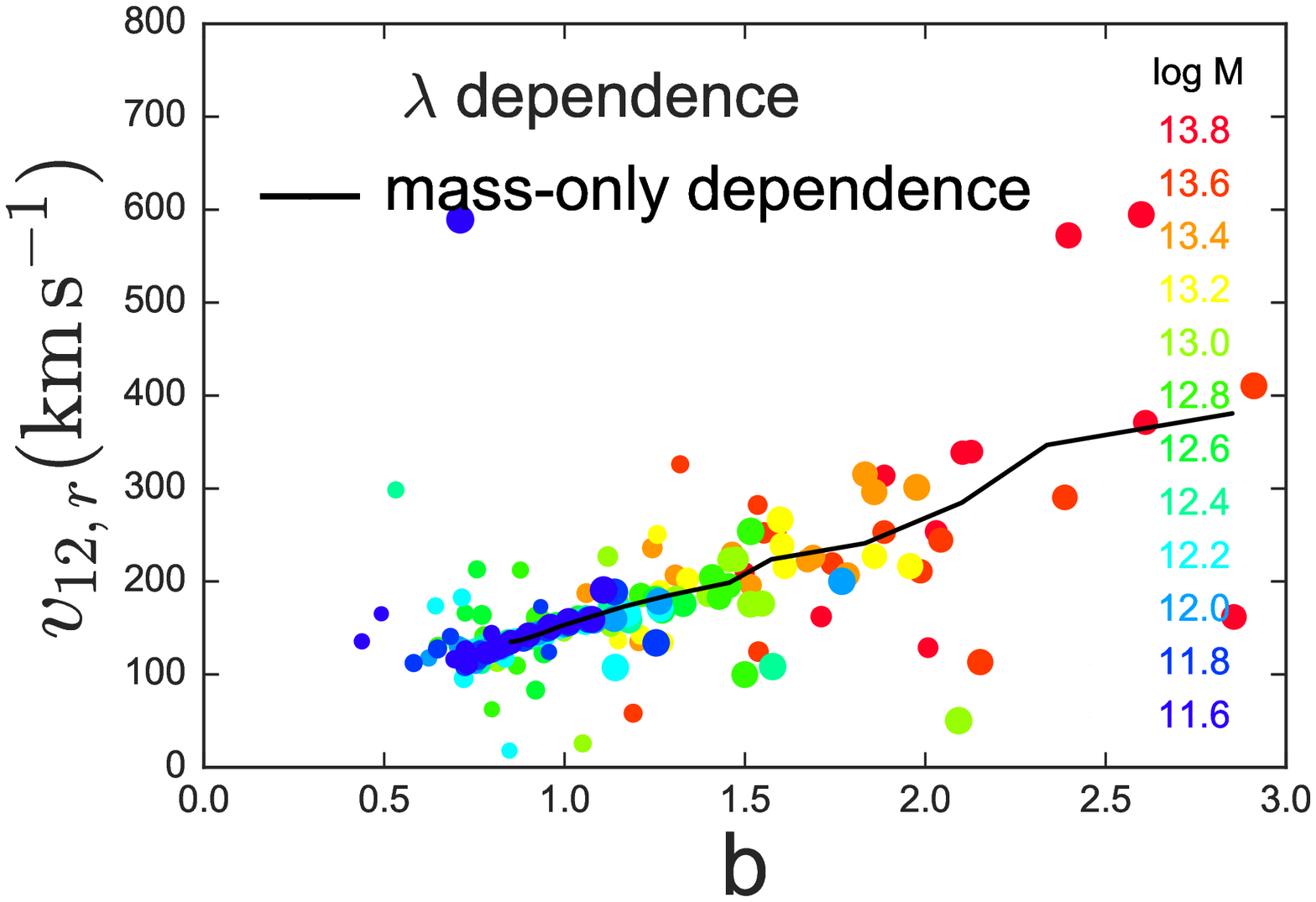}
    \end{subfigure}
    \hfill
    \begin{subfigure}[h]{0.24\textwidth}
        \centering
        \includegraphics[width=\textwidth]{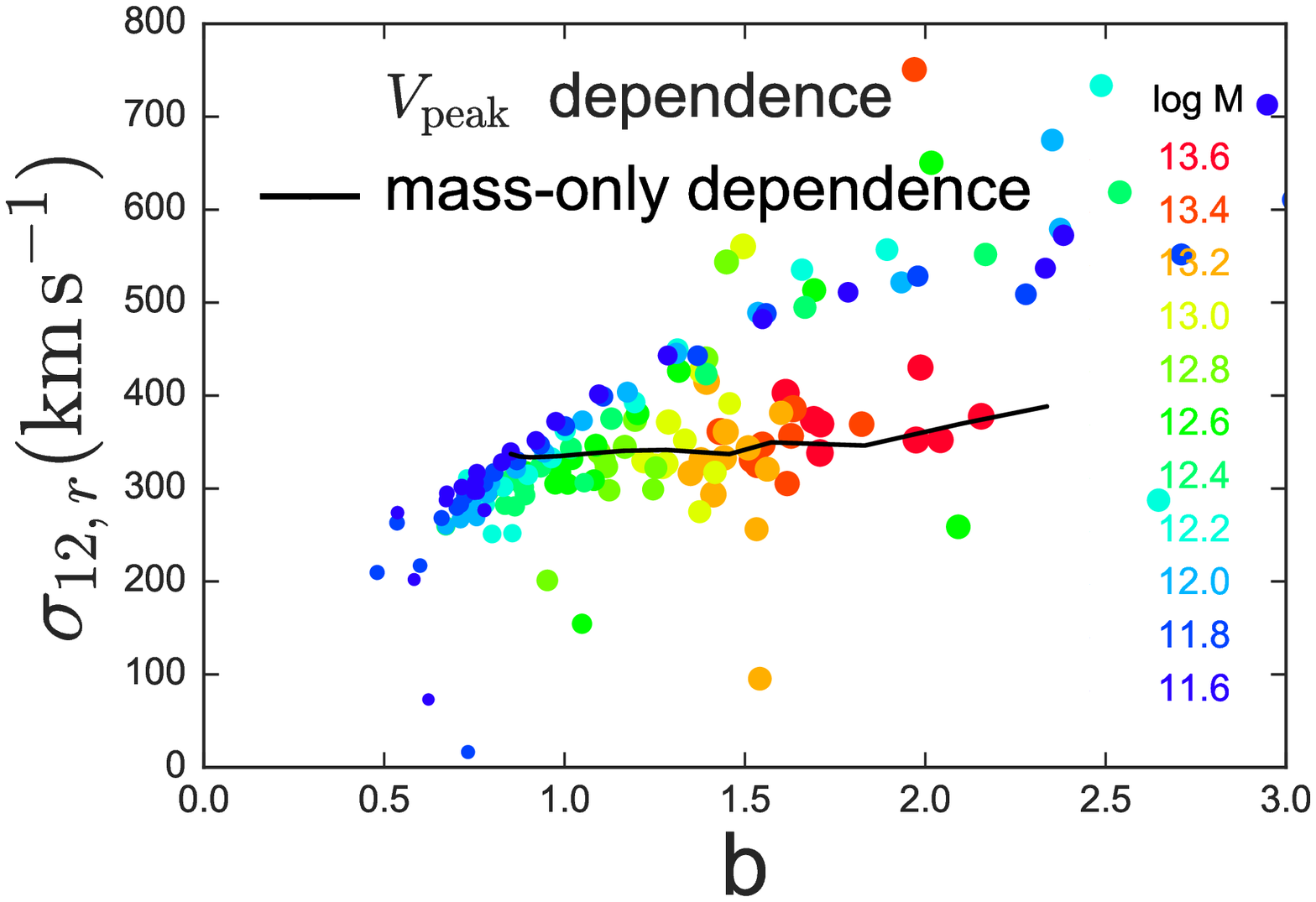}
    \end{subfigure}
    \hfill
    \begin{subfigure}[h]{0.24\textwidth}
        \centering
        \includegraphics[width=\textwidth]{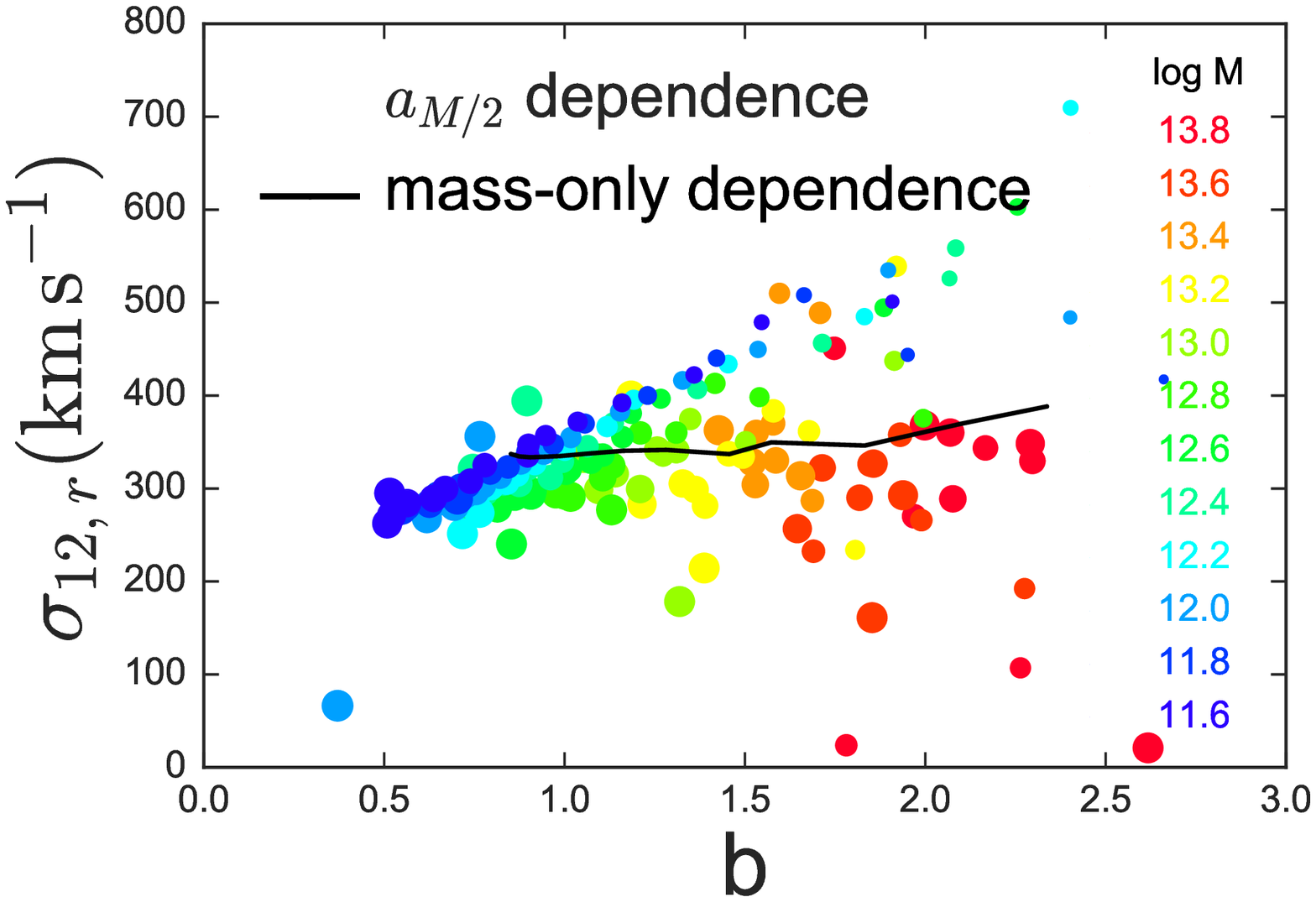}
    \end{subfigure}
    \hfill
    \begin{subfigure}[h]{0.24\textwidth}
        \centering
        \includegraphics[width=\textwidth]{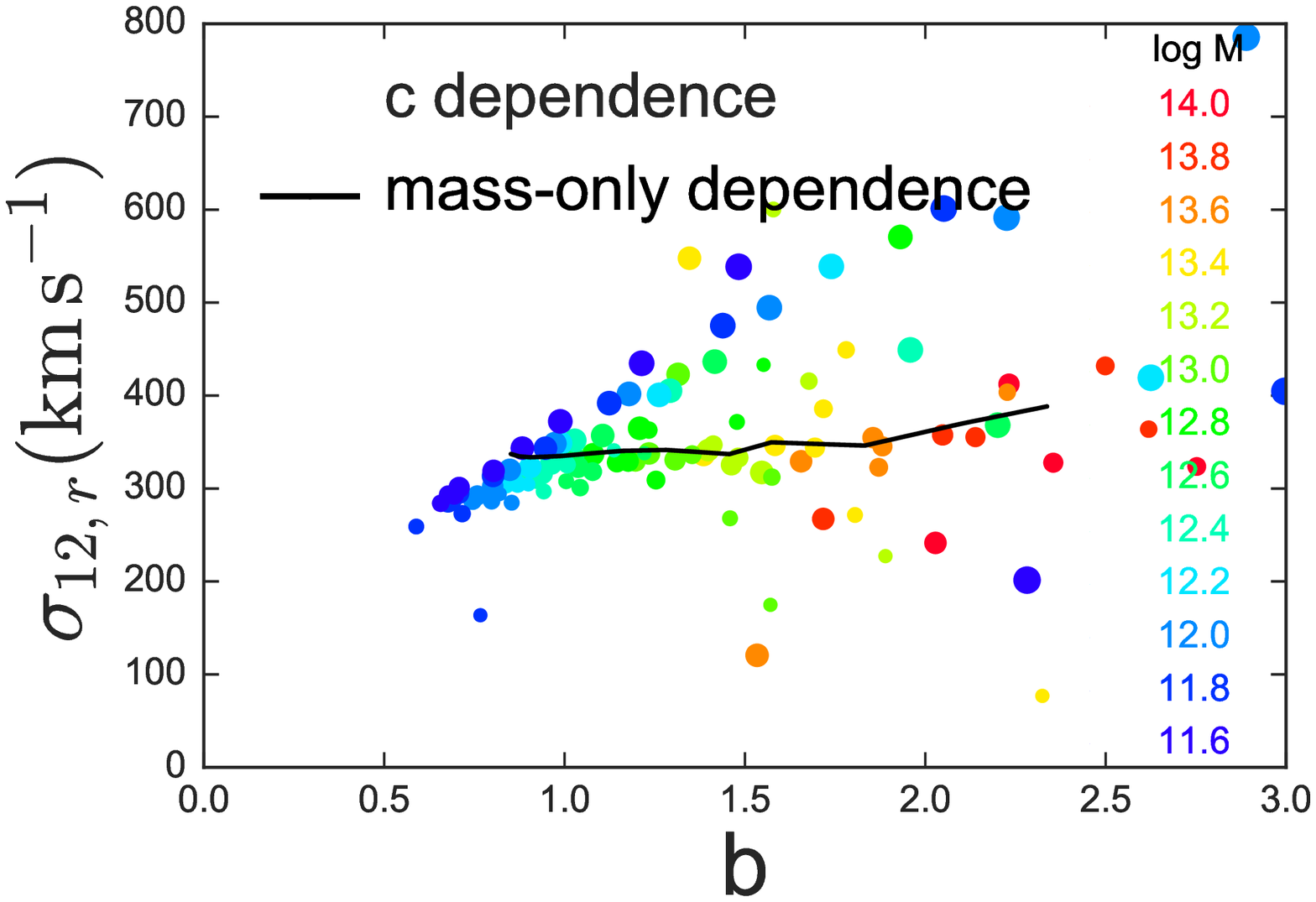}
    \end{subfigure}
    \hfill
    \begin{subfigure}[h]{0.24\textwidth}
        \centering
        \includegraphics[width=\textwidth]{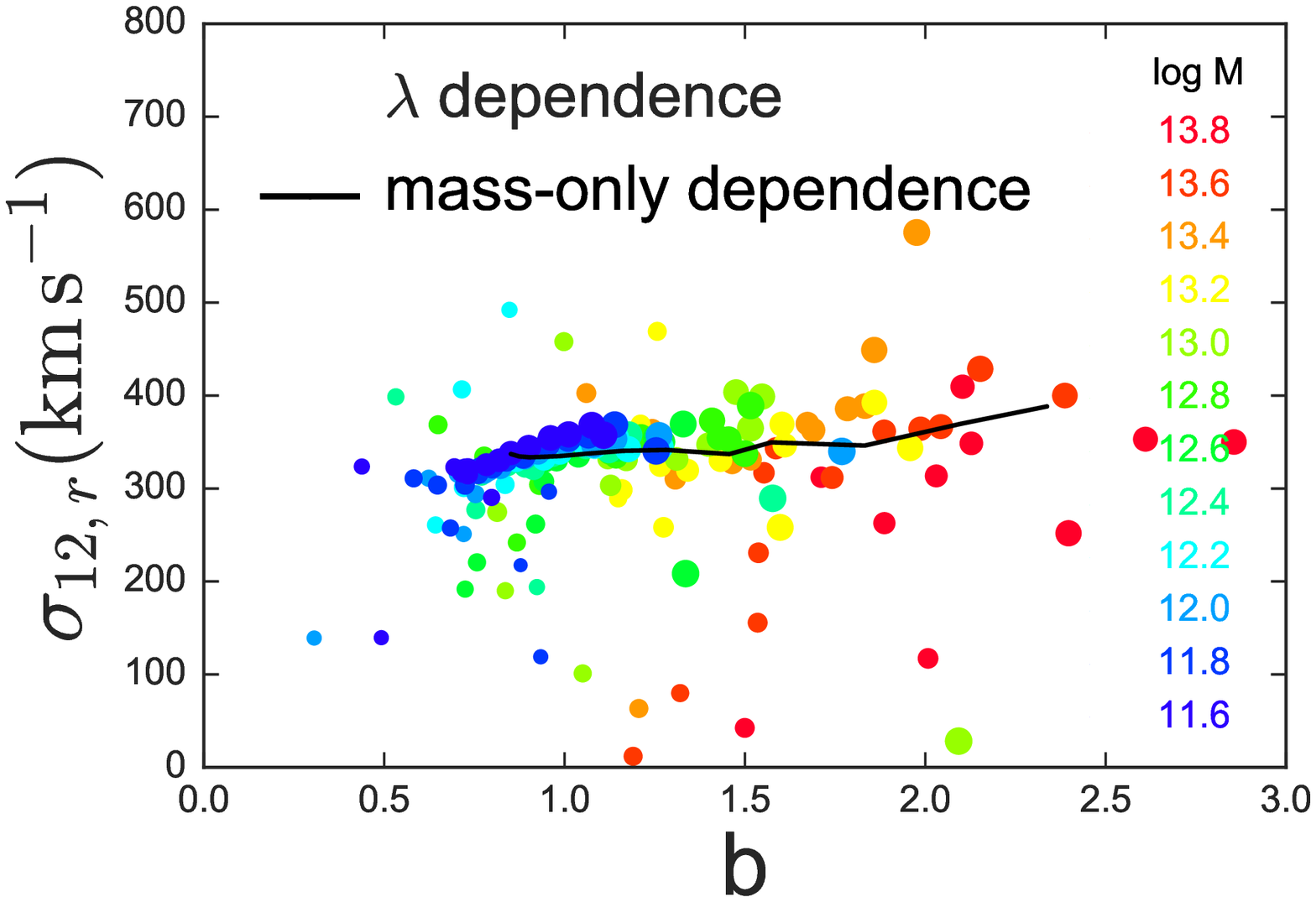}
    \end{subfigure}

    \hfill
    \begin{subfigure}[h]{0.24\textwidth}
        \centering
        \includegraphics[width=\textwidth]{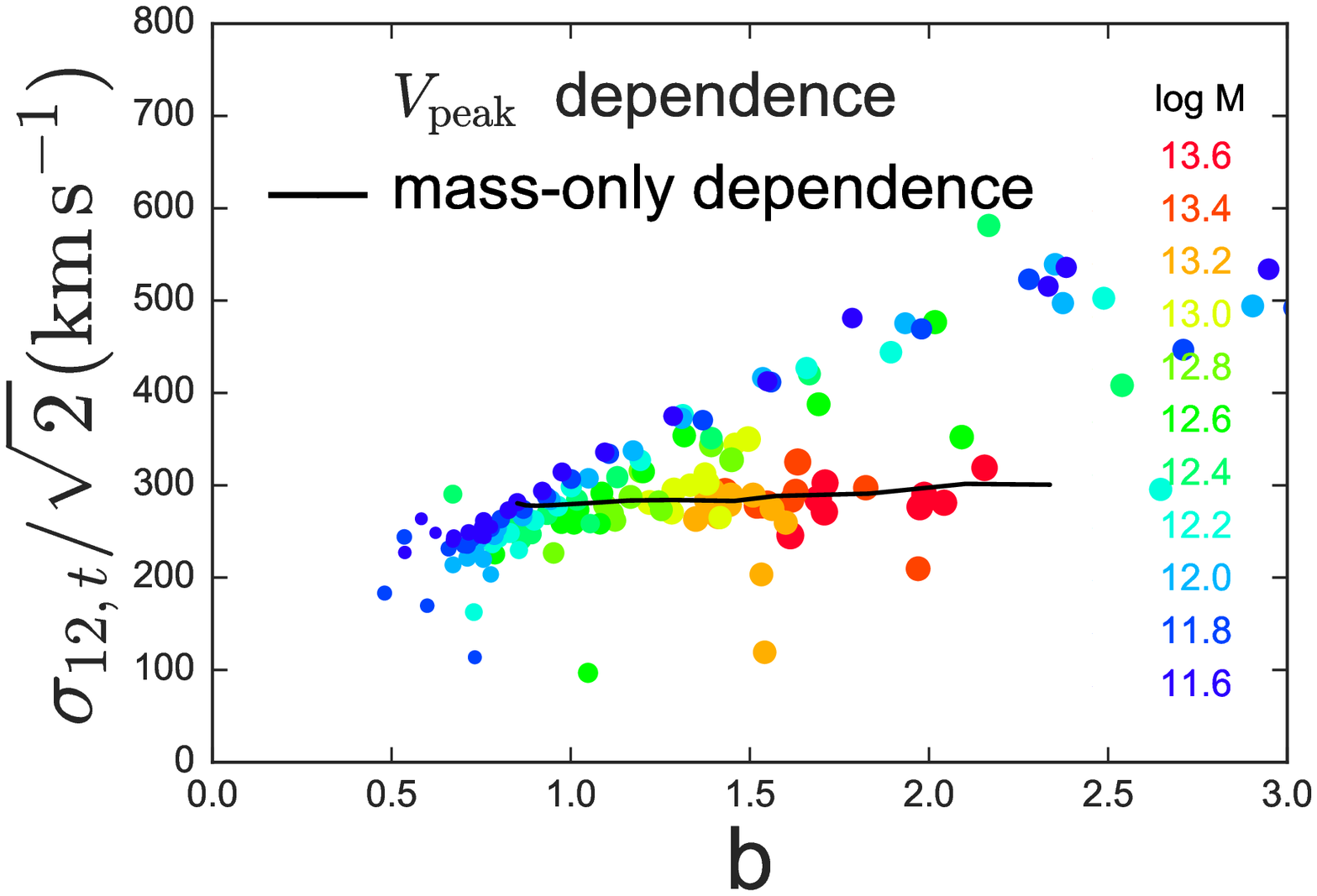}
    \end{subfigure}
    \hfill
    \begin{subfigure}[h]{0.24\textwidth}
        \centering
        \includegraphics[width=\textwidth]{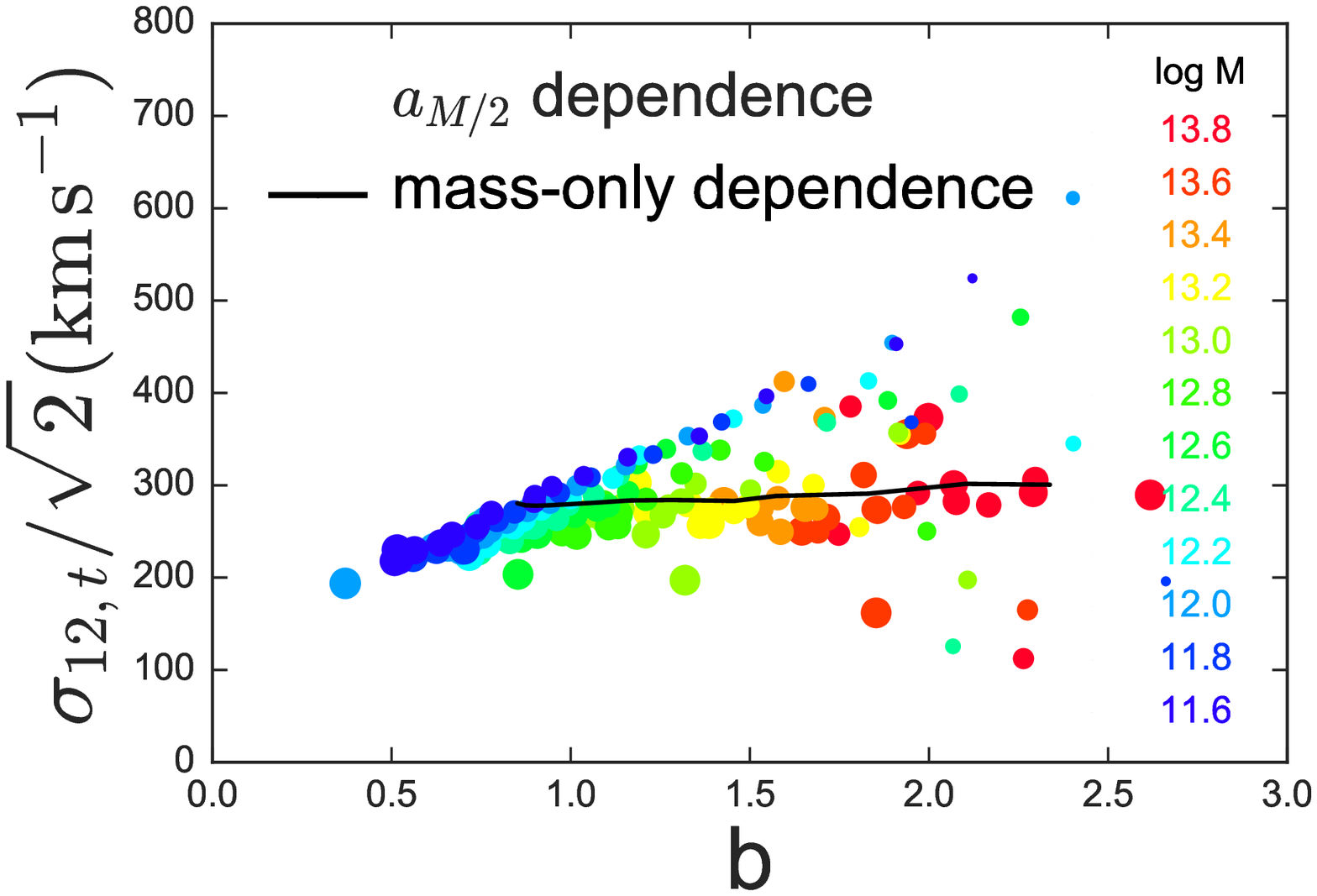}
    \end{subfigure}
    \hfill
    \begin{subfigure}[h]{0.24\textwidth}
        \centering
        \includegraphics[width=\textwidth]{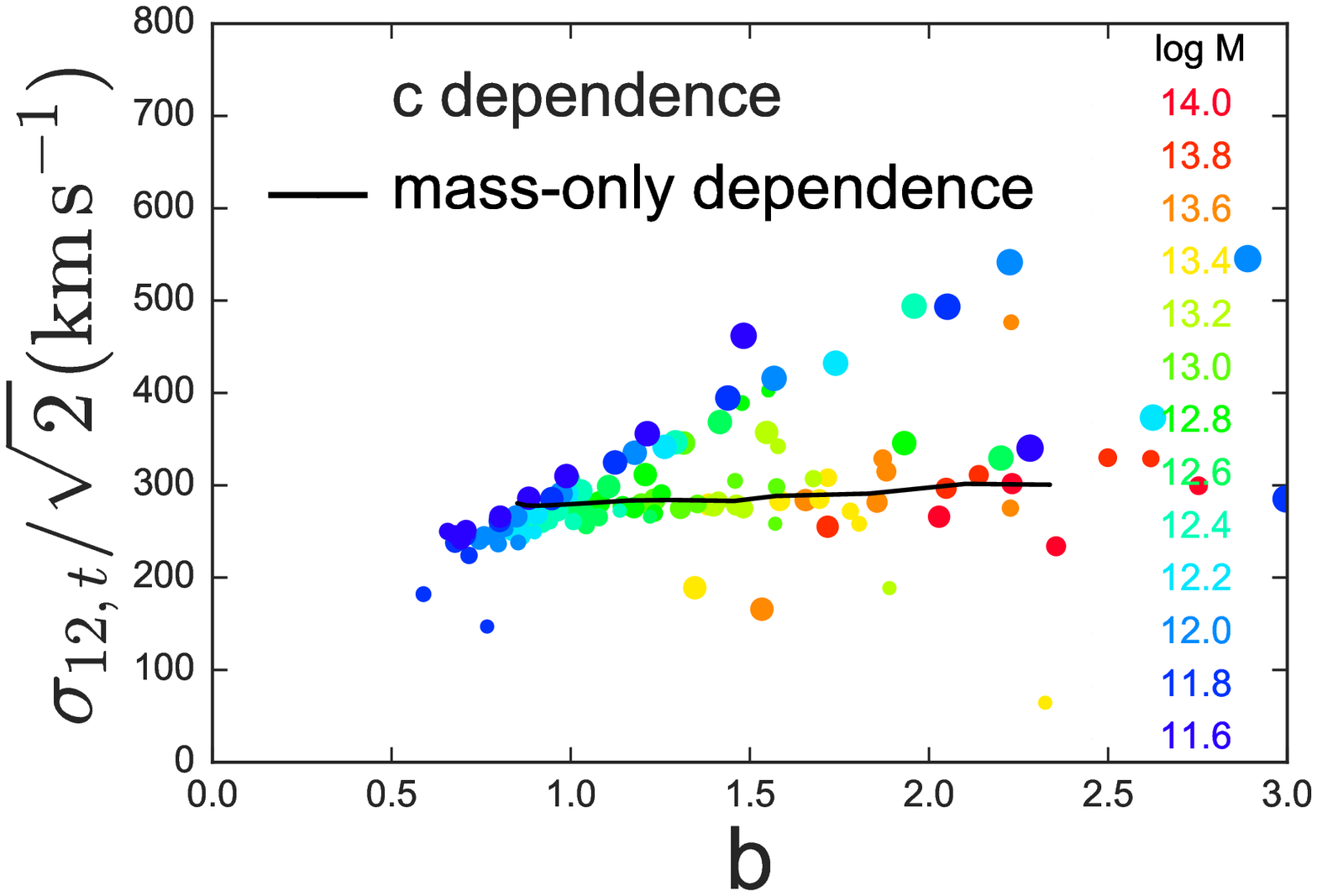}
    \end{subfigure}
    \hfill
    \begin{subfigure}[h]{0.24\textwidth}
        \centering
        \includegraphics[width=\textwidth]{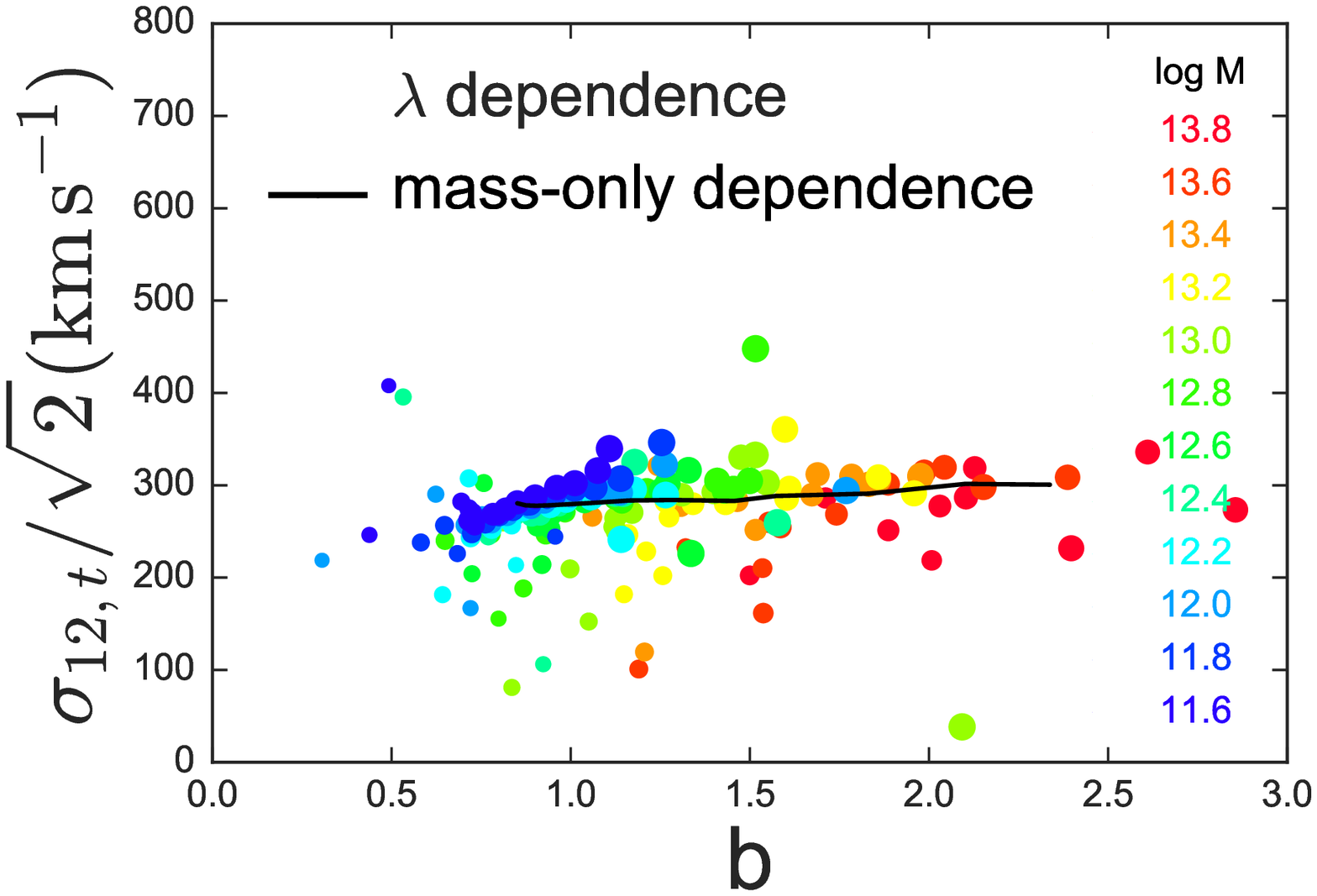}
    \end{subfigure}
\caption{Relation between pairwise velocity and velocity dispersions and halo bias. 
Top, middle, and bottom panels show pairwise radial 
velocity, pairwise radial velocity dispersion, and pairwise transverse 
velocity dispersion versus halo bias, respectively. In each panel, the points are 
colour-coded by halo mass and their size indicates the value of the corresponding
assembly variable. The pairwise velocity and velocity dispersions are evaluated at 
$r\sim 10\hinvMpc$. The solid curve is computed from mass-only dependence of halo 
clustering.
}
\label{fig:v_b}
\end{figure*}

For the joint dependence of pairwise velocity and velocity dispersions on halo mass
and each assembly variable, we plot $v_{\rm 12,r}$, $\sigma_{\rm 12,r}$, and
$\sigma_{\rm 12,t}/\sqrt{2}$ at the scale $r\sim 5\hinvMpc$ in Fig.~\ref{fig:vel_m_A}.
The trend is similar to the joint dependence for halo bias 
(Fig.\ref{fig:joint_dep}), and at fixed mass, more strongly clustered haloes display higher
pairwise velocity and velocity dispersions.

\subsection{Relation between spatial clustering and pairwise velocity under assembly effect}

The overall pattern of pairwise velocity and velocity dispersions in Fig.~\ref{fig:vel_m_A} 
shows a clear correlation with that of halo bias in Fig.\ref{fig:joint_dep}. With the joint
effect of halo mass and assembly, more strongly clustered haloes tend to move faster with
higher velocity dispersions. Such a correlation motivates us to further study the relation
between spatial clustering and halo velocity field.

Under the assumption that halo pairs are conserved during evolution, a relation 
can be established between halo pairwise velocity and spatial clustering. On large 
scales, halo pairwise velocity is found to be proportional to halo bias 
\citep[e.g.][]{Sheth01b,Zhang04}, $v_{\rm 12,r} \propto b$. This would have interesting 
implications for the assembly effect. For example, in the low halo mass regime, if all
older haloes originate from stripped massive haloes so that their stronger clustering
is a manifestation of that of more massive haloes, their pairwise velocity should also
follow that of the massive haloes. We would expect a tight correlation between $v_{\rm 12,r}$
and $b$ for all subsets of haloes (divided according to mass and assembly history).

In the top panels of Fig.~\ref{fig:v_b}, the values of $v_{\rm 12,r}$ around 
$r\sim 10\hinvMpc$ and halo bias $b$ are plotted for halo subsamples in bins of halo 
mass and assembly variables. 
 
In each panel, the colour indicates the halo mass and the size of the points denotes the magnitude of the assembly variable. We note that the magnitude of the assembly bias is not the focus here. At fixed mass it is simply related to the value of the bias factor in a way seen in Fig.~\ref{fig:joint_dep}. The plot is meant to show the $v_{\rm 12,r}$--$b$ sequence induced by assembly effect within each halo mass bin (represented by one colour) and as a collection from different mass bins.

In each panel, the curve 
is the relation derived from the dependence on halo mass only, $v_{\rm 12,r}(M)$ 
versus $b(M)$, which is close to the expected linear relation. On average, 
the $v_{\rm 12,r}$--$b$ relation from the points tracks 
the curve quite well. At fixed mass, more strongly clustered haloes originated from 
assembly effect tends to have pairwise velocity shifted toward the curve (e.g. the 
blue points with the lowest mass). However, the magnitude of the shift is not large 
enough to make the points fall on top of the curve. 

That is, within each halo mass bin, the $v_{\rm 12,r}$--$b$ relation from assembly effect shows deviations from the mass-only curve. In the $v_{\rm 12,r}$--$b$ relation as a whole from different mass bins, the assembly effect causes the scatter around the mass-only relation.
The deviation and scatter from the expected $v_{\rm 12,r}$--$b$ relation implies that the environment plays a 
role more complicated than our naive expectation.

In the middle and bottom panels of Fig.~\ref{fig:v_b}, the pairwise radial and 
transverse velocity dispersions show a considerably large scatter around the 
mass-only curve. The mass-only curve is almost flat, which means that
for haloes selected by mass the pairwise velocity dispersions are nearly independent 
of halo mass. Such a behaviour can be explained in linear theory by associating haloes 
with smoothed perturbation peaks \citep[e.g.][]{Bardeen86,Sheth01}. Once haloes are
split by their assembly history, the pairwise velocity dispersions no longer follow 
the mass-only curve, especially for low mass haloes. For example, the velocity dispersions 
for the lowest mass haloes (blue points) from assembly effect almost monotonically 
increase with increasing bias, substantially deviating from the mass-only curve. Low-mass 
haloes formed earlier (or with higher \Vpeak\ or higher concentration) can have 
1D pairwise velocity dispersions as high as 600--800 ${\rm km\, s^{-1}}$, much higher 
than the mass-only value ($\sim$300 ${\rm km\, s^{-1}}$). Interestingly they are even 
higher than those of massive haloes. This is consistent with those haloes being in 
denser environment, where larger pairwise velocity dispersions are expected 
\citep{Sheth01}. The other possible origin of the high velocity dispersion is that
some low-mass haloes are stripped haloes ejected from massive haloes.

The above results indicate that assembly effect in the kinematics of haloes, in 
particular in the pairwise velocity dispersions, does not follow a simple description
and that the environment can play an important role in shaping it. 

As the assembly bias for low mass haloes is shown to be connected to their tidal environment \citep[e.g.][]{Hahn09,Borzyszkowski17,Paranjape18}, studying the dependence of pairwise velocity statistics on the tidal environment (e.g. characterised by the tidal tensor) would be a useful step to help understand the assembly effect on the halo kinematics.

The assembly effect on halo kinematics can affect the redshift-space clustering of
haloes, raising the possibility of using redshift-space clustering to detect assembly 
effect with galaxy clustering. We analyse the redshift-space two-point correlation 
function and the multiple moments (monopole, quadrupole, and hexadecapole) of halo 
samples in mass and assembly variable bins. We find that the assembly effect on 
redshift-space clustering is mainly through the bias factor and that halo kinematics 
only introduce a minor effect. Similar to the scale-dependence case 
(\S~\ref{sec:scale_dep}), the halo assembly effect may be hard to be revealed in practice
from redshift-space clustering, but it deserves a further, detailed study.

%5
\section{Summary and Discussion}
\label{sec:conclusion}

Using dark matter haloes identified in the MultiDark MDR1 simulation, we 
investigate the joint dependence of halo bias on halo mass and halo assembly
variables and the assembly effect on halo kinematics.

For the halo assembly variables (peak maximum circular velocity \Vpeak, 
halo formation scale factor \ahalf, halo concentration $c$, and halo spin 
$\lambda$) considered in this work, the joint dependence of halo bias on 
halo mass and each assembly variable can be largely described as that halo
bias increases outward from a point of global minimum in the plane of
mass and assembly variable. All previous results of halo assembly bias measured
for certain percentiles of a given halo assembly variable can be inferred from
the above dependence and the distribution of assembly variable. We explore the
possibility of finding an effective halo variable to minimise the assembly
bias by using a combination of halo mass and spin. While an effective halo mass
constructed through the combination absorbs the dependence of assembly bias 
on halo spin, at fixed effective mass, halo bias still depends on other
assembly variables. The investigation indicates that assembly bias is 
multivariate in nature and that it is unlikely for one halo variable to 
absorb every aspect of the assembly effect.

From studying the joint dependence of halo bias on two assembly variables at
fixed halo mass, we find that it is not necessarily true to predict the
trend of assembly bias for one assembly variable solely based on that in
the other assembly variable and their correlation. It only becomes possible 
if the gradient in halo bias follows the correlation direction. Whether
the gradient and correlation directions align with each other relies on which
two assembly variables to choose. It also depends on halo mass -- with respect 
to the correlation direction, the gradient direction rotates as halo mass varies, 
showing the non-trivial nature of assembly bias.

We also study the kinematics of haloes under the assembly effect by dividing haloes
according to halo mass and assembly variables. In general, more strongly clustered 
haloes have higher pairwise radial velocity and higher pairwise velocity dispersions.
For low-mass haloes showing higher bias caused by assembly effect, the pairwise radial 
velocity tends to approach that of massive haloes of similar clustering amplitude,
while the pairwise velocity dispersions can be substantially higher than those of
the massive haloes. The results supports the picture that the evolution of low mass 
haloes is influenced by the surrounding environment, especially the tidal field, 
and that some of low mass haloes could be ejected haloes around massive haloes. 
However, we do not find a simple description for the relation between halo kinematics 
and spatial clustering under assembly effect.

The assembly bias for low mass haloes ($\Mh\la\Mnl$) has a substantial scale
dependence, showing as a drop on scales below $\sim 3\hinvMpc$. The scale 
dependence of assembly bias and the assembly effect on halo kinematics can 
potentially provide an approach to identify assembly effects in galaxy clustering 
data through the shape of the scale-dependent galaxy bias and redshift-space 
distortions. However, the effect can be subtle, which can be masked by the one-halo 
term of galaxy clustering and the scatter between the galaxy properties and
halo assembly. Further study is needed to see how and whether this approach 
works with high precision galaxy clustering data by incorporating the description 
of galaxy-halo connection with assembly effect included.

\section*{Acknowledgements}

We thank Kyle Dawson for useful discussions and the anonymous referee for constructive comments. The support and resources from the Center for High Performance Computing at the University of Utah are gratefully acknowledged.
The CosmoSim database used in this paper is a service by the Leibniz-Institute for Astrophysics Potsdam (AIP). The MultiDark database was developed in cooperation with the Spanish MultiDark Consolider Project CSD2009-00064. The authors gratefully acknowledge the Gauss Centre for Supercomputing e.V. (www.gauss-centre.eu) and the Partnership for Advanced Supercomputing in Europe (PRACE, www.prace-ri.eu) for funding the MultiDark simulation project by providing computing time on the GCS Supercomputer SuperMUC at Leibniz Supercomputing Centre (LRZ, www.lrz.de).

%%%%%%%%%%%%%%%%%%%%%%%%%%%%%%%%%%%%%%%%%%%%%%%%%%

%%%%%%%%%%%%%%%%%%%% REFERENCES %%%%%%%%%%%%%%%%%%

%%%%%%%%%%%%%%%%%%%%%%%%%%%%%%%%%%%%%%%%%%%%%%%%%%

%%%%%%%%%%%%%%%%% APPENDICES %%%%%%%%%%%%%%%%%%%%%

\appendix
\section{Mass resolution effect on halo bias and pairwise velocity statistics}
\label{sec:appendix}

\begin{figure*}
	\centering
	\begin{subfigure}[h]{0.48\textwidth}
		\includegraphics[width=\textwidth]{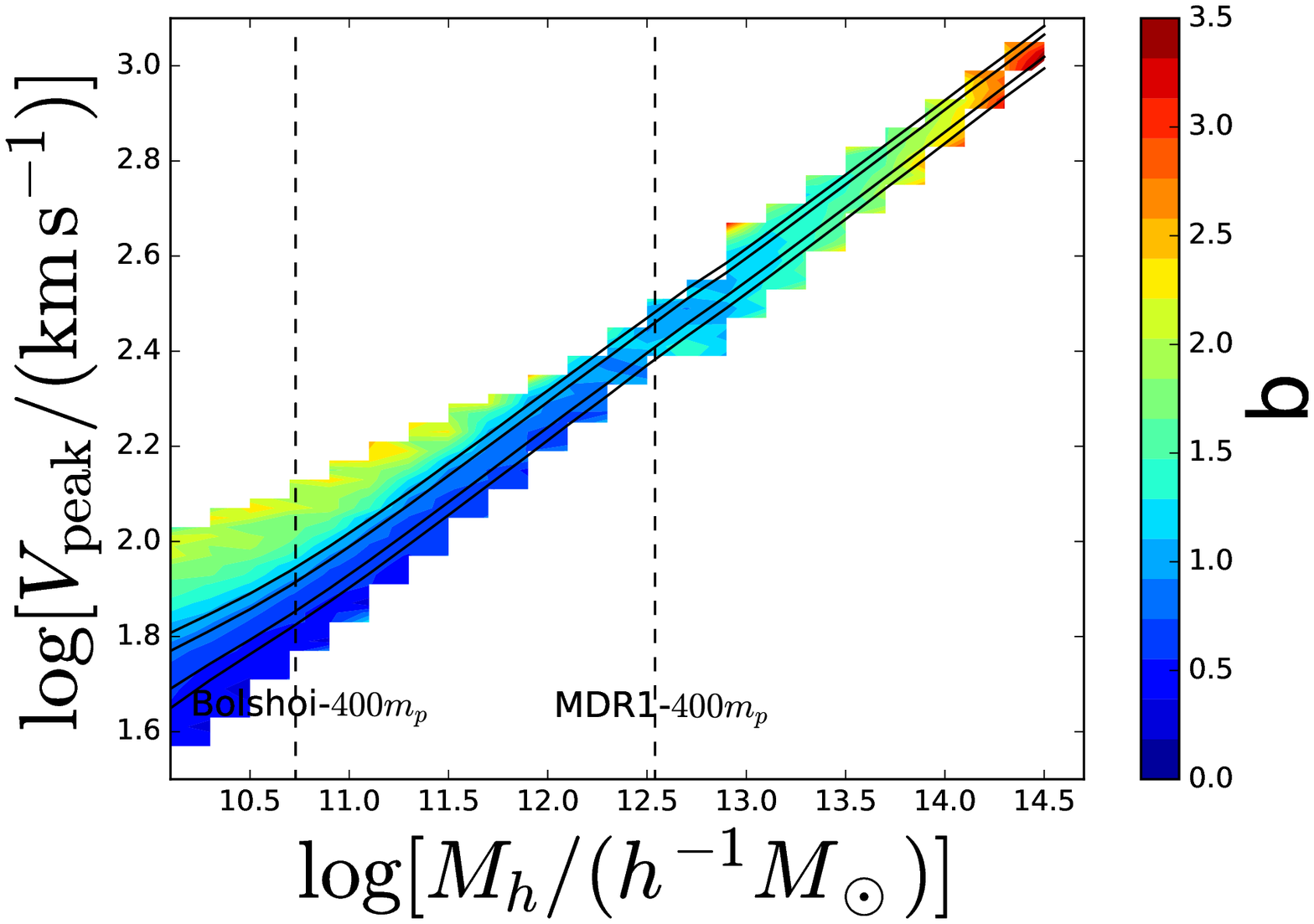}
	\end{subfigure}
	\hfill
	\begin{subfigure}[h]{0.48\textwidth}
                \includegraphics[width=\textwidth]{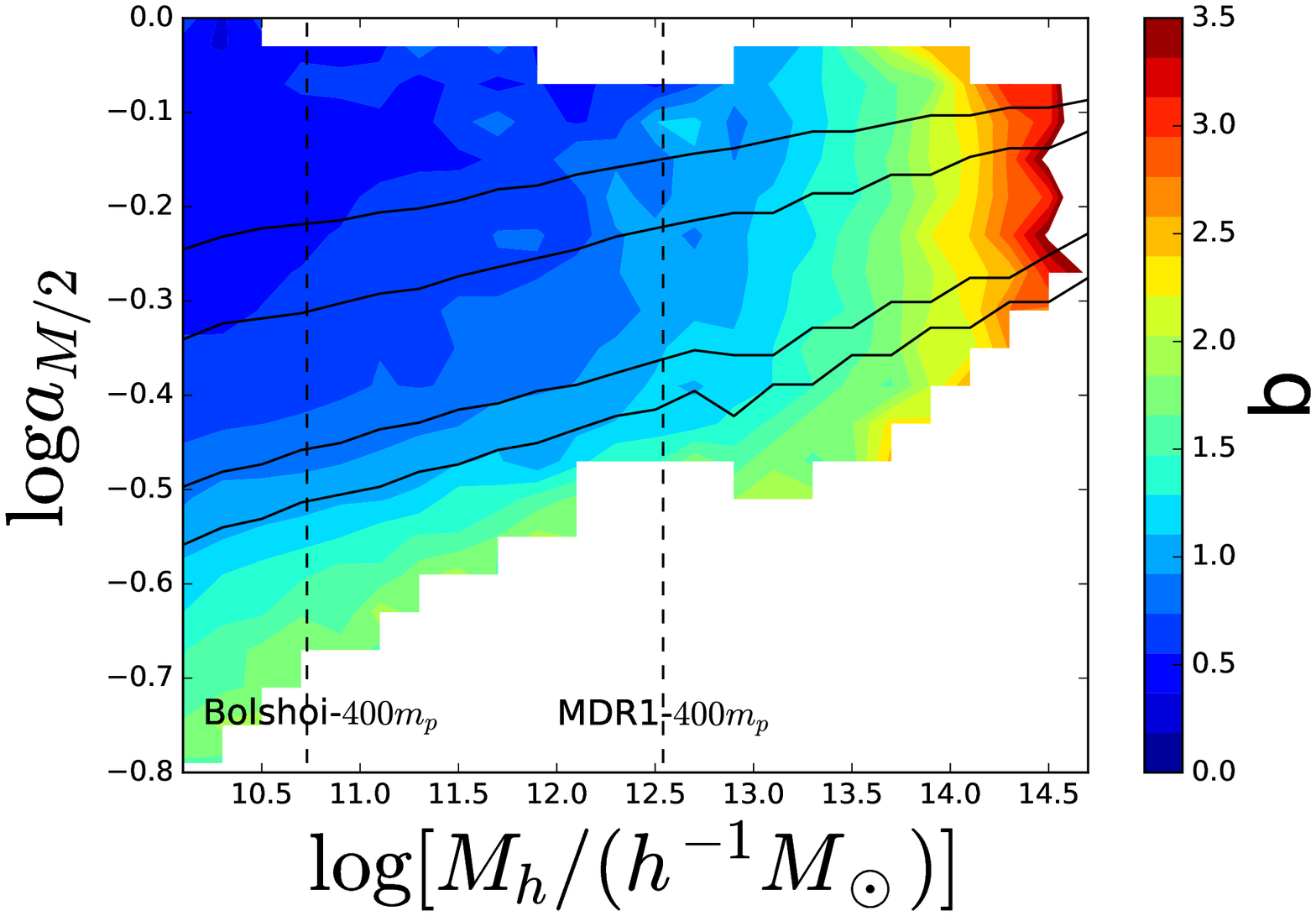}
	\end{subfigure}
	\hfill
	\begin{subfigure}[h]{0.48\textwidth}
                \includegraphics[width=\textwidth]{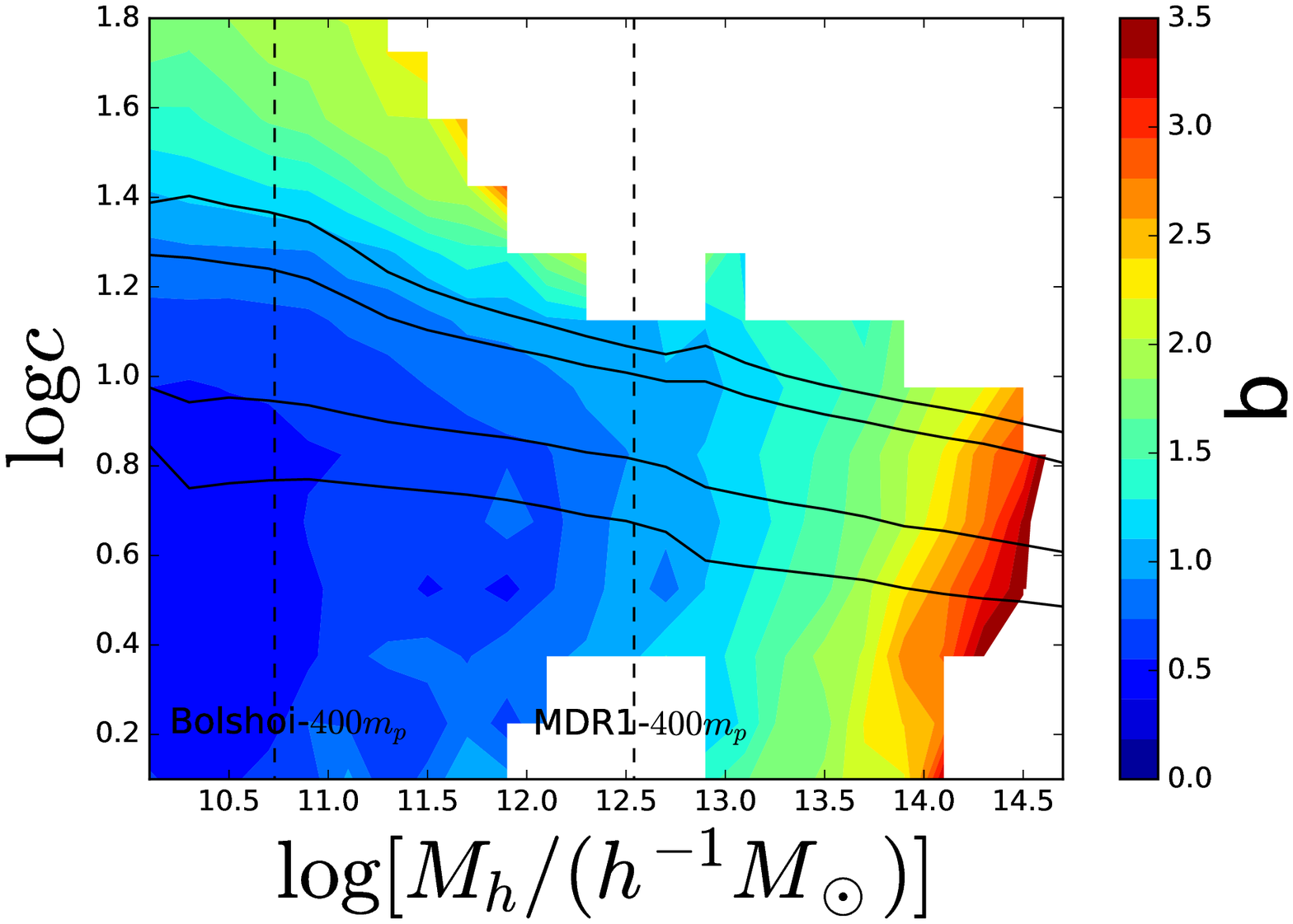}
	\end{subfigure}
	\hfill
	\begin{subfigure}[h]{0.48\textwidth}
                \includegraphics[width=\textwidth]{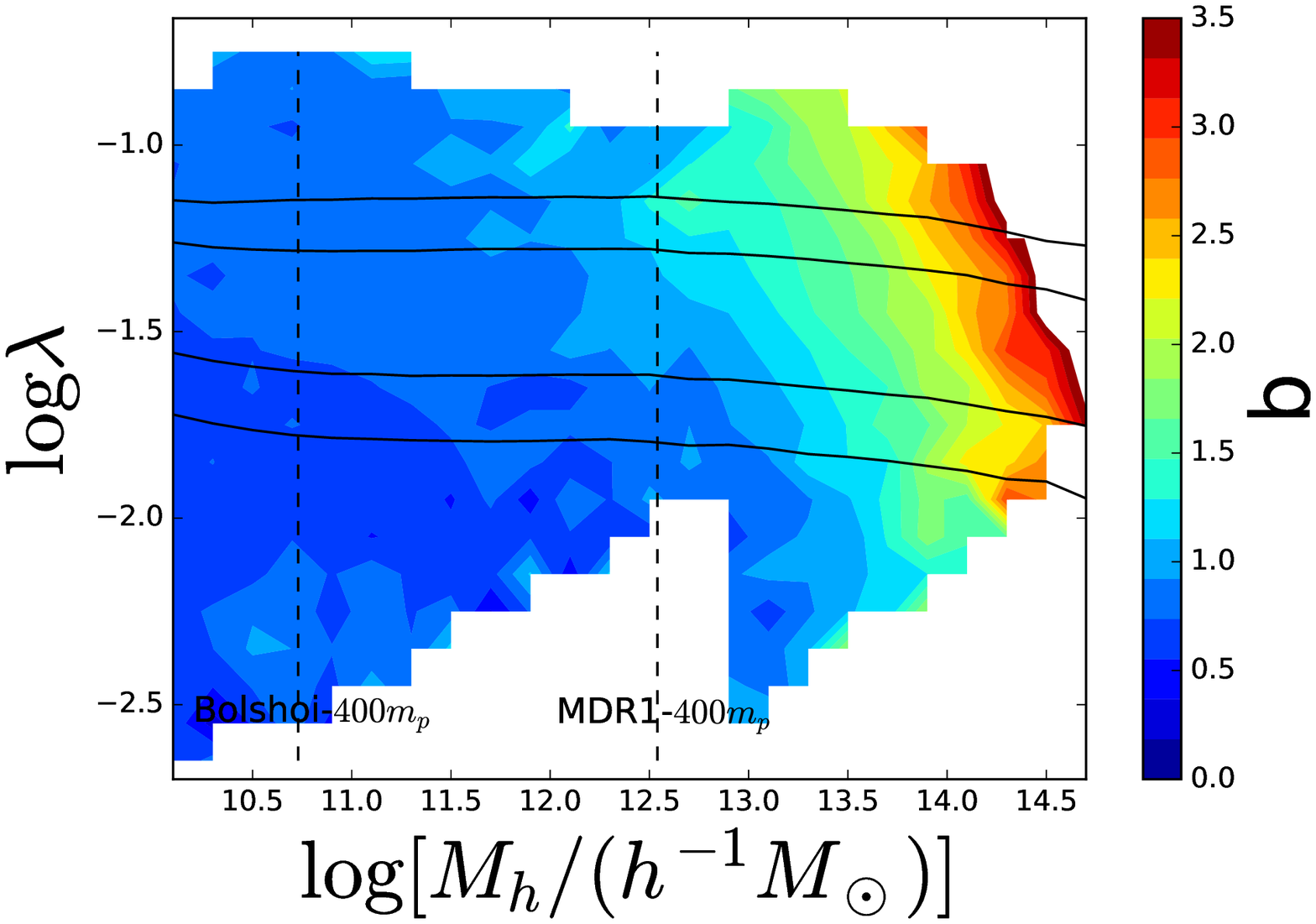}
	\end{subfigure}
	\hfill
\caption{Same as Fig.~\ref{fig:joint_dep}, but the Bolshoi simulation is used for $\log[\Mh/(\hinvMsun)]<12.8$ and MDR1 simulation for higher mass. The two vertical dashed lines in the each panel indicate the masses of haloes with 400 particles in the two simulations.
}
\label{fig:joint_dep_bolshoi}
\end{figure*}

\begin{figure*}
    \centering
    \begin{subfigure}[h]{0.24\textwidth}
        \centering
        \includegraphics[width=\textwidth]{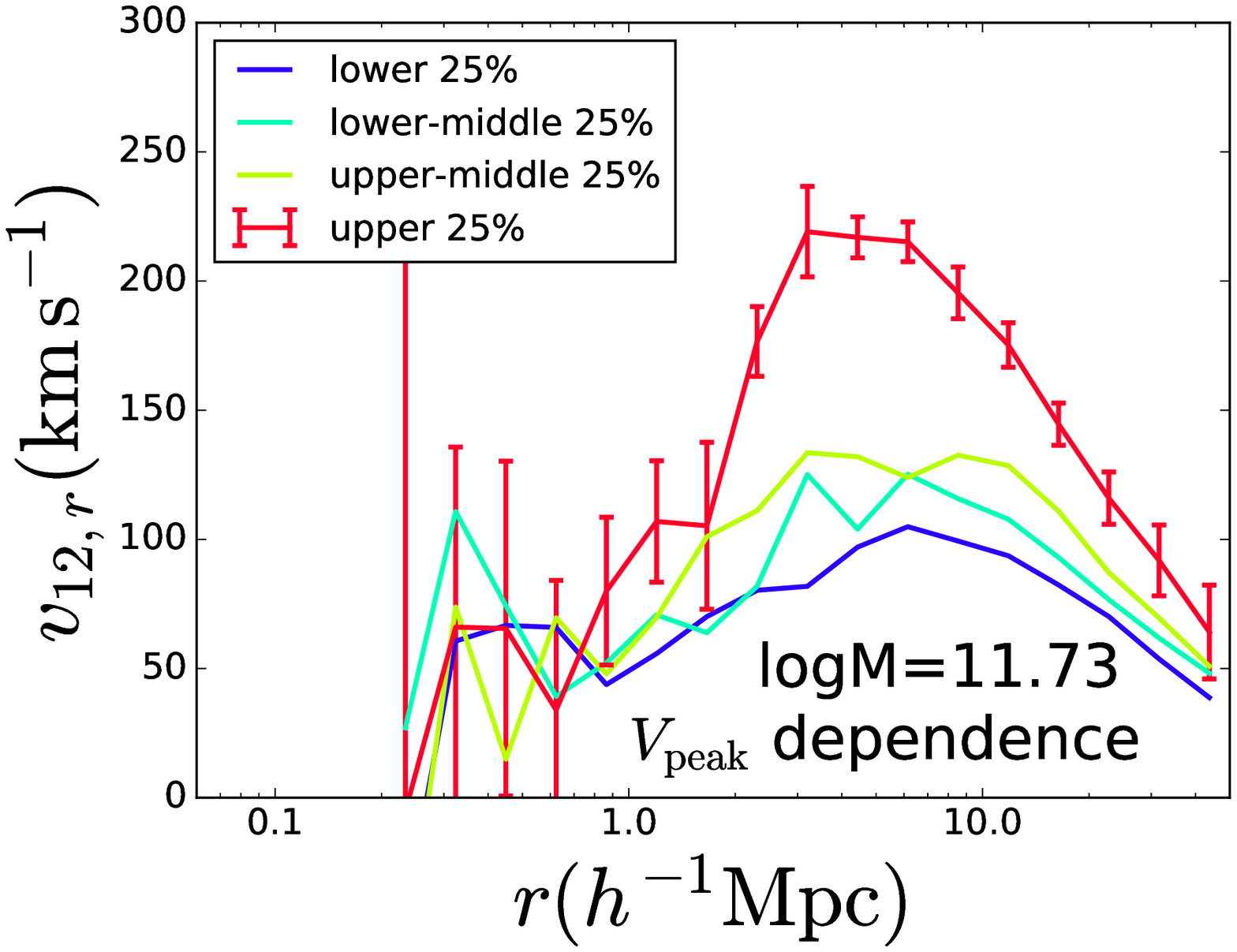}
    \end{subfigure}
    \hfill
    \begin{subfigure}[h]{0.24\textwidth}
        \centering
        \includegraphics[width=\textwidth]{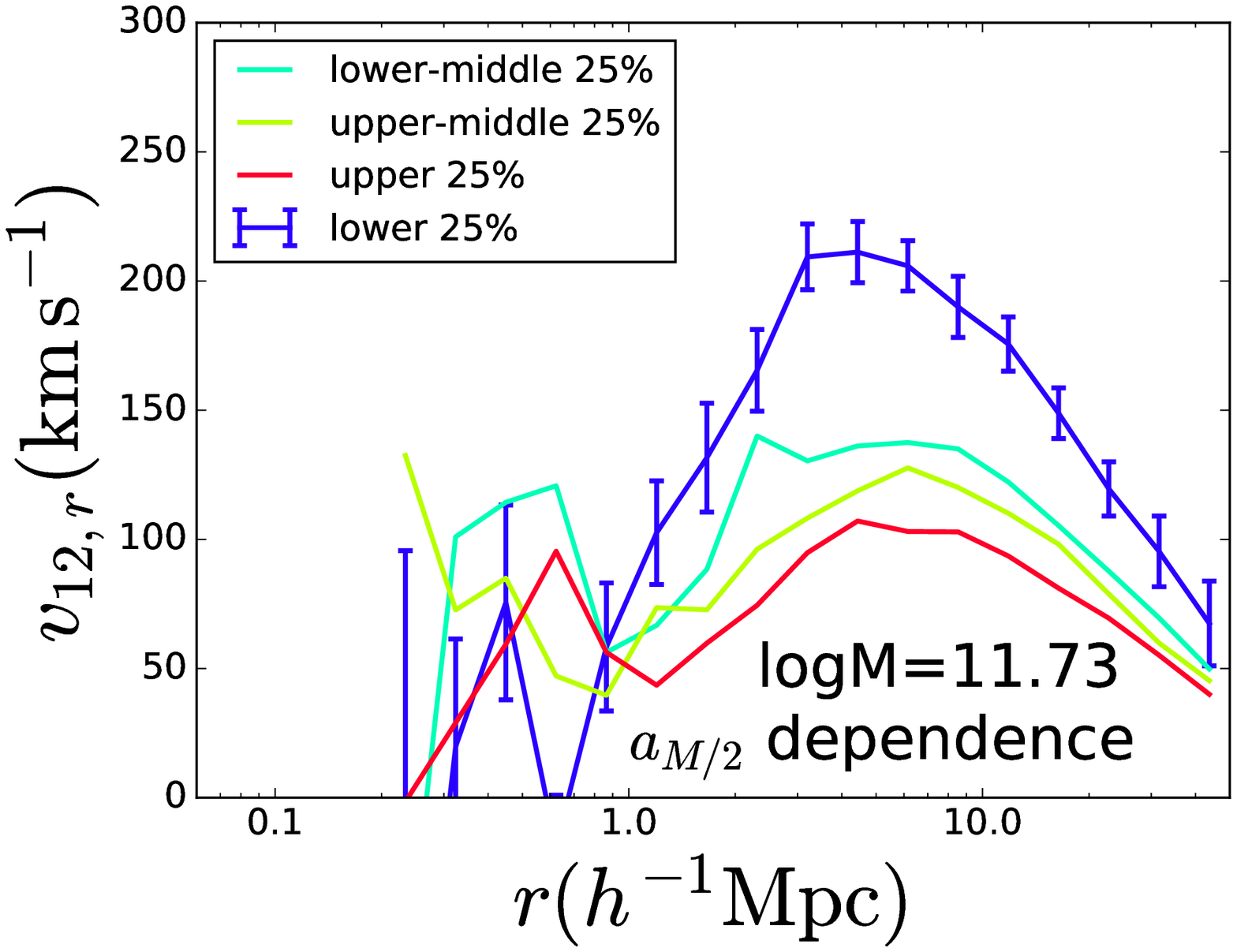}
    \end{subfigure}
	\hfill
    \begin{subfigure}[h]{0.24\textwidth}
        \centering
        \includegraphics[width=\textwidth]{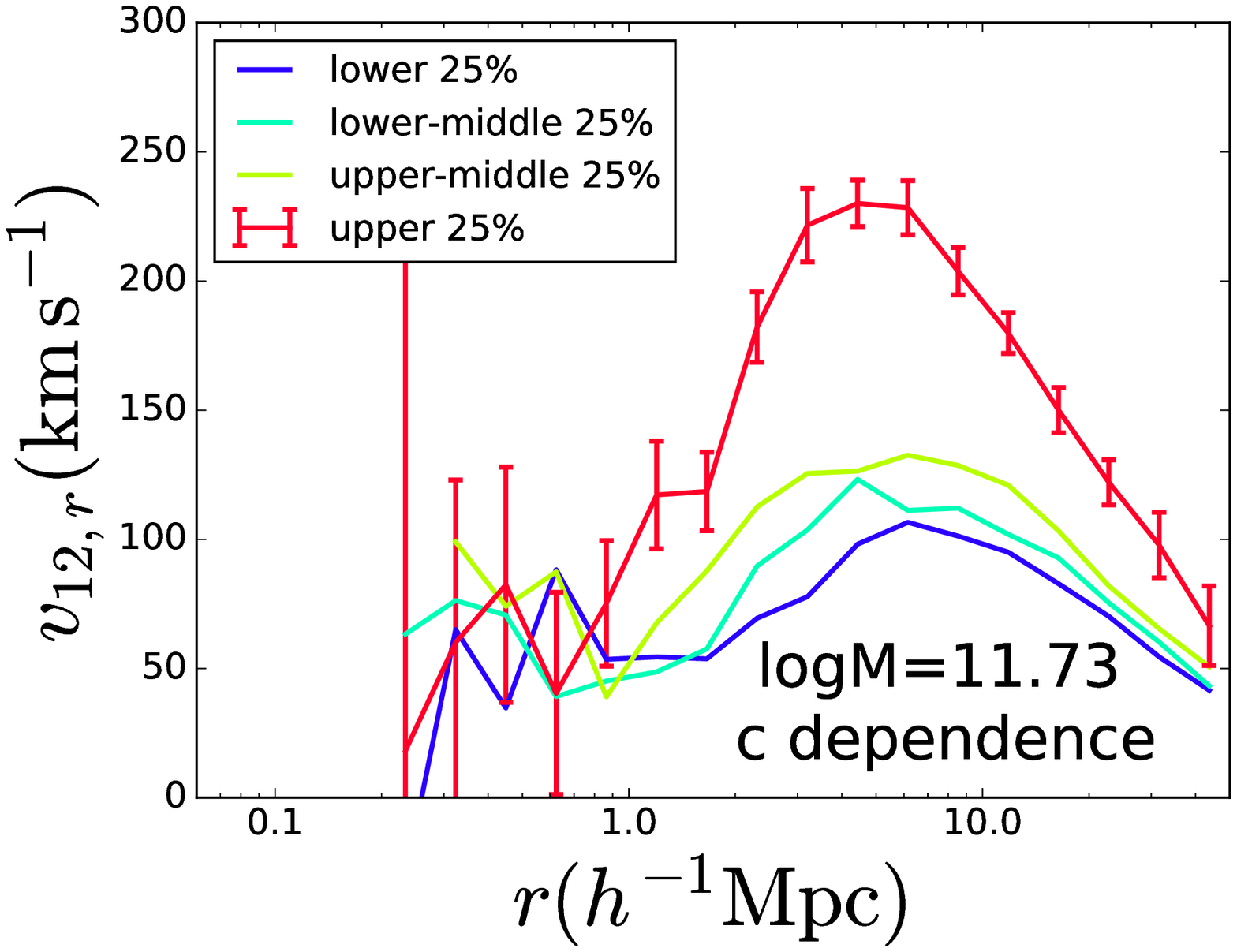}
    \end{subfigure}
	\hfill
    \begin{subfigure}[h]{0.24\textwidth}
        \centering
        \includegraphics[width=\textwidth]{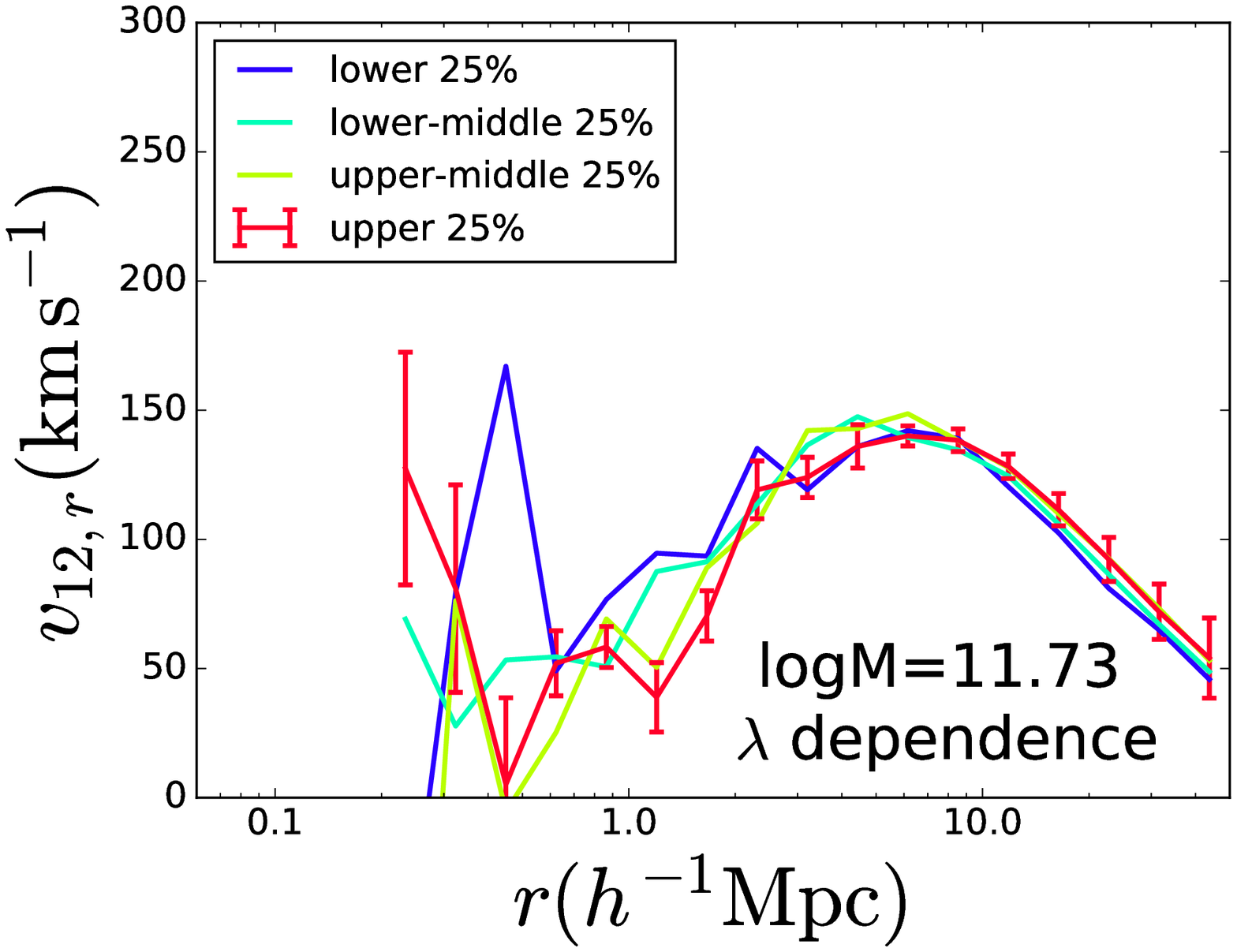}
    \end{subfigure}
	\hfill
    \begin{subfigure}[h]{0.24\textwidth}
        \centering
        \includegraphics[width=\textwidth]{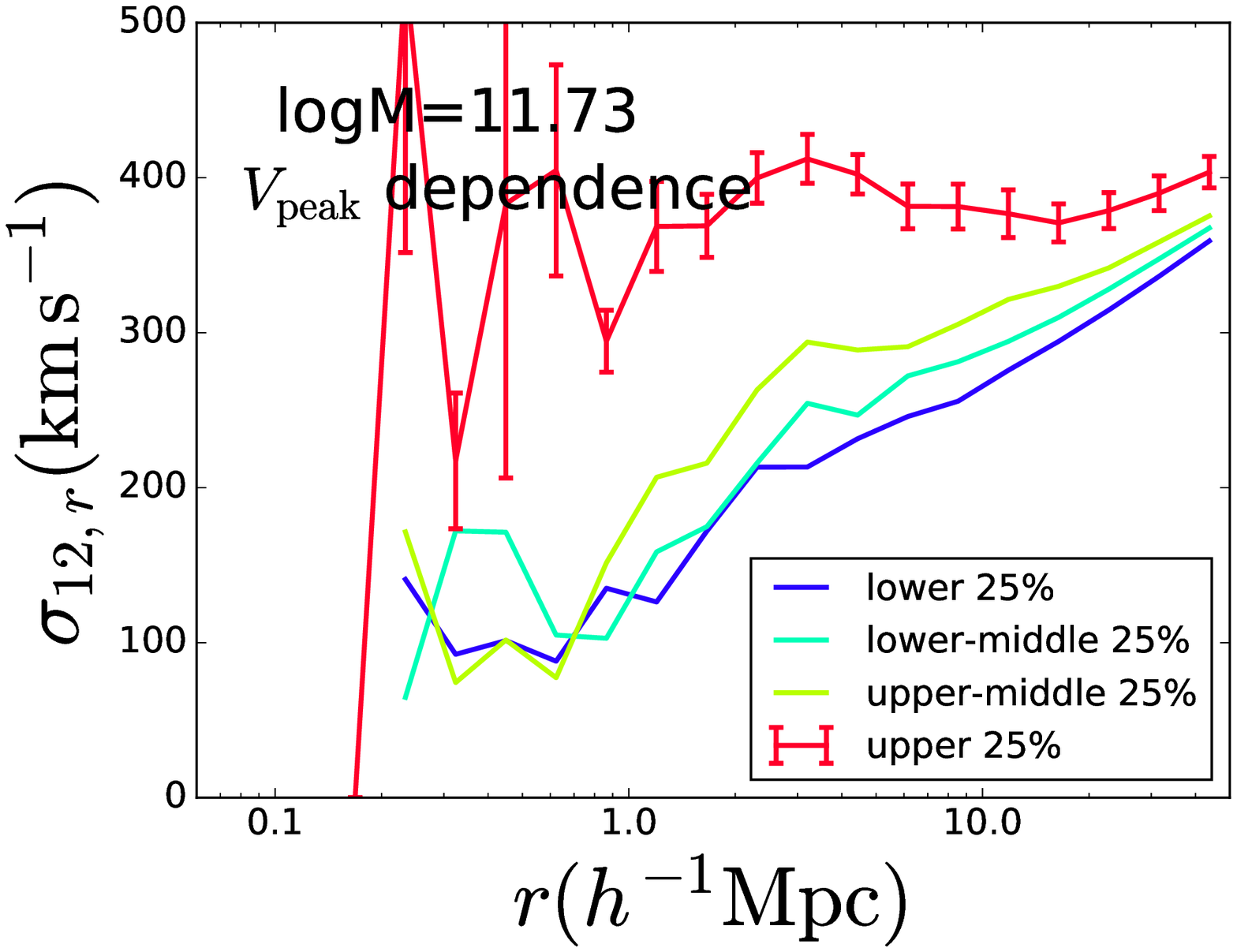}
    \end{subfigure}
	\hfill
    \begin{subfigure}[h]{0.24\textwidth}
        \centering
        \includegraphics[width=\textwidth]{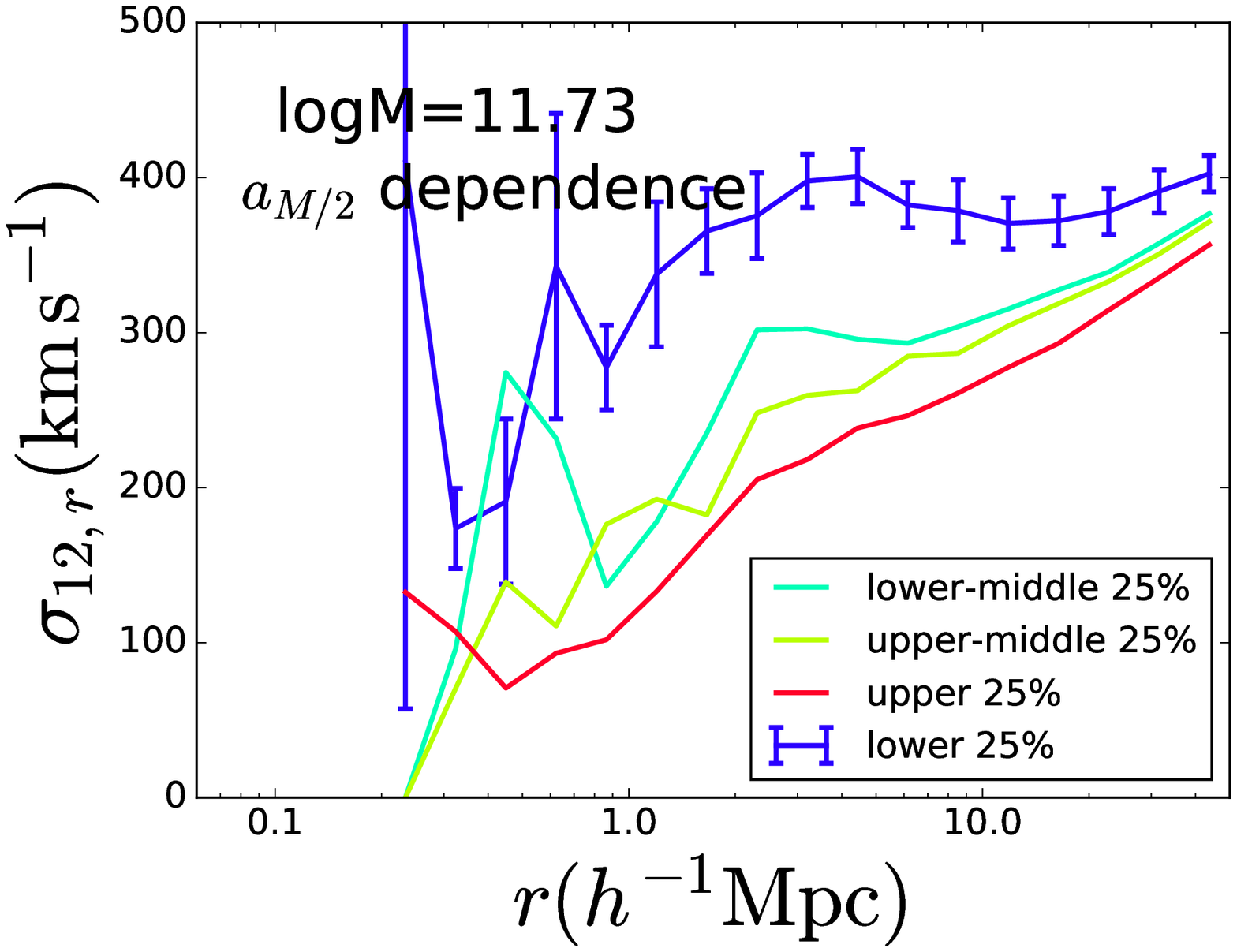}
    \end{subfigure}
	\hfill
    \begin{subfigure}[h]{0.24\textwidth}
        \centering
        \includegraphics[width=\textwidth]{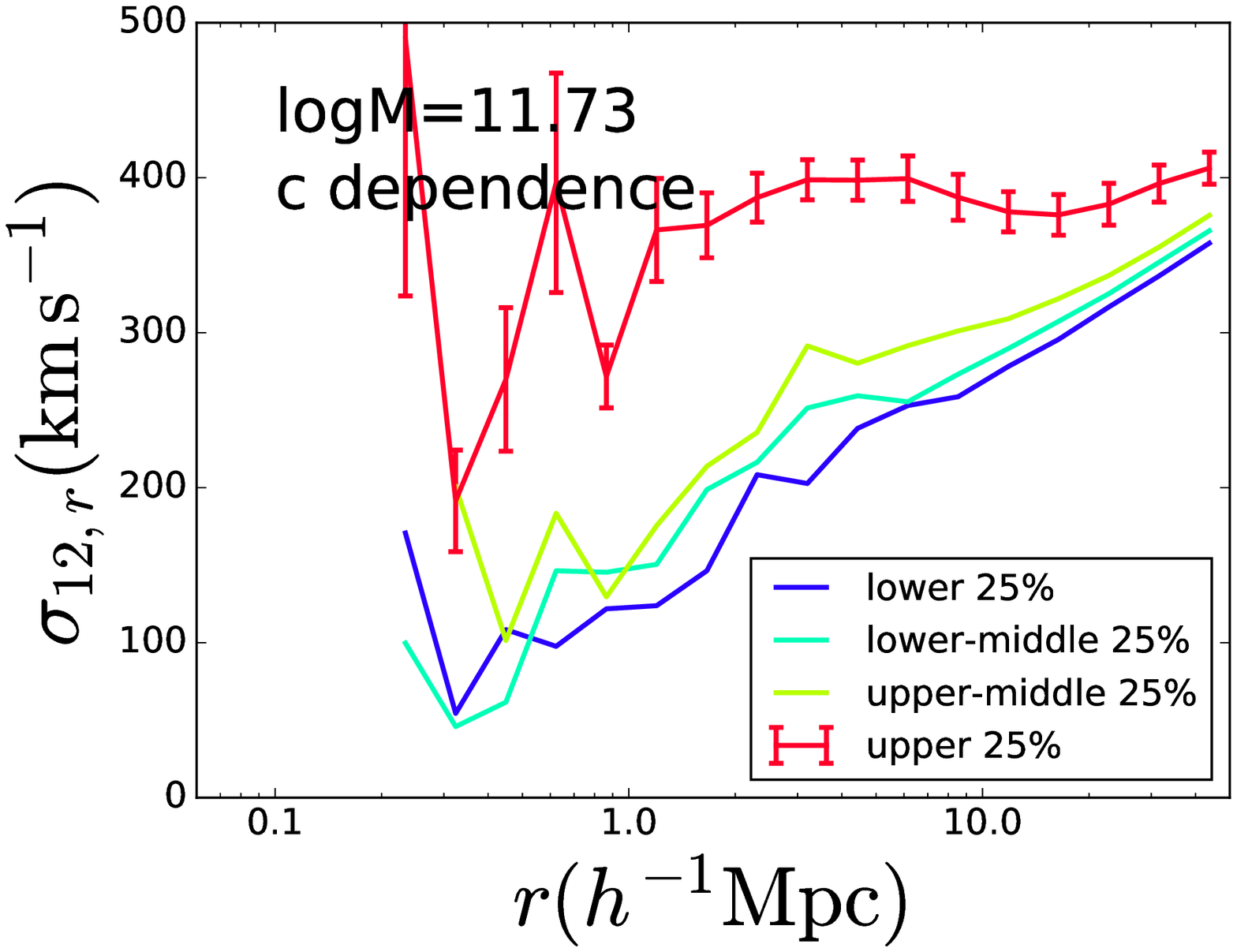}
    \end{subfigure}
	\hfill
    \begin{subfigure}[h]{0.24\textwidth}
        \centering
        \includegraphics[width=\textwidth]{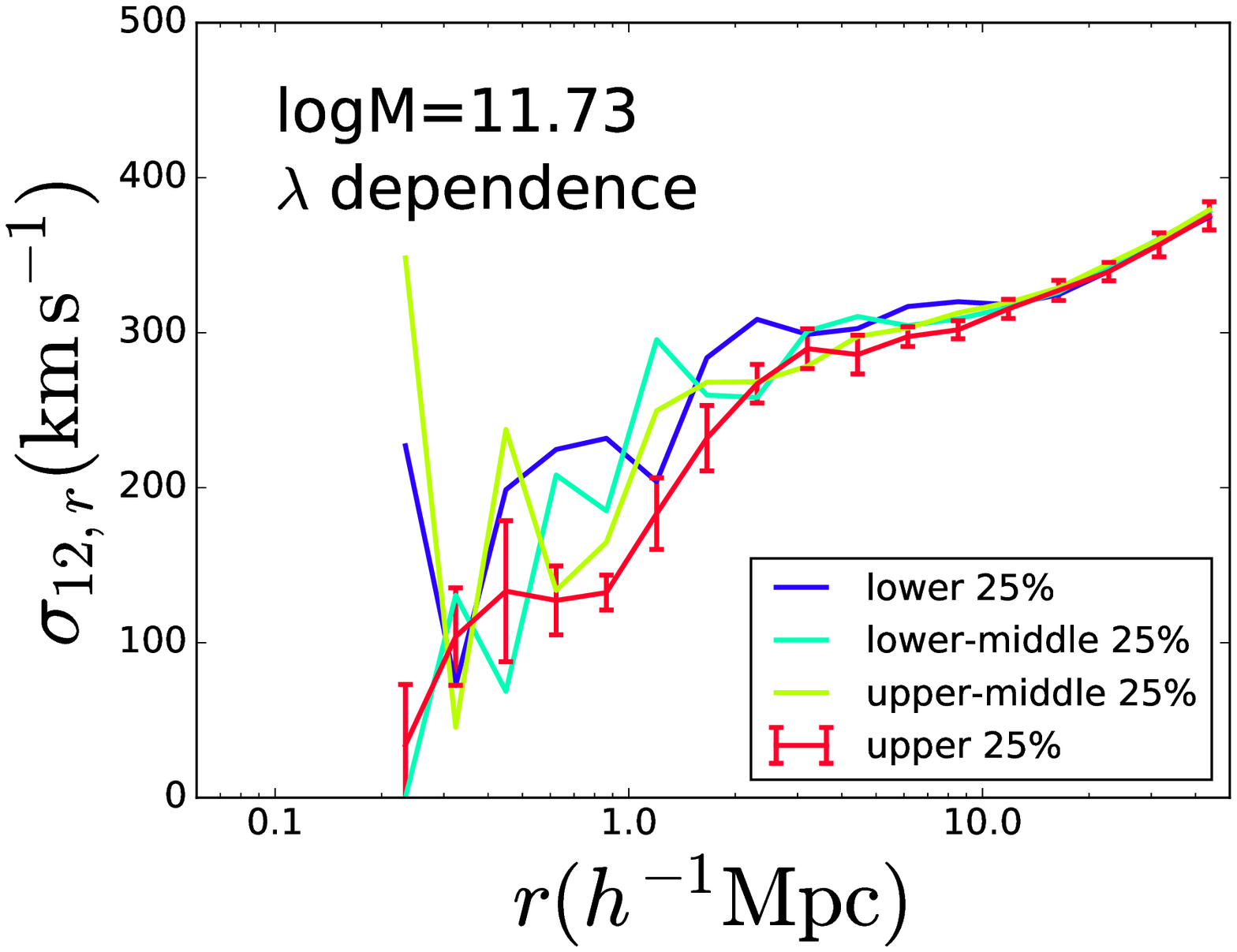}
    \end{subfigure}
    \begin{subfigure}[h]{0.24\textwidth}
        \centering
        \includegraphics[width=\textwidth]{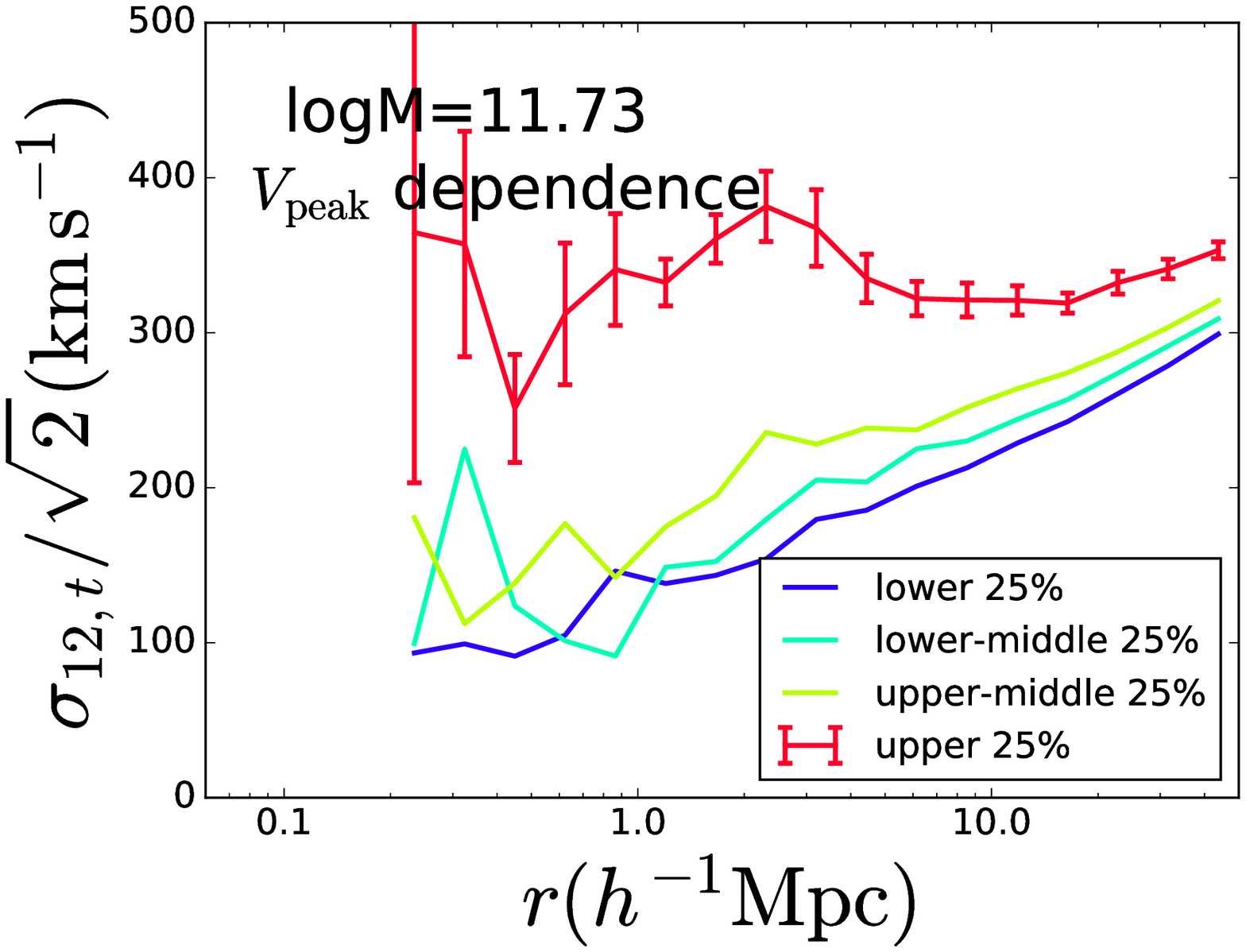}
    \end{subfigure}
    \hfill
    \begin{subfigure}[h]{0.24\textwidth}
        \centering
        \includegraphics[width=\textwidth]{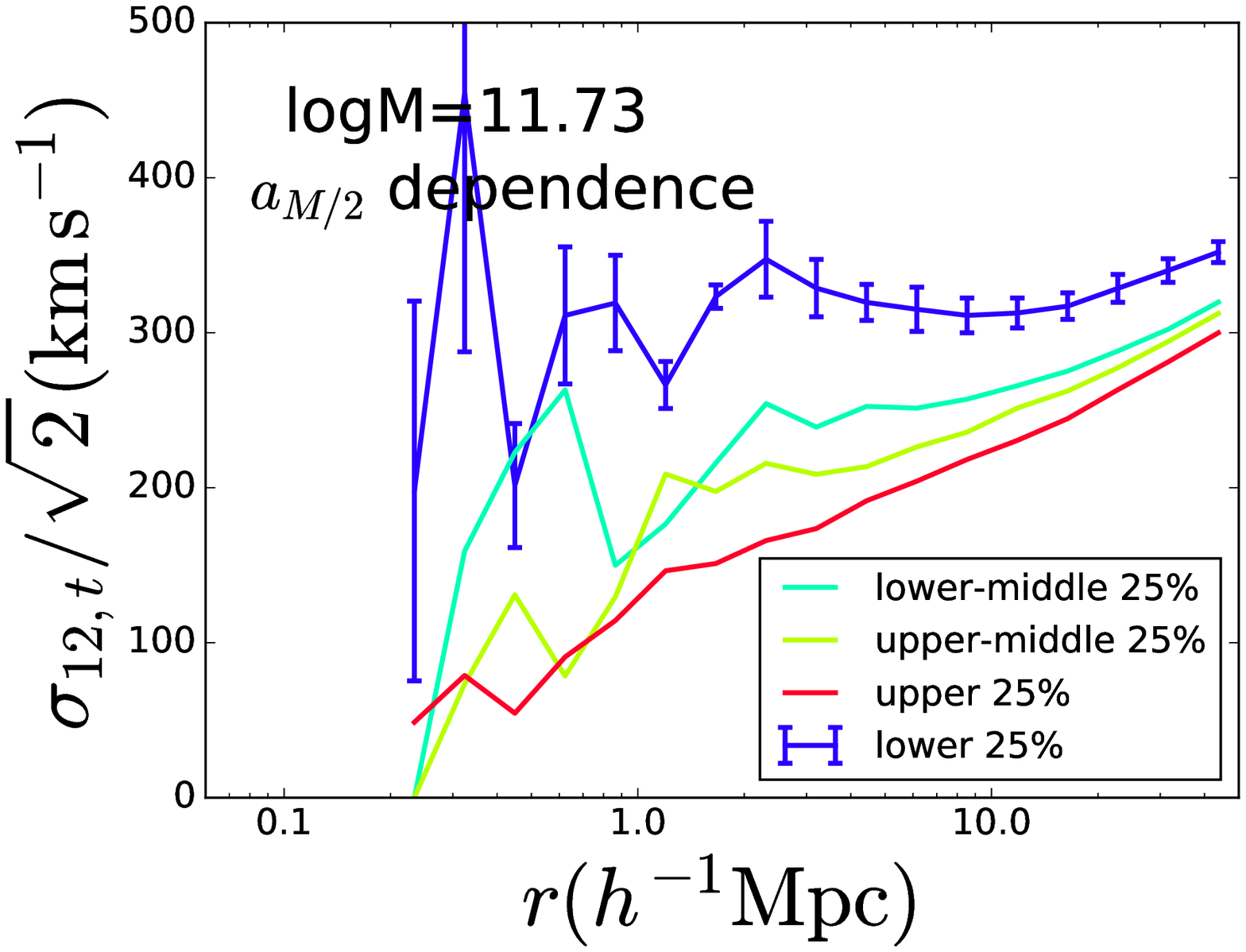}
    \end{subfigure}
	\hfill
    \begin{subfigure}[h]{0.24\textwidth}
        \centering
        \includegraphics[width=\textwidth]{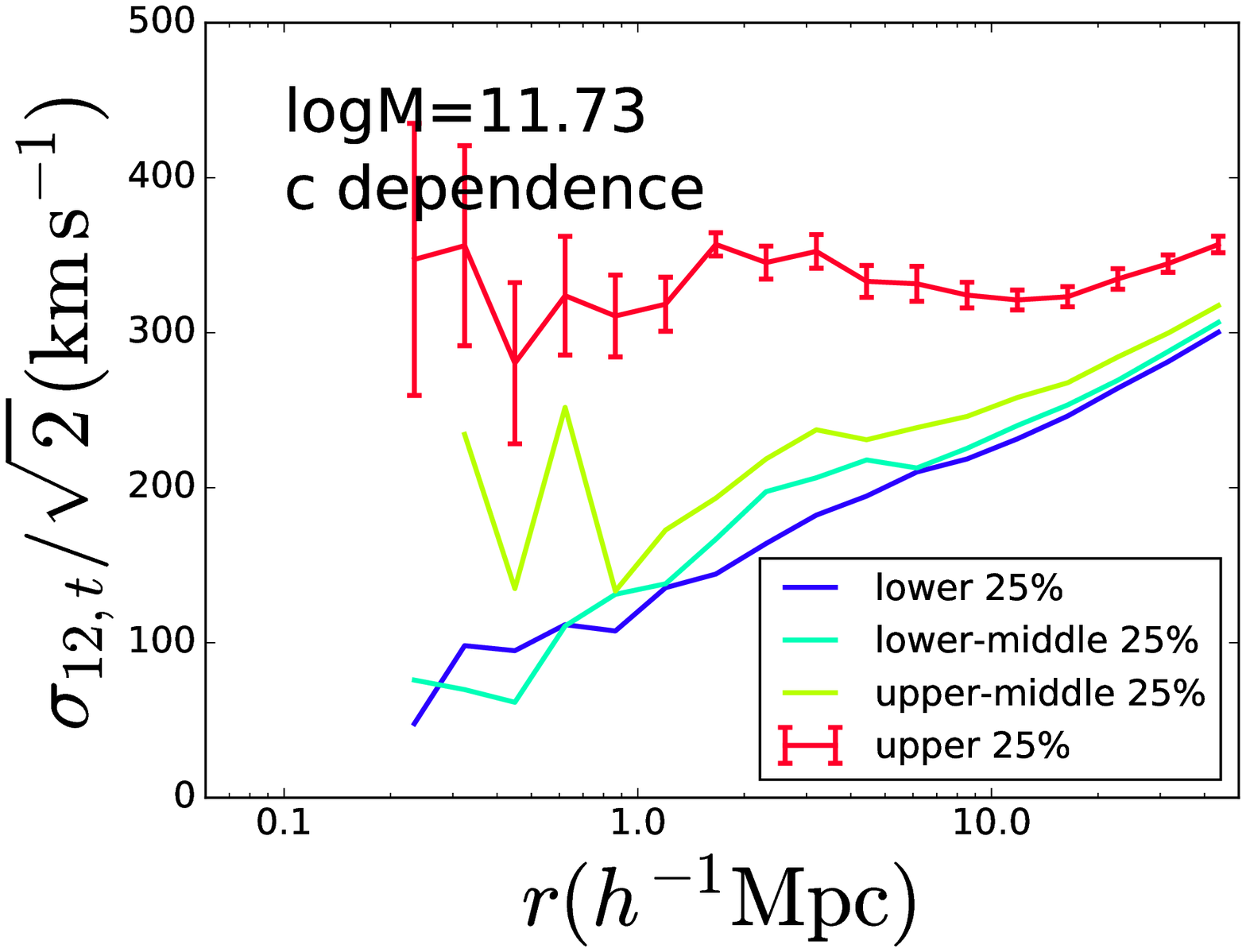}
    \end{subfigure}
	\hfill
    \begin{subfigure}[h]{0.24\textwidth}
        \centering
        \includegraphics[width=\textwidth]{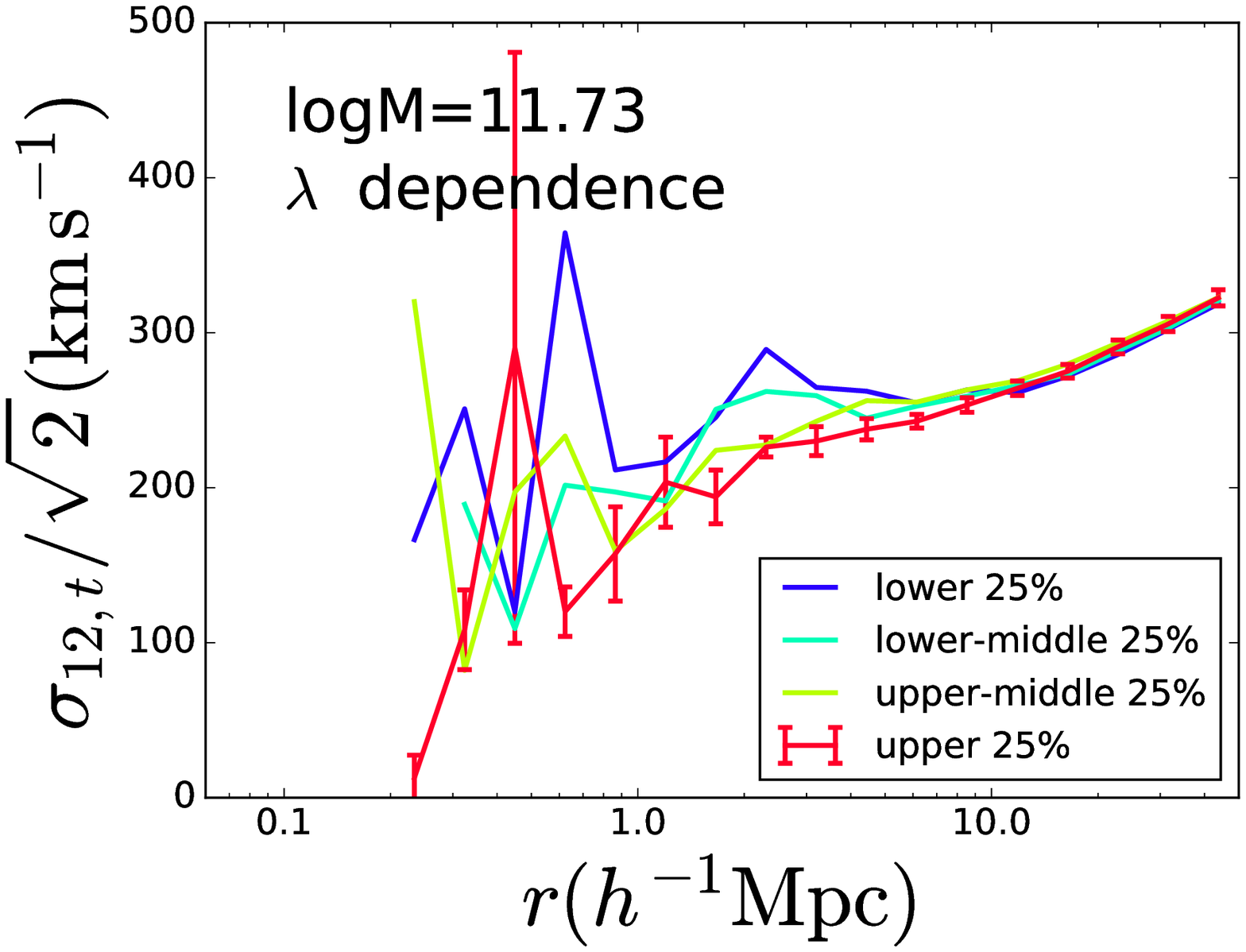}
    \end{subfigure}
\caption{
Same as Fig.~\ref{fig:vel_assembly_percentile}, but haloes from the Bolshoi simulation is used for the analysis.}
\label{fig:vel_assembly_percentile_bolshoi}
\end{figure*}

\begin{figure*}
    \centering
    \begin{subfigure}[h]{0.24\textwidth}
        \centering
        \includegraphics[width=\textwidth]{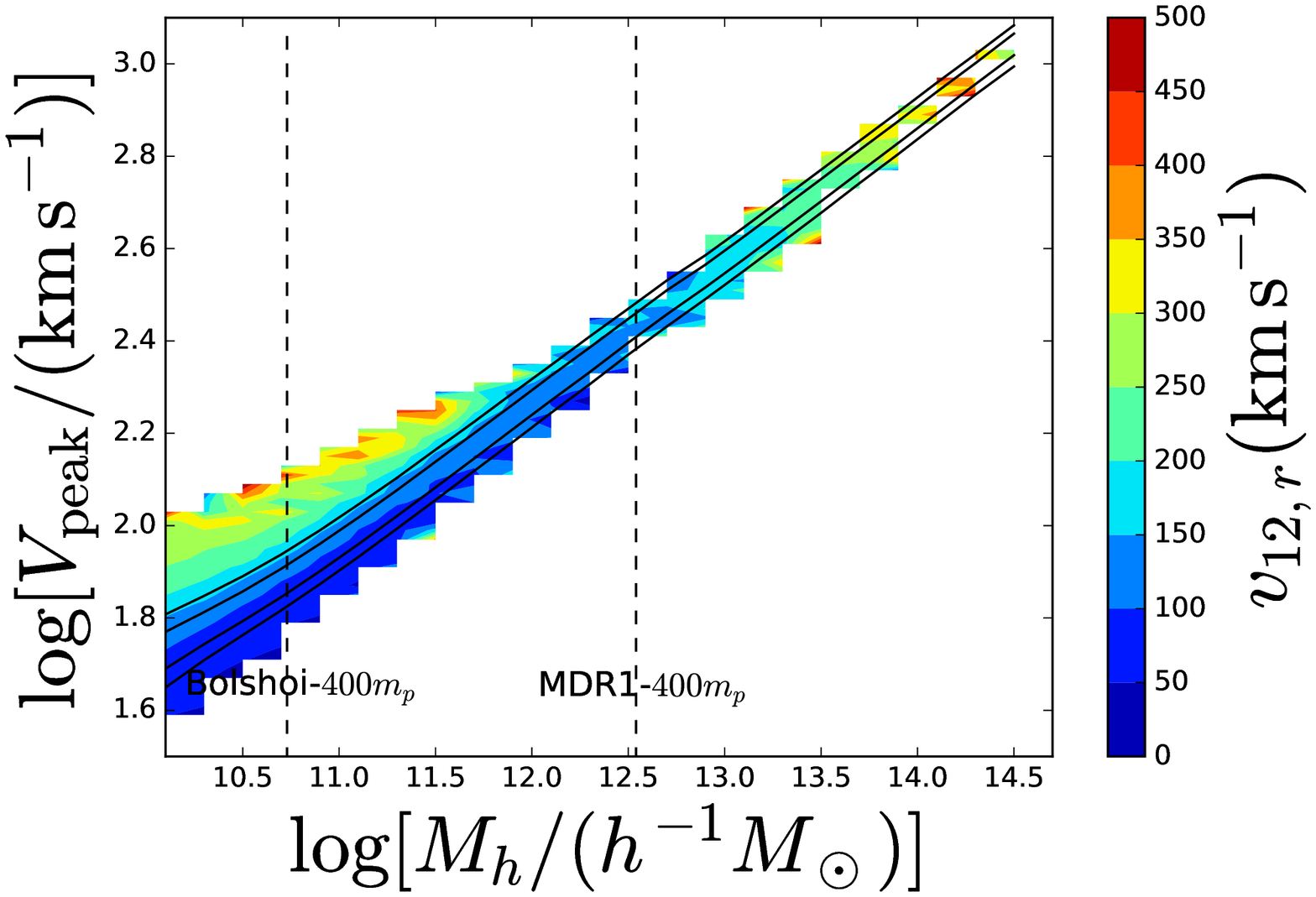}
    \end{subfigure}
    \hfill
    \begin{subfigure}[h]{0.24\textwidth}
        \centering
        \includegraphics[width=\textwidth]{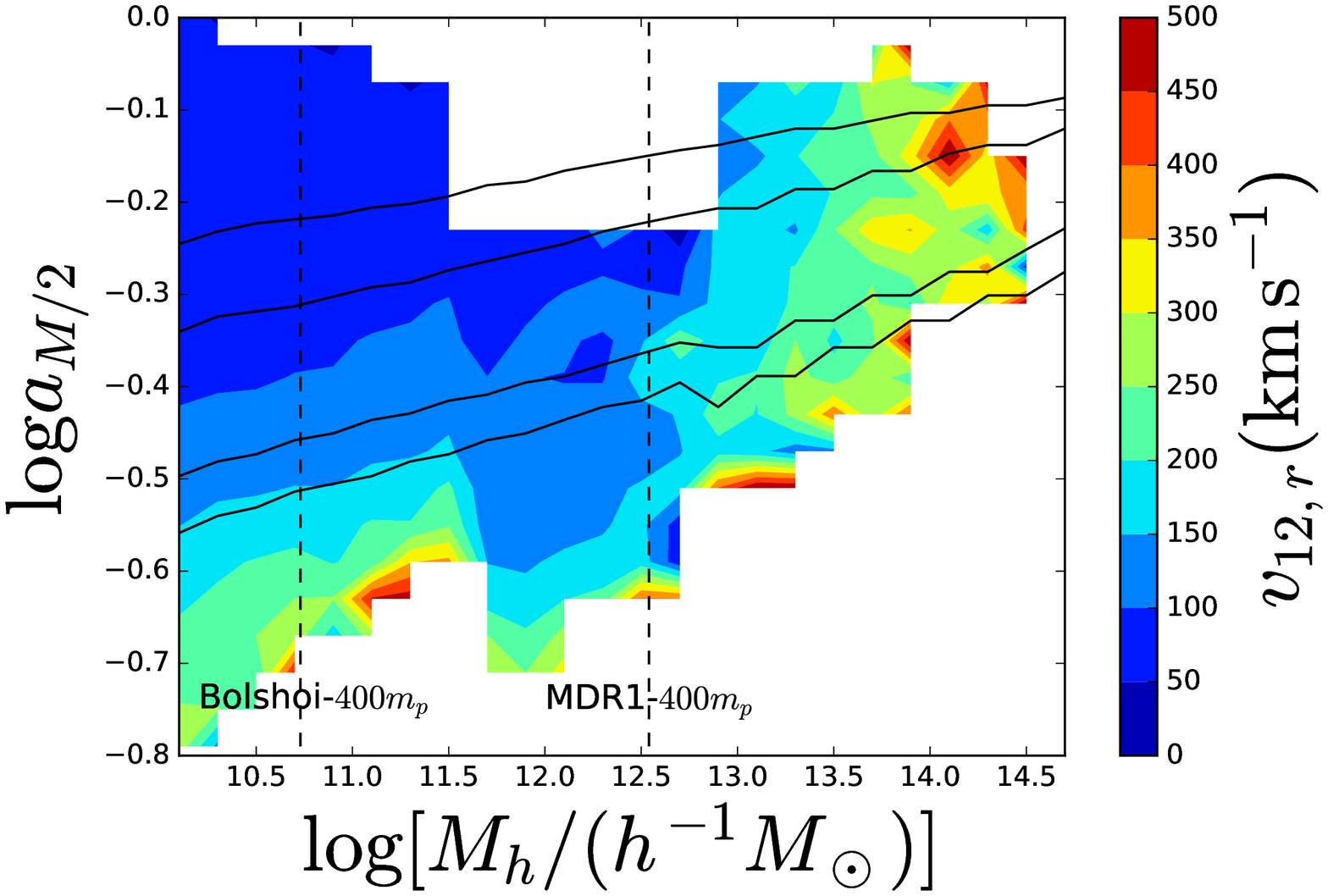}
    \end{subfigure}
    \hfill
    \begin{subfigure}[h]{0.24\textwidth}
        \centering
        \includegraphics[width=\textwidth]{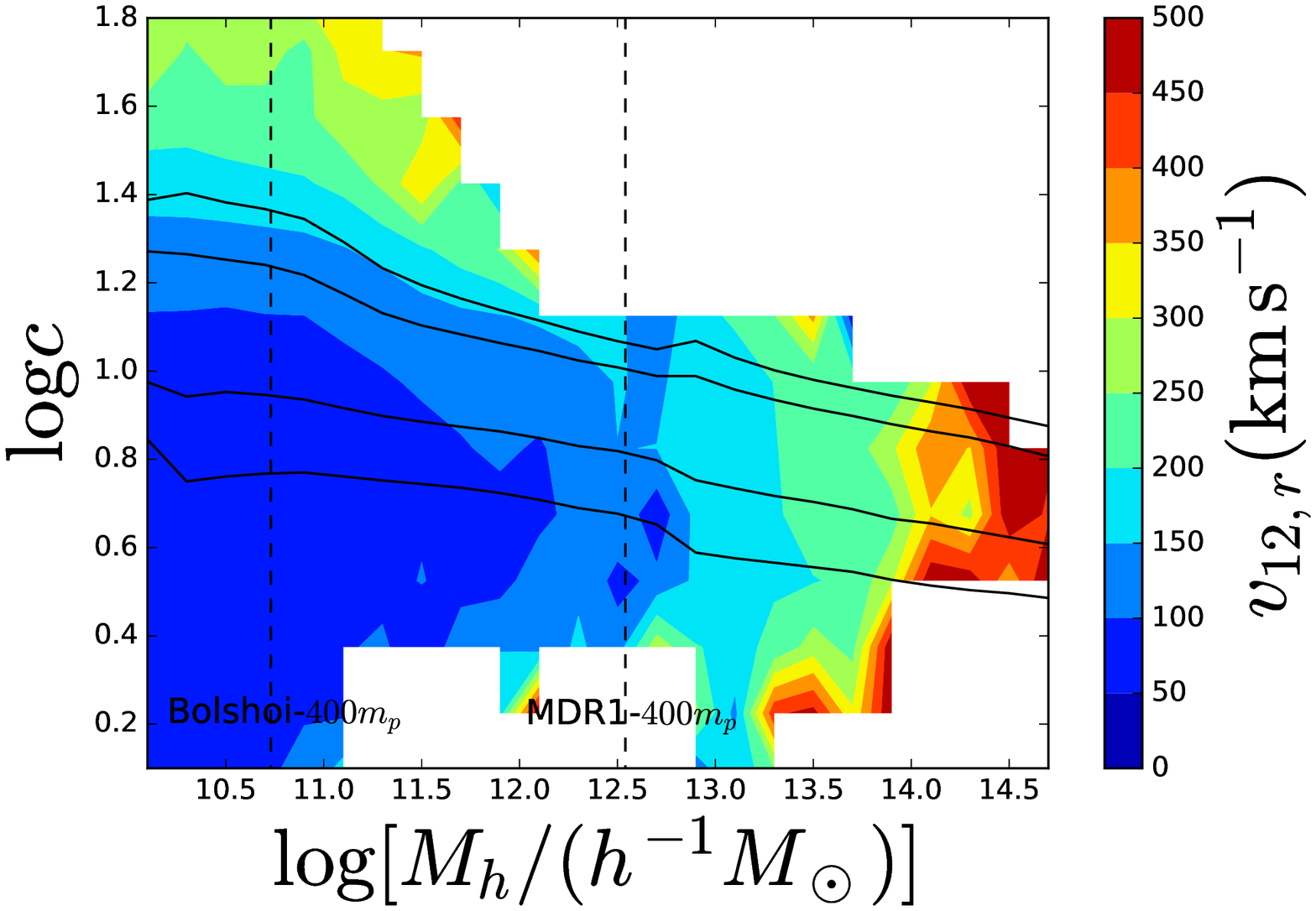}
    \end{subfigure}
    \hfill
    \begin{subfigure}[h]{0.24\textwidth}
        \centering
        \includegraphics[width=\textwidth]{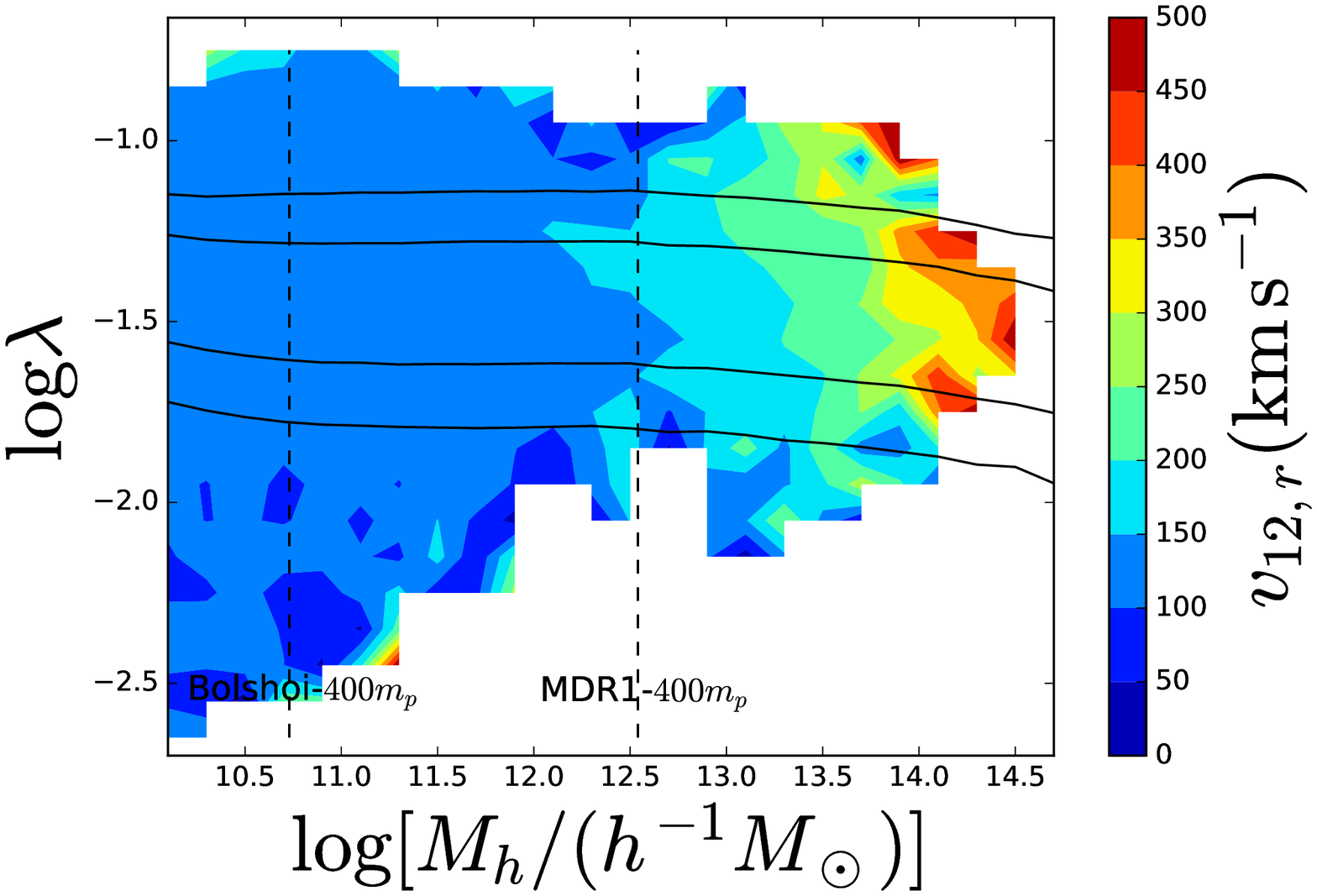}
    \end{subfigure}
    \hfill
    \begin{subfigure}[h]{0.24\textwidth}
        \centering
        \includegraphics[width=\textwidth]{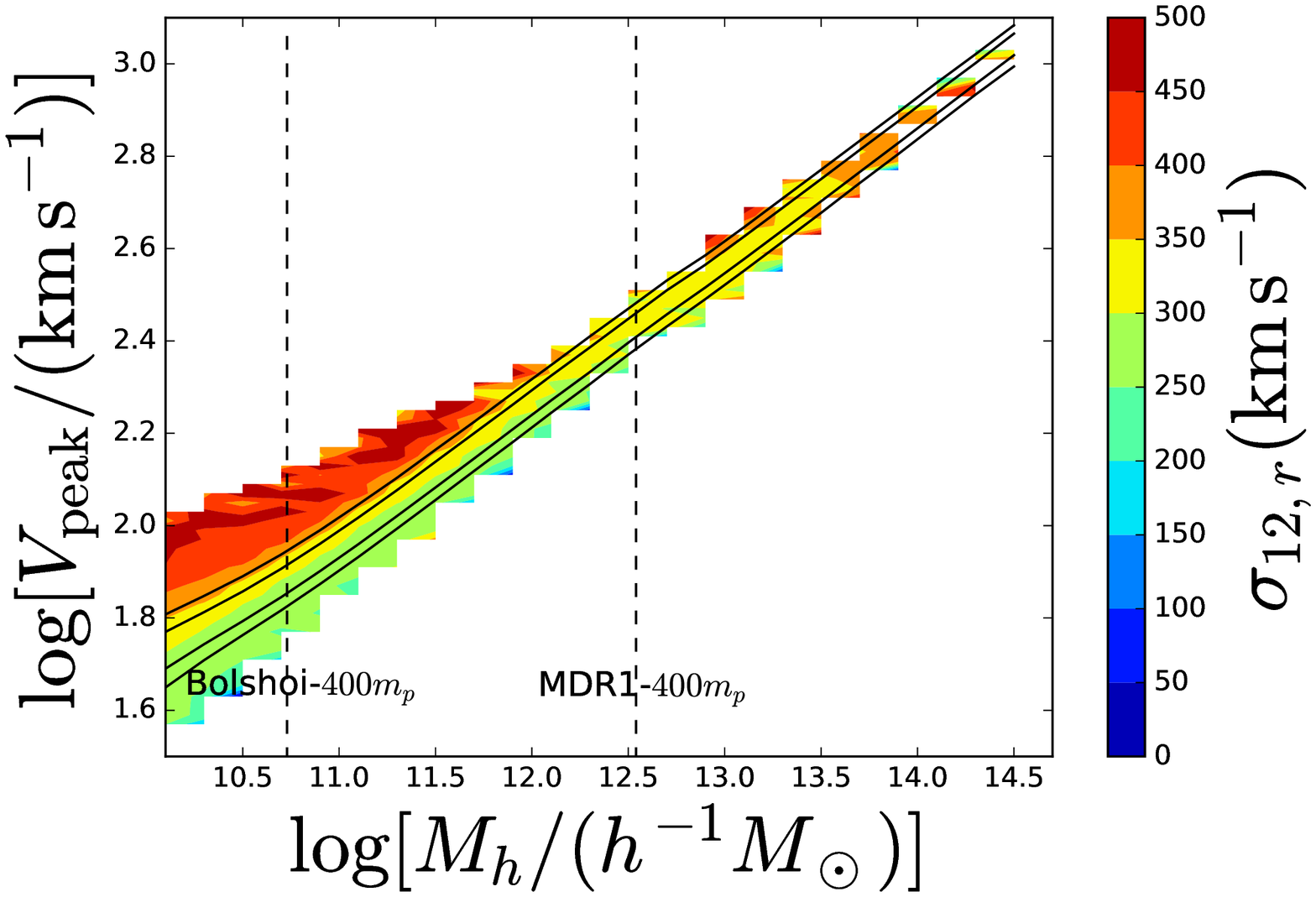}
    \end{subfigure}
    \hfill
    \begin{subfigure}[h]{0.24\textwidth}
        \centering
        \includegraphics[width=\textwidth]{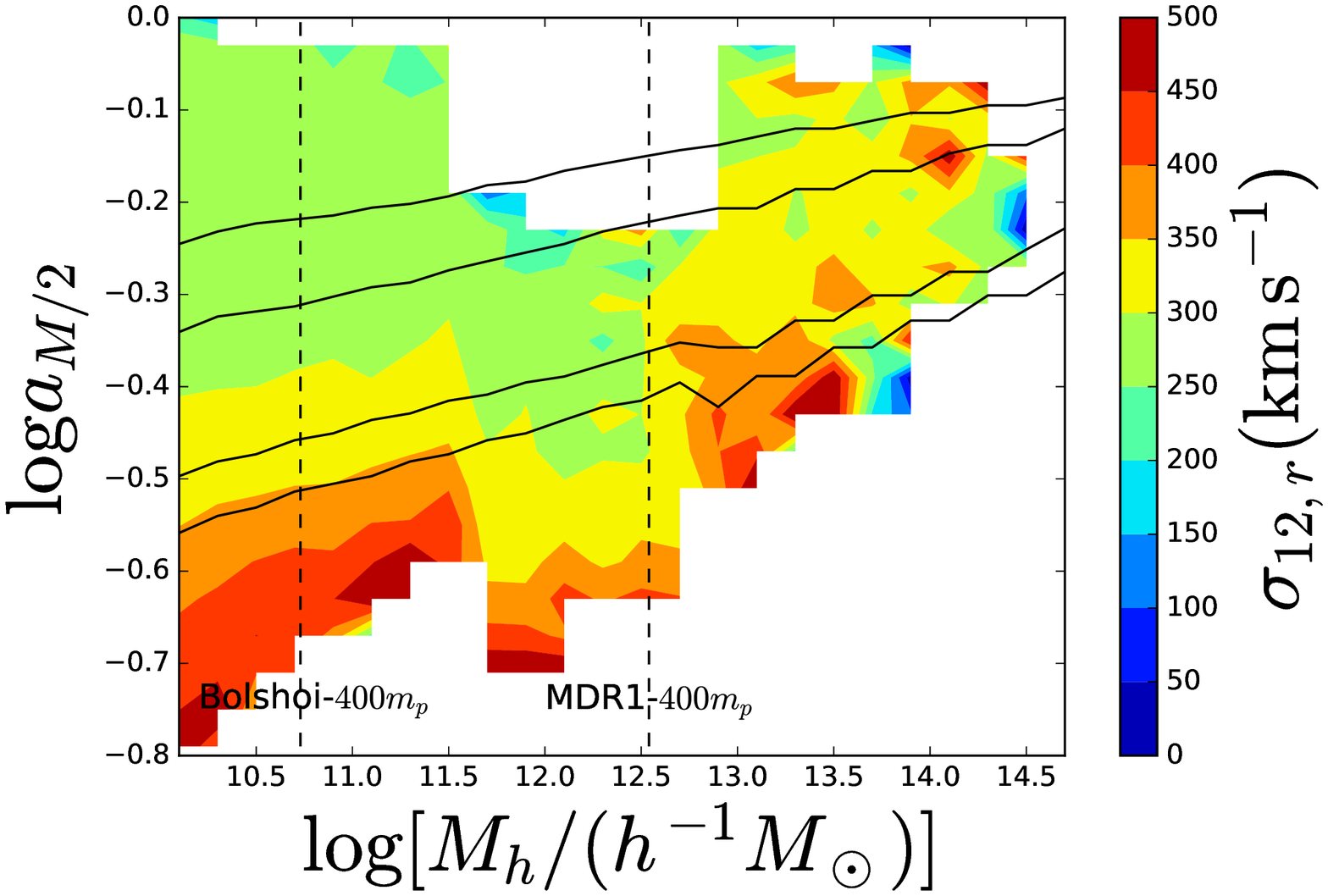}
    \end{subfigure}
    \hfill
    \begin{subfigure}[h]{0.24\textwidth}
        \centering
        \includegraphics[width=\textwidth]{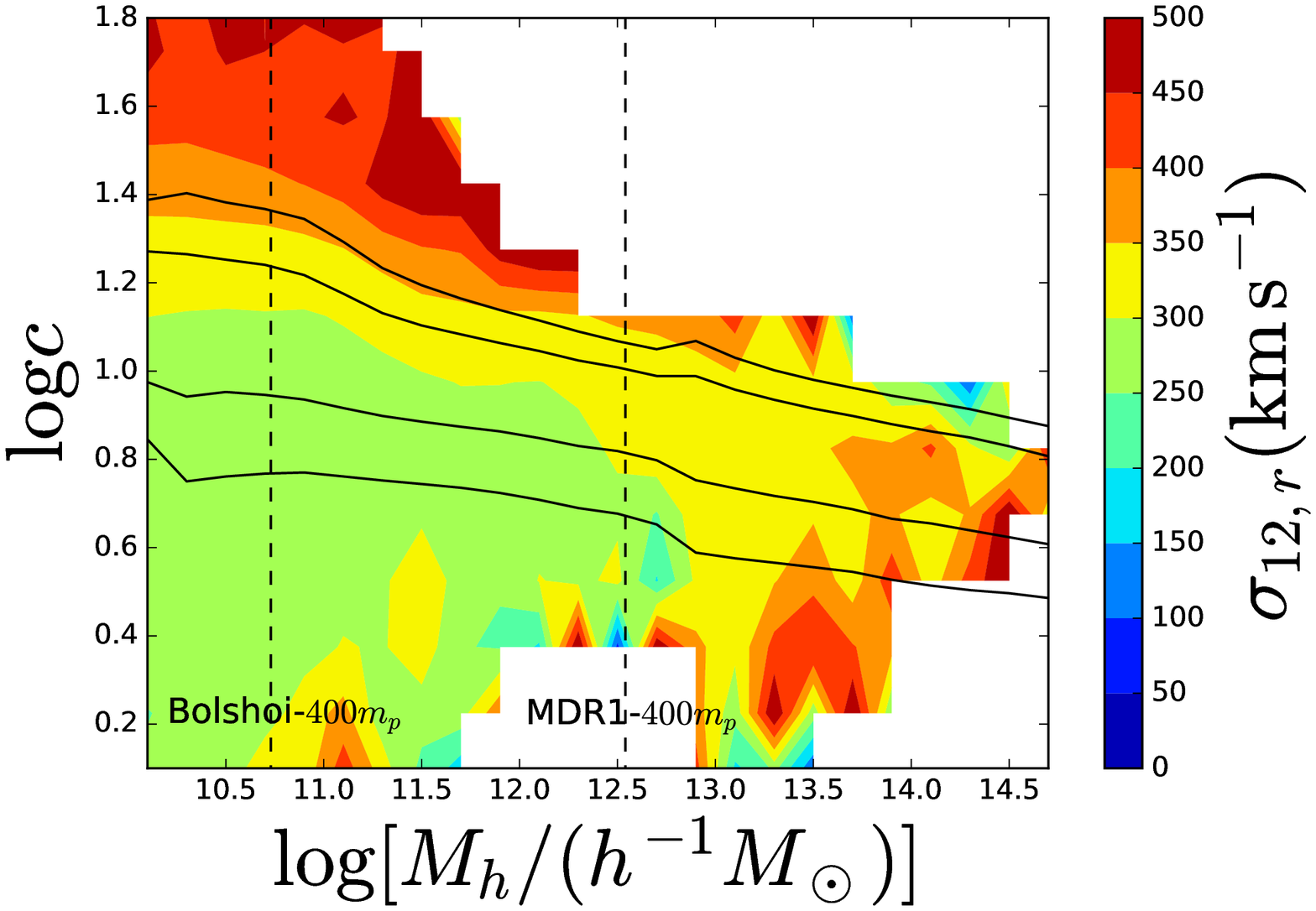}
    \end{subfigure}
    \hfill
    \begin{subfigure}[h]{0.24\textwidth}
        \centering
        \includegraphics[width=\textwidth]{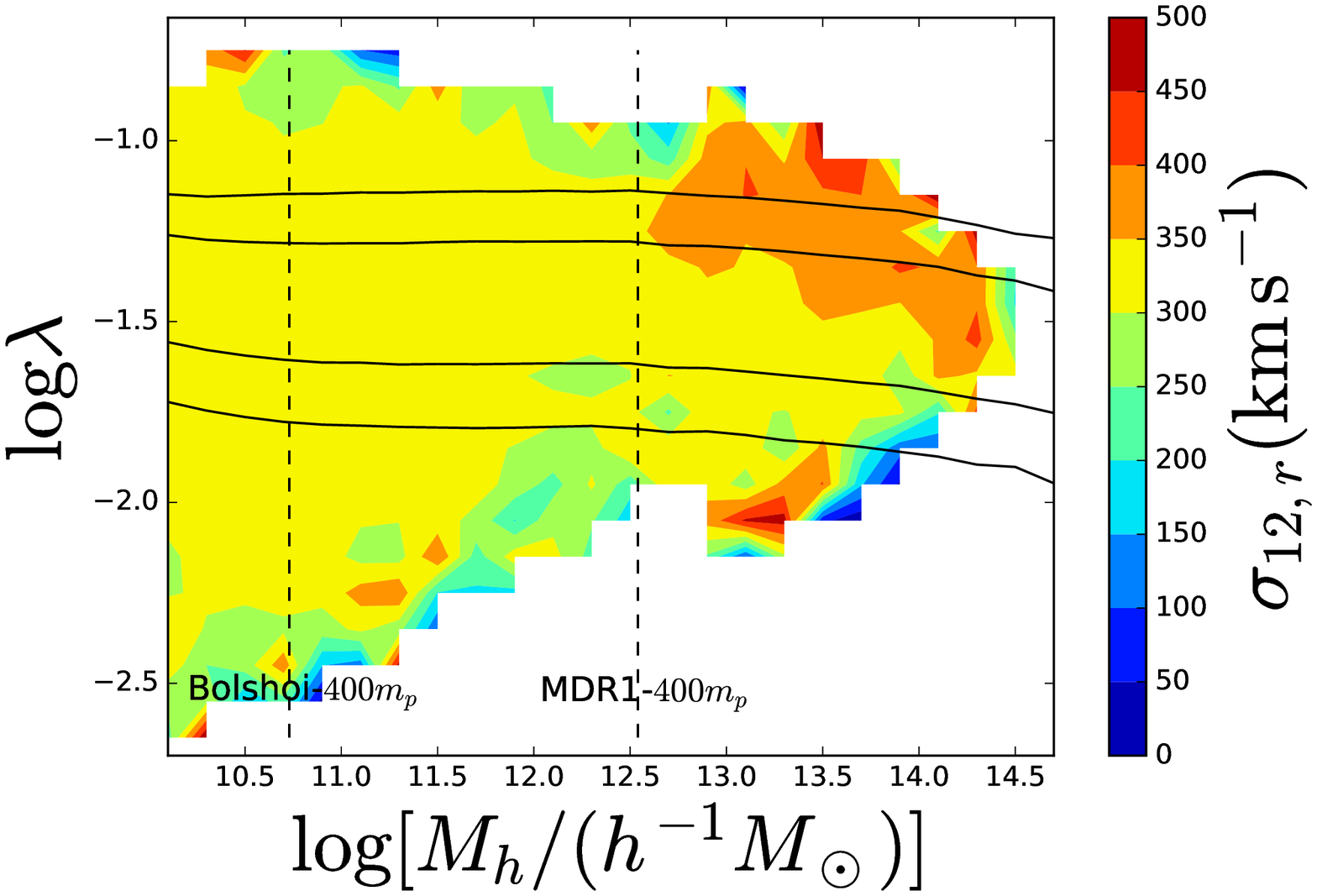}
    \end{subfigure}

    \hfill
    \begin{subfigure}[h]{0.24\textwidth}
        \centering
        \includegraphics[width=\textwidth]{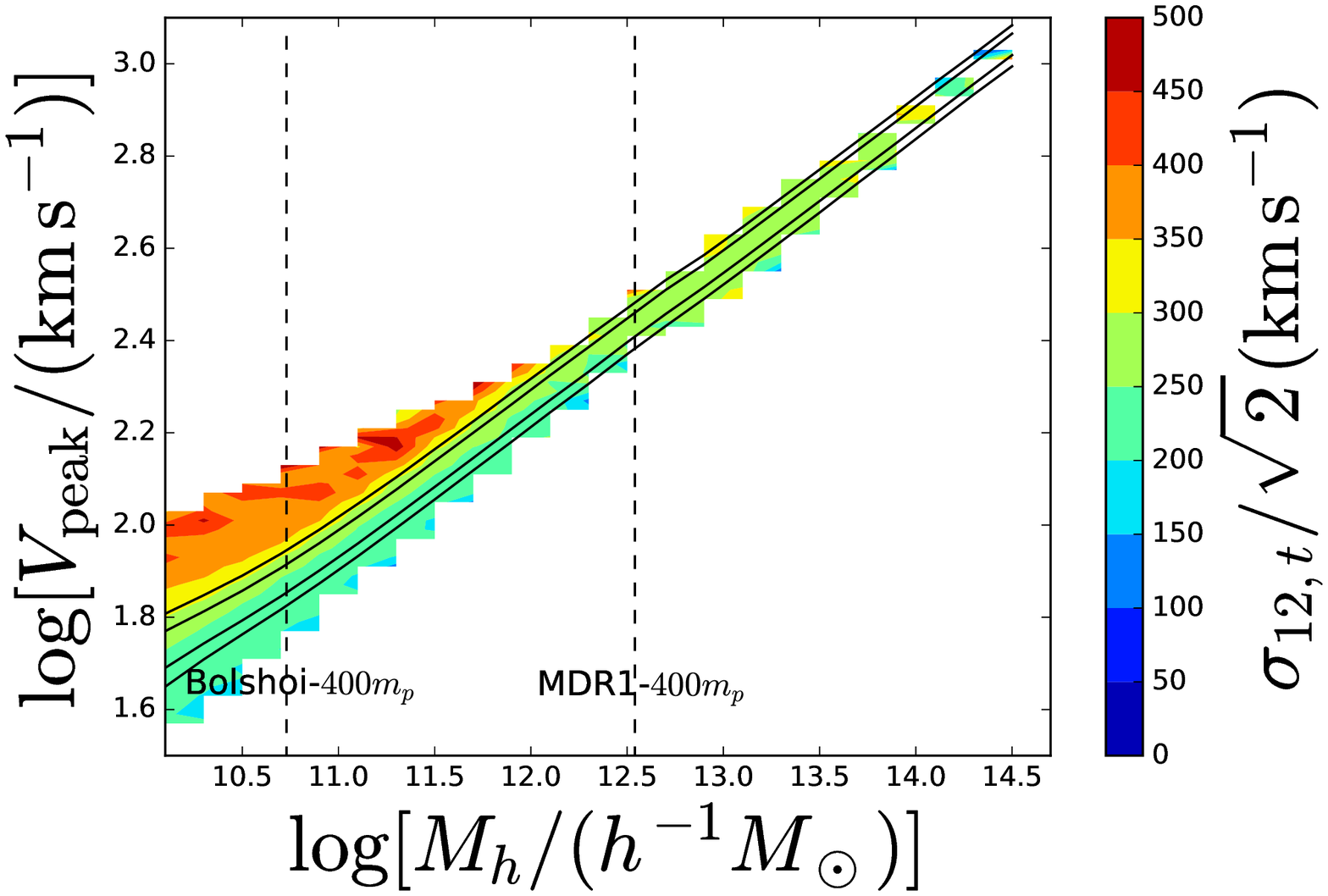}
    \end{subfigure}
    \hfill
    \begin{subfigure}[h]{0.24\textwidth}
        \centering
        \includegraphics[width=\textwidth]{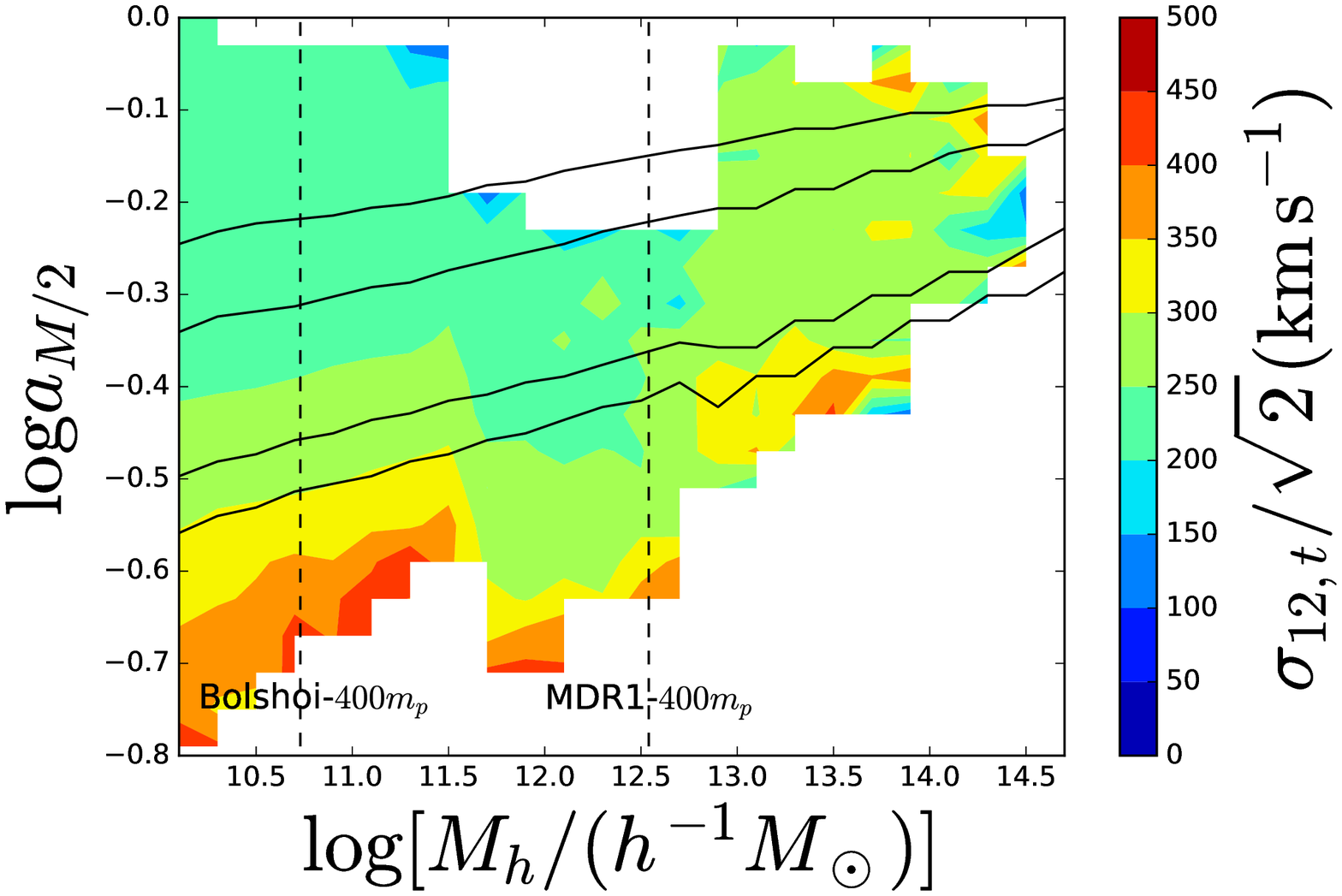}
    \end{subfigure}
    \hfill
    \begin{subfigure}[h]{0.24\textwidth}
        \centering
        \includegraphics[width=\textwidth]{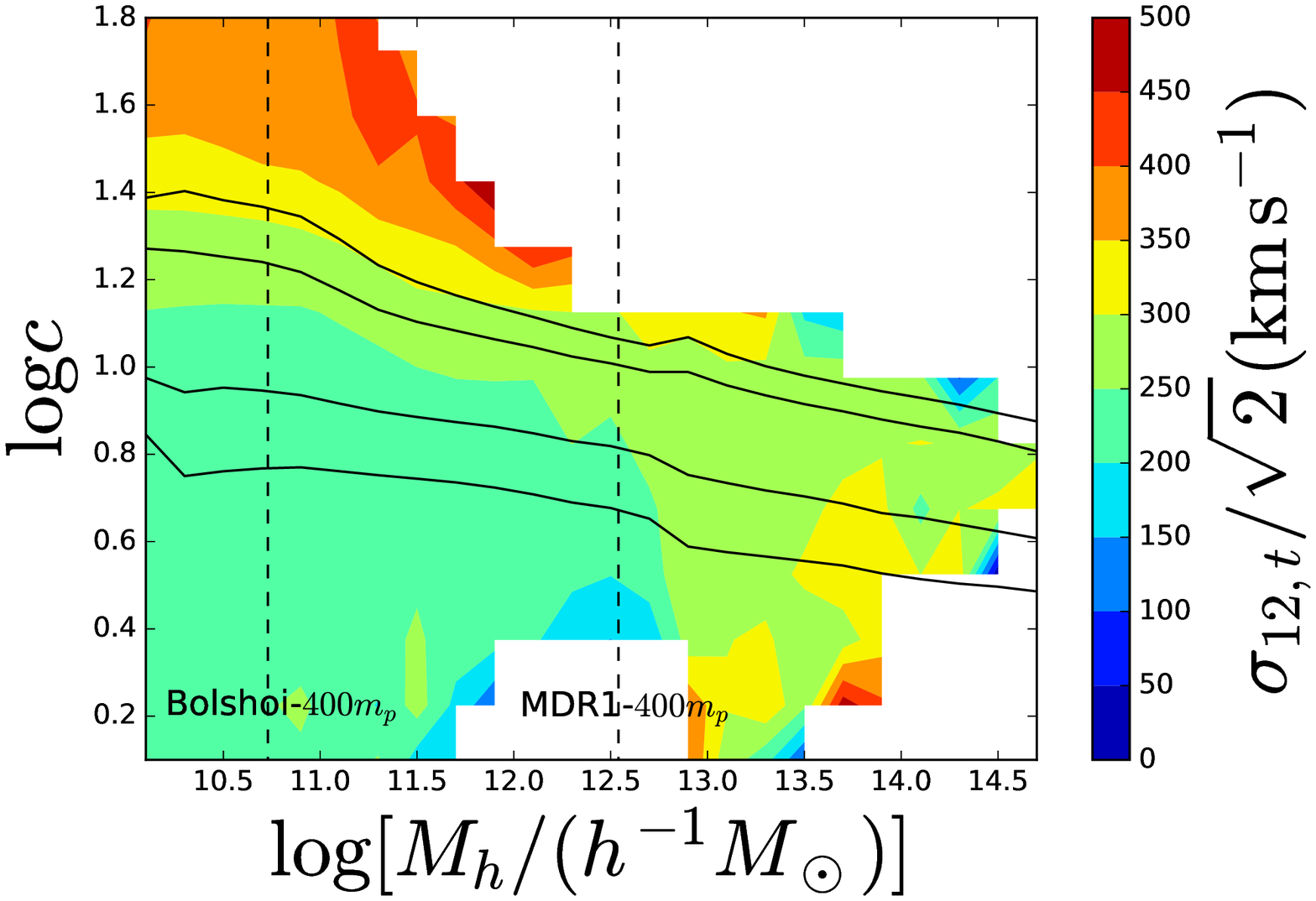}
    \end{subfigure}
    \hfill
    \begin{subfigure}[h]{0.24\textwidth}
        \centering
        \includegraphics[width=\textwidth]{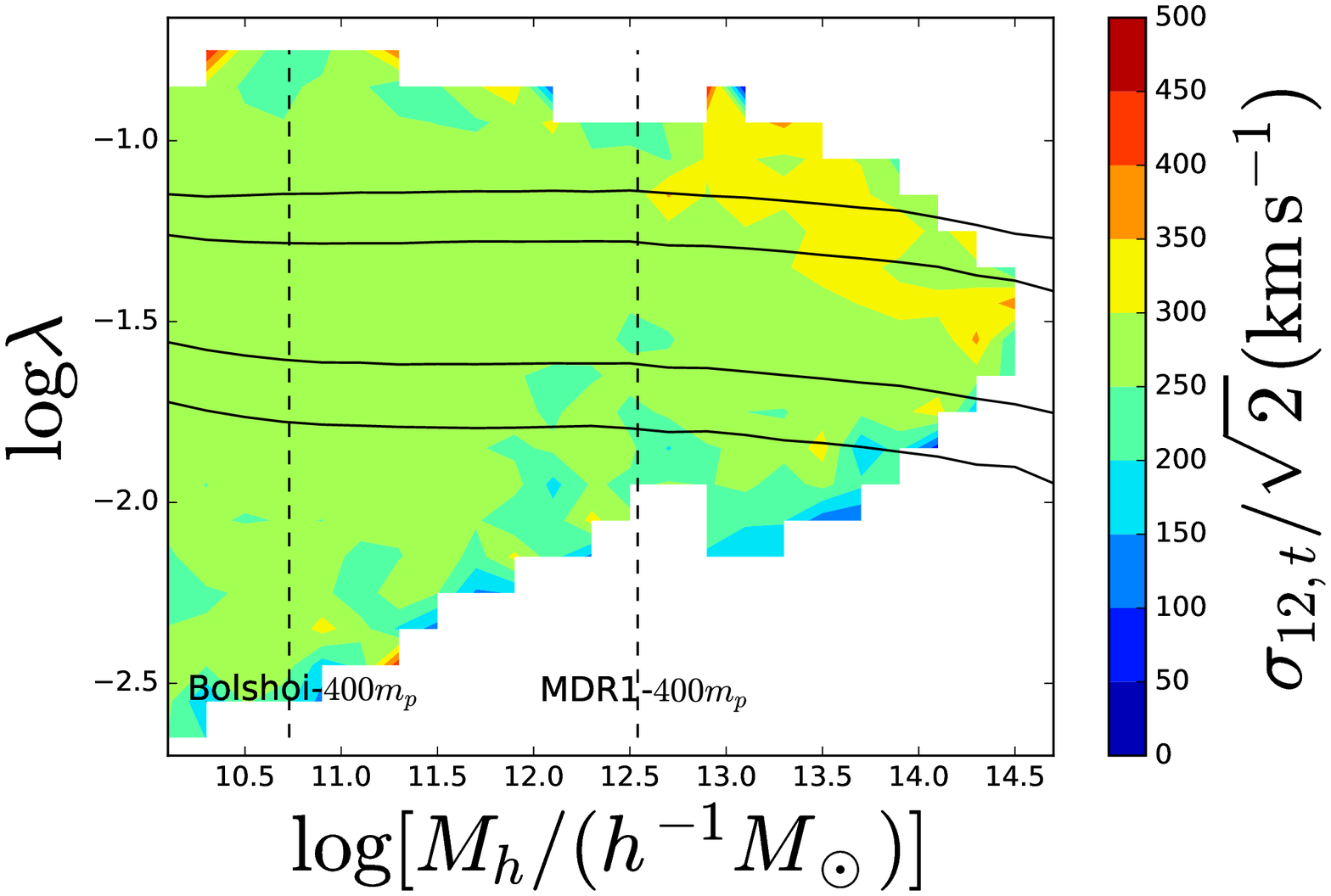}
    \end{subfigure}
\caption{
Same as Fig.~\ref{fig:vel_m_A}, but the Bolshoi simulation is used for $\log[\Mh/(\hinvMsun)]<12.8$ and MDR1 simulation for higher mass. The two vertical dashed lines in the each panel indicate the masses of haloes with 400 particles in the two simulations.
}
\label{fig:vel_m_A_bolshoi}
\end{figure*}

In the main analysis in this paper, we primarily use haloes from the MDR1 simulation, which has particle mass of $8.721\times 10^9\hinvMsun$. \citet{Paranjape17a} show that for haloes with less than 400 particles, halo concentration may not be determined accurately, usually lower than the true value (their Fig.2; also see Fig.9 of \citealt{Trenti10}). The distribution of other assembly variables (such as spin) for haloes of low number of particles can also be affected (e.g. Fig.9 in \citealt{Trenti10} and Fig.3 in \citealt{Benson17}). Such a numerical effect would potentially affect the pattern of the dependence of halo bias and velocity statistics on assembly variables for halos of mass below $3.5\times 10^{12}\hinvMsun$ in the MDR1 simulation. To test the effect, we make use of the Bolshoi simulation, with the same cosmology but a 64 times higher mass resolution. The mass of haloes with 400 particles in the Bolshoi simulation is $5.5\times 10^{10}\hinvMsun$, well below the minimum mass in our analysis.

First, we perform similar calculations as in Fig.~\ref{fig:joint_dep}, but using haloes in the Bolshoi simulation for $\log[\Mh/(\hinvMsun)]<12.8$. The results are presented in Fig.~\ref{fig:joint_dep_bolshoi}. By comparing with those in Fig.~\ref{fig:joint_dep}, we see that the trend in the joint dependence of halo bias on halo mass and one assembly variable from the Bolshoi simulation is similar to that from the MDR1 simulation in the mass range considered in the paper. The test assures that we can present the main results based on MDR1 simulation. The large volume of MDR1 compared to Bolshoi has the advantage of making the halo bias calculation less noisy and enabling the inclusion of a larger range in assembly variable at fixed mass, which is clearly seen when comparing Fig.~\ref{fig:joint_dep} and Fig.~\ref{fig:joint_dep_bolshoi}.
For the above reason, the results on bias (except for those in Section~\ref{sec:Meff}, see below) from the MDR1 simulation are shown in the paper. We caution, however, that for investigations or applications that require accurate values of halo bias, the mass resolution effect needs to be accounted for.

For the investigation with effective halo mass in Section~\ref{sec:Meff}, we need to combine the Bolshoi and MDR1 simulation. This is not mainly driven by the above numerical effect, but the need for low mass haloes. As can be seen from the left panel of Fig.~\ref{fig:Meff}, for effective halo mass in the range $\log[\Meff/(\hinvMsun)]\sim$ 11--12.5, we need to reach haloes with high spin and low mass (below $\log[\Mh/(\hinvMsun)]\sim$11.6), which are covered incompletely in the MDR1 simulation. A combination of MDR1 and Bolshoi is therefore necessary for the effective mass analysis (see Section~\ref{sec:Meff} for details).

Then we analyse the pairwise velocity statistics of haloes of $\log[\Mh/(\hinvMsun)]\sim$11.73 in the Bolshoi simulation. Fig.~\ref{fig:vel_assembly_percentile_bolshoi} shows the dependence of pairwise velocity and velocity dispersion on assembly variable, which is to be compared to Fig.~\ref{fig:vel_assembly_percentile} with the MDR1 simulation. For the cases with \Vpeak, \ahalf, and $c$ (the left three columns), the trends from the Bolshoi simulation well track those from the MDR1 simulation. There are slight amplitude shifts for the corresponding curves from the two simulations, which can be partly explained by the sample variance effect (i.e. difference in the long-wave length modes in the two simulations). The mass resolution may also contribute. Imagine that we have a simulation with the same box size and initial condition as Bolshoi but with MDR1 mass resolution. For Bolshoi haloes of a given concentration, those of the corresponding MDR1-resolution haloes would have a scatter \citep[e.g.][]{Paranjape17a,Trenti10}. That is, for the same percentile, haloes from the two simulations would not have the exact correspondence, and the pairwise velocity amplitude is expected to be slightly different. For the above three assembly variables, the pattern of the dependence of pairwise velocity statistics on the value of assembly bias from the Bolshoi simulation is similar to that from the MDR1 simulation, as can be seen by comparing Fig.~\ref{fig:vel_m_A_bolshoi} and Fig.~\ref{fig:vel_m_A}.

For the case with halo spin $\lambda$, the pairwise velocity statistics from the percentile analysis seem to be different between the MDR1 and Bolshoi simulations (right panels of Fig.~\ref{fig:vel_assembly_percentile} and Fig.~\ref{fig:vel_assembly_percentile_bolshoi}). With the MDR1 simulation, there is a clear trend of higher pairwise velocity and velocity dispersion for haloes of higher spin, although it is substantially weaker than those  with other three assembly variables. With the Bolshoi simulation, however, we do not see such a clear trend (also see the corresponding panels of Fig.~\ref{fig:vel_m_A_bolshoi}). If what Bolshoi suggests is close to the truth, given the expected scatter in determining the spin parameter for a low-resolution simulation \citep[e.g.][]{Trenti10,Benson17}, we would not expect to see any trend with the MDR1 simulation. However, here we have the opposite results. We note that the error bars with the Bolshoi simulation analyses are substantially larger than those with the MDR1. It is possible that the sample variance effect in the Bolshoi simulation masks the spin dependence, i.e. the noise level is too high to reveal the intrinsically weak trend. The other possibility is that with low-resolution simulations the spin measurement is more easily affected by environment and hence halo assembly, making the scatter mentioned above not random. Further investigations using simulations of the same initial condition and different resolutions, like those in \citet{Trenti10}, would help  resolve the issue of the spin-dependence of the pairwise velocity statistics.

To summarise, from comparing the halo statistics with MDR1 and Bolshoi simulations, for most results presented in this paper, the MDR1 simulation works well in revealing the patterns related to assembly variables, which has the benefit of better statistics from the much larger volume.
For the effective halo mass analysis, we have to include the Bolshoi simulation to reach haloes of much lower mass for completeness consideration. For halo pairwise velocity statistics, the spin case appears to be different between the two simulations, weak trend with the MDR1 simulation and the lack of trend with the Bolshoi simulation. The cause of the apparent discrepancy is not entirely clear, which could be noise in the Bolshoi simulation or some unknown environment-dependent systematic effect in halo spin measurement. In the main text, we choose to present the spin results from the MDR1 simulation, but put the caveat that there is an apparent difference compared to those from the Bolshoi simulation.

%We would like to investigate whether mass resolution of simulation will lead to any problem in section 3.1. Here we replace the haloes of mass below  $\log[\Mh/(\hinvMsun)]<12.8$ with Bolshoi haloes. Bolshoi has the same cosmology with MDR1 Multidark, but smaller boxsize and higher mass resolution, particle mass  $1.35\times 10^8\hinvMsun$. With this mass resolution, haloes in lowest mass bin in section 3.1. have about 3000 particles, instead of 50 particles. This also means we can extend our lower limit of mass range.

%By comparing mass range $11.6<\log[\Mh/(\hinvMsun)]<12.8$ Fig.~\ref{fig:joint_dep} and Fig.~\ref{fig:joint_dep_bolshoi}, we can see the results from section 3.1. still hold, since the main trend of contours are hold. But the parts from Bolshoi appear more noisy, caused by the smaller volume (fewer haloes). So for the analysis other than effective mass (section 3.2), we would like to keep using MDR1. 
%In section 3.2, due to the incomplete coverage of spin parameter using only MDR1 data, we combine the two data togeter.
%%%%%%%%%%%%%%%%%%%%%%%%%%%%%%%%%%%%%%%%%%%%%%%%%%

\bsp	% typesetting comment
\label{lastpage}

\begin{thebibliography}{99}

\bibitem[Bardeen et al.(1986)]{Bardeen86}
Bardeen J.~M., Bond J.~R., Kaiser N., \& Szalay A.~S.\ 1986, \apj, 304, 15 

%\bibitem[Bett et al.(2007)]{Bett07} 
%Bett P., Eke V., Frenk C.~S., et al.\ 2007, \mnras, 376, 215 

\bibitem[Behroozi et al.(2013)]{Behroozi13} 
Behroozi P.~S., Wechsler R.~H., \& Wu H.-Y.\ 2013, \apj, 762, 109 

\bibitem[Benson(2017)]{Benson17}
Benson, A.~J.\ 2017, \mnras, 471, 2871 

\bibitem[Berlind \& Weinberg(2002)]{Berlind02}
Berlind A.~A., \& Weinberg D.~H.\ 2002, \apj, 575, 587

%\bibitem[Berlind et al.(2003)]{Berlind03}
%Berlind A.~A., Weinberg D.~H., Benson A.~J., et al.\ 2003, \apj, 593, 1

\bibitem[Birnboim \& Dekel(2003)]{Birnboim03} 
Birnboim Y., \& Dekel A.\ 2003, \mnras, 345, 349 

%\bibitem[Blake et al.(2008)]{Blake08}
%Blake C., Collister A., \& Lahav O.\ 2008, \mnras, 385, 1257

\bibitem[Blake et al.(2011)]{Blake11}
Blake C., Kazin E.~A., Beutler F., et al.\ 2011, \mnras, 418, 1707,

\bibitem[Bond et al.(1991)]{Bond91}
Bond J.~R., Cole S., Efstathiou G., \& Kaiser N.\ 1991, \apj, 379, 440

\bibitem[Borzyszkowski et al.(2017)]{Borzyszkowski17} Borzyszkowski, M., Porciani, C., Romano-D{\'{\i}}az, E., \& Garaldi, E.\ 2017, \mnras, 469, 594 

%\bibitem[Brown et al.(2008)]{Brown08}
%Brown M.~J.~I., Zheng Z., White M., et al.\ 2008, \apj, 682, 937

%\bibitem[Bullock et al.(2002)]{Bullock02}
%Bullock J.~S., Wechsler R.~H., \& Somerville R.~S.\ 2002, \mnras, 329, 246

\bibitem[Busch \& White(2017)]{Busch17}
Busch P., \& White S.~D.~M.\ 2017, \mnras, 470, 4767

\bibitem[Chaves-Montero et al.(2016)]{Chaves16}
Chaves-Montero J., Angulo R.~E., Schaye J., et al.\ 2016, \mnras, 460, 3100

\bibitem[Colless(1999)]{Colless99}
Colless M.\ 1999, Philosophical Transactions of the Royal Society of London
Series A, 357, 105,

%\bibitem[Cooray \& Sheth(2002)]{Cooray02}
%Cooray A., \& Sheth R.\ 2002, Phys.\ Rept., 372, 1

%\bibitem[Cooray(2006)]{Cooray06}
%Cooray A.\ 2006, \mnras, 365, 842

%\bibitem[Coupon et al.(2012)]{Coupon12}
%Coupon J., Kilbinger M., McCracken H.~J.\ 2012, \aap, 542, 5

\bibitem[Croft et al.(2012)]{Croft12} 
Croft R.~A.~C., di Matteo T., Khandai N., et al.\ 2012, \mnras, 425, 2766 

\bibitem[Dalal et al.(2008)]{Dalal08}
Dalal N., White M., Bond J.~R., \& Shirokov A.\ 2008, \apj, 687, 12

\bibitem[Dawson et al.(2016)]{Dawson16}
Dawson K.~S., Kneib J.-P., Percival W.~J., et al.\ 2016, \aj, 151, 44,

%\bibitem[Diemand et al.(2007)]{Diemand07}
%Diemand J., Kuhlen M., \& Madau P.\ 2007, \apj, 667, 859

\bibitem[Eisenstein et al.(2011)]{Eisenstein11}
Eisenstein D.~J., Weinberg D.~H., Agol E., et al.\ 2011, \aj, 142, 72,

\bibitem[Faltenbacher \& White(2010)]{Faltenbacher10}
Faltenbacher A., \& White S.~D.~M.\ 2010, \apj, 708, 469

\bibitem[Gao et al.(2004)]{Gao04} 
Gao L., White S.~D.~M., Jenkins A., Stoehr F., \& Springel V.\ 2004, \mnras, 355, 819 

\bibitem[Gao et al.(2005)]{Gao05}
Gao L., Springel V., \& White S.~D.~M.\ 2005, \mnras, 363, L66

\bibitem[Gao \& White(2007)]{Gao07} 
Gao L., \& White S.~D.~M.\ 2007, \mnras, 377, L5 

\bibitem[Garaldi et al.(2018)]{Garaldi18} 
Garaldi E., Romano-D{\'{\i}}az E., Borzyszkowski M., \& Porciani C.\ 2018, \mnras, 473, 2234 

%\bibitem[Guo et al.(2014)]{Guo14}
%Guo H., Zheng Z., Zehavi I., et al.\ 2014, \mnras, 441, 2398

%\bibitem[Guo et al.(2015a)]{Guo15a}
%Guo H., Zheng Z., Zehavi I., et al.\ 2015a, \mnras, 446, 578

%\bibitem[Guo et al.(2015b)]{Guo15c}
%Guo H., Zheng Z., Zehavi I., et al.\ 2015c, \mnras, 453, 4368

%\bibitem[Guo et al.(2016)]{Guo16a}
%Guo H., Zheng Z., Behroozi P.~S., et al.\ 2016, \mnras, 459, 3040

\bibitem[Guo et al.(2017)]{Guo17} 
Guo H., Li C., Zheng Z., et al.\ 2017, \apj, 846, 61

\bibitem[Hahn et al.(2007)]{Hahn07} 
Hahn O., Carollo C.~M., Porciani C., \& Dekel A.\ 2007, \mnras, 381, 41 

\bibitem[Hahn et al.(2009)]{Hahn09} 
Hahn O., Porciani C., Dekel A., \& Carollo C.~M.\ 2009, \mnras, 398, 1742 

%\bibitem[Hamana et al.(2006)]{Hamana06}
%Hamana T., Yamada T., Ouchi M., Iwata I., \& Kodama T.\ 2006, \mnras, 369, 1929

\bibitem[Han et al.(2018)]{Han18}
Han J., Li Y., Jing Y.~P., Nishimichi T., Wang W., \& Jiang C.\ 2018, arXiv:1802.09177

\bibitem[Harker et al.(2006)]{Harker06}
Harker G., Cole S., Helly J., Frenk C., \& Jenkins A.\ 2006, \mnras, 367, 1039

\bibitem[Hearin(2015)]{Hearin15}
Hearin A.~P.\ 2015, \mnras, 451, L45

%\bibitem[Jenkins et al.(2001)]{Jenkins01} 
%Jenkins A., Frenk C.~S., White S.~D.~M., et al.\ 2001, \mnras, 321, 372 

%\bibitem[Jing et al.(1998)]{Jing98}
%Jing Y.~P., Mo H.~J. \& B{\"o}rner G.\ 1998, \apj, 494, 1,

%\bibitem[Jing(1998)]{Jing98a} 
%Jing Y.~P.\ 1998, \apjl, 503, L9 

\bibitem[Jing(1999)]{Jing99} 
Jing Y.~P.\ 1999, \apjl, 515, L45 

\bibitem[Jing et al.(2007)]{Jing07} 
Jing Y.~P., Suto Y., \& Mo H.~J.\ 2007, \apj, 657, 664 

\bibitem[Kere{\v s} et al.(2005)]{Keres05} 
Kere{\v s} D., Katz N., Weinberg D.~H., \& Dav{\'e} R.\ 2005, \mnras, 363, 2

\bibitem[Klypin et al.(2011)]{Klypin2011}
Klypin A. A., Trujillo-Gomez S., \& Primack J.\ 2011, \apj, 740, 102

\bibitem[Lazeyras et al.(2017)]{Lazeyras17}
Lazeyras T., Musso M., \& Schmidt F.\ 2017, \jcap, 3, 059

%\bibitem[Lee et al.(2006)]{Lee06}
%Lee K., Giavalisco M., Gnedin O.~Y., et al.\ 2006, \apj, 642, 63

%\bibitem[Li et al.(2008)]{Li08}
%Li Y., Mo H.~J., \& Gao L.\ 2008, \mnras, 389, 1419

\bibitem[Lin et al.(2016)]{Lin16}
Lin Y.-T., Mandelbaum R., Huang Y.-H., et al.\ 2016, \apj, 819, 119

%\bibitem[Ludlow et al.(2009)]{Ludlow09}
%Ludlow A.~D., Navarro J.~F., Springel V., et al.\ 2009, \apj, 692, 931

\bibitem[Macci{\`o} et al.(2007)]{Maccio07} 
Macci{\`o} A.~V., Dutton A.~A., van den Bosch F.~C., et al.\ 2007, \mnras, 378, 55 

\bibitem[Mao et al.(2018)]{Mao18} 
Mao Y.-Y., Zentner A.~R., \& Wechsler R.~H.\ 2018, \mnras, 474, 5143 

%\bibitem[Mao et al.(2017)]{Mao17} 
%Mao Y.-Y., Zentner A.~R., \& Wechsler R.~H.\ 2017, arXiv:1705.03888

\bibitem[McEwen \& Weinberg(2016)]{McEwen16}
McEwen J.~E., \& Weinberg D.~H.\ 2016, arXiv:1601.02693

\bibitem[Miyatake et al.(2016)]{Miyatake16}
Miyatake H., More, S., Takada M., et al.\ 2016, Physical Review Letters, 116, 041301

\bibitem[Mo \& White(1996)]{Mo96} 
Mo H.~J., \& White S.~D.~M.\ 1996, \mnras, 282, 347 

%\bibitem[Musso et al.(2018)]{Musso18} 
%Musso M., Cadiou C., Pichon C., et al.\ 2018, \mnras, 476, 4877 

%\bibitem[Musso et al.(2017)]{Musso17}
%Musso M., Cadiou C., Pichon C., et al.\ 2017, arXiv:1709.00834

\bibitem[Paranjape \& Padmanabhan(2017)]{Paranjape17a}
Paranjape A., \& Padmanabhan N.\ 2017, \mnras, 468, 2984

\bibitem[Paranjape et al.(2018)]{Paranjape18} 
Paranjape A., Hahn O., \& Sheth R.~K.\ 2018, \mnras, 476, 3631 

%\bibitem[Peacock \& Smith(2000)]{Peacock00}
%Peacock J.~A., \& Smith R.~E.\ 2000, \mnras, 318, 1144

%\bibitem[Phleps et al.(2006)]{Phleps06}
%Phleps S., Peacock J.~A., Meisenheimer K., \& Wolf C.\ 2006, \aap, 457, 145

\bibitem[Prada et al.(2012)]{Prada12} 
Prada F., Klypin A.~A., Cuesta A.~J., Betancort-Rijo J.~E., \& Primack J.\ 2012, \mnras, 423, 3018 

\bibitem[Press \& Schechter(1974)]{Press74} 
Press W.~H., \& Schechter P.\ 1974, \apj, 187, 425 

%\bibitem[Riebe et al.(2013)]{Riebe13}
%Riebe K., Partl A.~M., Enke H., et al.\ 2013, Astronomische Nachrichten, %334, 691 

\bibitem[Salcedo et al.(2018)]{Salcedo18} 
Salcedo A.~N., Maller A.~H., Berlind A.~A., et al.\ 2018, \mnras, 475, 4411 

%\bibitem[Salcedo et al.(2017)]{Salcedo17}
%Salcedo A.~N., Maller A.~H., Berlind A.~A., et al.\ 2017, %arXiv:1708.08451

%\bibitem[Sandvik et al.(2007)]{Sandvik07} 
%Sandvik H.~B., M{\"o}ller O., Lee J., \& White S.~D.~M.\ 2007, \mnras, %377, 234 

%\bibitem[Scoccimarro et al.(2001)]{Scoccimarro01}
%Scoccimarro R., Sheth R.~K., Hui L., \& Jain B.\ 2001, \apj, 546, 20

%\bibitem[Seljak(2000)]{Seljak00}
%Seljak U.\ 2000, \mnras, 318, 203

%\bibitem[Sheth \& Tormen(1999)]{Sheth99} 
%Sheth R.~K., \& Tormen G.\ 1999, \mnras, 308, 119 

%\bibitem[Sheth et al.(2001a)]{Sheth01a} 
%Sheth R.~K., Mo H.~J., \& Tormen G.\ 2001, \mnras, 323, 1 

\bibitem[Sheth et al.(2001b)]{Sheth01b}
Sheth R.~K., Diaferio A., Hui L., \& Scoccimarro R.\ 2001, \mnras, 326, 463 

\bibitem[Sheth \& Diaferio(2001)]{Sheth01} 
Sheth R.~K., \& Diaferio A.\ 2001, \mnras, 322, 901 

%\bibitem[Sheth \& Tormen(2004)]{Sheth04}
%Sheth R.~K., \& Tormen G.\ 2004, \mnras, 350, 1385

\bibitem[Shi et al.(2015)]{Shi15} 
Shi J., Wang H., \& Mo H.~J.\ 2015, \apj, 807, 37 

%\bibitem[Skibba et al.(2015)]{Skibba15}
%Skibba R.~A., Coil A.~L., Mendez A.~J., et al.\ 2015, \apj, 807, 152

\bibitem[Sunayama et al.(2016)]{Sunayama16} 
Sunayama T., Hearin A.~P., Padmanabhan N., \& Leauthaud A.\ 2016, \mnras, 458, 1510 

\bibitem[Tinker et al.(2008)]{Tinker08} 
Tinker J., Kravtsov A.~V., Klypin A., et al.\ 2008, \apj, 688, 709-728 

\bibitem[Trenti et al.(2010)]{Trenti10} 
Trenti, M., Smith, B.~D., Hallman, E.~J., Skillman, S.~W., \& Shull, J.~M.\ 2010, \apj, 711, 1198 

%\bibitem[Vakili \& Hahn(2016)]{Vakili16}
%Vakili M., \& Hahn C.~H.\ 2016, arXiv:1610.01991

%\bibitem[van den Bosch et al.(2003)]{Bosch03}
%van den Bosch F.~C., Mo H.~J., \& Yang X.\ 2003, \mnras, 345, 923

%\bibitem[Villarreal et al.(2017)]{Villarreal17}
%Villarreal A.~S., Zentner A.~R., Mao Y.-Y., et al.\ 2017, \mnras, 472, %1088

\bibitem[Wang et al.(2009)]{Wang09} 
Wang H., Mo H.~J., \& Jing Y.~P.\ 2009, \mnras, 396, 2249 

\bibitem[Wang et al.(2011)]{Wang11} 
Wang H., Mo H.~J., Jing Y.~P., Yang X., \& Wang Y.\ 2011, \mnras, 413, 1973 

\bibitem[Wechsler et al.(2002)]{Wechsler02} 
Wechsler R.~H., Bullock J.~S., Primack J.~R., Kravtsov A.~V., \& Dekel A.\ 2002, \apj, 568, 52 

\bibitem[Wechsler et al.(2006)]{Wechsler06} 
Wechsler R.~H., Zentner A.~R., Bullock J.~S., Kravtsov A.~V., \& Allgood B.\ 2006, \apj, 652, 71 

\bibitem[Wetzel et al.(2014)]{Wetzel14} 
Wetzel A.~R., Tinker J.~L., Conroy C., \& van den Bosch F.~C.\ 2014, \mnras, 439, 2687 

\bibitem[White \& Rees(1978)]{White78} 
White S.~D.~M., \& Rees M.~J.\ 1978, \mnras, 183, 341 

%\bibitem[White et al.(2007)]{White07}
%White M., Zheng Z., Brown M.~J.~I., Dey A., \& Jannuzi B.~T.\ 2007, \apj, 655, L69

%\bibitem[White et al.(2011)]{White11}
%White M., Blanton M., Bolton A., et al.\ 2011, \apj, 728, 126

\bibitem[Yang et al.(2003)]{Yang03}
Yang X., Mo H.~J., \& van den Bosch F.~C.\ 2003, \mnras, 339, 1057

%\bibitem[Xu et al.(2016)]{Xu16}
%Xu H., Zheng Z., Guo H., Zhu J., \& Zehavi I.\ 2016, \mnras, 460, 3647

%\bibitem[Xu et al.(2018)]{Xu18}
%Xu H., Zheng Z., Guo H., Zu Y., Zehavi I., \& Weinberg D.~H. \ 2018, arXiv:1801.07272

%\bibitem[Yang et al.(2005)]{Yang05}
%Yang X.~H., Mo H.~J., Jing Y.~P., \& van den Bosch F.~C.\ 2005

\bibitem[York et al.(2000)]{York00}
York D.~G., Adelman J., Anderson J.~E., Jr., et al.\ 2000, \aj, 120, 1579,

\bibitem[Zehavi et al.(2005)]{Zehavi05}
Zehavi I., Zheng Z., Weinberg D.~H., et al.\ 2005, \apj, 630, 1

\bibitem[Zehavi et al.(2011)]{Zehavi11}
Zehavi I., Zheng Z., Weinberg D.~H., et al.\ 2011, \apj, 736, 59

%\bibitem[Zehavi et al.(2018)]{Zehavi18}
%Zehavi I., Contreras S., Padilla N., et al.\ 2018, \apj, 853, 84 

%\bibitem[Zehavi et al.(2017)]{Zehavi17}
%Zehavi I., Contreras S., Padilla N., et al.\ 2017, arXiv:1706.07871

\bibitem[Zentner et al.(2014)]{Zentner14}
Zentner A.~R., Hearin A.~P., \& van den Bosch F.~C.\ 2014, \mnras, 443, 3044,

\bibitem[Zentner et al.(2016)]{Zentner16}
Zentner A.~R., Hearin A., van den Bosch F.~C., Lange J.~U., \& Villarreal A.\ 2016, arXiv:1606.07817,

\bibitem[Zhang \& Jing(2004)]{Zhang04}
Zhang H.-Y., \& Jing Y.-P.\ 2004, \cjaa, 4, 507 

\bibitem[Zheng et al.(2002)]{Zheng02} 
Zheng Z., Tinker J.~L., Weinberg D.~H., \& Berlind A.~A.\ 2002, \apj, 575, 617 

%\bibitem[Zheng(2004)]{Zheng04}
%Zheng Z.\ 2004, \apj, 610, 61

\bibitem[Zheng et al.(2005)]{Zheng05}
Zheng Z., Berlind A.~A., Weinberg D.~H., et al.\ 2005, \apj, 633, 791

%\bibitem[Zheng \& Weinberg(2007)]{Zheng07}
%Zheng Z., \& Weinberg D.~H.\ 2007, \apj, 659, 1

%\bibitem[Zheng et al.(2009)]{Zheng09}
%Zheng Z., Zehavi I., Eisenstein D.\ J., Weinberg D.\ H., \& Jing Y. 2009, \apj, 707, 554

\bibitem[Zhu et al.(2006)]{Zhu06}
Zhu G., Zheng Z., Lin W.~P., et al.\ 2006, \apjl, 639, L5

%\bibitem[Zu \& Mandelbaum(2018)]{Zu18} 
%Zu Y., \& Mandelbaum R.\ 2018, \mnras, 476, 1637 

%\bibitem[Zu \& Mandelbaum(2017)]{Zu17}
%Zu Y., \& Mandelbaum R.\ 2017, arXiv:1703.09219

\bibitem[Zu et al.(2017)]{Zu17a}
Zu Y., Mandelbaum R., Simet M., Rozo E., \& Rykoff E.~S.\ 2017, \mnras, 470, 551

\end{thebibliography}
\end{document}